\documentclass[12pt,a4paper,reqno,oneside]{amsbook}

\usepackage{amsmath}
\usepackage{amssymb}
\usepackage{slashed}
\usepackage{pstricks,pst-node,pst-plot}
\numberwithin{figure}{chapter}

\renewcommand{\(}{\left(}
\renewcommand{\)}{\right)}
\renewcommand{\[}{\left[}
\renewcommand{\]}{\right]}
\newcommand{\itbd}[1]{\mbox{\boldmath $#1$}}

\renewcommand{\d}{\partial}
\newcommand{\f}[2]{\frac{#1}{#2}}
\newcommand{\df}[2]{\dfrac{#1}{#2}}
\newcommand{\ph}{\phantom{\mu}}
\newcommand{\pd}[3]{\left(\frac{\partial #1}{\partial #2}\right)_{#3}}
\newcommand{\cal}{\mathcal}

\newcommand{\normal}{\hspace{-1mm}:\hspace{-1mm}}
\newcommand{\la}{\langle}
\newcommand{\ra}{\rangle}
\newcommand{\bra}[1]{\la #1 |}
\newcommand{\ket}[1]{| #1 \ra}
\newcommand{\beq}{\begin{equation}}
\newcommand{\eeq}{\end{equation}}
\newcommand{\beqa}{\begin{eqnarray}}
\newcommand{\eeqa}{\end{eqnarray}}
\newcommand{\beg}{\begin{gather}}
\newcommand{\eeg}{\end{gather}}
\newcommand{\nol}{\nonumber \\ }

\begin{document}

\pagenumbering{roman}
\begin{titlepage}
\vspace{3cm}
\begin{center}
 \Large\textbf{Classical and Quantum Radiation Reaction}
  \end{center}
  \vspace{2cm}
  \begin{center}
   \textbf{Giles D. R. Martin}
    \end{center}
    \vspace{1cm}
   \begin{center}
   \textbf{A Thesis Submitted for the Degree of PhD}
   \end{center}
   \vspace{1cm}
   \begin{center}
   \textbf{University of York\\ Department of Mathematics}
   \end{center}
   \vspace{1cm}
   \begin{center}
   \textbf{July 2007}
\end{center}
\end{titlepage}
\newpage
\textbf{Abstract:} This thesis reports on work undertaken in
comparing the effects of the phenomenon of radiation reaction in
classical and quantum theories of electrodynamics. Specifically, it
is concerned with the prediction of the change in position of a
particle due to the inclusion of the self-force in the theory. We
calculate this position shift for the classical theory, treating
radiation reaction as a perturbation in line with the reduction of
order procedure. We calculate the contributions to the position
shift in the $\hbar\to 0$ limit of quantum field theory to order
$e^2$ in the coupling, the order of the classical self-force. These
calculations contain the emission and forward scattering one loop
processes of quantum electrodynamics. The quantum calculations are
completed for the case of a particle represented by a scalar field
wave packet and then for a particle represented by the Dirac spinor
field. We additionally give an alternative derivation of the scalar
results using the interpretation of radiation reaction via a Green's
function decomposition, in order to explain and contrast the results
achieved.
\newpage
\tableofcontents
\newpage
\listoffigures
\chapter*{Acknowledgements}
I would like to thank my supervisor Atsushi Higuchi for his guidance
throughout my three years at York, under which this work was
completed, and for his expert help in the many hours of conversation
and discussion on the work and theory.

I would also like to thank the Department of Mathematics at the
University of York for welcoming me and creating a friendly and
educating atmosphere in which to live and work. I would like to
thank the members of the Mathematical Physics Group within the
department for their hospitality and in particular Chris Fewster for
his willingness to discuss and explain any theory of interest.
Further thanks go to the graduate students in the department for
their friendship, discussions, seminars and cakes and in particular
Calvin Smith for our frequent whiteboard aided investigations. In
addition, I would like to thank the University of York itself for
supporting me during my studies with a University Studentship and
with the facilities for study and community.

Finally, I would like to thank my family for their encouragement to
me in following this path of study and I am forever indebted to
Jenni Karley for her support throughout my study and her patience
during the writing of this thesis.
\newpage
\chapter*{Author's declaration}
I declare that the work contained in the thesis is original. Chapter
\ref{introchapter}, Chapter \ref{Semiclassical} sections
\ref{wkbsection} and \ref{semiclassscalarsection} and Appendices
\ref{IntHam} and \ref{diracrefs} are reviews. Chapter \ref{Semiclassical}
 section \ref{semiclassspinorsection}, Chapters
 \ref{classicalshiftchapter}, \ref{scalarshiftchapter}, \ref{greenschapter},
 \ref{spinorshiftchapter} and \ref{summarychapter} and Appendix \ref{spinoridappendix}
are my original work done in collaboration with my supervisor, Dr
Atsushi Higuchi. The work contained in Chapters
\ref{classicalshiftchapter}, and \ref{scalarshiftchapter} is reported in \cite{HM23} and
\cite{HM4}. The work in Chapter \ref{greenschapter} is reported in
\cite{HM5}.
\newpage
\pagenumbering{arabic}

\mainmatter

\chapter{Introduction}\label{introchapter}
\begin{quote}
In this chapter we introduce the work presented in this thesis. We
introduce the background theory of relevance, the work to be
presented and define the classical and quantum theoretical models to
be used.
\end{quote}

\section{Overview}
The concept that an accelerated particle radiates is one the of most
widely known and used phenomena from the theory of classical
electrodynamics. It is thus ironic that the theory of the process
and the mechanism behind it is in fact one of the least understood
and most debatable areas of classical theory. In truth, there is no
real consensus over the correct interpretation of the theory, or
even exactly which theory is the correct one to interpret. The
problems stem from the attempt to describe the effect that the
emission of such radiation would have on the particle itself -
\emph{radiation reaction}. That radiation is in fact produced by
various systems involving the acceleration of charged particles is
an observable experimental fact. The phenomenon is one of the most
frequently employed in electromagnetism, for example in the
production of radio waves from antennas.
\footnote{Curiously, the plural antennae is used for biological
appendages, whereas antennas is the use for equipment sending and
receiving electromagnetic waves.} The radiation itself carries away
energy-momentum which must consequently affect the particle's motion
via recoil in order to conserve the energy-momentum of the system.
Thus radiation reaction alters the equations of motion of a charged
particle. This is, of course, fundamental to our understanding as
the equations of motion for a system are one of the most basic
underpinnings of a theory. Nevertheless, the effect is rarely
considered (or even taught in undergraduate courses). The usual
focus in classical theory is either the study of the fields given
the motion of a charged particle, or the motion of a charged
particle given some external field(s). The problem of radiation
reaction, on the other hand, is one of the effect on the motion of
the particle of its
\emph{own} field, hence the frequently employed alternative name, \emph{self-force}. The
lack of attention to this effect is possibly due to a combination of
factors including,
\begin{itemize}
\item The effect of radiation reaction is very small for most
purposes; sufficiently small to be ignored.
\item The unresolved and/or debatable problems alluded to above which
prevent the presentation of a consistent theory on a concrete
footing.
\item Classical electrodynamics is no longer considered to be the
fundamental theory, having been superseded by quantum
electrodynamics.
\end{itemize}
The focus of this project is a comparison of the effects of
radiation reaction in both the classical and quantum
electrodynamics' theories. In this way, we hope to gain further
understanding of how radiation reaction is treated within these
theories and how this treatment differs. The fundamental nature of
the effects of adding radiation reaction to a model, as one must do
to obtain a realistic model, means that a fuller understanding of
the nature of radiation reaction is essential. Indeed, it is not
only in the theory of electrodynamics that we are presented with
this problem and much current research is, at the time of writing,
focused on radiation reaction problems in classical
gravity.\footnote{We shall return to this subject briefly later
(Sec.\ref{curvedandgrav}).}

In the coming sections of this introductory chapter we present a
description of the background theory of relevance to the study of
radiation reaction and the origin of the work presented here.  This
work is based on the calculation of the `position shift', the change
in position due to radiation reaction, as a measure of the effect of
radiation reaction and from section
\ref{modelsection} we then explain the models to be used, detailing
the choice of calculations to be performed. We introduce the
classical model and the conventions and definitions for the quantum
field theoretic models on which we shall base our investigations.
The body of work that forms this thesis is then split into relevant
chapters as follows: In Chapter
\ref{Semiclassical}, we introduce and calculate the semiclassical
approximations for use in describing the quantum fields in our
calculations and in Chapter
\ref{classicalshiftchapter} we calculate the position shift in the
classical theory of electrodynamics. In Chapter
\ref{scalarshiftchapter}, we then proceed with the calculations using
the quantum scalar electrodynamics, calculating the contributions to
the position shift and comparing the $\hbar \to 0$ limit with the
results from the classical theory. Chapter \ref{greenschapter} then
gives an alternative derivation for some of the results for the
scalar field by using the Green's function decomposition description
of radiation reaction in order to gain further understanding and
interpret the previously obtained results. Chapter
\ref{spinorshiftchapter} then repeats our calculations for the
canonical quantum electrodynamics model based on the Dirac spinor
field. The appendices include definitions and calculations which are
used and referred to within the main body of the text.

\section{Radiation from moving charges}
Before considerations are made of our theory of radiation reaction,
it would be timely to remind ourselves of some of the basic theory
concerning radiation from accelerated charges in flat
spacetime.\footnote{This explanation is intended as a reminder for
those familiar with the theories quoted. For a more in depth
discussion
\cite{Jackson} is a good place to start.} We recall that in the
absence of incoming fields, we may write the $4$-vector
electromagnetic potential $A$ generated by the motion of a charged
particle in terms of the retarded Green's function $G_-$ and the
particle's $4$-current $j$:
\begin{equation}
A^\mu(x) = \int d^4x' G^{\ph\mu}_{-\ph\nu}(x-x') j^{\nu}(x')\,,
\end{equation}
with $G^{\ph\mu}_{-\ph\nu}=\delta^\mu_\nu G_-$ and where our metric
signature is represented by $g_{\mu\nu}={\rm diag}\,(+1,-1,-1,-1)$.
The units are chosen so that $c=1$, where $c$ is the speed of light.
We also let the electromagnetic field satisfy the Lorentz gauge
condition $\partial_\alpha A^\alpha = 0$. The current is given by
\begin{equation}
j^\nu(x) = e \int d\tau \f{d x^\nu}{d\tau} \delta^4 (x-X(\tau))\,,
\end{equation}
where $X(\tau)$ is the space-time trajectory of the particle.  The
point particle nature of the theory is represented by the delta
function point distribution. This is technically the retarded
potential, with the advanced potential being analogously generated
from the advanced Green's function. We canonically choose the
retarded solution due to our wish to look at propagation forward in
time, which can be seen more explicitly below.

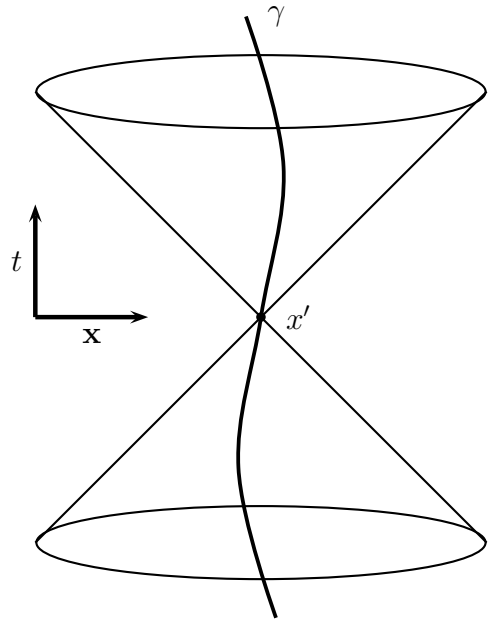
\begin{figure}
\begin{center}
\begin{pspicture}(-3.5,-4)(4,4.5)
\pscurve[linewidth=0.5mm]{-}(0.2,-4)(-0.3,-2)(0,0)(0.3,2)(-0.2,4)
 \psline(-3,-3)(3,3)
 \psline(-3,3)(3,-3)
 \psellipse(0,-3)(3,0.5)
 \psellipse(0,3)(3,0.5)
 \rput(0.5,0){$x'$}
 \rput(0.2,4){$\gamma$}
 \psdots(0,0)
 \psline[linewidth=0.6mm]{->}(-3,0)(-3,1.5)
 \psline[linewidth=0.6mm]{->}(-3,0)(-1.5,0)
 \rput(-3.25,0.75){$t$}
 \rput(-2.25,-0.25){${\bf x}$}
\end{pspicture}
\end{center}
\caption{Light cones of a point on a world line $\gamma$.} \label{lightcone1}
\end{figure}

\begin{figure}
\begin{center}
\begin{pspicture}(-4,-4)(4,4)
 \pscurve[linewidth=0.5mm]{-}(-1.5,-4)(-2,-2)(-1.6,0)(-1.4,1.4)(-1.9,4)
 \psdots(-2,-2)(-1.4,1.4)
 \psline(-3,-3)(3,3)
 \psline(-3,3)(3,-3)
 \psellipse(0,-3)(3,0.5)
 \psellipse(0,3)(3,0.5)
 \psdots(0,0)
 \rput(0.5,0){$x$}
 \rput(-2.8,-1.75){$X(\tau_-)$}
 \rput(-2.3,1.15){$X(\tau_+)$}
\end{pspicture}
\end{center}
\caption{The intersection of the light cone of point $x$ with the world line of the particle,
 at the retarded and advanced points.} \label{lightcone2}
\end{figure}
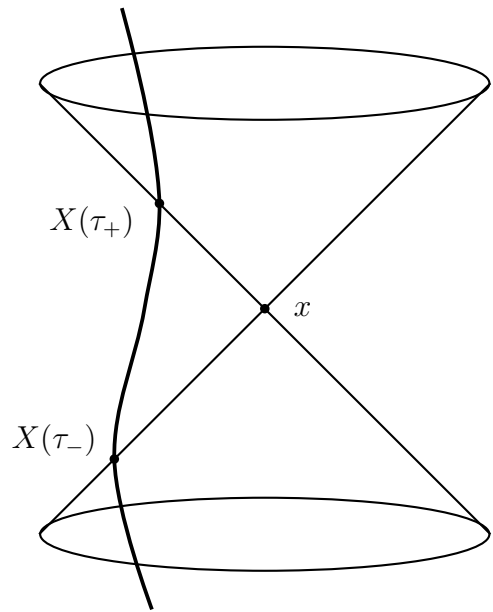

For the electromagnetic field, the Green's functions are the
fundamental solutions to the wave equation
\begin{equation}
\square G (x,x') = \delta^4(x-x')\,,
\end{equation}
where the translation invariance means that the solutions depend
only on $x-x'$, hence $G(x,x')=G(x-x')$. By utilizing the resultant
algebraic equation for the Fourier transform, the solutions can be
written in the integral form
\begin{equation}
G(x-x') = - \int \f{d^4 k}{(2\pi)^4} \f{e^{ik\cdot(x-x')}}{k^2}\,.
\end{equation}
\begin{figure}
\begin{center}
\begin{pspicture}(-4,-3)(4,3)
\psline[linewidth=0.5mm]{<->}(-4,0)(4,0)
\psline[linewidth=0.5mm]{<->}(0,-3)(0,3)
\psline[linewidth=0.3mm](-2.2,0.2)(-1.8,-0.2)
\psline[linewidth=0.3mm](-2.2,-0.2)(-1.8,0.2)
\psline[linewidth=0.3mm](2.2,0.2)(1.8,-0.2)
\psline[linewidth=0.3mm](2.2,-0.2)(1.8,0.2)
\psline[linewidth=0.3mm, linestyle=dashed, arrowsize=8pt]{->}(-4,1)(-2,1)
\psline[linewidth=0.3mm, linestyle=dashed, arrowsize=8pt]{->}(-2,1)(2,1)
\psline[linewidth=0.3mm, linestyle=dashed](2,1)(4,1)
\psline[linewidth=0.3mm, linestyle=dashed, arrowsize=8pt]{->}(-4,-1)(-2,-1)
\psline[linewidth=0.3mm, linestyle=dashed, arrowsize=8pt]{->}(-2,-1)(2,-1)
\psline[linewidth=0.3mm, linestyle=dashed](2,-1)(4,-1)
\rput(-2.0,1.5){Retarded}
\rput(-2.0,-1.5){Advanced}
\end{pspicture}
\caption{The contours used by the Retarded and Advanced Green's
functions avoiding the poles (X) in the $k_0$ integration.}
\label{Gcontours}
\end{center}
\end{figure}
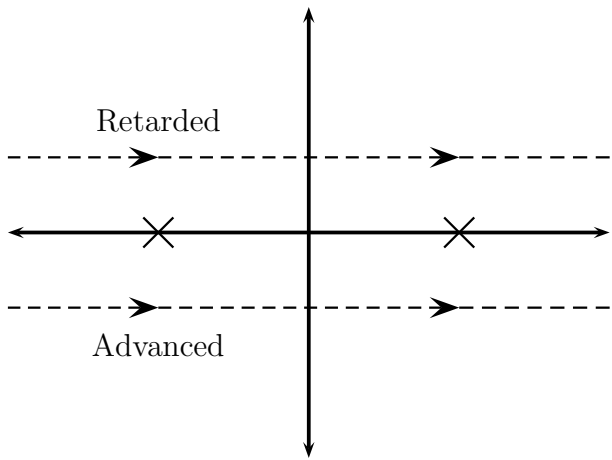
The singularities from the cone $k^2=0$ are dealt with via deforming
the contour of integration. The different Green's functions,
resulting from the alternative boundary conditions applied to the
wave equation, generate the different contours.\footnote{The
difference between the choices must, in the end, correspond to a
solution to the homogeneous equation.} The retarded/advanced Green's
functions are generated using the contours that travel above/below
the poles of the $k_0$ integration viz
\begin{equation}
G_{\mp}(x-x') =  - \int \f{d^4 k}{(2\pi)^4}
\f{e^{ik\cdot(x-x')}}{(k_0 \pm i\varepsilon)^2-{\bf k}^2} \,,
\end{equation}
where the limit $\varepsilon\to 0^+$ is assumed (see Fig.
\ref{Gcontours}). This gives
\begin{equation}
G_{\mp}(x-x') = \f{1}{2\pi} \theta(\pm (t-t'))\delta((x-x')^2) \,.
\end{equation}
where $\theta$ is the Heaviside step function. This expression shows
clearly that the support for the retarded and advanced Green's
functions lies on the future and past light cones respectively of
the particle at $x'$~\footnote{That is, $x$ lies on the future/past
light cone of $x'$ and consequently, $x'$ lies on the past/future
light cone of $x$.}, as is expected from causality reasoning. Fig.
\ref{lightcone1} shows these light cones for $x'$ on a particle
trajectory given by $\gamma=X(\tau)$.
The potential at $x$ is generated by the point on the world line of
the trajectory that intersects $x$'s \emph{past} light cone, which
is known as the retarded point. The proper time of the particle at
this intersection is known as the retarded time, which we label
$\tau_-$ here. Likewise for the advanced potential, we have the
intersection with $x$'s \emph{future} light cone, at the advanced
time $\tau_+$. This is represented by Fig.
\ref{lightcone2}.

Returning to our electromagnetic potential generated by the moving
charged particle, we have
\begin{equation}
A^\mu (x) = \f{e}{2\pi} \int d\tau
\theta\(x^0-X^0(\tau)\)\delta\(\(x-X(\tau)\)^2\)\f{d X^\mu}{d\tau}
\,,
\end{equation}
Labelling $X(\tau_\pm)=X_\pm$ as the advanced/retarded points on the
particle trajectory for $x$, and $\dot{X}_\pm=dX/d\tau(\tau_\pm)$ as
the world-velocities at those points, we can solve the integral and
write
\begin{equation}
A^\mu(x) = \f{e}{4\pi}
\f{\dot{X}^\mu_-}{\dot{X}_-^\nu\(x-X_-\)_\nu} \,.
\end{equation}
These are commonly known as the Li\'{e}nard-Wiechert potentials and
are the usual starting place in textbooks and electrodynamics
courses for the analysis of the radiation of a moving charge. This
expression is fully covariant and can be considered to be the
relativistic generalisation of the Coulomb potential.\footnote{In
the frame in which $\dot{X}_-=(1,0,0,0)$ we obtain the Coulomb
potential.} The most common exposition of the potentials and
radiation involve a non-covariant form written in terms of the
quantity ${\bf r}={\bf x}-{\bf X}_-$, which we name the radial
vector and where due to the null separation of the points the
magnitude can be given similarly by $r=|{\bf r}|=x^0-X^0_-$. Hence,
\begin{align}
A^0 &= \f{e}{4\pi}\f{1}{(r-{\bf r}\cdot{\bf v}_-)} \,, \\
{\bf A} &= \f{e}{4\pi} \f{{\bf v}_-}{(r-{\bf r}\cdot{\bf v}_-)}
\,,
\end{align}
where ${\bf v}_-=d{\bf X}/dt(\tau_-)$. The magnetic and electric
fields from this potential can then be given by
\begin{align}\label{LWB}
{\bf B} &= \f{{\bf r}\times{\bf E}}{r} \,, \\
{\bf E} &= \f{e}{4\pi} \f{({\bf r}-r{\bf v}_-)(1-{\bf v}_-^2)+{\bf
r}\times \[ ({\bf r}-r{\bf v}_-)\times \dot{\bf v}_-(t) \]}{(r-{\bf
r}\cdot{\bf v}_-)^3} \,, \label{LWE}
\end{align}
with $\dot{\bf v}_-(t) = d{\bf v}_-/dt$. It is worth recalling from
the definitions that all terms on the right hand side are evaluated
at the retarded time $\tau_-$. The expressions in (\ref{LWB}) and
(\ref{LWE}) can each be separated into two terms, representing the
so called `velocity fields', which do not depend on the
acceleration, and the `acceleration fields', which do. Introducing
the notation ${\bf n}={\bf r}/r$, we rewrite the ${\bf E}$ field for
example, and obtain
\begin{equation} \label{LWEfield}
{\bf E} = \f{e}{4\pi} \[ \f{({\bf n}-{\bf v}_-)(1-{\bf
v}^2_-)}{(1-{\bf n}\cdot{\bf v}_-)r^2} + \f{{\bf n}\times\[({\bf
n}-{\bf v}_-)\times
\dot{\bf v}_-\]}{(1-{\bf n}\cdot{\bf v}_-)r} \] \,.
\end{equation}
The first term, the velocity field, can be seen to be an inverse
square field, thus effectively a static Coulomb type field. The
second term is the acceleration field, which we see has the inverse
radial dependence one would expect from a radiation field. For this
field we can also confirm that both ${\bf E}$ and ${\bf B}$ are
transverse to the radial vector. That our interpretation of the
velocity field as a Coulomb type field is valid can be confirmed by
consideration of the situation in which the particle undergoes
uniform motion, hence when the acceleration field naturally
vanishes. The remaining contribution should be the Lorentz
transformation of the static Coulomb field. This can indeed be shown
to be the case. Therefore, this transformation to or from the static
frame implies that the velocity field should not affect the motion
of the particle\footnote{Indeed, one does not expect a static
particle to move due to its own Coulomb field.} nor consequently
cause any modifications to the equations of motion. We thus arrived
at the conclusion that the acceleration of a charged particle
induces a radiation field, i.e. the emission of radiation from the
particle.

Having predicted that an accelerated charged particle will radiate,
one is naturally inclined to ask `how much radiation is expected?'
Now the energy flux ${\cal E}$ across a sphere of finite size and
centered on the particle is given in terms of the Poynting vector,
viz
\begin{align}
\f{d{\cal E}}{dt} &= \int_S d{\bf S}\cdot({\bf E}\times{\bf B})
\nol
 &= \int_S d\Omega ({\bf r}\times{\bf E})^2 \,.
\end{align}
The last line demonstrates that the energy flux is positive. For the
non-relativistic case, i.e. at small velocities, the electric
acceleration field contribution is the standard dipole field
\begin{equation}
{\bf E}_a = \f{e}{4\pi r}{\bf n}\times \({\bf n}\times\dot{\bf v}_-
\) \,.
\end{equation}
The radiated power\footnote{Given that we are now considering the
situation at the particle itself, the time in question is in fact
the retarded time, and the `retarded' subscript is henceforth
redundant.} $P$ in this limit is known as the Larmor formula:
\begin{equation}\label{Larmorformula}
P=\f{d{\cal E}}{dt} = \f{2}{3} \f{e^2}{4\pi} \dot{\bf v}^2 \,.
\end{equation}
The relativistic generalization, written in terms of the
energy-momentum $4$-vector $p$ of the particle, is given by
\begin{equation}\label{relLarmor}
P=-\f{2}{3}\f{e^2}{4\pi m^2}\(\f{dp^\mu}{d\tau}\f{dp_\mu}{d\tau}\)
\,.
\end{equation}

\section{Radiation Reaction}
So far we have looked only at those areas usually considered, namely
the effect of the interaction with external fields on a particle
(the acceleration of our moving charge) and more chiefly above, the
fields produced by such a moving charge and the consequent
radiation. What we have yet to consider is of course the effect of
this radiation on the motion of the charge. That is, what are the
effects on the motion of the particle of its own fields? Put another
way, what are the self-interaction effects? As previously mentioned,
the radiation carries away energy-momentum and thus one expects the
particle's energy-momentum to be affected and hence its motion. We
have said that this classical radiative correction is frequently
neglected, and one of the main reasons given was that the effect is
very small. The approximation is justified provided that the energy
concerned is small in comparison with the typical energies of the
problem under consideration. Let us consider a period of interaction
of time $T$ and with typical resultant acceleration $a$. The energy
of the emitted radiation ${\cal E}_{\rm r}$ is, from above, of order
\begin{equation}
{\cal E}_{\rm r} \sim \f{2}{3}\f{e^2}{4\pi}a^2 T \,.
\end{equation}
The change in the particle's energy ${\cal E}_{\rm p}$ is by
comparison of order
\begin{equation}
{\cal E}_{\rm p} \sim m a^2 T^2 \,.
\end{equation}
The demand that ${\cal E}_{\rm p} \gg {\cal E}_{\rm r}$ leads to the
relation for the interaction time period
\begin{equation}\label{taue}
T \gg \tau_e = \f{2}{3} \f{e^2}{4\pi m} \,,
\end{equation}
where we have defined the characteristic radiation time $\tau_e$. We
note that $\tau_e=2 r_e/3$ where $r_e$ is the classical electron
radius and consequently, $\tau_e$ is of the order of the time taken
for light-signals to travel this distance.\footnote{Recall our units
$c=1$.} For time scales in excess of this period, the corrections
can be justifiably ignored. It is evident that this set of larger
time scales effectively contains all classical phenomena. Indeed,
for lower scales, we would expect to have reached to limit of
validity of the classical theory and expect quantum effects to
become important.

\subsection{Abraham-Lorentz-Dirac Force}
Below we shall present a description of the canonical classical
theory of radiation reaction, the alteration to the equations of
motion given by Abraham-Lorentz-Dirac force, and describe some of
the associated pathologies. A classical point particle moving under
the influence of some external (non-zero) potential, producing a
force $F_{\rm ext}$, is accelerated and the equations of motion
given by
\begin{align} \label{newtoneom}
m\f{d^2 x^\mu}{d\tau^2} = F^\mu_{\rm ext} \,,
\end{align}
where $x^\mu$ are the space-time coordinates of the particle at
proper time $\tau$. A charged particle emits radiation when
accelerated in such a potential and as stated above this will affect
the motion of the particle and thus the equations of motion. We
could consider the correction as the effect of the addition of an
extra force $F_{\rm R}$ on the right hand side of (\ref{newtoneom}):
\begin{equation}
m\frac{d^2x^\mu}{d\tau^2} = F_{\rm ext}^\mu + F_{\rm R}^\mu \,.
\label{LD}
\end{equation}
We call this additional force the \emph{radiation reaction force}.
Considering for a moment the non-relativistic approximation, as
described by the Larmor emission power (\ref{Larmorformula}), we
note that there are certain restrictions on $F_{\rm R}$. Given that
when there is no acceleration, there should be no radiation, and
thus no reaction to it, $F_{\rm R}$ should vanish if $\dot{\bf
v}=0$. In addition, the only parameter available to use is the
characteristic time, hence it is likely to feature in the force. In
fact, it is likely to feature at first order, given that the power
radiated is of order $e^2$, in common with $\tau_e$, and that
furthermore a sign change on the charge cannot alter the result. One
approach is to demand that the work done over the period of
interaction is the negative of the total energy radiated i.e.
\begin{equation}
\int_T dt \, {\bf F}_{\rm R} \cdot {\bf v} = -\int_T dt \, m\tau_e \dot{\bf
v}^2 \,.
\end{equation}
Integrating the right hand side by parts, then given the assumption
that either periodic motion or that $\dot{\bf v}\cdot{\bf v}=0$ at
the end points of the interaction period, the surface term vanishes
and the remainder leads to the conclusion
\begin{equation}
{\bf F}_{\rm R} = m \tau_e \ddot{\bf v} = \f{2}{3} \f{e^2}{4\pi}
\ddot{\bf v} \,,
\end{equation}
where $\ddot{\bf v}=d^2{\bf v}/dt^2=d^3{\bf x}/dt^3$. This reasoning
leads to what is commonly referred to as the Abraham-Lorentz
equation of motion:
\begin{equation}\label{ALforce}
m\dot{\bf v} = {\bf F}_{\rm ext} + m\tau_e \ddot{\bf v} \,,
\end{equation}
(See \cite{Abraham} and \cite{Lorentz}). The resultant equation of
motion is different to that which one usually encounters in
mechanics due to the presence of third-order differential terms. The
inclusion of such a term would imply that a third initial condition
would be needed in addition to the position and velocity. It is
indeed this type of term that is the source of the debate over the
physical correctness and interpretation of this theory. The problem
remains when we remove our un-physical non-relativistic
approximation.

\begin{figure}
\begin{center}
\begin{pspicture}(-4,-1)(4,4)
 \psline{->}(-4,0)(4,0) \psline{->}(0,-0.5)(0,4)
 \pscurve[linewidth=0.5mm]{-}(-4,0)(-3,0)(-2,0.1)(-1,0.5)(0,2)
 \psline[linewidth=0.5mm]{-}(0,2)(4,2)
 \uput[l](0,3){$a$} \uput[d](4,0){$t$} \uput[d](0,-0.5){$t_0$}
 \psframe[linestyle=none, fillstyle=vlines](0,0)(4,4)
\end{pspicture}
\end{center}
\caption{Pre-acceleration of a charged particle.} \label{preacceleration}
\end{figure}
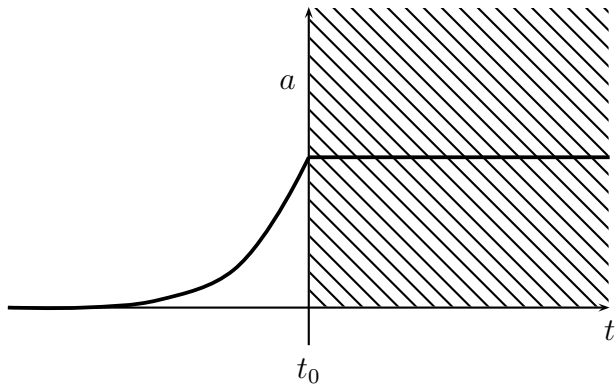

The fully relativistic generalization of the radiation reaction
force was first obtained by Dirac, using the local conservation of
energy and momentum \cite{Dirac}. The Abraham-Lorentz-Dirac force is
the canonical model of radiation reaction in classical
electrodynamics, commonly referred to as the Lorentz-Dirac force
\begin{equation}\label{LD4force}
F^\mu_{\rm LD} = \frac{2\alpha_c}{3}\left[
\frac{d^3x^\mu}{d\tau^3} + \frac{dx^\mu}{d\tau}\left(\frac{d^2
x^\nu}{d\tau^2} \frac{d^2x_\nu}{d\tau^2}\right)\right] \,,
\end{equation}
where we define $\alpha_c \equiv e^2/4\pi$. Due to the orthogonality
of the world-velocity and its proper time derivative
\begin{equation}
\f{dx^\mu}{d\tau}\f{d^2x_\mu}{d\tau^2}=0 \,,
\end{equation}
the Lorentz-Dirac force is often alternatively written
\begin{equation}\label{LD4force1}
F_{\rm R}^\mu=F^\mu_{\rm LD}
\equiv\frac{2\alpha_c}{3}\left[\delta^\mu_\nu
 - \frac{dx^\mu}{d\tau}\frac{d
x_\nu}{d\tau}
\right]\frac{d^3x^\nu}{d\tau^3} \,.
\end{equation}
The non-relativistic (\ref{ALforce}) is the result in the special
Lorentz frame which is momentarily co-moving with the particle. In
both cases, we see the presence of a third-derivative term, usually
referred to as the Schott term. Not only is this type of
differential equation fundamentally different to that which is
expected in dynamical motion, it also presents us with problematic
un-physical solutions. Using the non-relativistic case for
simplicity, we rearrange as an inhomogeneous differential equation
\begin{equation}
m\(\dot{v} - \tau_e \ddot{v} \) = {\bf F}_{\rm ext} \,.
\end{equation}
Now, in the homogenous case, i.e. in the absence of any external
force, the above equation presents us with two possible solutions
\begin{align}
\dot{v} (t) = \begin{cases} 0 \\ \dot{v}(0) e^{t/\tau_e}
\end{cases} \,.
\end{align}
The second solution is referred to as a `runaway' solution. It would
involve a particle effectively accelerating under its own radiation
reaction and is not physically acceptable, let alone an observed
phenomenon. Additionally, the reader may note that it breaks the
boundary conditions imposed during the above derivation which caused
the annihilation of the surface term. In order to restrict ourselves
to physical solutions, we must add these boundary conditions and in
particular demand that should ${\bf F}_{\rm ext}\to 0$ as
$t\to\infty$ then $\dot{\bf v}$ should also vanish. With the
addition of these conditions, we may produce an integro-differential
form of the Lorentz-Dirac equation, free from the troublesome
higher-derivative induced runaways:
\begin{equation}\label{integrodiff}
m\dot{\bf v}(t) = \int^\infty_0 dt' e^{-t'} {\bf F}_{\rm ext}
(t+\tau_e t') \,.
\end{equation}
Unfortunately, this version of the equation is plagued by an
alternative problem: pre-acceleration. Consider the case in which
the external force is `switched on' at $t=0$ i.e.
\begin{align}
{\bf F}_{\rm ext}  \begin{cases} =0 & \text{if $t<0$} \\ \neq 0
&\text{if $t \geq 0$} \end{cases} \,.
\end{align}
The reader will note that (\ref{integrodiff}) implies that the
acceleration of the particle is not zero for $t<0$, but instead
begins at a time of order $-\tau_e$. Hence the particle accelerates
\emph{before} the force is applied. This situation is represented in
Fig. \ref{preacceleration}. We note that again, the timescale with
which we find ourselves concerned is $\tau_e$.
Given the previous discussion on the size of the radiative
correction, we note that this timescale is beyond the expected
validity of classical electrodynamics, and thus the pre-acceleration
effect can be considered classically unobservable.

Whilst the `small' nature of the correction may be a good enough
reason to ignore the problems, or indeed the entire phenomenon, for
most practical purposes it is hardly satisfying from a fundamental
theoretical perspective. If we wish to obtain a proper understanding
of the dynamics then we must a) consider radiation reaction and b)
attempt to understand and ultimately solve the problems with the
current theory. One conclusion we can take from the current
situation is that the Lorentz-Dirac theory is at most only
approximately accurate. From our discussion, we note that it appears
that this accuracy extends only as long as the radiative correction
is small. In this regime, one can then treat radiation reaction as a
perturbative correction to the motion of the particle. If one
proceeds as such, then the problematic third derivative term,
treated as a perturbation, can be considered as referring to the
acceleration along the original trajectory (rather than the
perturbed self-interacting trajectory) viz
\begin{equation}
m\f{d a^\mu}{d\tau} = \f{dF_{\rm ext}^\mu}{d\tau} \,.
\end{equation}
Substituting this relation back into (\ref{LD4force1}), the
equations of motion
 give a second order differential equation,
\begin{equation}\label{LDreduce}
m\f{d^2x^\mu}{d\tau^2} = F^\mu_{\rm ext}+
\frac{2\alpha_c}{3m}\left[\delta^\mu_\nu
 - \frac{dx^\mu}{d\tau}\frac{d
x_\nu}{d\tau}
\right]\frac{dF^\nu_{\rm ext}}{d\tau} \,.
\end{equation}
Consequently, the treatment of radiation reaction as a perturbation
is akin to a reduction of order process on the differential
equations of the motion. With the, dynamically standard, second
order equation we are free from run-away and pre-acceleration
solutions. In (\ref{LDreduce}), we now have a pathology-free theory
for radiation reaction in classical electrodynamics which we can
proceed to use, subject to the constraints mentioned. It is in fact
this reduced order type of radiation reaction force that is most
commonly utilized for calculations involving the self-interaction
with electromagnetic and also other fields. The perturbative
treatment of the radiative corrections is also familiar as the main
calculation tool for quantum electrodynamics (QED). QED is most
frequently dealt with using perturbation theory, leading to the
ubiquitous Feynman diagrams representing the perturbation expansion
terms. This similarity in treatment is one of the motivations for
the focus of this work.

The problems with the full theory imply a problem with the starting
assumptions. Whilst classical mechanics is of course now known to be
based on erroneous grounds\footnote{One ought really to say
`inaccurate grounds' as, along with most modern developments in
theory, whilst incorrect, calling the theory false masks the fact
that it is remarkably successful in most regimes and also that any
new theory must reproduce the results of classical mechanics within
their range of validity.}, due to the need for quantization, we can
also query the legitimacy of the point particle model for example,
just as extended models are proposed as alternatives in quantum
theories. Over the years, a number of different alternative models
have been proposed, usually by the addition of either further
constraints or changes to the fundamental model. In \cite{FordOC1},
Ford and O'Connell drop the point particle model. Using a particular
electron structure type\footnote{Following Feynman, Ford and
O'Connell use a form factor $\Omega^2/(\Omega^2+\omega^2)$ dependent
on a cut-off frequency $\Omega$ and where $\omega$ is the typical
frequency of the external force.} they show that the pathologies in
the non-relativistic radiation reaction are due to the
point-particle assumption. Under certain restrictions, they
reproduce the reduced order equation from this direction.

So far, we have merely stated the Lorentz-Dirac force as Dirac's
relativistic generalization to the Abraham-Lorentz force. Dirac
derived this equation using considerations of local energy-momentum
\cite{Dirac}. More explicitly, enclosing the world line of the
particle with a `world-tube', Dirac calculated the energy-momentum
flux of the electromagnetic field. Incidently, the shape of the
world-tube is irrelevant, provided that the end surfaces are the
same. The generalised space-time formulations of Gauss' theorem can
be utilized to show that the flux over a deformed tube is the same
as the original.\footnote{\cite{Poisson} demonstrates the equality
in this context during the discussion of Dirac's derivation. The
derivation in this paper differs slightly from the original, but
follows the same basic idea.} By conservation of momentum, the
change in the particles momentum can then be deduced as the
balance.\footnote{Technically, one of Dirac's postulates is that the
change in the mechanical energy-momentum of the particle is balanced
by that for the electromagnetic field.} Following such a line of
reasoning will lead one to the Lorentz-Dirac equation. However, to
achieve this result the reader will have to make a modification to
the physical mass by subtracting the (infinite) contribution from
the rate of change of the bound energy-momentum. Thus Dirac found
that he had to renormalise the mass by subtracting the infinite
contribution of the electro-static self-energy $m_{\rm em}$, viz
\begin{equation}\label{renormLD}
ma^\mu = \f{2}{3} \alpha_c \( \dot{a}^\mu - a^2 u^\mu\)
\end{equation}
where $m=m_0+m_{\rm em}$.~\footnote{The expression for the
particle's energy-momentum is also not as straight forward as
$m_0a^\mu$, but consideration must be given to the end surfaces of
the world-tube. These complications are detailed in \cite{Poisson}
and in more detail \cite{Teitel} and are not as important to the
main discussion here.} Lorentz had also had to perform a similar
manipulation for the Abraham-Lorentz force. This is another
similarity with quantum field theory and in fact many students of
theoretical physics are likely to come across the concept of
renormalisation in quantum electrodynamics \emph{before} the
classical theory. This procedure naturally begs the question
concerning the similarities and differences between the classical
and quantum renormalisations and thus provides a further motivation
for the work presented here.

The separation of the electrostatic contribution from the radiative
contribution was noted in the discussion above on the
Li\'{e}nard-Wiechert potentials. Here we mentioned that the
electrostatic, short-range contribution was effectively the
generalisation of the Coulomb force. Considering the Coulomb force
at the (point) particle, one sees why this contribution is infinite.
From this perspective, it is also clearer why one would wish this
self-energy to be considered part of the left-hand side of
(\ref{renormLD}), as part of the mass, rather than on the right.

\subsection{Green's function decomposition}
An alternative derivation of the radiation reaction force is
motivated by the singular self-energy contribution. We wish to
consider the interaction of the particle with its own field. Now,
the particle's field can be generated from the retarded Green's
function. The reader will recall that the action of the wave
operator on the retarded Green's function is to produce a delta
function
\begin{equation}
\square G_- (x,x') = \delta^4(x-x') \,,
\end{equation}
where the distribution is singular at $x=x'$ i.e. at the particle
itself. This Green's function was used to generate the
electromagnetic Li\'{e}nard-Wiechert potentials, which in turn were
shown to have a singular contribution and a regular, radiative
contribution. This derivation is based on a similar decomposition to
the retarded Green's function.

The theory of electromagnetism, contained in Maxwell's equations, is
time-symmetric. The process of radiation reaction however is not a
time-symmetric process; whilst the emission of radiation from a
particle would transform on time reversal to the absorption, the
self-energy contributions ought to be the same i.e. time symmetric.
We note that in choosing the retarded potential, we broke the
time-reversal symmetry of the theory, in order to accommodate our
`time arrow' . Starting from the time-symmetric theory, the opposite
choice, of the advanced Green's function, could technically have
been made.  We thus re-introduce the symmetry by taking the linear
combination of Green's functions
\begin{equation}\label{GSing}
G_S=\f{1}{2} \[ G_- + G_+ \] \,,
\end{equation}
which is a solution to the \emph{inhomogeneous} wave equation.
Alternatively, we have the antisymmetric combination
\begin{equation}\label{GReg}
G_R=\f{1}{2} \[ G_- - G_+ \] \,.
\end{equation}
These two Green's functions form a decomposition of the retarded
$G_-$.~\footnote{$G_-=G_R+G_S$\,.} Now, as stated above, $G_S$
solves the inhomogeneous wave equation
\begin{equation}
\square G_S = \delta^4 \,,
\end{equation}
whilst $G_R$ solves the homogeneous equation
\begin{equation}
\square G_R = 0 \,.
\end{equation}
The singular nature of the retarded potential is thus entirely
contained within the field generated $G_S$. With reference to the
Li\'{e}nard-Wiechert potentials, we would consequently hope to
assign this contribution as the singular self-energy. Indeed, it can
be shown that the singular field
\begin{equation}
A^\mu_S (x) = \int d^4 x' G^\mu_{S\nu'}(x,x') j^{\nu'}(x') \,,
\end{equation}
does \emph{not} affect the motion of the particle (see
\cite{Poisson2}). We thus consider this to indeed be the self-energy
Coulomb-like time-symmetric contribution.\footnote{The time symmetry
means that there should be equal amounts of incoming and outgoing
radiation and thus should not affect the motion.} The remaining
contribution $G_R$, which generates a regular field and is not
time-symmetric, we now postulate as the `radiative' Green's function
solution which is responsible for radiation reaction.\footnote{We
note here that it is the behaviour at the particle's worldline of
the \emph{fields} generated by $G_{R,S}$ that is regular or
singular.}

The action of the particle's (retarded) field on itself is therefore
split into an infinite correction to the mass, generated by $G_S$,
and the remainder $G_- - G_S=G_R$ acts on the particle to produce
the radiation reaction force. Explicitly, the radiative field $A_R$
is given by
\begin{equation}
A^\mu_R (x) = \int d^4 x' G^\mu_{R\nu'}(x,x') j^{\nu'}(x') \,.
\end{equation}
The field tensor $F^R$ acting on the particle is then
\begin{equation}\label{FmunuR}
F^R_{\mu\nu} = \nabla_\mu A_\nu^R - \nabla_\nu A_\mu^R \,,
\end{equation}
and the force is given by the Lorentz force from this field tensor
leading to the equations of motion which, with the external field
already acting on the particle, are
\begin{equation}\label{LDeqR}
ma_\mu = F_\mu^{\rm ext} + eF^R_{\mu\nu}u^\nu \,.
\end{equation}
Using this postulated source for the radiation reaction field, the
above equation of motion gives the Lorentz-Dirac force
\cite{Poisson}.

\subsection{Curved Space and Gravitational Radiation
Reaction}\label{curvedandgrav}

Much recent work on classical radiation reaction has been
concentrating on the motion in curved space. Here we briefly mention
some interesting extensions to curved space, and to the
self-interaction of other fields. In these comments we follow
Poisson's excellent review article
\cite{Poisson2} on
\emph{Radiation reaction of point particles in curved space}, to
which the reader is referred for a detailed pedagogic introduction.
The main references for the curved space work in this subsection use
the metric signature $(-+++)$, thus for consistency with these works
we shall temporarily adopt this signature for this (and only this)
subsection.\footnote{This temporary change should help the reader
should they wish to consult the references on this short aside for
more information. For the main body of our work the signature is
$(+---)$ due to its ease of use in particle physics.} The extension
of the Lorentz-Dirac equation to curved space was originally given
by DeWitt and Brehme in 1960
\cite{DeWitt}.\footnote{This paper actually contains a mistake,
corrected by Hobbs in 1968
\cite{Hobbs}, leading to the absence of a term containing the Ricci
tensor in the final equations of motion.} The Green's function
method for the derivation can be extended to the solutions to the
wave equation in curved space. If, as is usually the case, the
space-time is globally hyperbolic\footnote{That is, the space-time
admits a Cauchy surface - a space-like 3-surface through which every
inextendible causal curve in the space-time manifold passes exactly
once.} then there exist unique advanced and retarded solutions to
the wave equation. However, these Green's functions have additional
features compared to their Minkowski space cousins when considering
the support. Recall that in flat space, the support of Green's
functions was \emph{on} the light cones. In curved space we have the
possibility of interaction between the radiation and the curvature -
scattering off the curvature - leading to the propagation of
electromagnetic waves at speeds \emph{up to and including} the speed
of light. With respect to the Green's functions, this means that the
support is extended within the light cones as well as on them. For
example, the retarded field, generated from the retarded Green's
function, is now dependent on the entire history of the world-line
of the particle, up to and including the retarded point. Similarly,
the advanced field is dependent on the entire future of the world
line, after and including the advanced point. As in the flat space
calculations, we note that the retarded solution is singular on the
world line of the particle. Following the method previously
utilized, we wish to remove this singularity\footnote{Consequently
renormalising the mass.}, before applying the particle's field to
the particle itself. Again, we could note that the retarded Green's
function's singularity is contained entirely within the symmetric
$(G_- + G_-)/2$. Proceeding to subtract this contribution as before,
we obtain the equation of motion for a point particle in curved
space undergoing electromagnetic radiation reaction. This method was
followed by DeWitt and Brehme\footnote{That is, they use the same
singular Green's function. The details of their working are based on
a definition of the `direct' and `tail' contributions to the Green's
function as those with support on and within the light cone
respectively.} to obtain
\begin{align}
m\f{D u^\mu}{d\tau} &= F^\mu_{\rm ext} + \alpha_c \( \delta^\mu_\nu
+ u^\mu u_\nu
\) \( \f{2}{3m} \dot{a}^\nu + \f{1}{3}R^\nu_{\ph\lambda}u^\lambda \)
\nol & \quad + 2e^2u_\nu \int^{\tau_-}_{-\infty}
\nabla^{[\mu}G^{\nu]}_{-\lambda'}(X(\tau),X(\tau'))u^{\lambda'}d\tau'
\,.
\end{align}
The last term is often referred to as the tail term and contains the
mentioned dependence on the past history of the world line of the
particle. The integral is cut-off at $\tau'=\tau_- - 0^+$ to avoid
the singular behaviour of the retarded potential. In flat space,
this equation collapses to the Lorentz-Dirac equation. Recalling
that this equation is also based on the point particle description,
we note the continued presence of the third derivative term and the
need for a reduction of order process, or something else, in order
to make the description physical.

So far, the extra features of the Green's functions appear only to
have manifested themselves in the presence of the tail term. A
difficulty is faced, however, in the interpretation of the
decomposition of the retarded solution. The combination
\begin{equation}
G_{\rm sym}=\f{1}{2} \(G_-+G_+\) \,,
\end{equation}
has the necessary properties that it is symmetric and solves the
inhomogeneous wave equation, thus the field that it generates
contains the singularity of the retarded field $A_-$. If we again
postulate that the remainder of the field $A_-$ is responsible for
the self-force, then although it is indeed regular, this combination
has support within both future and past light cones. The appeal of
this approach is that there is no support at space-like separation
for the arguments which is in keeping with a field theory
perspective. However, taking the effect on the world line itself,
then
\begin{equation}
G_--G_{\rm sym} = \f{1}{2}\(G_--G_+\) \,,
\end{equation}
is dependent on the entire past and future history of the world
line, which is somewhat problematic from a causal perspective if we
are to then interpret the resultant field as on radiative field
acting on the particle. The key to solving this problem, identified
by Detweiler and Whiting in 2003 \cite{Whiting}, is the recognition
that although the symmetric combination $G_{\rm sym}$ does indeed
contain the singularity, it is not unique in this respect. This
non-uniqueness is part of the reason that we stated in the flat
space description that the use of $G_--G_{\rm sym}$ as wholly
responsible for the self-force was \emph{postulated}. We are free to
add any solution of the homogeneous wave equation to $G_{\rm sym}$
and the result will remain a solution of the inhomogeneous equation.
This then is how we proceed. The additional homogeneous solution is
defined precisely to solve the causal issues present in the choice
of $G_{\rm sym}$. We note additionally, that we must ensure that
this solution is also symmetric, otherwise we shall affect the
symmetric property of the resultant singular solution. We therefore
define $H(x,x')$ such that it is equal to the advanced Green's
function $G_+$ when $x$ is in the chronological past of $x'$ and, by
symmetry, agrees with $G_-$ when $x$ is in the chronological future
of $x'$. Subtracting this solution from $G_{\rm sym}$, we define the
curved space singular Green's function as
\begin{equation}
G_S = \f{1}{2} \( G_-+G_+-H\) \,.
\end{equation}
This function has support at spatially separated points, and the
resulting
\begin{equation}
G_R = G_--G_S \,,
\end{equation}
is dependent on the history of the world line up to and including
the
\emph{advanced} time $\tau_+$. Whilst it is somewhat counterintuitive
to use a result with apparent dependence at spatial separation, we
recall that the decomposition is used only in calculating the effect
of the field on the particle itself, hence on the world line (where
the separation is zero). It should be stressed that this choice of
Green's function decomposition does not actually affect the
resulting equation for the radiation reaction force. It does
however, put the interpretation of the Green's function
decomposition on a more physically reasonable footing by providing a
physical field which can act on the particle.

This treatment of the self-interaction, by identification and
subtraction of the singular component of the field, can be extended
to other forces as well. Recent work has included the calculation of
the self-force for a scalar charge in curved space by Quinn in 2000
\cite{Quinn}. In this case, instead of interacting with a vector field,
as is the case for electromagnetism, the particle is coupled to a
spin-zero scalar field. The equations of motion for a particle with
charge $q$ are
\begin{equation}
ma^\mu = q\(g^{\mu\nu}+u^\mu u^\nu \)\nabla_\nu \Phi \,,
\end{equation}
where the scalar field $\Phi$ emitted by the particle satisfies the
wave equation
\begin{equation}
\( \square - \xi R \) \Phi = - \mu(x) \,,
\end{equation}
where $\xi$ is a constant measuring coupling to the
curvature\footnote{The constant $\xi$ here is arbitrary, however the
most commonly picked values are the minimally coupled scalar field
with $\xi=0$ and the conformally invariant $\xi=1/6$.}, and where
$\mu(x)$ is the charge density given by
\begin{equation}
\mu(x) = q \int_\gamma d\tau\, \delta_4(x,X) \,,
\end{equation}
on the world line $\gamma=X(\tau)$ and $\delta_4(x,X)$ is the
invariant generalisation of the Dirac delta function
\begin{equation}
\delta_4(x,x') = \f{\delta(x-x')}{\sqrt{-g}} \,.
\end{equation}
The combination of these equations adds an extra feature to the
dynamics in curved space: If one derives the above equations of
motion from a variational principle, then the inertial mass must be
time-dependent. Specifically,
\begin{equation}
\f{dm}{d\tau} = -qu^\mu\nabla_\mu \Phi \,.
\end{equation}
Subtracting the singular Green's function from the particle's field,
and adding the self interaction to the equations of motion, Quinn
found\footnote{Quinn's results \cite{Quinn} were for the minimally
coupled scalar field and were extended by Poisson to arbitrary $\xi$
for his review
\cite{Poisson2}.}
\begin{align}
ma^\mu &= F^\mu_{\rm ext} \nol & \quad +
\f{q^2}{4\pi}\(\delta^\mu_\nu +u^\mu u_\nu\) \[ \f{1}{3m}
\dot{a}^\nu + \f{1}{6} R^\nu_{\ph\lambda}u^\lambda +
\int^{\tau_-}_{-\infty} d\tau'
\nabla^\nu G_-
(X(\tau),X(\tau')) \] \\
\f{dm}{d\tau} &= -\f{1}{12}\f{q^2}{4\pi}\(1-6\xi\)R - \f{q^2}{4\pi}
u^\mu \int^{\tau_-}_{-\infty} d\tau' \nabla_\mu G_-
(X(\tau),X(\tau')) \,.
\end{align}
The reader can note the similarities, and differences, between the
scalar and vector self-force expressions.

Having looked at scalar and vector (electromagnetic) fields, the
next type of potential field of interest is that of gravitational
radiation. The emission of gravitational radiation is one of the
predictions of general relativity and one which is of current
interest at the time of writing. A number of detectors have recently
been built hoping to receive signals from the gravitational waves
reaching earth. From our current perspective, we see the possibility
of a point mass interacting with its own gravitational field. We
approach this problem analogously to the scalar and vector cases, by
considering a point mass and its potential field. As we are
considering the gravitational field, the field of the mass will be
the perturbation of the space-time that it induces. The idea of a
point mass poses some difficulties within general relativity, not to
mention the usual difficulties in the consideration of the
non-linear equations of motion. However, provided we keep the
perturbations produced by the mass small, which we would in any case
wish to do given the previous discussions, then we can proceed. We
thus consider the case of a small mass moving in a background
space-time $g_{\mu\nu}$, which here is analogous to the charged
particle moving in the external potential. The unperturbed path is
then a geodesic of $g$. We assume that $g_{\mu\nu}$ is a solution to
the vacuum Einstein equations. We then treat the mass as a
perturbation $h$ to this background metric and use the mass itself
as the expansion parameter to produce the full perturbed space-time
$f_{\mu\nu}$
\begin{equation}
f_{\mu\nu}=g_{\mu\nu}+h_{\mu\nu} +{\cal O}(m^2) \,.
\end{equation}
As the coupling to gravity, the mass is effectively the charge in
this context. In the background space-time, the equations of motion
are
\begin{align}
a^\mu = -\f{1}{2}\(g^{\mu\nu} + u^\mu u^\nu\)\(2h_{\nu\lambda ;
\kappa} - h_{\lambda\kappa ; \nu} \)u^\lambda u^\kappa \,,
\end{align}
using the `;' notation for covariant differentiation. Now, the
potential field which we use is not actually that of the
perturbation, but rather the trace-reversed tensor $\gamma$ defined,
akin to the Einstein tensor from the Ricci, as
\begin{align}
\gamma_{\mu\nu} = h_{\mu\nu} - \f{1}{2} \(
g^{\lambda\kappa}h_{\lambda\kappa}\) g_{\mu\nu} \,,
\end{align}
where the reader will note the use of the background metric for
contraction in keeping with the perturbation approximation. These
trace-reversed potentials then satisfy the wave equation
\begin{align}
\square \gamma ^{\mu\nu} + 2 R^{\ph\mu\ph\nu}_{\lambda\ph\kappa}
\gamma^{\lambda\kappa} = -16\pi T^{\mu\nu} \,,
\end{align}
where $T^{\mu\nu}$ is the stress-energy tensor of the point mass.
Following calculations with these potentials, the original $h$
fields can be obtained by trace-reversing again. These equations are
the appropriate counterparts for the spin two graviton to the scalar
(spin zero) and vector (spin one) potential cases. Subtraction of
the singular field from the perturbation leaves the regular, or
radiative, field $h^R$ given by
\begin{align}
h^R_{\mu\nu ; \lambda} = -4m \( u_{(\mu} R_{\nu)\kappa \lambda \rho}
+ R_{\mu \kappa \nu \rho} u_\lambda \) u^\kappa u^\rho + h^{\rm
tail}_{\mu\nu\lambda} \,,
\end{align}
with the tail term given by
\begin{align}
h^{\rm tail}_{\mu\nu\lambda} = 4m \int^{\tau_-}_{-\infty}d\tau'\,
\nabla_\lambda \( G_{-\mu\nu\mu'\nu'} - \f{1}{2} g_{\mu\nu}
G_{-\ph\lambda\mu'\nu'}^\lambda \)u^{\mu'}u^{\nu'} \,.
\end{align}
The retarded Green's function here is that for the wave equation for
$\gamma$, hence the trace-reversed presence here. The equations of
motion are then
\begin{align}
a^\mu = -\f{1}{2}\(g^{\mu\nu} + u^\mu u^\nu\)\(2h^{\rm
tail}_{\nu\lambda \kappa} - h^{\rm tail}_{\lambda\kappa\nu}
\)u^\lambda u^\kappa \,,
\end{align}
with only the tail term remaining. These equations were found in
1997 by Mino, Sasaki and Tanaka \cite{MiSaTa} and using different
methods reproduced by Quinn and Wald \cite{QuWa}. Consequently they
go by the name `MiSaTaQuWa equations'. The variety of methods used
removes some of the difficulties in the analysis of a point mass.
Given that the unperturbed path was the geodesic of the background
space-time, from a general relativity perspective, one would
naturally ask about the geometric properties of the new path. From
the analysis and interpretation of the regular Green's function by
Detweiler and Whiting \cite{Whiting}, we have already noted that the
scalar and electromagnetic charges would move under the influence of
the combination of the original external field and the particle's
radiative field. Thus extending to this case,
\cite{Whiting} gives us the interpretation of the new path as the
geodesic of the space-time with metric
\begin{align}
f^R_{\mu\nu}=g_{\mu\nu}+h^R_{\mu\nu} \,,
\end{align}
which remains a solution to the vacuum Einstein equations. Work on
gravitational radiation reaction uses the above equations, with
attempts to calculate the tail terms, for such situations as small
black holes and orbits in the Schwarzchild (black hole) metric. One
of the aims of such work is to produce a description of the motion
in such extreme circumstances and consequently accurately predict
the gravitational radiation that the new detectors hope to detect.
Should the process work, then we would ultimately obtain a new type
of telescope for probing some of the more extreme gravitational
events in the cosmos.

The equations of motion for gravitational radiation reaction appear
initially somewhat different in form from those for the other
fields. Some of this difference is because the above ignores some
extra difficulties. Namely, the above equations are produced using
the Lorentz gauge condition $\gamma^{\mu\nu}_{\ph ;
\nu}=0$ and are \emph{not} gauge invariant. Under coordinate
transformation of the background coordinates using a smooth field of
order $m$, $x^\mu \to x^\mu + \xi^\mu$, the change in the particles
acceleration is given by the `gauge acceleration'
\begin{align}
\delta a[\xi]^\mu = \(\delta^\mu_\nu +u^\mu u_\nu\)\(\f{D^2
\xi^\nu}{d\tau^2} + R^\nu_{\ph\lambda\kappa\rho}u^\lambda \xi^\kappa
u^\rho \) \,.
\end{align}
The consequences of this, such as the possible gauging away of the
self-force accelerations, should indicate the need to add into
consideration the full metric perturbation in order to obtain
gauge-invariant observables. As we are only giving a brief overview
of extensions to the radiation reaction problem here, we shall not
go into anymore detail but refer the interested reader to the
literature quoted. From this aside, we now return to considerations
of flat-space electromagnetic radiation reaction.

\section{Quantum Theory}
\begin{figure}
\begin{center}
\begin{pspicture}(-4,-4)(4,4)
\parametricplot[linewidth=1.2pt,plotstyle=curve, linestyle=dashed]%
{360}{850}{t cos  t mul 360 div t sin  t mul 360 div}
\pscurve[linewidth=1.2pt]{->}(2,0)(2.25,0.25)(2.75,-0.25)(3.25,0.25)(3.75,-0.25)(4.25,0.25)(4.5,0)
\pscurve[linewidth=1.2pt]{->}(-1.5,0)(-1.75,0.25)(-2.25,-0.25)(-2.75,0.25)(-3.25,-0.25)(-3.75,0.25)(-4,0)
\pscircle[linewidth=0.8pt, fillstyle=solid, fillcolor=gray](0,0){0.8}
\pspolygon[linewidth=1.2pt](0,2.15)(0,2.35)(0.2,2.25)
\pspolygon[linewidth=1.2pt](0,-1.65)(0,-1.85)(-0.2,-1.75)
\end{pspicture}
\caption{An electron in a classical atom would radiate, losing energy, and spiral into the nucleus. It is thus unstable.}
\label{classatom}
\end{center}
\end{figure}
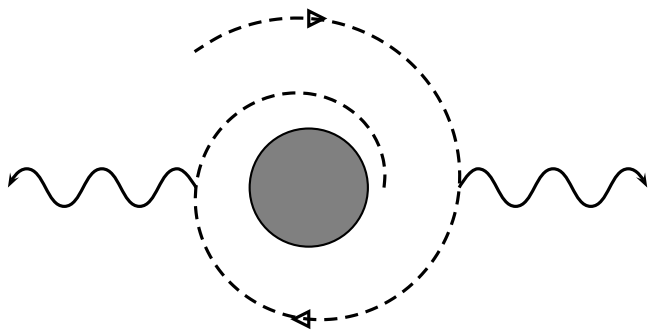

Classical electrodynamics is no longer considered to be the most
fundamental theory, but is currently superseded by quantum
electrodynamics, or to use the more common acronym, QED. The
classical theory is however very successful within limits and forms
the basis on which we normally construct the quantum theory, as with
many other classical theories. Classical Electromagnetism, unifying
two of the fundamental forces of nature, was one of the great
success stories of 19th century science. It was in the study of
radiation that the cracks began to appear. The ultraviolet
catastrophe\footnote{The ultraviolet catastrophe is also known as
the Rayleigh-Jeans catastrophe.} is usually given as the example of
this, whereby classical electromagnetism predicted that a black body
at thermal equilibrium would emit radiation with infinite power.
This is demonstrably false, with the problem occurring in the short
wavelength (hence ultraviolet) region. The well-known solution was
Max Planck's quantum hypothesis - that the radiation was emitted
only in discrete `quanta' of energy, which Einstein suggested be
used to address the issue. Einstein also used the hypothesis to
solve another classical problem relating to radiation: the
photoelectric effect. The issue of electromagnetic radiation was
thus one at the focus of the early work on quantum theory. Another
example frequently presented as a way of showing the successes of
early quantum theory is that concerning the structure of the atom
and one which is related to our main consideration. After
Rutherford's experiments providing the evidence for the positive
nucleus model of the atom\footnote{Instead of, say, the plum-pudding
model.}, the orbit style view of the atom, in which the electrons
circled the nucleus, like the Newtonian motion of the planets to the
sun\footnote{Using circled in a more liberal sense to include
elliptical motion.}, was the classical model of the motion of the
electrons. This model appears to be a fairly good analogy until one
considers the oft-ignored radiation reaction and considers the
motion of the particle itself. As the electron is continuously
accelerated, although with the acceleration vector changing to
always point to the nucleus, the theory predicts that it will emit
radiation. This would mean that the system would lose energy and
thus the prediction is that the electron would spiral into the
nucleus, consequently rendering all classical atoms inherently
unstable (see Fig. \ref{classatom}).
This is of course another effect which is (thankfully) demonstrably
incorrect. The extension of the quantum hypothesis to the energy
spectrum of the atom, thus allowing only certain stable energy
levels, quickly gives very accurate predictions. After these
beginnings, the full theory of quantum electrodynamics was gradually
developed. This has in turn become one of the success stories of
20th century science, and one frequently stated to be the most
accurate theoretical model of all time - so far at least.

\begin{figure}
\begin{center}
\begin{pspicture}(-4,-1)(4,2)
\psline[linewidth=0.5mm]{->}(-7,0)(-5,0)
\psline[linewidth=0.5mm](-5.1,0)(-3,0)

\psline[linewidth=0.5mm]{->}(-2,0)(-1,0)
\psline[linewidth=0.5mm]{->}(-1.1,0)(1,0)
\psline[linewidth=0.5mm](0.9,0)(2,0)
\pscurve[linewidth=0.5mm]{*-}(0,0)(0,0.5)(0.5,0.5)(0.5,1.0)(1.0,1.0)
(1.0,1.5)(1.5,1.5)

\psline[linewidth=0.5mm]{->}(3,0)(5,0)
\psline[linewidth=0.5mm](4.9,0)(7,0)
\pscurve[linewidth=0.5mm]{*-*}(3.75,0)(3.75,0.5)(4.25,0.5)(4.25,1.0)(4.7,1.0)
(5.0,1.25)(5.3,1.0)(5.75,1.0)(5.75,0.5)(6.25,0.5)(6.25,0)
\end{pspicture}
\caption{The first three types of Feynman diagrams for QED
representing the perturbation expansion up to order $e^2$.}
\label{Feyexp}
\end{center}
\end{figure}
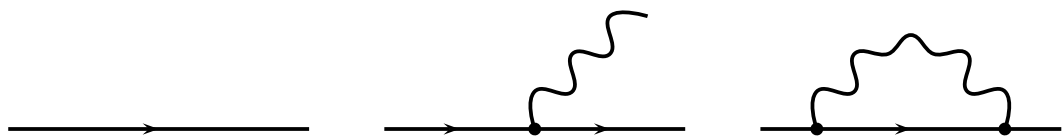

QED is usually studied using the techniques of perturbation theory,
in which the interactions between the particle and electromagnetic
fields are expanded as a series in terms of the coupling between
them. In Feynman diagrams, the first three terms of the perturbative
expansion of the scattering amplitude are given in Fig. \ref{Feyexp}
and are the basic diagrams usually considered when learning about
interactions in quantum field theory.\footnote{Along with the
particle creation/annihilation diagrams, which are really the same
process as the emission diagram in Fig. \ref{Feyexp}.}
These diagrams are for the perturbation theory up to order $e^2$,
which the reader may recall is the order of the classical radiation
reaction. We have a process corresponding to the emission of
radiation (the first order interaction). The strength of this
contribution to the perturbation calculation will be dependent on
the physical situation i.e. the classical external forces
accelerating the particle.\footnote{By classical here we mean that
such forces are treated non-perturbatively. Should an external force
be added in perturbatively, then the contributions would have to be
shown in the diagrams by additional boson lines.} The one-loop
contribution represents the most basic self-energy process in
quantum field theory. It is the self-interaction with a virtual
photon emitted and absorbed by the particle itself which we shall
also at times term the forward scattering.\footnote{We recall that
these diagrams are representative of the contributing terms in the
perturbation expansion rather, as is sometimes mistakenly thought,
the actual physical process.} This represents the contribution of
the particle interacting with its own electromagnetic field. The
contribution is infinite, as are a number of other self-interaction
type processes in QED, hence the reason for the necessity of
renormalisation. At this level, renormalisation consists of removing
the infinite self-energy contribution in order to obtain finite
answers. There is more than one equivalent way in which to achieve
this and we shall use the counter-term method in this work as it
fits the calculation best. The methods are however all equivalent to
redefining the mass.\footnote{For higher order perturbation terms,
one would also need to renormalise the field etc.}

In the above paragraph, whilst we gave a brief overview of the basic
perturbation contributions, the aim was in fact to word the
descriptions in terms similar to those that we have been using to
describe classical radiation reaction. Some of the similarities in
the way in which we deal with both theories should hopefully now be
apparent. In both cases, we use a perturbation expansion in terms of
the coupling for many calculations.\footnote{However, the reasons
for using the expansion are different.} In both cases we can split
the processes up into emission and self-energy interactions. In
addition, in both the classical and quantum theories it is necessary
to subtract an infinite contribution corresponding to the
self-energy via a renormalisation of the mass. There are of course
also differences; otherwise we would not need to replace classical
electrodynamics with the quantum theory in the first place. These
similarities and differences are then a further motivation for both
the work contained here and the details of the models chosen. The
above sections of the introduction should be kept in mind when we
introduce and justify the model and calculations to be performed.

\section{Origins of the present work}
This work is based on the initial results and models given in
\cite{Higuchi2}. In these papers, Higuchi looks at the process of
radiation reaction in quantum mechanics and the non-relativistic
approximation in quantum field theory for comparison with the
results of classical Abraham-Lorentz-Dirac theory. The papers look
at the calculation of the change in position due to radiation
reaction, which is labelled the position shift.\footnote{As the
model used later is based on that from \cite{Higuchi2}, we shall not
go into detail now, but rather ask the reader to wait until the next
section where the extended model for this work is presented.} The
comparison is then made between the predicted value of this quantity
for the classical theory with the classical limit of the
non-relativistic approximation of the first order interactions of
quantum field theory for a charged scalar field. The results are
that the predictions agree, thus supporting the idea of the
Lorentz-Dirac theory as the appropriate classical limit for quantum
theory. It is on this base that we build the work presented here.
Our aim is to compare the classical and quantum theories of
radiation reaction in order to gain a further understanding of the
effect in both. Given the debates over the interpretation of the
Lorentz-Dirac theory with the associated problems and possible
solutions as detailed above, a comparison at the level of the
classical limit of the more fundamental QED is also useful. The next
sections detail the models used and calculations to be presented
along with justifications for the choices made. The models are based
on those in \cite{Higuchi2}, but extended to a fully relativistic
theory, to the spinor field, and also to considerations of the
second order interaction at order $e^2$. The previous sections have
detailed the background theories with which we are concerned; the
following sections detail and introduce fully the current research
on which this work reports.

\section{The Model} \label{modelsection}
We wish to compare the effects of radiation reaction in the
classical and quantum electrodynamics theories. Possibly the most
fundamental effect of radiation reaction is to change the equations
of motion. These equations are in turn simply differential equations
to be solved for the position of the particle. Consequently, the
observable effect due to the existence of radiation reaction, is a
(possible) change in the measured position of the particle. We
therefore choose to make this observable the measured effect which
we shall investigate. To be more precise, we wish to measure the
change in position of the particle due to radiation reaction. This
rather unwieldy description we give the name the \emph{position
shift}.

We now require a model involving radiation reaction in which to make
our measurement of the position shift. The reader will recall that
the canonical set-up used in the perturbation theory of quantum
field theory is the situation in which the fields are regarded to be
free at future and past temporal infinity, with the perturbative
interaction in between. The particle interpretation is in fact
dependent on the states being non-interacting at temporal infinity
(past and future). We have a situation in which a free particle
enters from past infinity, interacts with the other fields (in our
case, the electromagnetic field) and then leaves as a free particle
to future infinity. At this point we remind the reader that in the
classical theory, the reduction of order procedure, as carried out
on the Lorentz-Dirac equation, is equivalent to treating the
Lorentz-Dirac force as a perturbation. The two theories we wish to
compare are consequently both best represented by the above
description of the quantum interaction model. We therefore choose
the following: Let the particle travel in a potential which is
constant in the asymptotic regions and non-constant for some finite
region in between. Only in the non-constant potential region will
the particle experience acceleration and thus radiation reaction.
Having given some explanation for the choice of such a model, we now
proceed to define precisely the model used for this work.

Let the potential $V$ be dependent on one of the spacetime indices,
say $x^a$. The potential is chosen to be equal to $V_0={\rm const.}$
for $x^a<X^a_1$ and equal to $0$ for $x^a>X^a_2$ for some
$X^a_2>X^a_1$. The acceleration is thus non-zero only in the region
$X^a_1\leq x^a \leq X^a_2$. The choice of $V=0$ for the final region
is made for simplicity (if it were not, we could simply redefine the
potential so that it was). Let us define the three regions as
\begin{align}
{\cal M}_- &= \{ x | x^a<X^a_1 \} \nol {\cal M}_I &= \{ x |
X^a_1\leq x^a
\leq X^a_2 \} \nol
{\cal M}_+ &= \{ x | x^a>X^a_2 \} \,.
\end{align}
With $x^a=x^0$, i.e. a time dependent potential, it is clear that
the particle will start in the region with $V_0$, enter the region
of acceleration and thenceforth finish in the region with $V=0$. For
$x^a$ spatial, we require the initial and final momenta to be
positive in the $x^a$ direction to achieve the same set-up. The only
assumption here is that there is no turning point in that
coordinate, something which in fact we shall require later in any
case. \footnote{Strictly speaking we have only so far assumed the
weaker condition that the number of turning points is not odd.}

\begin{figure}
\begin{center}
\begin{pspicture}(0.5,-1)(9,4)
 \psline{->}(0.5,0)(9,0) \psline{->}(8.5,-1)(8.5,4)
 \psline[linewidth=0.8mm](0.5,3)(3,3)
 \psline[linewidth=0.8mm](6,0)(8.8,0)
 \uput[r](8.5,3.5){$V$} \uput[d](9,0){$x^a$}
 \psline[linestyle=dotted](3,0)(3,4)
 \psline[linestyle=dotted](6,0)(6,4)
 \uput[u](1.5,3){$V(x^a)=V_0$} \uput[u](7,0){$V(x^a)=0$}
 \uput[d](3,0){$X^a_1$} \uput[d](6,0){$X^a_2$}
 \psline{<->}(3,2)(6,2)
 \uput[u](4.5,2){Acceleration}
\end{pspicture}
\end{center}
\caption{The potential $V(x^a)$ and period of acceleration.}
\label{potentialmodel}
\end{figure}
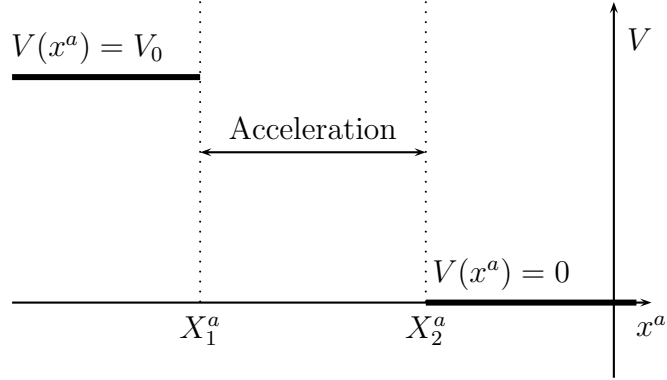

We wish to analyse the effect of the radiation reaction which takes
place in the region of acceleration as measured by the position
shift. It makes sense that the measurement takes place outside the
region itself. We note that in the non-interacting region,
representing the quantum field by a free field is an approximation
which becomes more valid as we move further from the interaction. We
thus state that the position shift is measured far enough into the
later asymptotic region so that the plane wave approximation for the
quantum mode function is accurate. Now, the position shift will be
measured by comparing the position of a non-radiation particle, a
control particle, to one undergoing radiation reaction. Again, for
the sake of simplicity, we define the coordinate such that the
control particle is at the origin at the time of the measurement.
This has the added bonus that the position shift is simply the
position of the radiating particle at the point of measurement. The
positions $X^a_{1,2}$ are thus negative in this coordinate system.
Fig. \ref{potentialmodel} represents the model graphically. The
choice of $V_0>0$ is made here simply for the purpose of the
graphical representation.

Now, whilst we wish to treat the radiation reaction effects as a
perturbation, we have no particular need to treat the potential as
such. In terms of the quantum theory, the potential is treated as a
so called classical potential (i.e.  non-perturbatively). The
potential $V$ is simply the source of the external force which
causes the particle to interact with its
\emph{own} field. The latter interaction is the one treated perturbatively.

\section{Scalar Field}
In this section we set up the quantum field theory model of the
charged scalar field. We give the appropriate definitions of the
field and the conventions and notation which we shall employ in the
further discussion. The Lagrangian density ${\cal L}$ for the free
complex scalar field is given by
\begin{equation} \label{KGlagrangian}
{\cal L} = \partial_\mu\varphi^\dagger \partial^\mu\varphi -
(m/\hbar)^2\varphi^\dagger \varphi \,.
\end{equation}
From this Lagrangian, the conserved current is given by
\begin{equation}\label{scalarcurrent}
j_\mu = \f{i}{\hbar}:\varphi^\dagger
\stackrel{\leftrightarrow}{\partial_\mu} \varphi : \,,
\end{equation}
where
$\stackrel{\leftrightarrow}{\partial_\mu}=\stackrel{\rightarrow}{\partial_\mu}
-\stackrel{\leftarrow}{\partial_\mu}$. The colons $:\,\,:$ represent
the normal ordering process which orders creation operators on the
left of annihilation operators to ensure that the vacuum expectation
value of the current $j_\mu$ vanishes. The zeroth component of the
conserved current is the charge density given by
\begin{equation}\label{scalardensity}
\rho(x) = \f{i}{\hbar}:\varphi^\dagger
\stackrel{\leftrightarrow}{\partial_t} \varphi : \,,
\end{equation}
which we shall have need of in order to calculate the expectation
values of the free state.

The equation of motion for a free charged scalar field $\varphi$ is
the Klein-Gordon equation
\beq \label{KG}
\( \hbar^2\Box + m^2 \) \varphi = 0 \,,
\eeq
where $\Box=\partial^\mu\partial_\mu$ is the d'Alembertian operator
and $m$ is naturally the mass of the field. In the absence of
coupling to another field, $\varphi$ is expanded via a Fourier
decomposition to give
\begin{equation}
\varphi(x) = \hbar \int \frac{d^3{\bf p}}{2p_0(2\pi\hbar)^3}
\left[ A({\bf p})\Phi_{\bf p}(x) + B^\dagger({\bf p}) \bar{\Phi}_{\bf
p}^\dagger(x)\right] \,. \label{scalar_expansion}
\end{equation}
In this expansion, $\Phi_{\bf p}(x)$ is the mode function i.e. a
solution to the field equation for $\varphi(x)$ (\ref{KG}).
Similarly, $\bar{\Phi}_{\bf p}$ is a solution to the field equation
for $\varphi^\dagger(x)$, which for the free field is again
(\ref{KG}), thus $\bar{\Phi}_{\bf p}(x)=\Phi_{\bf
p}(x)$.~\footnote{The introduction for the notation $\bar{\Phi}_{\bf
p}$ whilst seemingly superfluous here, shall be needed shortly.}
Modeling the field as a plane wave we substitute
\beq
\Phi_{\bf p}(x) = e^{-i p \cdot x/\hbar} \,.
\eeq
$A^\dagger({\bf p})$ and $B^\dagger({\bf p})$ are the creation
operators for the positive and negative charged particles, with
$A({\bf p})$, $B({\bf p})$ the respective annihilation operators.
The quantisation of the field is given by the commutation relations
\begin{align}\label{scalarcommutation}
\[ A({\bf p}), A^\dagger({\bf p'})\] &= \[ B({\bf p}), B^\dagger({\bf p'})\]
= 2p_0 (2\pi\hbar)^3 \delta^3({\bf p}-{\bf p'}) \,,
\end{align}
with all other commutation relations set to zero. At this point we
remind the reader of our conventions as set out above. The measure
$d^3{\bf p}/p_0$ is a Lorentz invariant element of phase space due
to the mass-shell condition $p_0^2={\bf p}^2+m^2$. Our convention
involves the constant multiplication to give the factors $2p_0
(2\pi\hbar)^3$ in both the measure's denominator and in the
commutation relations. Note the presence of the $\hbar$'s both here
and as an overall multiplier in the field in
(\ref{scalar_expansion}). Their presence is of course frequently
omitted in discussions due to the use of natural units ($\hbar=1$).
However, whilst very useful for most particle physics discussions,
such a unit system is not conducive to the analysis and
investigation of the classical, i.e. $\hbar\to 0$, limit which we
shall later wish to perform and hence their inclusion.

The Poincar\'{e} invariance of Minkowski space can be employed to
define a unambiguous vacuum state $\ket{0}$ for the scalar field,
given by the condition $A({\bf p})\ket{0}=B({\bf p})\ket{0}=0$. The
successive applications of the creation operators then build up the
Fock space for scalar field with the appropriate particle
interpretation.

The free scalar field is used to model the particle in terms of an
incoming and outgoing wave packet. We represent the initial state $|
i\rangle$ by
\begin{equation}\label{istate}
| i\rangle = \int \frac{d^3{\bf p}}{\sqrt{2p_0}(2\pi\hbar)^3} f({\bf
p})A^\dagger({\bf p})|0\rangle \,,
\end{equation}
where the function $f({\bf p})$ is sharply peaked about a given
momentum $\overline{\bf p}$. For our later use, we require that $f$
is sufficiently sharply peaked such that we can approximate $|f({\bf
p})|^2$ by $(2\pi\hbar)^3 \[ \delta({\bf p} - \overline{\bf p}) +
{\cal O}(\hbar^2) \]$. The normalisation of the operators
$A^\dagger({\bf p})$ is such that the condition $\langle
i\,|\,i\rangle = 1$ leads to
\begin{equation}\label{inorm}
\int\frac{d^3{\bf p}}{(2\pi\hbar)^3}|f({\bf p})|^2 = 1 \,.
\end{equation}
This shows that the function $f({\bf p})$ can heuristically be
regarded as the one-particle wave function in the momentum
representation.

So far, we have dealt only with the free fields i.e. in the absence
of the potential. Before considering the interaction between the
scalar and electromagnetic fields that will contribute to the
radiation reaction process, we must include the external potential
$V$ in the presence of which the interaction of interest will take
place. As previously stated, we shall treat this external potential
non-perturbatively i.e. we shall not expand in orders of $V$. The
inclusion is most easily achieved by substitution of the derivatives
as follows:
\begin{equation}
\partial_\mu \varphi \to D_\mu \varphi = \(\partial_\mu + \f{i}{\hbar}
V_\mu\) \varphi \,,
\end{equation}
where the $V_\mu$ are the spacetime components of the potential $V$.
In the presence of the potential, the Lagrangian density becomes
\begin{equation} \label{KGlagrangianV}
{\cal L} = (D_\mu\varphi)^\dagger D^\mu\varphi -
(m/\hbar)^2\varphi^\dagger \varphi \,.
\end{equation}
The field equations can be similarly obtained to give
\begin{align}
\(\hbar^2 D^\mu D_\mu + m^2\) \varphi &= 0 \label{KGV} \\
\(\hbar^2 D^{\dagger\mu} D^\dagger_\mu + m^2\) \varphi^\dagger &=
0 \,.
\label{KGVdagger}
\end{align}
In this case we note that the equations are no longer the same;
$D^\dagger_\mu \neq D_\mu$. Writing the field in the Fourier mode
expansion as before
\begin{equation}
\varphi(x) = \hbar \int \frac{d^3{\bf p}}{2p_0(2\pi\hbar)^3}
\left[ A({\bf p})\Phi_{\bf p}(x) + B^\dagger({\bf p}) \bar{\Phi}_{\bf
p}^\dagger(x)\right] \,.
\end{equation}
The mode functions $\Phi_{\bf p}(x)$ and $\bar{\Phi}_{\bf p}(x)$ are
solutions to the non-free field equations for $\varphi(x)$ in
(\ref{KGV}) and $\varphi^\dagger(x)$ in (\ref{KGVdagger})
respectively. The difference, arising from the $i$ in $D_\mu$, is of
course the charge difference between the particle and antiparticle
modes.\footnote{The potential has been added using minimal
substitution, as the (perturbative) electromagnetic field will be,
and thus has a charge coupling in a similar manner.} When analyzing
the field in ${\cal M}_I$ we shall use the semiclassical expansions
for the mode functions, which are detailed in Chapter
\ref{Semiclassical}. The commutation relations for the scalar field
creation and annihilation operators are those detailed above in
(\ref{scalarcommutation}) for field in the `free' regions.

The next step in the construction of our model is to add the
interaction between the scalar and electromagnetic fields, without
which there will be no radiation reaction. With the inclusion of
coupling to the electromagnetic field, we write the Lagrangian
density as
\begin{equation}\label{KGEMLagrangian}
{\cal L} = [\(D_\mu + ieA_\mu/\hbar\)\varphi]^\dagger
\(D^\mu +ieA^\mu/\hbar\)\varphi - (m/\hbar)^2\varphi^\dagger \varphi -
\frac{1}{4}F_{\mu\nu}F^{\mu\nu} - \frac{1}{2}(\partial_\mu
A^\mu)^2 \,,
\end{equation}
where $A_\mu$ is the electromagnetic potential and
$F_{\mu\nu}=\partial_\mu A_\nu- \partial_\nu A_\nu$ is the
electromagnetic field tensor. The last term in the Lagrangian
density is the Lagrange multiplier representing the choice of the
Lorentz gauge condition $\partial_\mu A^\mu=  0$. The choice of the
prefactor of $1/2$ on this term is known as the Feynman gauge and is
made in order to simplify the photon propagator.\footnote{See for
example \cite{Itzykson} for further theoretical details.} We proceed
as per the scalar field to give the expansion of the electromagnetic
potential in terms of the plane wave solutions viz
\begin{equation}
A_\mu(x) = \int\frac{d^3{\bf k}}{2k(2\pi)^3} \left[ a_\mu({\bf k})
e^{-ik\cdot x} + a^\dagger_\mu({\bf k})e^{ik\cdot x} \right] \,.
\end{equation}
We make use of the notation $k=|{\bf k}|$. Due to the massless
nature of the photons, with $k^\mu k_\mu=0$ and thus $k=k_0$, we
will use $k$ and $k_0$ interchangeably depending on the emphasis
required at the time. The quantisation is given by the commutation
relations for the photon creation and annihilation operators
\begin{equation}\label{EMcommutation}
\left[ a_\mu({\bf k}),a^\dagger_\nu({\bf k}')\right] =
-g_{\mu\nu}(2\pi)^32\hbar k\delta^3({\bf k}-{\bf k}') \,.
\end{equation}
Notice that the scalar field $\varphi$ is expanded in terms of the
momentum ${\bf p}$ whereas the electromagnetic field $A_\mu$ is
expanded in terms of the wave number ${\bf k}$. We adopt this
convention because the vectors ${\bf p}$ and ${\bf k}$ are regarded
as classical rather than ${\bf p}/\hbar$, the wave number of the
scalar particle, and $\hbar{\bf k}$, the momentum of the
electromagnetic field.

We are now in a position to turn our attention to the interaction
and evolution of the wave packet taking place during the
acceleration period in ${\cal M}_I$. The evolution of the state is
modeled by perturbation theory. We are interested in terms up to
second order in the coupling, i.e. $e^2$, and consequently need to
consider the first two orders in the interaction. The evolution from
an initial state $\ket{i}$, written in terms of the interaction
Hamiltonian is to second order given by the map
\begin{equation}\label{evol}
\ket{i}
\mapsto \ket{i} -\frac{i}{\hbar}\int d^4 x
{\cal H}_I(x)\ket{i} + \(\f{-i}{\hbar}\)^2 \int d^4x\,d^4x' T \[
{\cal H}_I(x){\cal H}_I(x')\]\ket{i} \,,
\end{equation}
where $T$ is the time ordering operator. The interaction Hamiltonian
density is obtained from the interaction Lagrangian to give
\begin{equation}\label{Hamiltonian}
{\cal H}_I(x) = \frac{ie}{\hbar}A_\mu \normal\left[ \varphi^\dagger
D^\mu
\varphi - (D^\mu \varphi)^\dagger \varphi\right]\normal
 + \frac{e^2}{\hbar^2}\sum_{i=1}^3 A_iA_i
\normal\varphi^\dagger\varphi\normal \,,
\end{equation}
where $D_\mu\equiv \partial_\mu + iV_\mu/\hbar$ as before. We have
normal-ordered the scalar-field operators to drop the vacuum
polarization diagram automatically. Note that the second term is
different from what might be na\"{i}vely expected, viz
$-(e^2/\hbar^2)A_\mu A^\mu
\,\normal\varphi^\dagger\varphi\normal\,\,$. This difference is
due to the presence of interaction terms involving $\dot{\varphi}$
or $\dot{\varphi}^\dagger$ in the Lagrangian density.\footnote{The
derivation of the interaction Hamiltonian is detailed in appendix
\ref{IntHam}.}

\begin{figure}
\begin{center}
\begin{pspicture}(-4,-1)(4,2)
\psline[linewidth=0.5mm]{->}(-7,0)(-5.5,0)
\psline[linewidth=0.5mm](-5.6,0)(-4,0)

\psline[linewidth=0.5mm]{->}(-3.4,0)(-2.65,0)
\psline[linewidth=0.5mm]{->}(-2.7,0)(-1.15,0)
\psline[linewidth=0.5mm](-1.2,0)(-0.4,0)
\pscurve[linewidth=0.5mm]{*-}(-1.9,0)(-1.9,0.4)(-1.5,0.4)(-1.5,0.8)(-1.1,0.8)
(-1.1,1.2)(-0.7,1.2)

\psline[linewidth=0.5mm]{->}(0.4,0)(1.9,0)
\psline[linewidth=0.5mm](1.8,0)(3.4,0)
\pscurve[linewidth=0.5mm]{*-*}(0.9,0)(0.9,0.4)(1.3,0.4)(1.3,0.8)(1.7,0.8)
(1.9,1)(2.1,0.8)(2.5,0.8)(2.5,0.4)(2.9,0.4)(2.9,0)

\psline[linewidth=0.5mm]{->}(4,0)(4.75,0)
\psline[linewidth=0.5mm]{->}(4.65,0)(6.25,0)
\psline[linewidth=0.5mm](6.15,0)(7,0)
\pscurve[linewidth=0.5mm]{*-}(5.5,0)(5.8,0.3)(6.1,0.2)(6.3,0.4)(6.1,0.6)(6.3,0.8)(6.1,1.0)(5.8,0.9)
(5.5,1.2)(5.2,0.9)(4.9,1.0)(4.7,0.8)(4.9,0.6)(4.7,0.4)(4.9,0.2)(5.2,0.3)(5.5,0)

\end{pspicture}
\caption{The Feynman diagrams for scalar QED
representing the perturbation expansion up to order $e^2$.}
\label{scalardiags}
\end{center}
\end{figure}
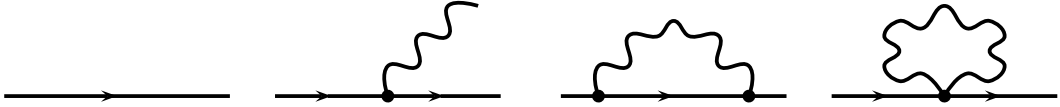

In addition to the standard one-loop QED process, for scalar QED we
also have the contribution where the start and end of the loop are
at the same point. In the Feynman diagrams, this is present by the
vertex with two photon and two scalar propagators, sometimes known
as a seagull vertex, and must be remembered if working from the
Feynman rules.\footnote{For most of the later work, we shall be
starting from the operators, and so this contribution will come out
of the works on its own. We only use the Feynman rules here for the
free-field calculations of the mass counter-terms.} The processes
contributing to order $e^2$ from the above interaction Hamiltonian
are, in diagrammatic form given in Fig. \ref{scalardiags}. The last
two diagrams in Fig. \ref{scalardiags} jointly give the first
non-trivial contribution to the one-particle irreducible Green's
function with two external lines, also known as the self energy,
which is divergent.

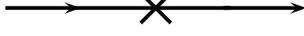
\begin{figure}
\begin{center}
\begin{pspicture}(-2,-1)(2,1)
\psline[linewidth=0.5mm]{->}(-2,0)(-1,0)
\psline[linewidth=0.5mm](-1.1,0)(1,0)
\psline[linewidth=0.5mm]{->}(0.9,0)(2,0)
\psline[linewidth=0.5mm](-0.2,0.2)(0.2,-0.2)
\psline[linewidth=0.5mm](-0.2,-0.2)(0.2,0.2)
\end{pspicture}
\caption{The mass counter term contribution to the propagator.}
\label{renormcross}
\end{center}
\end{figure}

We consequently now come to the renormalisation process to deal with
the divergences. To the required order in $\hbar$ for the
calculations that shall follow we shall only require the
renormalisation of the mass. As we shall be dealing with the
contributions from the interaction Hamiltonian terms to the position
shift, the natural method of renormalisation will be using the mass
counter-term. The contribution of the counter-term, which is of
course infinite by definition, can then be added the our results.
The counter-term takes the form of the addition to the Lagrangian of
\begin{equation}
\delta {\cal L} = \f{\delta m^2}{\hbar^2} \varphi^\dagger \varphi
\,,
\end{equation}
where the $\hbar^2$ is needed due to our field conventions for the
scalar field. The mass counter-term is local, i.e. has no momentum
dependence, and is designed to cancel the divergences from the
one-loop diagrams for the free-field.\footnote{For the scalar field
calculations we shall refer jointly to last two diagrams in Fig.
\ref{scalardiags} as the forward scattering or one loop process.}
The calculation here is thus the standard quantum scalar field
theory renormalisation for which the reader is referred to the
literature for a full introduction.\footnote{The author recommends,
for example
\cite{Itzykson}, \cite{Ryder} and \cite{Mandl}.} This standard nature is emphasized
due to the fact that in the presence of the potential, the general
quantum field theory calculations are \emph{not} standard free field
QED, hence the Feynman rules are not used there. The propagator is
modified to remove these divergent contributions via the subtraction
of the self-energy
\begin{equation}
\f{i}{p^2-m^2} \mapsto  \f{i}{p^2-m^2-\Sigma(p)} \,,
\end{equation}
where $m$ here is the bare mass. Rewritten as an expansion
\begin{align}
\f{i}{p^2-m^2-\Sigma(p)} &=  \f{i}{p^2-m^2} + \f{i}{p^2-m^2} \[ -i\Sigma(p) \] \f{i}{p^2-m^2}
\nol & \quad +\f{i}{p^2-m^2}\[ -i\Sigma(p) \]\f{i}{p^2-m^2}\[ -i\Sigma(p)
\]\f{i}{p^2-m^2}+
\ldots \,,
\end{align}
it is easy to see that this operation then effectively adds a
further Feynman diagram to the perturbation expansion, given in Fig.
\ref{renormcross}.
This perturbation contribution can be read straight from the Feynman
rules for the scalar field. Using our conventions we obtain, with
$K=\hbar k$ for the photon momentum,
\begin{align}\label{scalarselfenergy}
-i\Sigma(p) &= \int\f{d^4 K}{(2\pi)^4} \left\{
\(-\f{ie}{\hbar}\) \left[ p_\mu + (p_\mu - K_\mu)\right]
\right. \nol
& \qquad \left. \times \f{i}{(p-K)^2 - m^2 + i\epsilon}
\f{-i\hbar g^{\mu\nu}}{K^2 + i\epsilon}
\( -\f{ie}{\hbar}\) \[ p_\nu + \( p_\nu - K_\nu \) \] \right. \nol & \qquad \left. +\(-\f{ie}{\hbar}\)^2
 \f{i\hbar \delta^\mu_\mu}{K^2+i\epsilon} \right\}
 \nol
& = -i e^2\int \f{d^4 K}{(2\pi)^4 i}
\left\{ \f{(2p-K)^2}{\[ (p-K)^2 - m^2+i\epsilon\] \[ K^2 + i\epsilon\]}
 - \f{4}{K^2+i\epsilon} \right\} \,.
\end{align}
The function $\Sigma$ is divergent and we need to regularise it,
e.g. by dimensional regularisation. Then $\delta m^2$ is chosen (as
a function of the regularising parameter) to cancel the divergence
and ensure that
\begin{equation}
\delta m^2 - \Sigma(p)|_{p^2=m_P^2} \to 0
\end{equation}
as the regulator is removed, where $m_P$ is the physical mass.

We have now introduced the conventions and definitions for the main
components of our quantum theoretic model of the complex scalar
field.

\section{Spinor Field}\label{spinordefnsection}
In this section we give our definitions and conventions for the
quantum model for a spinor field. The spinor field is the spin $1/2$
field $\psi$ satisfying the first order Dirac equation
\begin{equation}\label{Diraceqn}
\(i\slashed{\partial}-m\)\psi=0 \,.
\end{equation}
In the above we have made use of the Feynman `slash' notation i.e.
for the covariant vector $A_\mu$, $\slashed{A}:=\gamma^\mu A_\mu$.
The gamma matrices $\gamma_\mu$, by virtue of the fact that $\psi$
must also satisfy the Klein-Gordon equation, are subject to the
relation $\{\gamma_\mu,\gamma_\nu\}=2g_{\mu\nu}$. In this work we
shall make use of the Dirac representation for the $\gamma$-matrices
(also known as the standard representation and detailed in Appendix
\ref{diracrefs}). The Dirac equation is the equation of motion for the field with
Lagrangian
\begin{equation}
{\cal L} = i \hbar \bar{\psi} \slashed{\partial} \psi - m
\bar{\psi}\psi \,,
\end{equation}
where the barred spinors are defined in terms of $\gamma^0$ and the
Hermitian conjugate spinor by $\bar{\psi}=
\psi^\dagger \gamma^0$. From this Lagrangian, the
canonical momentum is given by $\pi(x) = i\psi^\dagger(x)$. Despite
the more complicated nature of spinors when compared with a scalar
field, the fact that the equations of motion are first order leads
to simpler expressions for most basic required quantities. The
zeroth component of the  current, the charge density, is given by
\begin{align}
j^0=\rho(x) = : \psi^\dagger(x) \psi(x) : \,,
\end{align}
with the usual normal ordering.

The free spinor field is expanded in the following way
\begin{equation}\label{spinorfield}
\psi(t,{\bf x}) = \int \f{d^3 p}{(2\pi\hbar)^3} \f{m}{p_0} \sum_\alpha
\[ b_\alpha({\bf p})\Phi_\alpha({\bf p}) +
d^\dagger_\alpha({\bf p})\Psi_\alpha({\bf p}) \] \,.
\end{equation}
The expansion includes the sum over $\alpha$, the spin index. The
spin $1/2$ field will have its spin along a particular axis in one
of two states, `up' and `down', which shall be represented in the
appropriate solution to (\ref{Diraceqn}). The different spin states
have their own creation and annihilation operators. These operators
satisfy the anticommutation relations
\begin{align} \label{anticomms}
\{ b_\alpha({\bf p}),b^\dagger_\beta({\bf p'})\} &= \{
d_\alpha({\bf p}),d^\dagger_\beta({\bf p'})\}=
\f{p_0}{m}(2\pi\hbar)^3\delta^3({\bf p}-{\bf
p'})\delta_{\alpha\beta} \nol
\{ b_\alpha({\bf p}),b_\beta({\bf p'})\} &=
\{ b^\dagger_\alpha({\bf p}),b^\dagger_\beta({\bf p'})\}= 0 \nol
\{ d_\alpha({\bf p}),d_\beta({\bf p'})\} &=
\{d^\dagger_\alpha({\bf p}),d^\dagger_\beta({\bf p'})\}= 0 \,.
\end{align}
We recall that quantisation of the spinor field uses the
anticommutation relations, as opposed the commutation relations, to
ensure that the energy of the field is positive definite. This also
means that when normal ordering one must be careful to take the
appropriate minus signs when swapping the order of fields. Here we
have used the multiple $p_0/m$ in the denominator of the measure and
the anticommutation relations.

From the scalar definitions, we recall that there was no need in the
free field case to distinguish between solutions of the field
equations for the field and its conjugate. However, the distinction
was important when adding the potential. The same should be
considered here. The conjugate of the Dirac equation gives the field
equations for the barred-conjugate field
\begin{equation}
\bar{\psi}(x) \(i\hbar \stackrel{\leftarrow}{\slashed{\partial}} +m \) = 0
\,,
\end{equation}
where the arrow indicates that $\partial$ acts on those terms to the
left (i.e. on the $\bar{\psi}(x)$ field here). We consequently
regard the mode function $\Phi$ as a solution for the Dirac equation
for $\psi$ and $\bar{\Psi}$ as a solution of the conjugate equation
for $\bar{\psi}$.~\footnote{Note that $\bar{\Psi}$ is the
barred-conjugate of the mode function in the decomposition
(\ref{spinorfield}), which must be a vector in the same vector space
as $\Phi$.} The latter designation is merely for emphasis; the mode
function $\Psi$ is still a solution to the Dirac equation. However,
we wish to emphasize that the mode functions are not conjugate to
each other. This is in keeping with the definitions of the scalar
mode functions and again, for the free field the distinction is
irrelevant. The functions $\Phi_\alpha({\bf p}),\Psi_\alpha({\bf
p})$ for the free field plane wave solution are given by
\begin{align} \label{freePhi}
\Phi^{\alpha}({\bf p})=& u_\alpha(p)e^{-ip\cdot x/\hbar} \\
\Psi^\alpha ({\bf p}) =& v_\alpha(p)e^{ip\cdot x/\hbar} \,. \label{freePsi}
\end{align}
From the Dirac equation, the spinors satisfy the equations
\begin{align}
(\slashed{p}-m)u(p)&=0 \\ (\slashed{p}+m)v(p)&=0 \,.
\end{align}
The spinors $u_\alpha(p),v_\alpha(p)$ are given by
\begin{align} \label{freeu}
u_\alpha(p) =& \sqrt{\df{p_0+m}{2m}} \begin{pmatrix} s_\alpha \\
\df{\itbd{\sigma}\cdot{\bf p}}{p_0+m} s_\alpha \end{pmatrix} \\
v_\alpha(p) =& \sqrt{\df{p_0+m}{2m}} \begin{pmatrix} \df{\itbd{\sigma}\cdot{\bf p}}{p_0+m} s_\alpha \\
 s_\alpha \end{pmatrix} \,. \label{freev}
\end{align}
For spin up/down along the $x^i$ axis, the vectors $s_\alpha$ are
the corresponding two eigenvectors of the spin matrix $\sigma^i$. To
simplify the notation, we shall make use of an Einstein convention
on the spin indices, for which we shall reserve the early-alphabet
Greek letters $\alpha,\beta,\gamma,\delta$. Thus
$b_\alpha\Phi^\alpha=\sum_{\alpha=1,2} b_\alpha \Phi_\alpha$. The
mid-alphabet Greek letters $\mu,\nu$ etc. will be reserved for the
spacetime indices which satisfy the usual Einstein convention with
space-time metric convention $(+---)$. Latin letters denote space
indices only.

We use the free field to model the wave packet in the
non-interacting regions. Similarly to before, we represent the
incoming wave packet as a distribution heuristically regarded as the
one-particle wave function in the momentum representation:
\begin{equation}\label{spinistate}
| i \rangle = \int\f{d^3 {\bf p}}{(2\pi\hbar)^3}\sqrt{\f{m}{p_0}}
f({\bf p}) b^\dagger_\alpha({\bf p}) |0\rangle \,,
\end{equation}
where $f$ is sharply peaked about the initial momentum in the region
${\cal M}_-$ and normalised via $\langle i|i\rangle=1$
\begin{equation}\label{normalisef}
\int \f{d^3 {\bf p}}{(2\pi\hbar)^3} f^*({\bf p})f({\bf p}) = 1 \,.
\end{equation}

Having considered the field in the asymptotic regions, we must now
consider the field in the interaction region ${\cal M}_I$ in the
presence of the potential $V$. We proceed in the same way as
previously by introducing the potential via the transformation of
the derivative
\begin{equation}\label{Vsubpsi}
\partial_\mu \psi \mapsto D_\mu \psi = \(\partial_\mu + \f{i}{\hbar}
V_\mu\) \psi \,.
\end{equation}
We again stress that we treat the potential non-perturbatively. The
Lagrangian is now
\begin{equation}
{\cal L} = i \hbar \bar{\psi} \gamma^\mu {D}_\mu \psi - m
\bar{\psi}\psi \,.
\end{equation}
The relative minus sign on V in the conjugate of (\ref{Vsubpsi})
ultimately represents the opposite charge of the antiparticle
solutions. Let $\slashed{D}=\gamma^\mu D_\mu$. The appropriate
equations of motion are now
\begin{align}
\(i\hbar\slashed{D} - m \) \psi &= 0 \\\intertext{and the conjugate gives}
\bar{\psi} \(i\hbar\stackrel{\leftarrow}{\slashed{D}}^\dagger + m \) &= 0
\,,
\end{align}
where the arrow indicates the differentiation of term to the left.
We note that
$\overline{(\slashed{D}\psi)}=\bar{\psi}\slashed{D}^\dagger$,
leading to the second equation. The mode functions in the
interacting region are now solutions of these two equations i.e.
\begin{align}
\(i\hbar\slashed{D} - m \)\Phi_\alpha (x) &= 0 \\
\bar{\Psi}_\alpha(x) \(i\hbar\stackrel{\leftarrow}{\slashed{D}}^\dagger + m \) &= 0 \,.
\end{align}

We add the electromagnetic field via minimal substitution as before,
which in this case gives
\begin{align}
\slashed{D} \to \slashed{D}+ie\slashed{A}/\hbar \,,
\end{align}
where once again the electromagnetic field has the expansion
\begin{equation}
A_\mu(x) = \int\frac{d^3{\bf k}}{2k(2\pi)^3} \left[ a_\mu({\bf k})
e^{-ik\cdot x} + a^\dagger_\mu({\bf k})e^{ik\cdot x} \right] \,,
\end{equation}
with the commutation relations
\begin{equation}
\left[ a_\mu({\bf k}),a^\dagger_\nu({\bf k}')\right] =
-g_{\mu\nu}(2\pi)^32\hbar k\delta^3({\bf k}-{\bf k}') \,.
\end{equation}
Most of what was said previously about the details of the EM field
applies equally here. The QED Lagrangian, in the presence of the
classical potential $V$, is given by
\begin{align}\label{QEDlag}
{\cal L} = i \hbar \bar{\psi} \gamma^\mu
\({D}_\mu+\f{ie}{\hbar}A_\mu \) \psi - m \bar{\psi}\psi
- \frac{1}{4}F_{\mu\nu}F^{\mu\nu} - \frac{1}{2}(\partial_\mu
A^\mu)^2 \,.
\end{align}
The interaction Lagrangian can be given from (\ref{QEDlag}) by
\begin{equation}
{\cal L}_I = -e\bar{\psi}\slashed{A}\psi \,.
\end{equation}
Unlike the scalar case, the switch to the Hamiltonian formulation is
straightforward and we find that the interaction Hamiltonian is
simply to negative of ${\cal L}_I$ viz
\begin{align}\label{spinorHamInt}
{\cal H}_I = e : \bar{\psi}\slashed{A}\psi : \,,
\end{align}
where we have added the normal ordering. The interaction Hamiltonian
is then substituted as appropriate in the evolution of the state.
The evolution in (\ref{evol}) is a general statement of perturbation
theory and thus relevant here:
\begin{equation}\tag{\ref{evol}}
\ket{i} \to \ket{i} -\frac{i}{\hbar}\int d^4 x
{\cal H}_I(x)\ket{i} + \(\f{-i}{\hbar}\)^2 \int d^4x\,d^4x' T \[
{\cal H}_I(x){\cal H}_I(x')\]\ket{i} \,.
\end{equation}

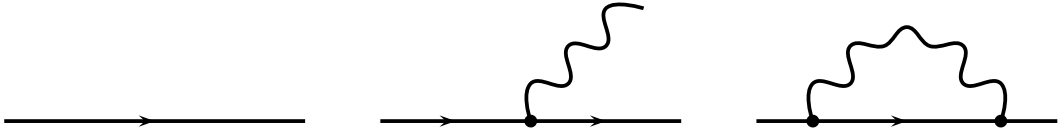
\begin{figure}
\begin{center}
\begin{pspicture}(-4,-1)(4,2)
\psline[linewidth=0.5mm]{->}(-7,0)(-5,0)
\psline[linewidth=0.5mm](-5.1,0)(-3,0)

\psline[linewidth=0.5mm]{->}(-2,0)(-1,0)
\psline[linewidth=0.5mm]{->}(-1.1,0)(1,0)
\psline[linewidth=0.5mm](0.9,0)(2,0)
\pscurve[linewidth=0.5mm]{*-}(0,0)(0,0.5)(0.5,0.5)(0.5,1.0)(1.0,1.0)
(1.0,1.5)(1.5,1.5)

\psline[linewidth=0.5mm]{->}(3,0)(5,0)
\psline[linewidth=0.5mm](4.9,0)(7,0)
\pscurve[linewidth=0.5mm]{*-*}(3.75,0)(3.75,0.5)(4.25,0.5)(4.25,1.0)(4.7,1.0)
(5.0,1.25)(5.3,1.0)(5.75,1.0)(5.75,0.5)(6.25,0.5)(6.25,0)
\end{pspicture}
\caption{~~The first three types of Feynman diagrams for QED
representing the perturbation expansion up to order $e^2$.}
\label{Feyexp2}
\end{center}
\end{figure}

We again look at the perturbation expansion contributions to order
$e^2$. As with the scalar case, we have null, emission and forward
scattering processes. For the spinor fields, which is of course
standard QED, the forward scattering does not contain the second
circular loop process seen as the last process in Fig.
\ref{scalardiags}, as there is no seagull vertex. We instead simply
have the three diagrams described in the introduction and given in
Fig. \ref{Feyexp}, which we repeat in this section (Fig.
\ref{Feyexp2}) to aid the reader.
The remaining one loop diagram is still divergent and the
contribution is subtracted via renormalisation in much the same way
as briefly described in the scalar field section. To order $e^2$ the
mass counter term adds to the Lagrangian the additional term
\begin{equation}
\delta {\cal L} = \delta m \bar{\psi} \psi \,.
\end{equation}
The counterterm $\delta m$ is again local, i.e. has no momentum
dependence. The spinor propagator, written in terms of the bare mass
$m$, is modified to remove the divergences from the one-loop
contribution via
\begin{align}
\f{i}{\slashed{p}-m} & \to  \f{i}{\slashed{p}-m} +
\f{i}{\slashed{p}-m}\[ -i\Sigma(p) \] \f{i}{\slashed{p}-m} \nol & \quad +
\f{i}{\slashed{p}-m}\[ -i\Sigma(p) \] \f{i}{\slashed{p}-m}\[ -i\Sigma(p) \] \f{i}{\slashed{p}-m} +
\ldots \nol
&= \f{i}{\slashed{p}-m-\Sigma(p)} \,.
\end{align}
The self-energy $\Sigma(p)$ is given here analogously to the
situation described for the scalar field and can similarly be
represented by an additional Feynman diagram contribution (see Fig.
\ref{renormcross}).\footnote{We are approaching these fields in a
somewhat reverse order by giving the `\emph{standard}' QED results
second, due to the order they are used in this work.} Using the
Feynman rules for standard (spinor) QED applied to the one loop
diagram, we obtain
\begin{equation}
\Sigma (p) = -\f{ie^2}{\hbar} \int \f{d^4 K}{(2\pi)^4}
\f{g_{\mu\nu}}{K^2+i\epsilon} \gamma^\mu
\f{1}{\slashed{p}-\slashed{K}-m+i\epsilon} \gamma^\nu \,.
\end{equation}
The self-energy is again divergent and may proceed as in the scalar
case to regularise it via dimensional regularisation. Similarly to
the previous case, the counterterm $\delta m$ is then chosen, as a
function of the regularising parameter, to cancel the divergence so
that as the regulator is removed we have
\begin{equation}
\delta m - \Sigma(p)|_{\slashed{p}=m_P} \to 0\,,
\end{equation}
where $m_P$ is the physical mass.

This concludes our introduction to the model that we shall use for
the quantum field theory description of the Dirac spinor field for
QED. An exhaustive or pedagogic introduction to quantum field theory
would be out of place here and the reader unfamiliar with the
canonical descriptions outlined above is referred the one of the
many textbooks, or indeed courses, designed specifically for that
purpose eg. \cite{Ryder},
\cite{Mandl} or
\cite{Itzykson}. On the other hand, the above two sections should now provide a
reader familiar with quantum field theory with an appropriate
reference for the definitions and conventions that are used in the
rest of this work.

\chapter{Semiclassical Approximation}\label{Semiclassical}
\begin{quote}
In this chapter we introduce and calculate the semiclassical
approximations to be used to model the scalar and spinor quantum
fields during their interaction with the classical potential.
\end{quote}

\section{Semiclassical and WKB approximations}\label{wkbsection}

In this section we have two purposes to keep in mind. Firstly, we
need to solve the field equations to find expressions for the mode
functions during the period of acceleration. Secondly, we aim to
take the classical limit i.e. the limit in which $\hbar \to 0$. It
is thus appropriate to use a semiclassical expansion, i.e. an
expansion in terms of $\hbar$, in order to obtain our mode function
solutions. Due to the nature of the model, we shall not however be
solving the equations exactly in terms of known quantities. This is
a simple consequence of the fact that we do not wish to constrain
the possible behaviour any more than is absolutely necessary. So far
little has been said of any possible constraints on the
acceleration. Most of the constraints that will become apparent are
in fact due to the semiclassical expansion detailed in this section.
In order for the expansion to be valid, and indeed found by the
following method, some restraints are necessary.

In order to set the scene before presenting the relevant
calculations, let us briefly recall some of the basic theory of
semiclassical expansions in quantum theory. The expansion of the
wave function in orders of $\hbar$ in quantum mechanics goes by the
name of the WKB approximation, named after Wentzel-Kramers-Brillouin
from their 1926 development of the method.\footnote{This is one of
those cases where multiple names are sometimes used in attempts to
credit the correct people. The WKB approximation is also known as
the WKBJ. The J is for Harold Jeffreys, who in 1923 developed the
general method of approximating linear, second-order differential
equations, including the later (1925) Schr\"{o}dinger equation.
Early quantum mechanics texts also use WBK, BWK, WKBJ and BWKJ.}
Strictly speaking, the WKB approximation is the expansion up to
order $\hbar$ of the solution to the Schr\"{o}dinger equation
\begin{equation}\label{Schrodinger}
i\hbar\f{\partial}{\partial t} \psi({\bf x},t) =  \[
-\f{\hbar^2}{2m} \nabla ^2 +V({\bf x}) \] \psi({\bf x},t) \,.
\end{equation}
The complex solution $\psi({\bf x},t)$ to (\ref{Schrodinger}),
rewritten in terms of some function $S$ as $e^{iS({\bf x},t)/\hbar}$
leads, with the assumption $\psi\neq 0$, to
\begin{equation}\label{WKB1}
-\f{\partial S}{\partial t} = \f{1}{2m} \( \nabla S \)^2
-\f{i\hbar}{2m} \nabla^2 S + V \,.
\end{equation}
The formal `classical limit', $\hbar\to 0$, gives the
Hamilton-Jacobi equation
\begin{equation}
-\f{\partial S}{\partial t} = \f{1}{2m} \( \nabla S \)^2 + V \,.
\end{equation}
In general, the semiclassical expansion is the expansion of $S$ in
terms of $\hbar$ viz
\begin{equation}
S=S_0+\hbar S_1+\hbar^2 S_2+\ldots \,.
\end{equation}
With the substitution of this expansion,the appropriate equation of
motion can then be solved order by order. For the Schr\"{o}dinger
equation we have, for $S_0$ and $S_1$,
\begin{align}
-\f{\partial S_0}{\partial t} &= \f{1}{2m} \( \nabla S_0 \)^2 + V \\
-\f{\partial S_1}{\partial t} &= \f{1}{2m} \[ -i\nabla^2 S_0 + 2
\itbd{\nabla}S_0 \cdot \itbd{\nabla}S_1 \] \,.
\end{align}
Note that the equation for $S_0$ is again the Hamilton-Jacobi
equation. As an example, consider the time-independent \emph{one
dimensional} Schr\"{o}dinger equation. For a wavefunction
proportional to $e^{-iEt/\hbar}$, we note that $S(x,t)=S(x)-Et$, and
so we may separate out the $e^{-iEt/\hbar}$ factor and consider the
semiclassical expansion as terms dependent on ${\bf x}$ only:
$S({\bf x}) = S_0({\bf x})+\hbar S_1({\bf x}) +\ldots$. The solution
up to order $\hbar$ in this expansion is
\begin{equation}
\psi(x,t) = \f{C}{\sqrt{p(x)}} \exp \[ \pm \f{i}{\hbar} \int p(x')
dx' \] e^{-iEt/\hbar} \,,
\end{equation}
where $p(x)=\sqrt{2m(E-V(x))}$ is the classical momentum of the
particle and $C$ is a constant.\footnote{The steps of this
calculation are nearly identical to those for scalar field which we
shall present fully later. As we present the Schr\"{o}dinger results
as a motivational example we have omitted the details here.} This
example shows explicitly the general restriction on the validity of
this approximation, namely that it breaks down when the classical
particle reaches a turning point, i.e. $p(x)=0$ above.

This conclusion can also be reached by analysis of the validity of
the approximation itself. In order for us to be justified in taking
the $\hbar$ expansion then the truncated series that we use must be
a good approximation.
 From the notation above, we
would require that $\hbar S_1$ be much smaller than $S_0$. From the
equation (\ref{WKB1}), we require that the $\hbar$ term be much
smaller than the other $\hbar^0$ terms. These general requirements
give in our current context the condition
\begin{equation}
|\(\nabla S\)^2| \gg \hbar |\nabla^2 S | \,.
\end{equation}
If we turn to our specific example of the time-independent
Schr\"{o}dinger equation, we obtain
\begin{equation}
p(x)^2 \gg \hbar \left| \f{d p(x)}{dx} \right| \,.
\end{equation}
Substituting the definition of $p(x)$ from above and rearranging, we
find
\begin{equation}
\left| \f{\hbar dV(x)/dx}{2(E-V(x))p(x)} \right| \ll 1 \,.
\end{equation}
As stated, we thus arrive at the same conclusion regarding the
validity conditions i.e. that the approximation breaks down at the
classical turning point $E=V$.~\footnote{Recall that $p(x)=0$ here
too.} One should however note that the approximation may still be
valid \emph{beyond} the classical turning point. This encapsulates
the fact that in quantum mechanics the probability amplitude need
not be zero in the classically forbidden regions, hence providing
for quantum phenomena such as quantum tunnelling. The above
condition, it should be recalled, is one for the approximation,
rather than the quantum wavefunction itself. That the WKB
approximation does not break down in the classically forbidden, yet
quantum-allowed, regions is an important point which demonstrates
that the $\hbar\to 0$ limit of the semiclassical expansion may still
contain quantum phenomena and thus can not technically be assumed to
be the classical limit, in the sense of producing the purely
classical theory. This limit is nevertheless frequently referred to
as the classical limit, and as we shall use this limit to compare
the quantum theory effects with those of the classical theory, it
shall be referred to as such here with the above caveat to be kept
in mind.

Having now reminded ourselves of the canonical semiclassical theory
for quantum mechanics, we can now turn our attention to the
approximations needed for the quantum model we have set out. We
start by looking at the semiclassical approximation for the scalar
field and then consider the same for the Dirac spinor field.

\section{Semiclassical Scalar
solutions}\label{semiclassscalarsection}

In this section we consider the scalar field solutions to the
Klein-Gordon field equations in the presence of the potential. In
the region of the acceleration of the particle, the mode functions
are these solutions to the field equations. We desire a
semiclassical expansion of the mode functions. Firstly, let us
consider the case of the time dependent (and space independent)
potential: $V=(0,{\bf V}(t))$ with the gauge choice $V_0=0$. In this
case we will have conservation of momentum. Firstly, we separate he
mode function as follows:
\beq
\Phi_{\bf p} \propto e^{\f{i}{\hbar} {\bf p} \cdot {\bf x}} \phi_{{\bf p}} (t) \,.
\eeq
The wave equation that is satisfied by $\Phi_{\bf p}$ is
\begin{multline}
\[\hbar^2 \d_t^2 + \(-i\hbar\d_x - V_x(t)\)^2
+\(-i\hbar\d_y - V_y(t)\)^2 \right. \\ \left. +\(-i\hbar\d_z -
V_z(t)\)^2+m^2\]\Phi_p=0 \,,
\end{multline}
thus the equation that must be satisfied by $\phi_{\bf p}(t)$ is
\beq\label{phitKG}
\[\hbar^2 \d_t^2 + \(p_x - V_x(t)\)^2
+\(p_y - V_y(t)\)^2+\(p_z - V_z(t)\)^2+m^2\]\phi_{\bf p}(t)=0 \,.
\eeq
We now wish to find the semiclassical expansion of this solution. In
the scalar case, we shall need to take only the first two terms of
an expansion in $\hbar$ in the exponential, which translates to
order $\hbar^0$ for the mode function, viz
\beq
\phi_{\bf p}(t) = \exp\[-\f{i}{\hbar}S^{(0)} - S^{(1)}\] \,.
\eeq
We substitute this expression into the wave equation and then equate
order by order. The first term, the $t$ differential gives
\begin{align}
&\hbar^2\d^2_t \exp\(-\f{i}{\hbar} S^{(0)}-S^{(1)}\) \nol
 &=
\hbar^2
\d_t \[\exp\(-\f{i}{\hbar}
S^{(0)}-S^{(1)}\)\[-\f{i}{\hbar}\d_tS^{(0)} - \d_tS^{(1)}\]\]\nol
 &= \left\{-\(\d_tS^{(0)}\)^2+2i\hbar\d_tS^{(0)}\d_tS^{(1)} -i\hbar\d^2_tS^{(0)} - \hbar^2\d^2_tS^{(1)}
 +\hbar^2\(\d_tS^{(1)}\)^2 \right\} \phi_{\bf p}(t) \,.
\end{align}
Thus the order $\hbar^0$ terms give an equation for $S^{(0)}$
\beq
\(\d_tS^{(0)}\)^2 =
\(p_x-V_x(t)\)^2+\(p_y-V_y(t)\)^2+\(p_z-V_z(t)\)^2+m^2 \,,
\eeq
which we can solve to give
\beq
S^{(0)} = \int^t_0 E_p(t')dt' \,,
\eeq
where
\beq
E_p(t) =
\sqrt{\(p_x-V_x(t)\)^2+\(p_y-V_y(t)\)^2+\(p_z-V_z(t)\)^2+m^2} \,.
\eeq
This is the classical energy of the particle.\footnote{We have
labelled the subscript as $p$ to distinguish which momentum this
energy is related too. It should be noted that $E_p(t)$ is dependent
on the vector ${\bf p}$.} The order $\hbar^1$ terms give
\beq
2\d_tS^{(0)}\d_tS^{(1)} = \d_t^2S^{(0)} \,,
\eeq
which has the solution
\beqa
S^{(1)} &=& \int^t_0 \f{\d_{t'}^2 S^{(0)}}{2\d_{t'} S^{(0)}} dt'
\nol
&=& \f{1}{2}\int d\(\d_tS^{(0)}\)\f{1}{\d_tS^{(0)}} \nol
 &=& \f{1}{2}\ln E_p(t) + \text{const.}
\eeqa
Thus the $t$-dependent part of the wave function is given by
\beqa
\phi_{\bf p}(t) &=& \f{C}{\sqrt{E_p}}e^{-i\int E_p dt/\hbar} \\
 &=&
 \sqrt{\f{p_0}{E_p(t)}}\exp\[-\f{i}{\hbar}\int^t_0E_p(\zeta)d\zeta
 dt'\] \,, \label{tdepwkb}
\eeqa
where $p_0=\sqrt{{\bf p}^2+m^2}$. The case of the potential
dependent on one of the spatial coordinates can be given by
considering one example. Here we choose a $z$ dependent potential
$V(z)=(V_t(z), V_x(z), V_y(z), 0)$ with the gauge choice $V_z=0$.
Again, we separate out the constituent parts of the mode function,
this time producing
\beq
\Phi_{\bf p} =  \phi_{{\bf
p}}(z) e^{-\f{i}{\hbar}\(p_0t-p_xx-p_yy\) } \,.
\eeq
The wave equation that is satisfied by $\Phi_{\bf p}$ is
\begin{multline}
\[-\(-i\hbar \d_t-V_t(z)\)^2 + \(-i\hbar\d_x - V_x(z)\)^2
\right. \\ \left. +\(-i\hbar\d_y -
V_y(z)\)^2+\(-i\hbar\d_z\)^2+m^2\]\Phi_p=0 \,,
\end{multline}
and consequently that for $\phi_{\bf p}(z)$ is
\beq
\[-(p_0-V_t(z))^2 + \(p_x - V_x(z)\)^2
+\(p_y - V_y(z)\)^2+\(-i\hbar\d_z\)^2+m^2\]\phi_{\bf p}(z)=0 \,.
\eeq
As with the previous case, we expand to order $\hbar^0$ overall, viz
\beq
\phi_{{\bf p}}(z) = \exp\[\f{i}{\hbar}S^{(0)} + S^{(1)}\] \,.
\eeq
Proceeding to analyse the solution order by order we note that the
$z$ differential gives
\begin{align}
&-\hbar^2\d^2_z \exp\(\f{i}{\hbar} S^{(0)}+S^{(1)}\) \nol
 &=-
\hbar^2
\d_z \[\exp\(\f{i}{\hbar}
S^{(0)}+S^{(1)}\)\[\f{i}{\hbar}\d_zS^{(0)} + \d_zS^{(1)}\]\]\nol
 &= -\left\{-\(\d_zS^{(0)}\)^2+2i\hbar\d_zS^{(0)}\d_zS^{(1)} +i\hbar\d^2_zS^{(0)} + \hbar^2\d^2_zS^{(1)}
 +\hbar^2\(\d_zS^{(1)}\)^2 \right\} \nol & \qquad \times \phi_{\bf p}(z)
 \,.
\end{align}
Thus the order $\hbar^0$ terms again produce an equation for
$S^{(0)}$
\beq
\(\d_zS^{(0)}\)^2 =
\(p_0-V_t(z)\)^2-\(p_x-V_x(z)\)^2-\(p_y-V_y(z)\)^2-m^2 \,,
\eeq
which solves to give
\beq
S^{(0)} = \int^z_0 \kappa_p(z')dz' \,,
\eeq
where
\beq
\kappa_p(z) =
\sqrt{\(p_0-V_t(z)\)^2-\(p_x-V_x(z)\)^2-\(p_y-V_y(z)\)^2-m^2} \,.
\eeq

The order $\hbar^1$ terms give
\beq
2\d_zS^{(0)}\d_zS^{(1)} =- \d_z^2S^{(0)} \,,
\eeq
which has the solution
\beqa
S^{(1)} &=& -\int^z_0 \f{\d_{z'}^2 S^{(0)}}{2\d_{z'} S^{(0)}} dz'
\nol
&=& -\f{1}{2}\int d\(\d_zS^{(0)}\)\f{1}{\d_zS^{(0)}} \nol
 &=& -\f{1}{2}\ln \kappa_p(z) + \text{const.}
\eeqa
Thus the $z$-dependent part of the wave function is given by
\begin{align}
\phi_{\bf p}(z) &= \f{C}{\sqrt{\kappa_p}}e^{i\int\kappa_p dz/\hbar} \nol
 &=
 \sqrt{\f{p_z}{\kappa_p(z)}}\exp\[\f{i}{\hbar}\int^z_0\kappa_p(\zeta)d\zeta
 \]  \,, \label{scalarsemispace}
\end{align}
with $p_z=\sqrt{p_0^2-p_x^2-p_y^2-m^2}$. The extension to potentials
dependent on $x$ or $y$ is straightforward and by simple
substitution, hence not repeated here.

\subsection{Antiparticle mode functions}
We recall that the antiparticle mode functions $\bar{\Phi}_{\bf
p}(x)$ are solutions to the wave equation
\begin{align}
\( \hbar^2 D^{\dagger\mu}D^\dagger_\mu +m^2\)\varphi^\dagger = 0
\,.
\end{align}
Transformation between the particle/antiparticle solutions is thus
accomplished by the transformation $V\to -V$. For the time-dependent
potential, we thus have
\begin{align}
\bar{\Phi}_{\bf p}(x) &= \bar{\phi}_{\bf p} (t) e^{i{\bf p}\cdot{\bf
x}/\hbar} \,, \\
\intertext{with}
\bar{\phi}_{\bf p}(t) &= \sqrt{\f{p_0}{E_{p_+}(t)}}\exp\[-\f{i}{\hbar}\int^t_0 E_{p_+}(\zeta)d\zeta
\] \,, \label{tdepwkb2}
\end{align}
where
\begin{equation}
E_{p_+}(t) = \sqrt{|{\bf p}+{\bf V}(t)|^2+m^2} \,.
\end{equation}

Similarly for the potential dependent on the spatial coordinate $z$
(for example) we have
\begin{align}
\bar{\Phi}_{\bf p}(x) &=  \bar{\phi}_{{\bf p}}(z) e^{-\f{i}{\hbar}\(p_0t-p_xx-p_yy\)
} \,,
\\ \intertext{with}
\bar{\phi}_{{\bf p}}(z) &=
\sqrt{\f{p_z}{\bar{\kappa}(z)}}\exp\[\f{i}{\hbar}\int^z_0\kappa_{p_+}(\zeta)d\zeta
\] \,,
\end{align}
where
\begin{equation}
\kappa_{p_+}(z) =
\sqrt{\(p_0+V_t(z)\)^2-\(p_x+V_x(z)\)^2-\(p_y+V_y(z)\)^2-m^2} \,.
\end{equation}

\section{Semiclassical Spinor
solutions}\label{semiclassspinorsection}
\subsection{Positive Energy Solution}

In a time-dependent potential the mode function can be split into
its space and time dependent parts:
\begin{equation}
\Phi(x) = \psi(t) e^{i{\bf p}\cdot{\bf x}/\hbar} \,,
\end{equation}
where we write the time-dependent component as
\beq
\psi(t) = \begin{pmatrix} \varphi \\ \chi \end{pmatrix} \exp \(
-\f{i}{\hbar}S\) \,,
\eeq
with the semiclassical expansion contained within the spinor term:
\beq \label{spinsemiexp}
\begin{pmatrix} \varphi \\ \chi \end{pmatrix} = \begin{pmatrix} \varphi^{(0)} \\ \chi^{(0)}\end{pmatrix}
+\hbar \begin{pmatrix} \varphi^{(1)} \\ \chi^{(1)} \end{pmatrix} +\hbar^2 \begin{pmatrix} \varphi^{(2)} \\
\chi^{(2)}
\end{pmatrix} + \ldots \,.
\eeq
This mode function must obey the Dirac equation with a
time-dependent potential. Defining ${\bf \tilde{p}} = {\bf p}-{\bf
V}(t)$, we have
\begin{gather}
i\hbar \partial_t \Phi(x) - \[ \itbd{\alpha}\cdot\(-i\hbar{\bf
\nabla}-{\bf V}(t)\) +\beta m \] \Phi(x) = 0 \,, \nol i\hbar \partial_t \psi(t) -
\[
\itbd{\alpha}\cdot{\bf \tilde{p}} +\beta m \] \psi(t) = 0 \,.
\end{gather}
Hence
\begin{multline}
i \hbar \begin{pmatrix} \dot{\varphi} \\ \dot{\chi} \end{pmatrix}
\exp\(-\f{i}{\hbar}S\) +i\hbar \begin{pmatrix} \varphi \\ \chi \end{pmatrix} \(-\f{i}{\hbar}\dot{S}\)\exp \(-\f{i}{\hbar}S\)
\\ - \[ \itbd{\alpha}\cdot{\bf \tilde{p}} +\beta m \]\begin{pmatrix} \varphi \\
\chi
\end{pmatrix} \exp \( -\f{i}{\hbar}S\) = 0 \,.
\end{multline}
Substituting the $\hbar$ expansion (\ref{spinsemiexp}) into this
equation, we obtain at lowest order
\begin{align}
\dot{S} \begin{pmatrix} \varphi^{(0)} \\ \chi^{(0)}\end{pmatrix} =
\[ \itbd{\alpha}\cdot{\bf \tilde{p}} +\beta m \] \begin{pmatrix} \varphi^{(0)} \\
\chi^{(0)}\end{pmatrix} \,.
\end{align}
Defining $E=\sqrt{{\bf \tilde{p}}^2+m^2}$, the eigenvalues of the
matrix $\itbd{\alpha}\cdot{\bf \tilde{p}} +\beta m$ are $\pm E$. In
this section we are considering the two `positive energy' mode
functions (i.e. those solutions corresponding to the $+E$
eigenvalue). We obtain
\beq
S = \int_0^t E(\xi)d\xi \,,
\eeq
and
\beq \label{spincomps}
\chi^{(0)} = \f{\itbd{\sigma}\cdot{\bf \tilde{p}}}{E+m}
\varphi^{(0)} \,.
\eeq
At higher orders, we have
\beq
i\begin{pmatrix} \dot{\varphi}^{(n)} \\
\dot{\chi}^{(n)}\end{pmatrix} +
\begin{pmatrix}
E-m & -\itbd{\sigma}\cdot{\bf \tilde{p}} \\
-\itbd{\sigma}\cdot{\bf \tilde{p}} & E+m \end{pmatrix}
\begin{pmatrix}
\varphi^{(n+1)} \\ \chi^{(n+1)}\end{pmatrix}= {\bf 0},  \ \
\forall n=0,1,2,3,\ldots \,.
\eeq
Multiplying both sides by
\beq
\begin{pmatrix}
E+m & \itbd{\sigma}\cdot{\bf \tilde{p}} \\
\itbd{\sigma}\cdot{\bf \tilde{p}} & E-m \end{pmatrix} \,,
\eeq
we obtain
\begin{align}
\begin{pmatrix} \label{ntspinor}
E+m & \itbd{\sigma}\cdot{\bf \tilde{p}} \\
\itbd{\sigma}\cdot{\bf \tilde{p}} & E-m \end{pmatrix} \begin{pmatrix} \dot{\varphi}^{(n)} \\
\dot{\chi}^{(n)} \end{pmatrix} = {\bf 0}, \ \  \forall n=0,1,2,3,\ldots
\,.
\end{align}
Thus for the lowest order case we can combine (\ref{ntspinor}) for
$n=0$ with (\ref{spincomps}) to find a differential equation for
$\varphi^{(0)}$, viz
\begin{align}
\dot{\varphi}^{(0)} =& -\f{\itbd{\sigma}\cdot{\bf \tilde{p}}}{E+m} \dot{\chi}^{(0)}
\nol
=& -\f{\itbd{\sigma}\cdot{\bf \tilde{p}}}{E+m} \f{d}{dt} \(
\f{\itbd{\sigma}\cdot{\bf \tilde{p}}}{E+m} \varphi^{(0)}\) \nol \intertext{which leads to}
\dot{\varphi}^{(0)} =&
-\f{\itbd{\sigma}\cdot{\bf \tilde{p}}}{2E} \f{d}{dt} \(
\f{\itbd{\sigma}\cdot{\bf \tilde{p}}}{E+m}\) \varphi^{(0)} \,,
\label{zerothdiff}
\end{align}
where we use $\itbd{\sigma}\cdot{\bf
\tilde{p}}\,\itbd{\sigma}\cdot{\bf \tilde{p}}={\bf \tilde{p}}^2 = E^2-m^2$. This
further becomes
\beq
\dot{\varphi}^{(0)}= \[ -\f{m\dot{E}}{2E(E+m)} - \f{i\itbd{\sigma}\cdot{\bf \tilde{p}}\times{\bf
\dot{\tilde{p}}}}{2E(E+m)} \] \varphi^{(0)} \,.
\eeq
With regards to the first term we note that
\beq
\f{d}{dt} \( \sqrt{\f{E+m}{2E}} \) = -\f{m\dot{E}}{2E(E+m)}
\sqrt{\f{E+m}{2E}} \,.
\eeq
We then treat the second term as a time-dependent perturbation. The
differential equation, due to the non-commutative nature of
matrices, does not simply give the exponential solution, but rather
the Taylor series expansion which can be rearranged to produce a
time-ordered product. The result is known as an ordered (or
path-ordered) exponential and we can thus write the spinor component
as
\beq \label{phizerosoln}
\varphi^{(0)} =  C \sqrt{\f{E+m}{2E}} \, T \( \exp \[ -i \int^t_0 d\tau
\f{\itbd{\sigma}\cdot\({\bf \tilde{p}}(\tau)\times{\bf
\dot{\tilde{p}}}(\tau)\)}{2E(\tau)(E(\tau)+m)} \] \)  s \,,
\eeq
where $s$ is a spin eigenstate at $t=0$, chosen normalised, $C$ a
constant and $T$ is the time-ordering operator. We note that the
exponential notation is a short hand to represent the series
expansion.

Defining
\beq \label{Udefn}
U(t) := T \( \exp \[ -i \int^t_0 d\tau
\f{\itbd{\sigma}\cdot\({\bf \tilde{p}}(\tau)\times{\bf
\dot{\tilde{p}}}(\tau)\)}{2E(\tau)(E(\tau)+m)} \] \) \,,
\eeq
we note that $U(t)$ is a unitary operator acting on $s$ that can be
considered as the time-evolution of the spin polarization. We define
\begin{equation} \label{Lambda}
\Lambda_p(t) = \f{\itbd{\sigma}\cdot{\bf
p}(t) \times \dot{\bf p}(t)}{\(E_p(t)+m\)} \,.
\end{equation}
Note that $\Lambda_p(t)$ is Hermitian and traceless. There are two
positive energy solutions. Thus $s$ is one of the two spin (up or
down) eigenstates. Define $s(t) = U(t) s$. The zeroth order term in
the spinor expansion is thus (up to a multiplicative constant)
\beq
\left( \begin{array}{c} \varphi^{(0)} \\ \chi^{(0)}
\end{array}\right)
= \sqrt{\frac{E+m}{2E}}\left( \begin{array}{c}
s_\alpha(t) \\
\df{\itbd{\sigma}\cdot{\bf \tilde{p}}}{E+m}
s_\alpha(t)\end{array}\right) \,.\label{zerothorderspinoru}
\eeq

\subsubsection{First order correction}

We now look at the first order term in the spinor expansion i.e. the
$\hbar$ correction term in the semiclassical expansion. We
consequently return to the full-order positive spinor equation
\beq
\left( \begin{array}{cc} E-m & - \itbd{\sigma}\cdot{\bf p} \\
- \itbd{\sigma}\cdot {\bf p} & E + m \end{array}\right)
\left( \begin{array}{c} \varphi \\ \chi \end{array}\right)
+ i\hbar\left( \begin{array}{c} \dot{\varphi} \\
\dot{\chi}\end{array}\right) = 0\,.  \label{exact2}
\eeq
For ease of notation let
\beq
\Sigma := \f{\itbd{\sigma}\cdot{\bf p}}{E+m} \,,
\eeq
where ${\bf p}$ and $E$ are time-dependent. Thus the zeroth order
spinor term is
\beq
\left( \begin{array}{c} \varphi^{(0)} \\ \chi^{(0)} \end{array}\right)
= \sqrt{\frac{E+m}{2E}}\left( \begin{array}{c}
s_\alpha(t) \\
\Sigma s_\alpha(t)\end{array}\right)\,.
\label{zeroth}
\eeq
Recall that the spinors $s_\alpha(t)$ satisfy $s_\alpha^\dagger (t)
s_\beta (t) = \delta_{\alpha\beta}$. Define the unitary matrix
$S(t)$ as follows:
\beq
S(t) \equiv \sqrt{\frac{E+m}{2E}}\left( \begin{array}{cc} U(t) &
- \Sigma U(t) \\
\Sigma U(t) & U(t) \end{array}
\right)\,.
\eeq
We note that
\beq
S^{-1}(t)\left( \begin{array}{cc} m & \itbd{\sigma}\cdot{\bf p} \\
\itbd{\sigma}\cdot{\bf p} & - m \end{array}\right)S(t)
= \left( \begin{array}{cc} E & {\bf 0} \\ {\bf 0} & - E\end{array}
\right)\,.
\eeq

Using this matrix, we change the representation of the spinors and
let
\beq
\left( \begin{array}{c} \varphi \\ \chi \end{array}\right)
= S(t) \left( \begin{array}{c} \tilde{\varphi} \\ \tilde{\chi}
\end{array}\right)\,.
\eeq
The positive energy spinor equation (\ref{exact2}) can be written in
this representation as
\beq
\left( \begin{array}{c} {\bf 0} \\ 2E \tilde{\chi} \end{array} \right)
+ i\hbar S^{-1}(t)\frac{d\ }{dt}S(t)\left( \begin{array}{c} \tilde{\varphi} \\
\tilde{\chi} \end{array}\right)
+ i \hbar\left( \begin{array}{c} \dot{\tilde{\varphi}} \\
\dot{\tilde{\chi}}
\end{array}\right) = \left( \begin{array}{c} {\bf 0} \\ {\bf
0}\end{array} \right)\,. \label{exact}
\eeq
We need to compute the matrix $S^{-1}(t)\dot{S}(t)$. Substituting
the solution for $\varphi^{(0)}$ (\ref{phizerosoln}) back into Eq.
(\ref{zerothdiff}) gives the relation
\beq
\frac{d\ }{dt}
\left\{ \sqrt{\frac{E+m}{2E}}\left(
\begin{array}{c} s_\alpha(t) \\
\Sigma s_\alpha(t) \end{array}
\right)\right\}
= \left( \frac{E+m}{2E}\right)^{3/2} \left(
\begin{array}{c} - \Sigma \dot{\Sigma} s _\alpha (t)
\\ \dot{\Sigma} s _\alpha (t) \end{array}\right)\,,
\eeq
from which we also obtain
\beq
\frac{d\ }{dt}
\left\{ \sqrt{\frac{E+m}{2E}}\left(
\begin{array}{c}
- \Sigma s_\alpha(t) \\
s_\alpha(t) \end{array}
\right)\right\}
= - \left( \frac{E+m}{2E}\right)^{3/2} \left(
\begin{array}{c} \dot{\Sigma} s_\alpha (t)  \\
 \Sigma \dot{\Sigma} s_\alpha (t)
\end{array}\right)\,.
\eeq
Hence, if we define a $2\times 2$ matrix $T(t)$ by
\beq
T(t) \equiv  \frac{E+m}{2E} U^\dagger(t) \dot{\Sigma} U(t)\,,
\eeq
then
\beq
S^{-1}(t)\frac{d\ }{dt}S(t) = \left( \begin{array}{cc} {\bf 0} & -
T(t) \\ T(t) & {\bf 0}\end{array}
\right)\,.
\eeq
Substituting the matrix into Eq.~(\ref{exact}) we obtain the two
equations
\beqa
\dot{\tilde{\varphi}} & = & T(t)\tilde{\chi}\,,\\
  i\hbar \dot{\tilde{\chi}} & = &
 - i\hbar T(t)\tilde{\varphi}-2E\tilde{\chi}\,.
\eeqa
Alternatively, using the semiclassical expansion (\ref{spinsemiexp})
we obtain
\beqa
\tilde{\chi}^{(n+1)} & = & -\frac{i}{2E}T(t)\tilde{\varphi}^{(n)} -
\frac{i}{2E}\dot{\tilde{\chi}}^{(n)}\,,
 \\
\dot{\tilde{\varphi}}^{(n+1)} & = & T(t)\tilde{\chi}^{(n+1)}\,.
\eeqa
In this representation, the zeroth-order solutions are somewhat
simpler: $\tilde{\chi}^{(0)}_\alpha = 0$ and
$\tilde{\varphi}^{(0)}_1 = \left( \begin{array}{c} 1 \\
0\end{array}\right)$ or $\tilde{\varphi}^{(0)}_2 =
\left( \begin{array}{c} 0 \\ 1 \end{array}\right)$. Thus,
\beq
\tilde{\chi}^{(1)}_\alpha = - \frac{i(E+m)}{4E^2}
\left( \begin{array}{c}
s_1^\dagger (t)
\dot{\Sigma} s_\alpha(t) \\
s_2^\dagger (t)
\dot{\Sigma} s_\alpha(t)
\end{array}\right)\,
\eeq
and
\beq
\dot{\tilde{\varphi}}^{(1)}_\alpha
= - \frac{i(E+m)^2}{8E^3}\left( \begin{array}{c} s_1^\dagger (t)
\dot{\Sigma}^2 s_\alpha(t) \\
s_2^\dagger (t)
\dot{\Sigma}^2 s_\alpha(t)
\end{array}\right)\,.
\eeq
Now
\beq
\frac{d\ }{dt}\Sigma
= \frac{1}{E+m}\left( \dot{\bf p}- \frac{\dot{E}}{E+m}{\bf p}\right)
\cdot\itbd{\sigma}\,.
\eeq
Then we obtain
\beq
\left( \dot{\Sigma} \right) ^2 = \frac{\dot{\bf p}^2 - \dot{E}^2}{(E+m)^2} =
-\frac{\dot{p}^2}{(E+m)^2}\,,
\eeq
where $\dot{p}^2=\dot{p}^\mu \dot{p}_\mu$. Hence
\beq
\tilde{\varphi}^{(1)}_1 = ig(t)\left( \begin{array}{c} 1 \\ 0
\end{array}\right)\,,\ \ \ \
\tilde{\varphi}^{(1)}_2 = ig(t)\left( \begin{array}{c} 0 \\
1\end{array}\right)\,,
\eeq
where $g(t)$ is a real function defined by
\beq
g(t)\equiv \int_{t_0}^t
\frac{\dot{p}(\tau)^2}{8E^3(\tau)}\,d\tau\,,
\eeq
with $t_0$ being a constant. Changing back to the standard
representation, the first-order spinor correction is
\begin{align}
\hbar \left( \begin{array}{c} \varphi^{(1)}_\alpha \\
\chi^{(1)}_\alpha \end{array}\right) &= \hbar S(t)\left( \begin{array}{c} \tilde{\varphi}^{(1)}_\alpha \\
\tilde{\chi}^{(1)}_\alpha \end{array}\right)
\nol
& =  i\hbar g(t)\left( \begin{array}{c} \varphi^{(0)}_\alpha \\
\chi^{(0)}_\alpha \end{array}\right)
- i\hbar \frac{(E+m)^{3/2}}{(2E)^{5/2}}
\left( \begin{array}{c} -\Sigma \dot{\Sigma} s_\alpha(t) \\
\dot{\Sigma} s_\alpha(t)
\end{array}\right)\,.\label{first}
\end{align}

The semiclassical expansion for the positive energy spinor can now
be written to order $\hbar$ as
\begin{align}
\Phi(x) &= C \sqrt{\f{E+m}{2E}} \[ \(1+i\hbar g(t)\)\begin{pmatrix} s_\alpha(t) \\
\Sigma s_\alpha(t) \end{pmatrix}  - i\hbar \frac{E+m}{(2E)^2}
\left( \begin{array}{c} -\Sigma \dot{\Sigma} s_\alpha(t) \\
\dot{\Sigma} s_\alpha(t)
\end{array}\right)\] \nol & \times \exp
\(-\f{i}{\hbar}\int^t_0 E(\xi)d\xi\) e^{i{\bf
p}\cdot{\bf x}/\hbar} \,.
\end{align}
Recall that the energy $E$ and the matrix $\Sigma$ are time
dependent. However, the ${\bf p}$ in the exponential is not. We can
choose the constant $C$ to achieve the desired normalisation:
$C=\sqrt{p_0/m}$. Rearranging, we obtain
\beq \label{wkbspinormodep}
\Phi(x) =  \[ \(1+i\hbar g(t)\) u_\alpha^{(0)} - i\hbar \frac{E+m}{(2E)^2}
\sqrt{\f{E+m}{2m}} \left( \begin{array}{c} -\Sigma \dot{\Sigma} s_\alpha(t) \\
\dot{\Sigma} s_\alpha(t)
\end{array}\right)\] \phi_{\bf p}(t) e^{i{\bf
p}\cdot{\bf x}/\hbar} \,,
\eeq
where
\begin{align}\label{wkbzerothu}
u_\alpha^{(0)} = \sqrt{\frac{E+m}{2m}}\left( \begin{array}{c}
s_\alpha(t) \\
\Sigma s_\alpha(t)\end{array}\right) \,,
\end{align}
can be considered the zeroth order spinor (for $V=\text{constant}$
it is the usual positive energy spinor), and
\begin{align}
\phi_{\bf p}(t) = \sqrt{\f{p_0}{E}} \exp
\(-\f{i}{\hbar}\int^t_0 E(\xi)d\xi\) \,,
\end{align}
is the time-dependent part of the WKB semiclassical expansion for
the scalar field. We also define
\begin{align}\label{wkbfirstu}
u_\alpha^{(1)}(p) &= i \hbar  g(t)
\sqrt{\f{E+m}{2m}}\begin{pmatrix} s_\alpha \\ \Sigma s_\alpha
\end{pmatrix} - i\hbar\f{E+m}{(2E)^2}\sqrt{\f{E+m}{2m}}
\begin{pmatrix} -\Sigma \dot{\Sigma} s_\alpha \\ \dot{\Sigma}
s_\alpha \end{pmatrix} \,,
\end{align}
as the first order spinor.

\subsection{Negative Energy Solution}

The Negative energy solutions are interpreted as the antiparticle
solutions and thus this time we look for solutions of the form
\begin{equation}
\Psi(x) = \[ \begin{pmatrix} \varphi^{(0)} \\ \chi^{(0)}\end{pmatrix}
+\hbar \begin{pmatrix} \varphi^{(1)} \\ \chi^{(1)} \end{pmatrix} +\hbar^2 \begin{pmatrix} \varphi^{(2)} \\
\chi^{(2)}
\end{pmatrix} + \ldots\] \exp \(
+\f{i}{\hbar}S\)  e^{-i{\bf p}\cdot{\bf x}/\hbar} \,.
\end{equation}
The mode function satisfies the conjugate Dirac equation, and as the
potential is the minimal substitution electromagnetic potential the
result is that the antiparticle has opposite charge. Relative to the
momentum operators we rewrite ${\bf V}(t)\to -{\bf V}(t)$. Due to
the sign change, we obtain
\begin{align}
i \hbar \begin{pmatrix} \dot{\varphi} \\ \dot{\chi} \end{pmatrix}
 -\begin{pmatrix} \varphi \\ \chi \end{pmatrix} \dot{S}
+ \[ \itbd{\alpha}\cdot\tilde{\bf p}_+ -\beta m \]\begin{pmatrix} \varphi \\
\chi
\end{pmatrix}  = 0 \,,
\end{align}
where this time we have $\tilde{\bf p}_+ = {\bf p}+{\bf V}(t)$. The
lowest order equation gives the eigenvector equation with eigenvalue
$\dot{S}=E_+$ where $E_+=\sqrt{\tilde{\bf p}_+^2+m^2}$ in keeping
with the positive energy solutions. The spinor equation is now
\begin{align}
i \hbar \begin{pmatrix}
\dot{\varphi} \\ \dot{\chi} \end{pmatrix}
 -\begin{pmatrix} E_++m & -\itbd{\sigma}\cdot\tilde{\bf p}_+ \\  -\itbd{\sigma}\cdot\tilde{\bf p}_+
 & E_+-m \end{pmatrix}   \begin{pmatrix} \varphi \\
\chi
\end{pmatrix}  = 0 \,. \label{exact3}
\end{align}
Comparing this equation with (\ref{exact2}) for the positive energy
solution we see that under the transformation $\hbar\to-\hbar$ and
$\varphi \leftrightarrow \chi$ they are the same. Thus the negative
energy solution can be written
\begin{align} \label{wkbspinormoden}
\Psi(x) &=  \[ \(1-i\hbar g(t)\) v_\alpha^{(0)} + i\hbar \frac{E_++m}{(2E_+)^2}
\sqrt{\f{E_++m}{2m}} \left( \begin{array}{c} \dot{\Sigma} s_\alpha(t) \\ -\Sigma \dot{\Sigma} s_\alpha(t)
\end{array}\right)\] \nol & \times \bar{\phi}^*_{\bf p}(t) e^{-i{\bf
p}\cdot{\bf x}/\hbar} \,,
\end{align}
where
\begin{align}\label{wkbzerothv}
v_\alpha^{(0)} = \sqrt{\frac{E_++m}{2m}}\left( \begin{array}{c}
\Sigma s_\alpha(t) \\ s_\alpha(t)
\end{array}\right) \,,
\end{align}
can be considered the zeroth order antiparticle spinor, and
\begin{align}
\bar{\phi}^*_{\bf p}(t) = \sqrt{\f{p_0}{E_+}} \exp
\(\f{i}{\hbar}\int^t_0 E_+(\xi)d\xi\) \,,
\end{align}
is the time-dependent part of the WKB semiclassical expansion for
the complex conjugate scalar field. Hence overall, the two solutions
are related by $\hbar\to-\hbar$, $\varphi \leftrightarrow
\chi$, ${\bf V}(t) \to -{\bf V}(t)$.

\chapter{Classical Position Shift}\label{classicalshiftchapter}
\begin{quote}
In this chapter we measure the effects radiation reaction in the
classical theory of electrodynamics via the calculation of the
position shift. We analyse the special case of linear acceleration
before deriving a more general description.
\end{quote}

\section{Linear Acceleration}

The case of linear acceleration simplifies matters considerably. Let
us orientate our coordinate system such that the direction of the
linear acceleration is along the $z$-axis. For the most part, we can
consider the system to be in 1+1 dimensions $(t,z)$. The reference
frame can naturally be shifted so that the perpendicular velocities
are zero. Before proceeding, we make a note of some simplifying
notation in the spirit of Newton: We use dot notation to represent
differentiation with respect to coordinate time $t$ and dash
notation to represent differentiation with respect to proper time
$\tau$. Thus
\begin{align}
\dot{z} &= \f{dz}{dt} &  z' = \f{d z}{d\tau} \,.
\end{align}
To enable the reader to easily follow the calculations and indeed
for ease of reproducing them, we give a number of simple identities
which are of use in this system. Firstly, in 1+1 dimensions the
relativistic gamma factor is defined, in our above notation, as
\begin{equation}
t'=\gamma = \f{1}{\sqrt{1-\dot{z}^2}} \,.
\end{equation}
The following are equalities between the dot and dash
representations
\begin{align}
\dot{\gamma} &= \gamma^3\ddot{z}\dot{z} \,, \\
t''=\gamma' &= \gamma^4\ddot{z}\dot{z} \,, \\
z' &= \gamma \dot{z} \,, \\
z'' &= \gamma^4 \ddot{z} \,, \\
z''' &= \gamma^5 \dot{\ddot{z}} + 4 \gamma^7 \ddot{z}^2 \dot{z}
\,.
\end{align}

Recall that the Lorentz-Dirac force is given by
\begin{equation}\tag{\ref{LD4force}}
F^\mu_{\rm LD} \equiv \frac{2\alpha_c}{3}\left[
\frac{d^3x^\mu}{d\tau^3} + \frac{dx^\mu}{d\tau}\left(\frac{d^2
x^\nu}{d\tau^2} \frac{d^2x_\nu}{d\tau^2}\right)\right]\,.
\end{equation}
In this system the expression for the force can be much simplified.
For the $z$ component one finds
\begin{align}
F_{\rm LD}^z &= \f{2\alpha_c}{3} \[ z''' +z' \[ (t'')^2-(z'')^2 \]
\] \nol
&= \f{2\alpha_c}{3} \[ \gamma^5 \dot{\ddot{z}} + 4 \gamma^7
\ddot{z}^2
\dot{z} +\gamma \dot{z} \[ \gamma^8\ddot{z}^2 \dot{z}^2 - \gamma^8
\ddot{z}^2 \] \] \nol
&= \f{2\alpha_c}{3} \[ \gamma^5 \dot{\ddot{z}} + 4 \gamma^7
\ddot{z}^2
\dot{z}-\gamma^7\ddot{z}^2\dot{z} \] \nol
&= \f{2\alpha_c}{3} \[ \gamma^5 \dot{\ddot{z}} + 3 \gamma^7
\ddot{z}^2
\dot{z} \] \nol
&= \f{2\alpha_c}{3} \gamma^2 \[ \gamma^3 \dot{\ddot{z}} + 3 \gamma^5
\ddot{z}^2
\dot{z} \] \nol
&= \f{2\alpha_c}{3} \gamma^2 d_t \(\gamma^3 \ddot{z} \) \,.
\end{align}
The $t$ component can similarly be given as
\begin{equation}
F_{\rm LD}^t =  \f{2\alpha_c}{3} \dot{z} \gamma^2 d_t \(\gamma^3
\ddot{z}
\) \,.
\end{equation}
Now, for linear acceleration in the potential $V(z)$, the external
force acting on the particle is given by
\begin{align}
 F^t_{\rm ext} &=-V'(z)\,dz/d\tau \,,  \nonumber \\
 F^z_{\rm ext} &= - V'(z)\,dt/d\tau \,, \nonumber \\
 F^x_{\rm ext} &= F^y_{\rm ext} = 0 \,.
\end{align}
The Lorentz-Dirac force can be similarly written as
\begin{align}
 F^t_{\rm LD} &= F_{\rm LD}\,dz/d\tau \nonumber \,, \\
 F^z_{\rm LD} &= F_{\rm LD}\,dt/d\tau \nonumber \,, \\
 F^x_{\rm LD} &= F^y_{\rm LD} = 0
\end{align}
where, using the more compact form found above, we have
\begin{equation}
F_{\rm LD}  \equiv \frac{2\alpha_c}{3}\gamma \frac{d\ }{dt}
(\gamma^3\ddot{z})\,.  \label{LDLD}
\end{equation}

\subsection{Space-dependent Potential}

In this section we explicitly calculate the position shift for
linear acceleration due to the potential $V(z)$, where $z$ is the
direction of the acceleration. We recall that the position shift is
the change in position due to radiation reaction. We also recall,
that we shall regard the radiation reaction force as a perturbation.
What this means in practice is that all quantities, such as
$\dot{z}$, $\ddot{z}$ and $\dot{\ddot{z}}$, in the equations
involving the radiation reaction force are evaluated using the
original unperturbed path given by $ma^\mu=F_{\rm ext}^\mu$. We
shall find the position shift to first non-trivial order in $F_{\rm
LD}$.

Suppose that, in the absence of radiation reaction, the particle
would be at $z=0$ at time $t=0$. This is the position of the
unperturbed particle obeying $ma^\mu=F_{\rm ext}^\mu$. The position
of the particle undergoing radiation reaction, and thus obeying
$ma^\mu=F_{\rm ext}^\mu+F_{\rm LD}^\mu$, is equal the position
shift, which we label $\delta z$. In the system with $V(z)$, the
calculation of $\delta z$ is facilitated by the observation that the
change in the total energy, $m\,dt/d\tau + V(z)$, is equal to the
work done by the Lorentz-Dirac force. We then find
\begin{eqnarray}\label{0}
 \int_{-\infty}^{t} F_{\rm LD}(t')\dot{z}(t') dt'
&=&\frac{d}{d\dot{z}}\frac{m}{\sqrt{1-\dot{z}^2}} \delta \dot{z}
 + V'(z)\delta z \nonumber \\
&=&  m\gamma^3\dot{z}^2\frac{d}{dt}\left(\frac{\delta
  z}{\dot{z}}\right)\,,
\end{eqnarray}
where we have used
\begin{equation}\label{vdash}
\frac{d\ }{dt}\left( m\gamma \dot{z}\right)
= m \gamma^3\ddot{z} = - V'(z) \,.
\end{equation}
This last line is to zeroth-order in $F_{\rm LD}$ as we explained
above. Rearranging and integrating, we obtain the position shift:
\begin{equation}
\delta z_{\rm LD} = \frac{v_0}{m}\int^0_{-\infty}
\left(\int^{t}_{-\infty}
F_{\rm
LD}\frac{dz}{dt'}dt'\right)\frac{1}{\gamma^3(t)[\dot{z}(t)]^2}dt
\,,\label{deltaz}
\end{equation}
where $v_0=\dot{z}(0)$ is the final velocity. The reader may note
that the outer integration limit is $t=0$, as is the time of
measurement for $v_0$, which is of course due to the fact that
$\delta z_{\rm LD}$ is the position shift at $t=0$.

Now, the current set-up, in which the unperturbed particle is at the
origin at the time of measurement, is naturally made for simplicity
and we indeed have complete freedom to do so by appropriate
definition of the coordinate system. However, it does encourage the
question as to what the position shift would be if this were not the
case, i.e. if $z=z_0\neq 0$ at $t=0$ as opposed to $z=0$. We assume
that $z_0$ is still in the final non-accelerated region and thus the
final velocity is still $v_0$. The result is that the time the
particle spends between the end of the acceleration and the
measurement at $t=0$ is lengthened by $t_0=z_0/v_0$. The effect is
the same as shifting the entire trajectory earlier in time by $t_0$.
Consequently, we may calculate the new position shift by using our
original trajectory and taking the measurement at $t=t_0$ instead of
$t=0$. The extra contribution to the position shift is thus
\begin{equation}\label{zextra}
\delta z_{\rm extra} =
\frac{v_0}{m}\int^{t_0}_{0}
\left(\int^{t}_{-\infty}
F_{\rm
LD}\frac{dz}{dt'}dt'\right)\frac{1}{\gamma^3(t)[\dot{z}(t)]^2}dt
\,,
\end{equation}
which is easily obtained with reference to the earlier comments
about the limits and the constant velocity. Within the new limits,
$t\in [0,t_0]$, we note that $\dot{z}(t)=v_0$ and
$\gamma(t)=\gamma_0 \equiv (1-v_0^2)^{-1/2}$ are constant and
$F_{\rm LD}=0$. We can interchange the order of integration to find
\begin{eqnarray}
\delta z_{\rm extra} &=&
\frac{v_0}{m}\int_{-\infty}^{0}
\left( \int^{t_0}_0 \frac{1}{\gamma^3_0 v_0^2}
dt'\right)F_{\rm LD}\frac{dz}{dt}dt\, \nonumber \\ &=&
-\frac{z_0}{m\gamma_0^3 v_0^2}E_{\rm em} \,, \label{extraaa}
\end{eqnarray}
where $E_{\rm em}$ is the energy emitted as radiation given by
\begin{eqnarray}
E_{\rm em} & = & - \int_{-\infty}^0 F_{\rm LD}\frac{dz}{dt}dt
\nonumber \\
& = & \frac{2\alpha_c}{3}\int_{-\infty}^0 (\gamma^3
\ddot{z})^2\,dt\,.
\label{Larmor}
\end{eqnarray}
This is the relativistic Larmor formula for one-dimensional motion.

The current form of the position shift (\ref{deltaz}), whilst useful
for the above comment, is somewhat more complicated than is
necessary. After interchanging the order of integration to obtain
\begin{equation}
\delta z_{\rm LD} = -\frac{v_0}{m}\int^0_{-\infty}
\left(\int^{t}_{0}\frac{1}{\gamma^3(t')[\dot{z}(t')]^2}dt'
\right)F_{\rm LD}\frac{dz}{dt}dt \,,
\label{deltazreorder}
\end{equation}
it can be simplified by noting that, for the space dependent
potential,
\begin{equation}\label{concrete}
\left(\frac{\partial z}{\partial p}\right)_t
= \frac{v_0}{m}\dot{z}(t)\int_0^t \frac{1}
{\gamma^3(t')\left[\dot{z}(t')\right]^2}\,dt' \,,
\end{equation}
where $p$ is the final momentum of the particle. This equation can
be demonstrated as follows. Since the energy is conserved, we have
\begin{align}\label{11}
  \sqrt{p^2+m^2}&=\sqrt{\(mz'\)^2 + m^2}+V(z) \nol
&= \f{m}{\sqrt{1-\dot{z}^2}}+V(z) \,,
\end{align}
and hence,
\begin{equation}\label{dzdteq}
\dot{z} =
\[1-m^2\(\sqrt{p^2+m^2}-V(z)\)^{-2}\]^{1/2}\,.
\end{equation}
By differentiating both sides with respect to $p$ with $t$ fixed,
and noting that
\begin{eqnarray}
p/\sqrt{p^2 + m^2} & = & v_0\,,\\
\sqrt{p^2 + m^2} - V(z) & = & m\gamma\,,
\end{eqnarray}
we obtain
\begin{equation}\label{gamma3}
\frac{d\ }{dt}\left(\frac{\partial z}{\partial p}\right)_t
= \frac{1}{m\gamma^3\dot{z}}
\left[v_0 - V'(z)\left(\frac{\partial z}{\partial p}\right)_t
\right]\,.
\end{equation}
By substituting the formula $V'(z) = - m\gamma^3 \ddot{z}$ (see
(\ref{vdash})) in (\ref{gamma3}) we find
\begin{equation}
\frac{d\ }{dt}
\left[\frac{1}{\dot{z}}\left(\frac{\partial z}{\partial p}\right)_t
\right] = \frac{v_0}{m\gamma^3\dot{z}^2}\,.
\end{equation}
Then by integrating this formula, remembering that $z=0$ at $t=0$
for all $p$, we arrive at (\ref{concrete}). Consequently, the
position shift can be written in the more compact form
\begin{equation}\label{dzLD}
\delta z_{\rm LD} = - \int_{-\infty}^0 dt\,
 F_{\rm LD}\left(\frac{\partial z}{\partial
p}\right)_t \,.
\end{equation}

\subsection{Time-dependent Potential}
The case of linear acceleration due to a time-dependent potential
can be analysed in a similar way to the previous exercise. We again
define the coordinate system such that the acceleration is in the
direction of the $z$-axis, but this time the potential is given by
$V(t)$. In the $V(z)$ case, we use the energy conservation, whereas
now we shall make use of the momentum conservation equation which
reads
\begin{equation}\label{momcons}
\frac{d\ }{dt}\left[ m\gamma \dot{z} + V(t)\right]
= F_{\rm LD} \,.
\end{equation}
The lack of symmetry between the two situations ($V(z)$ and $V(t)$)
is worth noting. In both situations we are measuring the change in
position at equal time, as opposed to the possible consideration of
the change in time for the same position. Retaining the same
measurement breaks some of the symmetry. Returning to the change in
momentum, we note that in the time-dependent case, the potential
\emph{at equal time} is the same for the particle in the presence or
absence of radiation reaction. The momentum conservation
(\ref{momcons}) thus leads to
\begin{equation}
\delta(m\gamma\dot{z}) =
m\gamma^3\,\frac{d\ }{dt}(\delta z) = \int_{-\infty}^t F_{\rm
LD}(t')\,dt'\,.
\end{equation}
Rearranging for $\delta z$ and interchanging the order of
integration as per the previous case, the position shift is given by
\begin{equation}\label{cshifttdep}
 \delta z_{\rm LD}  = -\int^0_{-\infty}
\(\int^t_0\frac{1}{m\gamma^3}dt'\)F_{\rm LD}dt\,.
\end{equation}
In line with the spatially dependent potential case, this can be
further simplified. In fact, we find that for the $t$-dependent
potential,
\begin{equation}\label{infact}
\(\frac{\partial z}{\partial p}\)_t
 = \int^t_0 \frac{1}{m\gamma^3}\, dt \,,
\end{equation}
where we again recall that this is for the unperturbed particle.
This is demonstrated as follows: The momentum conservation for this
particle in the $z$-direction reads
\begin{equation}
m\frac{dz}{d\tau} + V(t) = p\,.
\end{equation}
Hence, with the condition $z=0$ at $t=0$, we find
\begin{equation}
z = \int^t_0 \left\{ 1+\frac{m^2}{[p-V(t)]^2}\right\}^{-1/2}dt\,.
\end{equation}
By differentiating this expression with respect to $p$ and using
$p-V(t) = m\,dz/d\tau$ and $\sqrt{[p-V(t)]^2 + m^2} = m\,dt/d\tau$,
we indeed obtain (\ref{infact}).

Consequently, we note that the position shift can be written in the
same form as (\ref{dzLD}) before, namely
\begin{equation}
\delta z_{\rm LD} = - \int_{-\infty}^0 dt\,
 F_{\rm LD}\left(\frac{\partial z}{\partial
p}\right)_t \,.
\end{equation}

\section{Generalised Classical Position Shift}

The fact that both the $t$-dependent and $z$-dependent potentials
for linear acceleration lead ultimately to the same expression for
the classical position shift and in addition that this expression is
fairly simple, leads one to suspect a more general argument for this
formula. This is indeed the case as we now proceed to relate.

We now look at the full three spatial dimensional system. The system
is one where the total force acting on the particle is the sum of an
external force $F$ and an additional force $\gamma{\cal F}$, which
we intend to treat as a perturbation:
\begin{equation}
m\f{d^2x^i}{d\tau^2} = F^i + {\cal F}^i\f{dt}{d\tau} \,.
\end{equation}

\subsection{Homogeneous system}

As yet, we have said nothing about what these forces are. Let us
consider this system to be the result of a perturbation from a
Hamiltonian system i.e. that in the absence of the extra force,
${\cal F}=0$, the system is described by a Hamiltonian $H({\bf
x},{\bf p})$, where $({\bf x},{\bf p})$ are the generalised
coordinates and conjugate momenta. We shall refer to this system as
the homogeneous system. Hamilton's equations are given by
\begin{eqnarray}
\dot{x^i} &=& \f{\d H}{\d p^i} \,,\\
\dot{p^i} &=& -\f{\d H}{\d x^i} \,.
\end{eqnarray}
Consider a perturbation to the solution $({\bf x}, {\bf p})$ given
by $({\bf x}+{\bf \Delta x}, {\bf p} + {\bf \Delta p})$. We shall
refer to these perturbations to the path as the homogeneous
perturbations. The expansion to second order of the Hamiltonian of
the perturbed solution is given by
\begin{multline}\label{Hamexp}
H({\bf x}+{\bf \Delta x}, {\bf p} + {\bf \Delta p}) = H({\bf x},{\bf
p}) + \f{\d H}{\d x^i} \Delta x^i + \f{\d H}{\d p^i}\Delta p^i \\
 + \f{1}{2} \[ \f{\d^2 H}{\d x^i \d x^j} \Delta x^i
\Delta x^j + 2
\f{\d^2 H}{\d x^i \d p^j} \Delta x^i \Delta p^j + \f{\d^2 H}{\d p^i \d
p^j} \Delta p^i \Delta p^j \] \,.
\end{multline}

The equations for the homogeneous perturbations are given by
\begin{eqnarray}
\Delta \dot{x^i} &=& \Delta \f{\d H}{\d p^i} \nol
 &=& \f{\d^2 H}{\d x^j \d p^i} \Delta x^j + \f{\d^2 H}{ \d p^j \d p^i}
 \Delta p^j \,,\\
 \Delta \dot{p^i} &=& -\Delta \f{\d H}{\d x^i} \nol
  &=& - \f{\d^2 H}{ \d x^j \d x^i} \Delta x^j - \f{ \d^2 H}{\d p^j
  \d x^i} \Delta p^j \,.
\end{eqnarray}
Thus these can be seen to be generated by the equations
\begin{align}\label{Hamdeltaq}
\Delta \dot{x^i} &= \f{\d \bar{H}}{\d \Delta p^i} \,,\\ \label{Hamdeltap}
\Delta \dot{p^i} &= -\f{\d \bar{H}}{\d \Delta x^i} \,,
\end{align}
where the Hamiltonian $\bar{H}$ is given by the second order terms
in the expansion in (\ref{Hamexp}), viz
\begin{equation}
\bar{H} =  \f{1}{2} \[ \f{\d^2 H}{\d x^i \d x^j} \Delta x^i \Delta x^j +2
\f{\d^2 H}{\d x^i \d p^j} \Delta x^i \Delta p^j + \f{\d^2 H}{\d p^i \d
p^j} \Delta p^i \Delta p^j \] \,.
\end{equation}
This can be rewritten in terms of the matrices $A_{ij}$, $B_{ij}$
and $C_{ij}$, where $A$ and $C$ are symmetric, as follows
\begin{equation}
\bar{H} = \f{1}{2} \[ A_{ij}\Delta p^i \Delta p^j + 2 B_{ij} \Delta x^i
\Delta p^j + C_{ij}  \Delta x^i \Delta x^j \] \,.
\end{equation}
Thus the equations (\ref{Hamdeltaq}) and (\ref{Hamdeltap}) can be
written
\begin{align}
\Delta \dot{x^i} &= \f{\d \bar{H}}{\d \Delta p^i}  =  B_{ji} \Delta x^j + C_{ij} \Delta p^j \,,\\
\Delta \dot{p^i} &= -\f{\d \bar{H}}{\d \Delta x^i} =  - A_{ij} \Delta x^j - B_{ij} \Delta
p^j \,.
\end{align}
As a consequence we can deduce that the symplectic product of the
perturbations is conserved. Given two solutions $(\Delta X^i, \Delta
P^i)$, $(\Delta x^i,
\Delta p^i)$, the symplectic product is given by
\begin{equation}
\langle \Delta X^i, \Delta P^i | \Delta x^i,
\Delta p^i \rangle = \Delta X^i \Delta p^i - \Delta x^i \Delta P^i
\,.
\end{equation}
The time conservation is easily seen as follows:
\begin{align}
& \f{d}{dt} \( \Delta X^i \Delta p^i - \Delta x^i \Delta P^i \) \nol
&= \Delta \dot{X^i} \Delta p^i + \Delta X^i \Delta \dot{p^i} -
\Delta \dot{x^i} \Delta P^i  - \Delta x^i
\Delta \dot{P^i} \nol
 &= \( B_{ji}\Delta X^j + C_{ij}\Delta P^j \) \Delta p^i +
 \(-A_{ij}\Delta x^j - B_{ij} \Delta p^j \) \Delta X^i \nol
 & \quad - \( B_{ji}\Delta x^j + C_{ij}\Delta p^j \) \Delta P^i - \(-A_{ij}\Delta X^j - B_{ij} \Delta P^j \) \Delta
 x^i \nol
 &= \Delta X^j \Delta p^i \(B_{ji}-B_{ji}\) + \Delta x^j \Delta P^i
 \( -B_{ji}+B_{ij} \) \nol & \quad + \Delta P^j \Delta p^i \( C_{ij}-C_{ji}\) +
 \Delta x^j \Delta X^i \( -A_{ij}+A_{ji}\) \nol
 &= 0 \,,
\end{align}
where we have made use of the fact that $A_{ij}$ and $C_{ij}$ are
symmetric.

\subsection{Inhomogeneous system}

We now consider the system for which we add the additional force,
${\cal F}\neq 0$. We shall refer to this system as the inhomogeneous
system. Let $({\bf x},{\bf p}) = ({\bf x}_0(t),{\bf p}_0(t))$ be a
solution to the homogeneous system such that $({\bf x}_0(0),{\bf
p}_0(0)) = (0,{\bf p})$. This solution gives the classical
trajectory of a particle passing through the origin at $t=0$ with
momentum ${\bf p}$ in the absence of radiation reaction. We let
$({\bf x}_0(t)+\delta {\bf x}(t),{\bf p}_0(t)+\delta {\bf p} (t))$
be a solution to the inhomogeneous system to first order in ${\cal
F}$. The $(\delta {\bf x}(t),\delta {\bf p}(t))$, which we call the
inhomogeneous perturbations, are the perturbations to the classical
trajectory due to the addition of the radiation reaction force,
treated as a perturbation to first order. They have the property
that $((\delta {\bf x}(t),\delta {\bf p}(t)) \to (0,0)$ as $t\to
-\infty$ and will satisfy the equations
\begin{align}
\f{d}{dt} \delta x^i &= B_{ji} \delta x^j + C_{ij} \delta p^j \,, \nol
\f{d}{dt} \delta p^i &= -A_{ij} \delta x^j - B_{ij}\delta p^j +
{\cal F} \,.
\end{align}

\begin{figure}
\begin{center}
\begin{pspicture}(-4,-4)(1.5,1.5)
 \psline{->}(-4,-2)(1,-2) \psline{->}(-1,-4)(-1,1)
 \uput[r](-1,1){$t$} \uput[d](1,-2){$x^i$}
 \pscurve[linewidth=0.5mm]{*-*}(0,0)(-1.2,-1)(-1.8,-2)(-3,-3)
 \pscurve[linewidth=0.5mm]{*-*}(0,0)(-0.6,-1)(-1,-2)(-2,-3)
 \psline{->}(-3,-3.3)(-2,-3.3)
 \uput[d](-2.5,-3.3){$\Delta x^i_{(j)} (t;s)$}
 \psline[linestyle=dotted](-4,0)(0,0)
 \uput[u](-3,0){$t=s$}
\end{pspicture}
\end{center}
\caption{The world lines for the solutions $({\bf x}_0(t),{\bf p}_0(t))$, which passes through the origin, and
$({\bf x}_0+\Delta {\bf x}_{(j)}(t;s), {\bf p}_0 + \Delta {\bf
p}_{(j)} (t;s))$ for some $j$.}
\label{deltaxst}
\end{figure}
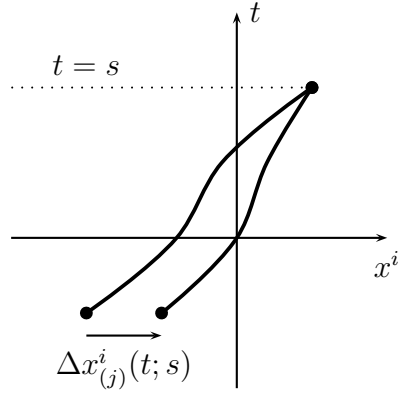

In order to solve these equations, we define a set of homogeneous
perturbations $(\Delta {\bf x}_{(j)} (t;s), \Delta {\bf p}_{(j)}
(t;s))$, with $j=1,2,3$ and $s \in \(-\infty,\infty\)$, by the
following initial conditions:
\begin{align}
\Delta x^i_{(j)} (s;s) &= 0 \,, \nol
\Delta p^i_{(j)} (s;s) &= \delta^i_j \,.
\end{align}
The solution $({\bf x}_0+\Delta {\bf x}_{(j)}(t;s), {\bf p}_0 +
\Delta {\bf p}_{(j)} (t;s))$ then represents the particle trajectory
which coincides with ${\bf x}_0(t)$ at time $t=s$, but which has
excess momentum solely in the $j$-direction at this time. This
trajectory is represented in Fig. \ref{deltaxst}.
With these solutions now defined we note that the solutions of the
coupled inhomogeneous equations can be given by
\begin{align}
\delta x^i &= \int^t_{-\infty} ds {\cal F}^j (s) \Delta
x^i_{(j)}(t;s) \,,
\nol
\delta p^i &= \int^t_{-\infty} ds {\cal F}^j (s) \Delta p^i_{(j)}
(t;s) \,.
\end{align}
where the index $j$ is summed over. The position shift due to the
additional force ${\cal F}$ can therefore be written as
\begin{equation}
\delta x^i = \int^0_{-\infty} dt {\cal F}^j (t) \Delta x^i_{(j)}
(0;t) \,.
\end{equation}
With final momentum at time $t=0$ as $p$ then, given the definition
of the $\Delta x^i_{(j)} (t;s)$ solutions, we can write
\begin{equation}
\( \f{\partial x^j}{\partial p^i} \)_t = \f{\Delta
x^j_{(i)}(t;0)}{\Delta p^i_{(i)}(0;0)} = \Delta x^j_{(i)}(t;0)
\,.
\end{equation}
Recall that the symplectic product of homogeneous perturbations is
conserved. Thus, by equating the symplectic products of the two
solutions $\(\Delta {\bf x}_{(i)} (t;s), \Delta p_{(i)}(t;s)\)$ and
$\(\Delta {\bf x}_{(j)} (t;u), \Delta p_{(j)}(t;u)\)$ at the times
$t=s$ and $t=u$ we have
\begin{multline}
\Delta {\bf x}_{(i)}(s;s) \cdot \Delta {\bf p}_{(j)}(s;u)
- \Delta {\bf x}_{(j)}(s;u) \cdot \Delta {\bf p}_{(i)}(s;s) \\ =
\Delta {\bf x}_{(i)}(u;s) \Delta {\bf p}_{(j)}(u;u)
- \Delta {\bf x}_{(j)} (u;u) \Delta {\bf p}_{(i)} (u;s) \,.
\end{multline}
This equation and the initial conditions that define these solutions
imply
\begin{equation}
\Delta x^i_{(j)} (s;u) = -\Delta x^j_{(i)} (u;s) \,.
\end{equation}
Of particular interest to us, we obtain $\Delta x^i_{(j)} (0;t) =
-\Delta x^j_{(i)} (t;0)$, which means that we can rewrite the
position shift as
\begin{equation}
\delta x^i = -\int^0_{-\infty} dt {\cal F}^j \( \f{\partial
x^j}{\partial p^i} \)_t \,.
\end{equation}
This is the same form of equation which we derived for the linear
acceleration and the Lorentz-Dirac force to first order. In that
case, we only have non-zero terms for $j=3$ with ${\cal F}^z={\cal
F}_{\rm LD}^z$ and $i=3$. We have thus derived a more general
relation for the classical position shift. We have assumed that the
position shift is due to an additional force taken as a
perturbation, to first order, to a Hamiltonian system.

The above conditions apply for the specific model we wish to
consider for the Lorentz-Dirac force in three dimensional motion.
Here we simply write ${\cal F}^i={\cal F}_{\rm LD}^i
\equiv \gamma^{-1} F_{{\rm LD}}^i$. Thus
\begin{equation}\label{gen3deofm}
m\f{d^2x^i}{d\tau^2} = F^i + {\cal F}^i_{\rm LD} \f{dt}{d\tau} \,.
\end{equation}
It would be useful to repeat here for the 3D case the conversion
from proper time variables to coordinate time variables as was done
for the 1D case at the beginning of the chapter. Using the dot
notation for differentiation with respect to $t$, as before, we
define ${\bf v}=\dot{\bf x}$ and ${\bf a}=\ddot{x}$ and note that
\begin{align}
\gamma &= \f{dt}{d\tau} = \f{1}{(1-\bf{v}\cdot{\bf v})^{1/2}} \\
\dot{\gamma} &= \f{{\bf v}\cdot{\bf a}}{\(1-{\bf v}^2\)^{3/2}} = \gamma^3 {\bf
v}\cdot{\bf a} \,,
\end{align}
and also
\begin{align}
\f{d x^i}{d\tau} &= \f{dt}{d\tau} \f{dx^i}{dt} = \gamma v^i \\
\f{d^2 x^i}{d\tau^2} &= \gamma \f{d}{dt} \( \gamma v^i \) \nol
&= \gamma \(\dot{\gamma}v^i + \gamma a^i \) \nol
&= \gamma^4 {\bf v}\cdot{\bf a} v^i + \gamma^2 a^i \\
\f{d^2 t}{d\tau^2} &= \gamma \dot{\gamma} = \gamma^4 {\bf
v}\cdot{\bf a} \,.
\end{align}
From these last two relations, we have
\begin{align}
\f{d^2x^\nu}{d\tau^2} \f{d^2 x_\nu}{d\tau^2} &= \(\gamma^4 {\bf
v}\cdot{\bf a}\)^2 - \(\gamma^4 {\bf v}\cdot{\bf a} {\bf v} +
\gamma^2 {\bf a} \)\cdot \( \gamma^4 {\bf v}\cdot{\bf a} {\bf v} + \gamma^2
{\bf a}  \) \nol &= \gamma^8 \({\bf v}\cdot{\bf a}\)^2-\gamma^8
\({\bf v}\cdot{\bf a}\)^2 {\bf v}^2 - 2 \gamma^6 \({\bf v}\cdot{\bf
a}\)^2 - \gamma^4 {\bf a}^2 \nol &= \gamma^8 \({\bf v}\cdot{\bf
a}\)^2 \gamma^{-2}- 2 \gamma^6 \({\bf v}\cdot{\bf a}\)^2 - \gamma^4
{\bf a}^2 \nol &= \gamma^6\({\bf v}\cdot{\bf a}\)^2- \gamma^4 {\bf
a}^2 \,.
\end{align}

Now, the Lorentz-Dirac force is given by
\begin{equation}\tag{\ref{LD4force}}
F^\mu_{\rm LD} \equiv \frac{2\alpha_c}{3}\left[
\frac{d^3x^\mu}{d\tau^3} + \frac{dx^\mu}{d\tau}\left(\frac{d^2
x^\nu}{d\tau^2} \frac{d^2x_\nu}{d\tau^2}\right)\right] \,.
\end{equation}
Thus the spatial components can be written
\begin{align}
F^i_{\rm LD} &= \frac{2\alpha_c}{3} \left[ \gamma \f{d}{dt} \(
\f{d^2 x^i}{d\tau^2} \) + \gamma v^i \left(\frac{d^2
x^\nu}{d\tau^2} \frac{d^2x_\nu}{d\tau^2}\right)\right] \,.
\end{align}
Substituting the relations above, we have
\begin{align}
F^i_{\rm LD} &= \frac{2\alpha_c}{3} \gamma \left[  \f{d}{dt} \(
 \gamma^4 {\bf v}\cdot{\bf a} v^i + \gamma^2 a^i \) +  v^i \left( \gamma^6\({\bf v}\cdot{\bf a}\)^2- \gamma^4
{\bf a}^2 \right)\right] \,,
\end{align}
and consequently,
\begin{align}\label{3dLD}
{\cal F}^i_{\rm LD} = \frac{2\alpha_c}{3}  \left[  \f{d}{dt} \(
 \gamma^4 {\bf v}\cdot{\bf a} v^i + \gamma^2 a^i \) +  v^i \left( \gamma^6\({\bf v}\cdot{\bf a}\)^2- \gamma^4
{\bf a}^2 \right)\right] \,.
\end{align}
It will be sufficient, and more useful, for our purposes to leave
this expression in its current form instead of rearranging or
evaluating it further.

Returning to the equations of motion (\ref{gen3deofm}), the external
force on the charged particle $F$, which causes the initial
acceleration in the first place, merely needs to be one derived from
a Hamiltonian, which is the case for most external forces that we
would consider. For example, the most natural external force on a
charged particle would be an electromagnetic Lorentz force. The
Lorentz force can be derived from the Hamiltonian
\begin{equation}
H = \sqrt{\( {\bf p} - e {\bf A}_{\rm ext}\)^2 +m^2} + e A_{\rm
ext}^0\,,
\end{equation}
where $A_{\rm ext}^\mu$ is the external electromagnetic potential.
In our previous notation, the potential is $V^\mu=eA_{\rm ext}^\mu$.
Henceforth we refer to
\begin{equation}\label{cshift}
\delta x^i_C = -\int^0_{-\infty} dt {\cal F_{\rm LD}}^j \( \f{\partial
x^j}{\partial p^i} \)_t \,.
\end{equation}
as the classical position shift (due to electromagnetic radiation
reaction as described by the Lorentz-Dirac force to first order).
This concludes our investigation using the classical theory and
provides us with the classical position shift, with which we can
compare the results of investigations using quantum field theory.

\chapter{Scalar Quantum Position Shift}\label{scalarshiftchapter}
\begin{quote}
In this chapter we derive the contributions to the position shift in
the quantum scalar electrodynamics model. We calculate the
contributions from the photon emission, forward scattering and
renormalisation counterterm perturbation effects and combine them to
compare the position shift in the $\hbar\to 0$ limit with that from
the classical theory.
\end{quote}

\section{Initial control state}\label{scalarinitialsection}

The initial state is given by the incoming wave packet in
(\ref{istate}) as
\begin{equation}
| i\rangle = \int \frac{d^3{\bf p}}{\sqrt{2p_0}(2\pi\hbar)^3} f({\bf
p})A^\dagger({\bf p})|0\rangle\,,
\end{equation}
with $f({\bf p})$ peaked about the initial momentum in the region
${\cal M}_-$. Let the potential satisfy $|V_0|<2m$, thus precluding
the possibility of scalar-particle pair creation. This would be a
vacuum process and thus not of interest to us in examining the
evolution of the particle under consideration. The `free' particle
that we wish to use as our control measurement does not interact
with the electromagnetic field via radiation reaction. Having passed
through the classical non-perturbative potential $V$ in the region
${\cal M}_I$, the final state can be considered as analogous to the
initial state, albeit with the wave packet peaked about the final
momentum in the region ${\cal M}_+$. Thus, we wish to find the
position expectation value for the particle in the above state
$\ket{i}$. Under the above restriction to the potential, if there is
only one particle in the state, then the probability density
coincides with the charge density $\rho(x)$ given in
(\ref{scalardensity}). The position expectation value is then given
by
\begin{equation}
\langle {\bf x} \rangle = \int d^3 {\bf x}\, {\bf x} \langle \rho (t,{\bf x'})
\rangle \,.
\end{equation}
The expectation value of the charge, and thus probability, density
for the initial state $\ket{i}$ is as follows
\begin{align}
\bra{i} \rho (t,{\bf x}) \ket{i} &= \bra{i} \f{i}{\hbar} :
\varphi^\dagger \stackrel{\leftrightarrow}{\partial_t} \varphi :
\ket{i} \nol
&=\int\frac{d^3{\bf p}'}{\sqrt{2p_0'}(2\pi\hbar)^3}\frac{d^3{\bf
p}}{\sqrt{2p_0}(2\pi\hbar)^3}\bra{0}   \nol & \times f^*({\bf p}')
A({\bf p}')
\f{i}{\hbar} :
\varphi^\dagger \stackrel{\leftrightarrow}{\partial_t} \varphi :  f({\bf p})A^\dagger({\bf p}) \ket{0}
\nol
&= \f{i}{\hbar} \hbar^2 \int\frac{d^3{\bf
p}'}{\sqrt{2p_0'}(2\pi\hbar)^3}\frac{d^3{\bf
p}''}{2p_0''(2\pi\hbar)^3}\frac{d^3{\bf
p}'''}{2p_0'''(2\pi\hbar)^3}\frac{d^3{\bf
p}}{\sqrt{2p_0}(2\pi\hbar)^3}
\nol
& \times
\langle 0| f^*({\bf p}') A({\bf p}')
\left[ A^\dagger({\bf p}'')\Phi_{{\bf p}''}^\dagger({\bf x},t)
 A({\bf p}''')\partial_t
\Phi_{{\bf p}'''}({\bf x},t)  \right. \nol & \left. -
 A^\dagger({\bf
p}'')\partial_t\Phi_{{\bf p}''}^\dagger({\bf x},t) A({\bf
p}''')\Phi_{{\bf p}'''}({\bf x},t)
\right]  f({\bf p})A^\dagger({\bf p})|0\rangle  \,, \label{rhoi1}
\end{align}
where we have made use of the lack of the pair creation to remove
the $B({\bf p})$ antiparticle creation/annihilation operators.
Proceeding, using the commutation relations set out in
(\ref{scalarcommutation}):
\begin{align}
\bra{i} \rho (t,{\bf x}) \ket{i} & = i\hbar \int\frac{d^3{\bf p}'}{\sqrt{2p_0'}(2\pi\hbar)^3}
\frac{d^3{\bf p}''}{2p_0''(2\pi\hbar)^3}
\frac{d^3{\bf p}'''}{2p_0'''(2\pi\hbar)^3}
 \frac{d^3{\bf p}}{\sqrt{2p_0}(2\pi\hbar)^3}\nol
& \times f^*({\bf p}') f({\bf p}) 2p''_0(2\pi\hbar)^3\delta({\bf
p}'-{\bf p}'') 2p'''_0(2\pi\hbar)^3\delta({\bf p}'''-{\bf p})\nol &
\times
\left[\Phi_{{\bf p}''}({\bf x},t)\partial_t
\Phi_{{\bf p}'''}({\bf x},t) - \partial_t\Phi_{{\bf p}''}({\bf
x},t)\cdot \Phi_{{\bf p}'''}({\bf x},t)\right]\nol & = i\hbar
\int\frac{d^3{\bf p}'}{\sqrt{2p_0'}(2\pi\hbar)^3}
 \int\frac{d^3{\bf p}}{\sqrt{2p_0}(2\pi\hbar)^3}
f^*({\bf p}') f({\bf p})\nol & \times \left[\Phi_{{\bf
p}'}^\dagger({\bf x},t)\partial_t \Phi_{{\bf p}}({\bf x},t) -
\partial_t\Phi_{{\bf p}'}^\dagger({\bf x},t)\cdot \Phi_{{\bf p}}({\bf
x},t)\right] \,. \label{rhoi2}
\end{align}
We are interested in making this measurement far into the
post-acceleration region ${\cal M}_+$, where we note that as
previously described we may write the mode functions as the
plane-wave $\Phi_{\bf p}(x) = e^{-ip\cdot x/\hbar}$. Hence, in terms
of the time-dependence of the mode function we have $\Phi_{\bf p}(x)
\propto e^{-ip_0t/\hbar}$. Substitution and a brief rearrangement
yield
\begin{align}
& \left. \bra{i} \rho (t,{\bf x}) \ket{i} \right|_{{\cal M}_+} \nol
&= \frac{1}{2} \int\frac{d^3{\bf p}'}{(2\pi\hbar)^3}
 \int\frac{d^3{\bf p}}{(2\pi\hbar)^3}
f^*({\bf p}') f({\bf p}) \left( \sqrt{\frac{p_0}{p_0'}} +
\sqrt{\frac{p_0'}{p_0}}\right)\Phi_{{\bf p}'}^*({\bf x},t)
\Phi_{{\bf p}}({\bf x},t) \,.
\end{align}
Recall that for the sake of simplicity we have defined our
coordinate system such that we shall be taking our measurements at
time $t=0$. Using the plane wave mode function, the position
expectation value in the direction $x^i(t)$ evaluated at $t=0$ is
\begin{align}
\langle x^i(0) \rangle & =  \frac{1}{2} \int d^3{\bf
x}\int\frac{d^3{\bf p}'}{(2\pi\hbar)^3}
 \int\frac{d^3{\bf p}}{(2\pi\hbar)^3}
f^*({\bf p}') f({\bf p}) \left( \sqrt{\frac{p_0}{p_0'}} +
\sqrt{\frac{p_0'}{p_0}}\right)x^i e^{i({\bf p}-{\bf p}')\cdot {\bf
x}/\hbar} \nol & =  - \frac{i}{2} \int\frac{d^3{\bf
p}'}{(2\pi\hbar)^3}
 \int\frac{d^3{\bf p}}{(2\pi\hbar)^3}
f^*({\bf p}') f({\bf p}) \left( \sqrt{\frac{p_0}{p_0'}} +
\sqrt{\frac{p_0'}{p_0}}\right) \nol &  \qquad \times \int d^3{\bf x} \hbar\partial_{p_i}
e^{i({\bf p}-{\bf p}')\cdot {\bf x}/\hbar} \,.
\end{align}
Integration of $p_i$ by parts and integration over ${\bf x}$
produces
\begin{align}
\langle x^i(0) \rangle & = \frac{i\hbar}{2}
 \int\frac{d^3{\bf p}'}{(2\pi\hbar)^3}
 \int\frac{d^3{\bf p}}{(2\pi\hbar)^3}
f^*({\bf p}') \frac{\partial\ }{\partial p_i}\left[f({\bf p}) \left(
\sqrt{\frac{p_0}{p_0'}} + \sqrt{\frac{p_0'}{p_0}}\right)\right]
\nol & \qquad \times (2\pi\hbar)^3\delta({\bf p}-{\bf p}')\nol
& =  \frac{i\hbar}{2}
 \int\frac{d^3{\bf p}}{(2\pi\hbar)^3}
 \int\frac{d^3{\bf p'}}{(2\pi\hbar)^3}(2\pi\hbar)^3\delta({\bf
p}'-{\bf p})
\nol
& \qquad \times f^*({\bf p}') \left[\frac{\partial\ }{\partial p_i}
f({\bf p})
\left( \sqrt{\frac{p_0}{p_0'}} + \sqrt{\frac{p_0'}{p_0}}\right) +
\f{f({\bf p})}{2p_0} \left( \frac{p_i}{\sqrt{p_0p_0'}} -
p_i\sqrt{\frac{p_0'}{p_0^3}}\right)\right] \nol & =
i\hbar\int\frac{d^3{\bf p}}{(2\pi\hbar)^3} f^*({\bf p})
\frac{\partial\ }{\partial p_i} f({\bf p})\nol & =
\frac{i\hbar}{2}
\int\frac{d^3{\bf p}}{(2\pi\hbar)^3} f^*({\bf p})
\stackrel{\leftrightarrow}{\partial}_{p_i} f({\bf p}) \,.
\label{nonradpev}
\end{align}
Due to the fact that the position expectation value is real, the
last line takes the real part of the previous one, thus restoring
some symmetry to the expression which was lost by the choice of
taking the derivative with respect to $p_i$, as opposed to $p_i'$,
of the exponential earlier. We recall that we have chosen to use the
remaining freedom in the choice of coordinate system to arrange the
wave packet $f({\bf p})$ such that the position expectation value
(\ref{nonradpev}) is equal to $0$, viz
\begin{equation}
\bra{i} x^i(0) \ket{i} = \frac{i\hbar}{2}
\int\frac{d^3{\bf p}}{(2\pi\hbar)^3} f^*({\bf p})
\stackrel{\leftrightarrow}{\partial}_{p_i} f({\bf p}) = 0 \quad \forall
i=1,2,3 \,. \label{posexpi}
\end{equation}
This formula henceforth represents the control against which the
position expectation value of the realistic particle whose state has
evolved through radiation reaction interactions can be compared. We
now duly turn our attention to this evolution.

\section{Final interacting state}

The `interacting' particle enters from ${\cal M}_-$ with the same
initial state as before, namely $\ket{i}$ with the wave packet
peaked about some initial momentum ${\bf p}_I$. During the
accelerations caused by the potential $V$ in the region ${\cal
M}_I$, the particle, unlike the previous case, is coupled to and
interacts with the electromagnetic field. This interaction results
in the possible emission of electromagnetic radiation and in
radiation reaction effects. Including such interactions to ${\cal
O}(e^2)$ in ${\cal M}_I$, the final out state in ${\cal M}_+$ can be
either a scalar particle, or a scalar particle and a photon. We
designate these two situations the zero-photon and one-photon
sectors respectively. In the one-photon sector, the probability
amplitude of the emission we, unsurprisingly, call the emission
amplitude. The zero-photon sector includes the possibility that the
particle does not interact with the electromagnetic field at all but
also the one-loop process, the amplitude of which we refer to as the
forward scattering amplitude. The Feynman diagrams representing the
one loop and emission interactions are presented in Figs.
\ref{oneloopfeynman} and \ref{emissionfeynman}.
\begin{figure}
\begin{center}
\begin{pspicture}(-4,-1)(4,2.5)
\psline[linewidth=0.5mm,linestyle=dashed]{->}(-3.5,0)(-3,0)
\psline[linewidth=0.5mm,linestyle=dashed]{->}(-3,0)(0,0)
\psline[linewidth=0.5mm,linestyle=dashed]{->}(0,0)(3,0)
\psline[linewidth=0.5mm,linestyle=dashed](3,0)(3.5,0)
\pscurve[linewidth=0.5mm]{*-*}(-2,0)(-2,.5)(-1.5,.5)%
(-1.4,1.2)(-.9,1.1)(-.8,1.6)(-.3,1.5)(0,2)(.3,1.5)(.8,1.6)(.9,1.1)%
(1.4,1.2)(1.5,.5)(2,.5)(2,0)
\rput(-3,-.5){${\bf p}$}
\rput(0,-.5){${\bf p}-\hbar{\bf k}$}
\rput(3,-.5){${\bf p}$}
\rput(0,2.4){$\hbar{\bf k}$}
\end{pspicture}
\caption{The one-loop diagram contributing to the
forward-scattering amplitude: the dashed and wavy lines represent
the scalar and photon propagators, respectively.}
\label{oneloopfeynman}
\end{center}
\end{figure}
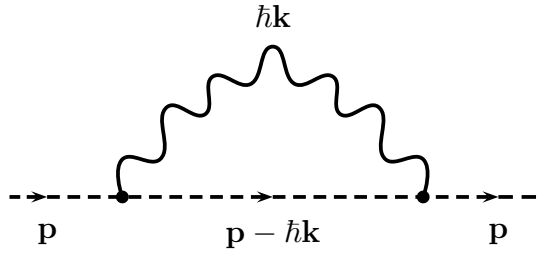
\begin{figure}
\begin{center}
\begin{pspicture}(-4,-1)(4,2.5)
\psline[linewidth=0.5mm,linestyle=dashed]{->}(-3.5,0)(-2,0)
\psline[linewidth=0.5mm,linestyle=dashed]{->}(-2,0)(0,0)
\psline[linewidth=0.5mm,linestyle=dashed]{->}(0,0)(2,0)
\psline[linewidth=0.5mm,linestyle=dashed]{->}(2,0)(3.5,0)
\pscurve[linewidth=0.5mm]{*-}(0,0)(0,0.5)(0.5,0.5)(0.5,1.0)(1.0,1.0)
(1.0,1.5)(1.5,1.5)(1.5,2.0)(2.0,2.0)
\rput(-2,-.5){${\bf p}$}
\rput(2,-.5){${\bf P}$}
\rput(0,1.5){$\hbar{\bf k}$}
\end{pspicture}
\caption{The one-photon emission diagram contributing to the
emission amplitude: the dashed and wavy lines represent the scalar
and photon propagators, respectively.}
\label{emissionfeynman}
\end{center}
\end{figure}
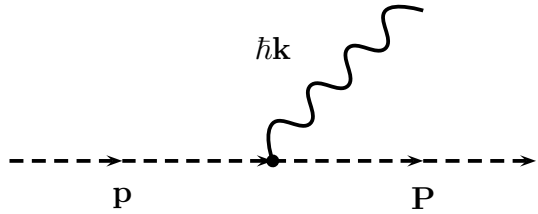
The two components of the final state can therefore be written (up
to ${\cal O}(e^2)$) as
\begin{align}
\ket{f}_{\rm for} &=  \int \f{d^3{\bf p}}{(2\pi\hbar)^3 \sqrt{2p_0}}
\[ 1+i{\cal F}({\bf p}) \] f({\bf p}) A^\dagger({\bf p}) \ket{0} \label{sffor} \\
\ket{f}_{\rm em} &=\f{i}{\hbar}
\int \f{d^3{\bf p}}{(2\pi\hbar)^3 \sqrt{2p_0}} \int \f{d^3{\bf k}}{2k_0(2\pi)^3} {\cal A}^\mu({\bf p},{\bf k})
a^\dagger_\mu({\bf k})f({\bf p})A^\dagger({\bf P})\ket{0} \,,
\label{sfem}
\end{align}
where ${\cal F}({\bf p})$ is the forward scattering amplitude and
${\cal A}^\mu({\bf p},{\bf k})$ the emission amplitude. The reader
is urged to note that the momentum of the final scalar particle
differs between the two terms due to the energy-momentum carried
away by the photon. In the case of a time-dependent potential, we
have conservation of momentum and thus ${\bf P}={\bf p}-\hbar{\bf
k}$. If the potential is dependent on one of the spatial directions
only, $x^3$ say, then ${\bf P}$ is determined by a combination of
energy conservation $\sqrt{{\bf p}^2+m^2} = \sqrt{{\bf P}+m^2}+\hbar
k$ and transverse-momentum conservation $p^i = P^i+\hbar k^i \,
(i=1,2)$.

Comparing these final states with the form of the initial state,
where the distribution $f({\bf p})$ was regarded as the one-particle
wave function in the momentum representation, let us define
\begin{align}\label{Fdefn}
F({\bf p}) & \equiv \left[ 1 + i{\cal F}({\bf p})
\right]f({\bf p})  \, \\
G^\mu({\bf p},{\bf k}) & \equiv {\cal A}^\mu({\bf p},{\bf k})f({\bf
p})\,. \label{Gdefn}
\end{align}
One can then heuristically regard the function $F({\bf p})$ as the
one-particle wave function in the zero-photon sector in the ${\bf
p}$-representation and the function $G^\mu({\bf p},{\bf k})$ as that
in the one-photon sector with a photon with momentum $\hbar {\bf k}$
in the ${\bf P}$-representation.

The full final state is simply the sum of the above $\ket{f} =
\ket{f}_{\rm for} +
\ket{f}_{\rm em}$.
The actual calculation of the forward scattering and emission
amplitudes will need to be completed using the mode functions of the
field in the region ${\cal M}_I$ and will depend upon the
circumstances there. We shall return to these calculations later
using the semiclassical approximation for the mode functions. In the
meantime we can obtain more general expressions for the position
expectation value of the final state in terms of these two
amplitudes. Later calculations of the amplitudes can then be
substituted into the position formulae. We proceed in much the same
way that we approached the position of the state $\ket{i}$. As there
is no cross term between the two states (\ref{sffor}) and
(\ref{sfem}), the final state density is the sum of the densities of
the above states.

\subsection{Zero Photon sector}
The final state density of the zero photon sector resulting from the
forward scattering is given by
\begin{align}
 _{\rm for}\bra{f} \rho(x) \ket{f}_{\rm for}  & = \int \f{d^3{\bf p'}}{(2\pi\hbar)^3 \sqrt{2p_0'}}
\f{d^3{\bf p}}{(2\pi\hbar)^3 \sqrt{2p_0}}
 \bra{0}
\[ 1-i{\cal F}^*({\bf p'}) \] f^*({\bf p'}) A^\dagger({\bf p'}) \nol
& \times
\f{i}{\hbar} :\varphi^\dagger
\stackrel{\leftrightarrow}{\partial_t} \varphi :
\[ 1+i{\cal F}({\bf p}) \] f({\bf p}) A^\dagger({\bf p}) \ket{0}
\,.
\end{align}
Comparison with the calculation pertaining to
$\ket{i}$~\footnote{See the first lines of (\ref{rhoi1}).} shows
that we have an identical situation after the substitution $f({\bf
p})
\to \[ 1+i{\cal F}({\bf p}) \] f({\bf p})$. As a consequence we can
write the position expectation value of the state $\ket{f}_{\rm
for}$ at time $t=0$ as
\begin{equation}
\langle x^i (0) \rangle_{\rm for} = \frac{i\hbar}{2}
\int\frac{d^3{\bf p}}{(2\pi\hbar)^3} \(\[ 1-i{\cal F}^*({\bf p}) \]f^*({\bf
p})\)
\stackrel{\leftrightarrow}{\partial}_{p_i} \(\[ 1+i{\cal F}({\bf p}) \] f({\bf
p})\) \,. \label{sfpevF}
\end{equation}
One could therefore regard this state as analogous to the initial
`free' state $\ket{i}$, but with the wave packet distribution
$F({\bf p}) \equiv \[ 1+i{\cal F}({\bf p}) \] f({\bf p})$, as
opposed to $f({\bf p})$. We note, however, that although $f({\bf
p})$ was arranged such that the position expectation value passed
through the origin, there is no reason to think the same would be
true of $F({\bf p})$. We expand out (\ref{sfpevF}) to order $e^2$.
${\cal F}$ is of order $e^2$ already and thus we ignore terms at
second order in the forward scattering.
\begin{align}
\langle x^i (0) \rangle_{\rm for} &=  \frac{i\hbar}{2}
\int\frac{d^3{\bf p}}{(2\pi\hbar)^3} f^*({\bf p})
\stackrel{\leftrightarrow}{\partial}_{p_i} f({\bf p}) \left\{ 1 +
i{\cal F}({\bf p}) - i{\cal F}^*({\bf p}) \right \} \nol & +
\frac{i\hbar}{2}
\int\frac{d^3{\bf p}}{(2\pi\hbar)^3} f^*({\bf p})f({\bf p})
\partial_{p_i} \[ i{\cal F}+i{\cal F}^* \] \nol
&= \frac{i\hbar}{2}
\int\frac{d^3{\bf p}}{(2\pi\hbar)^3} \( f^*({\bf p})
\stackrel{\leftrightarrow}{\partial}_{p_i} f({\bf p})\) \[ 1-2\Im
{\cal F}({\bf p}) \] \nol & -\hbar
\int\frac{d^3{\bf p}}{(2\pi\hbar)^3} |f({\bf p})|^2
\partial_{p_i} \Re {\cal F} \,. \label{posexpfor}
\end{align}
The reader will undoubtedly have noticed an expression similar to
the position expectation value for $\ket{i}$ present in the above.
We shall obviously return to this. However, before doing so it will
be advantageous to obtain a similar expression for the one photon
sector, with which we now proceed.

\subsection{One photon sector}
In common with the treatment of the forward scattering, we define
the position expectation value for the one photon sector of the
final state as
\begin{align}
 _{\rm em}\bra{f} \rho(x) \ket{f}_{\rm em}   & =-\f{i}{\hbar}\f{i}{\hbar}\int \f{d^3{\bf p'}}{(2\pi\hbar)^3 \sqrt{2p_0'}}
 \f{d^3{\bf p}}{(2\pi\hbar)^3 \sqrt{2p_0}}
 \f{d^3{\bf k'}}{2k_0'(2\pi)^3}  \f{d^3{\bf k}}{2k_0(2\pi)^3}
 \nol & \bra{0}
 A({\bf P'})a_\nu({\bf k'}){\cal A}^{\nu *}({\bf p'},{\bf k'})
f^*({\bf p'})
   \rho(x)  {\cal A}^\mu({\bf p},{\bf k})
f({\bf p})a^\dagger_\mu({\bf k})A^\dagger({\bf P})\ket{0} \,,
\end{align}
with ${\bf P'}$ defined analogously to ${\bf P}$. \footnote{i.e.
using the same conservation relations albeit with primed variables
instead.} Using the commutation relations for the electromagnetic
field, which we recall are given by
\begin{equation}\tag{\ref{EMcommutation}}
\left[ a_\mu({\bf k}),a^\dagger_\nu({\bf k}')\right] =
-g_{\mu\nu}(2\pi)^32\hbar k\delta^3({\bf k}-{\bf k}') \,,
\end{equation}
we can write
\begin{align}
\langle \rho \rangle_{\rm em} &= -\hbar \int \f{d^3{\bf k}}{2k_0(2\pi)^3}
\bra{0} C_\mu({\bf k}) \rho C^{\mu\dagger}({\bf k}) \ket{0} \,,
\label{rhoem1}
\end{align}
where
\begin{align}
C^{\mu\dagger}({\bf k}) \equiv \f{i}{\hbar} \int \f{d^3{\bf
p}}{\sqrt{2p_0}(2\pi\hbar)^3}  {\cal A}^\mu ({\bf p},{\bf k})f({\bf
p}) A^\dagger({\bf P}) \,. \label{Cmudefn}
\end{align}
In this form it is easier to see the similarities once again present
between the current calculation and that for the initial state. A
complication is the difference between ${\bf p}$ and ${\bf P}$,
which we shall now address.

\subsubsection{Space-dependent potential}
In the case of the potential dependent on one of the spatial
coordinates\footnote{The space-dependent case is more complicated
than the time dependent case. In this particular calculation it
turns out to be more advantageous to perform the complicated version
first.} - let us choose this coordinate to be $x^a$ - we note that
formally we have the transformation $d^3 {\bf p} =
\(\partial p_a / \partial P_a \)d^3 {\bf P}$. We denote the Jacobian $J_a=\(\partial p_a / \partial P_a \)$
and stress that this is not a sum as $a$ represents one specific
coordinate. We may rewrite $C^{\mu\dagger}$ as
\begin{align}
C^{\mu\dagger}({\bf k}) = \f{i}{\hbar}\int \f{d^3{\bf
P}}{\sqrt{2P_0}(2\pi\hbar)^3}  {\cal A}^\mu ({\bf p},{\bf k})f({\bf
p}) A^\dagger({\bf P}) \sqrt{\f{P_0}{p_0}}J_a \,.
\end{align}
Now, defining
\begin{align}
g^\mu\({\bf P},{\bf k}\) \equiv \f{i}{\hbar} {\cal A}^\mu \({\bf
p},{\bf k}\) f({\bf p}) \sqrt{\f{P_0}{p_0}} J_a \,,
\end{align}
we can rewrite (\ref{rhoem1}) as
\begin{align}
\langle \rho \rangle_{\rm em} &= -\hbar \int \f{d^3{\bf k}}{2k_0(2\pi)^3}
\nol &  \bra{0} \int \f{d^3{\bf P'}}{\sqrt{2P_0'}(2\pi\hbar)^3} g_\mu^*({\bf
P'},{\bf k}) A({\bf P'}) \rho(x) \int \f{d^3{\bf
P}}{\sqrt{2P_0}(2\pi\hbar)^3} g^\mu({\bf P},{\bf k}) A^\dagger({\bf
P}) \ket{0} \,.
\end{align}
Comparison with (\ref{rhoi1}) is now clear. The calculation for
$\ket{i}$ may then be followed for the position expectation value of
the state $\ket{f}_{\rm em}$ to give
\begin{align}
\langle x^i \rangle_{\rm em} &= -\f{i\hbar^2}{2} \int \f{d^3{\bf
k}}{2k_0(2\pi)^3} \f{d^3{\bf P}}{(2\pi\hbar)^3} \( g_\mu^*({\bf
P},{\bf k})
\stackrel{\leftrightarrow}{\partial}_{P_i} g^\mu({\bf P},{\bf k})
\) \,.
\end{align}
Returning to our original notation and using the symmetry of the
derivative operator $\stackrel{\leftrightarrow}{\partial}_{P_i}$, we
find
\begin{align}
\langle x^i \rangle_{\rm em} &= -\f{i}{2} \int \f{d^3{\bf
k}}{2k_0(2\pi)^3} \f{d^3{\bf P}}{(2\pi\hbar)^3} \nol & \qquad \times
\( {\cal A}_\mu^*({\bf p},{\bf k})f^*({\bf p})
\stackrel{\leftrightarrow}{\partial}_{P_i} {\cal A}^\mu({\bf p},{\bf k}) f({\bf p})\)
\f{P_0}{p_0}J_a^2 \,.
\end{align}
Converting the integration variable back from ${\bf P}$ to ${\bf p}$
and changing the $P_i$-derivative to a $p_i$-derivative we produce
\begin{align}
\langle x^i (0) \rangle_{\rm em} &= -\f{i}{2} \int \f{d^3{\bf
k}}{2k_0(2\pi)^3} \f{d^3{\bf p}}{(2\pi\hbar)^3} \nol & \qquad \times
\( {\cal A}_\mu^*({\bf p},{\bf k})f^*({\bf p})
\stackrel{\leftrightarrow}{\partial}_{p_i} {\cal A}^\mu({\bf p},{\bf k}) f({\bf p})\)
\f{P_0}{p_0}J_a\f{\partial p_i}{\partial
P_i} \,.
\end{align}
Separating out the emission amplitude and wave packet distribution
terms, the position expectation value for the emission state is
given by two terms:
\begin{align}
\langle x^i (0) \rangle_{\rm em}  & =
-\frac{i}{2} \int \frac{d^3{\bf p}}{(2\pi\hbar)^3} f^*({\bf
p})\stackrel{\leftrightarrow}{\partial}_{p_i} f({\bf p})
 \nol & \qquad \times \int\frac{d^3{\bf k}}{2k(2\pi)^3} {\cal A}^{\mu
*}({\bf p},{\bf k}){\cal A}_\mu({\bf p},{\bf k})
 \f{P_0}{p_0}J_a\f{\partial p_i}{\partial P_i}
\nol & \quad -
\frac{i}{2} \int \frac{d^3{\bf p}}{(2\pi\hbar)^3} | f({\bf
p})|^2
\int \frac{d^3{\bf k}}{2k(2\pi)^3}
{\cal A}^{\mu *}({\bf p},{\bf k})
\stackrel{\leftrightarrow}{\partial}_{p_i} {\cal A}_\mu({\bf p},{\bf
k}) \,. \label{posexpem}
\end{align}
We have dropped the factor $\(P_0/p_0\)J_a\(\partial p_i/\partial
P_i\)$ from the second term for the following reasons: The emission
amplitude ${\cal A}^\mu$ and its $p_i$-derivative are
both\footnote{This is demonstrated in section
\ref{scalaremissioncalcsection} in the calculation of the emission amplitude using the
semiclassical approximation (given in (\ref{amplitude}) or
(\ref{emamp})).} of order ${\hbar}^0$, and thus so is the second
term in the above expression as it is now written. To order
$\hbar^0$, $P_0=p_0$ and $\partial p_i / \partial P_i = 1$ for all
$i=1,2,3$ (including $a$), hence $J_a=1$ to lowest order.
Consequently we may replace these terms with unity at this order.
However, one needs to keep these factors in the first term of
(\ref{posexpem}), which is of order $\hbar^{-1}$.\footnote{This can
be seen from the previous footnote, on the emission amplitude, and
by comparison with the expression in (\ref{posexpi}).}

\subsubsection{Time-dependent potential}
Returning to the expression for the charge density/probability
density expectation value (\ref{rhoem1}), we proceed with the
simpler case of the time-dependent potential. The definition of
$C^{\mu\dagger}$ from (\ref{Cmudefn}) still holds, which we repeat
as
\begin{align}
C^{\mu\dagger}({\bf k}) \equiv \f{i}{\hbar} \int \f{d^3{\bf
p}}{\sqrt{2p_0}(2\pi\hbar)^3}  {\cal A}^\mu ({\bf p},{\bf k})f({\bf
p}) A^\dagger({\bf P}) \,. \tag{\ref{Cmudefn}}
\end{align}
The time-dependent potential is simpler because the conservation of
momentum means that we can write ${\bf P}={\bf p}-\hbar{\bf k}$.
Consequently we have $d^3{\bf P}=d^3{\bf p}$. In connection with the
previous workings we can write
\begin{align}
C^{\mu\dagger}({\bf k}) \equiv \int \f{d^3{\bf
P}}{\sqrt{2P_0}(2\pi\hbar)^3} g_t^\mu({\bf P},{\bf k})
A^\dagger({\bf P}) \,,
\end{align}
where this time the momentum distribution is given by
\begin{align}
g_t^\mu({\bf P},{\bf k}) \equiv \f{i}{\hbar}{\cal A}^\mu({\bf
p},{\bf k}) \sqrt{\f{P_0}{p_0}} \,.
\end{align}
The difference with respect to the previous case is the absence of
the $\partial p_a/\partial P_a$ factor. With reference to this we
see that we have
\begin{align}
\langle x^i \rangle_{\rm em} &= -\f{i\hbar^2}{2} \int \f{d^3{\bf
k}}{2k_0(2\pi)^3} \f{d^3{\bf P}}{(2\pi\hbar)^3} \( g_{t\mu}^*({\bf
P},{\bf k})
\stackrel{\leftrightarrow}{\partial}_{P_i} g_t^\mu({\bf P},{\bf k}) \)
\nol
&= -\f{i}{2} \int \f{d^3{\bf k}}{2k_0(2\pi)^3} \f{d^3{\bf
p}}{(2\pi\hbar)^3} \( {\cal A}_\mu^*({\bf p},{\bf k})f^*({\bf p})
\stackrel{\leftrightarrow}{\partial}_{p_i} {\cal A}^\mu({\bf p},{\bf k}) f({\bf p})\)
\f{P_0}{p_0}\f{\partial p_i}{\partial P_i}
\nol & =
-\frac{i}{2} \int \frac{d^3{\bf p}}{(2\pi\hbar)^3} f^*({\bf
p})\stackrel{\leftrightarrow}{\partial}_{p_i} f({\bf p})
 \int\frac{d^3{\bf k}}{2k(2\pi)^3} {\cal A}^{\mu
*}({\bf p},{\bf k}){\cal A}_\mu({\bf p},{\bf k})
 \f{P_0}{p_0}\f{\partial p_i}{\partial P_i}
\nol & \quad -
\frac{i}{2} \int \frac{d^3{\bf p}}{(2\pi\hbar)^3} | f({\bf
p})|^2
\int \frac{d^3{\bf k}}{2k(2\pi)^3}
{\cal A}^{\mu *}({\bf p},{\bf k})
\stackrel{\leftrightarrow}{\partial}_{p_i} {\cal A}_\mu({\bf p},{\bf
k}) \,. \label{posexpem2}
\end{align}
The previous arguments pertaining to the order of the two terms are
fully applicable here too. We see that the only difference between
the two is the removal of the $\partial p_a/\partial P_a$ type
factor when moving from $V(x^a)$ to $V(t)$.

Considering $\langle x^i \rangle_{\rm em}$ in (\ref{posexpem}) and
(\ref{posexpem2}), the ever observant reader will once again note
the similarity between the first term and the position expectation
value for $\ket{i}$. When this was noted for the forward scattering,
we delayed consideration of the factor until after the corresponding
emission calculation had been performed. We now return as promised
to consider these two terms.

\subsection{Normalisation and unitarity}
The final state $\ket{f}$ for the interacting particle is the sum of
the components $\ket{f}_{\rm em}$ and $\ket{f}_{\rm for}$. Whilst we
have already observed that there is no cross term, there is however
a connection to be made using the normalisation condition for
$\ket{f}$. Recalling the definition and normalisation of $\ket{i}$
and using the unitarity of time evolution, we find
\begin{eqnarray}
\langle f | f \rangle = 1
\end{eqnarray}
In other words, the normalisation of the two components must also
add to unity. For the forward scattering zero-photon sector, we have
\begin{align}
_{\rm for}\langle f | f \rangle_{\rm for} &= \int \f{d^3{\bf
p}}{\sqrt{2p_0}(2\pi\hbar)^3} \f{d^3{\bf
p'}}{\sqrt{2p_0'}(2\pi\hbar)^3} \nol & \qquad \bra{0} A({\bf p'})
f({\bf p'})\[ 1-i{\cal F}^*({\bf p'})\] \[ 1+i{\cal F}({\bf p})\]
f({\bf p}) A^\dagger \ket{0} \nol &= \int \f{d^3{\bf
p}}{(2\pi\hbar)^3} |f({\bf p})|^2 \[ 1-2\Im {\cal F}({\bf p})\]
+{\cal O}(e^4) \,. \label{forprob1}
\end{align}
The last term is added as a reminder that terms are taken to first
order in $e^2$. For the one-photon sector, the inner product
produces the emission probability ${\cal P}_{\rm em}$ viz
\begin{align}
_{\rm em}\langle f | f \rangle_{\rm em} &= {\cal P}_{\rm em} \nol &=
-\f{i}{\hbar}\f{i}{\hbar}\int \f{d^3{\bf p'}}{(2\pi\hbar)^3
\sqrt{2p_0'}}
 \f{d^3{\bf p}}{(2\pi\hbar)^3 \sqrt{2p_0}}
 \f{d^3{\bf k'}}{2k_0'(2\pi)^3}  \f{d^3{\bf k}}{2k_0(2\pi)^3}
 \nol & \quad \bra{0}
 A({\bf P'})a_\nu({\bf k'}){\cal A}^{\nu *}({\bf p'},{\bf k'})
f^*({\bf p'})
   \rho(x)  {\cal A}^\mu({\bf p},{\bf k})
f({\bf p})a^\dagger_\mu({\bf k})A^\dagger({\bf P})\ket{0} \nol &=
-\hbar
\int \f{d^3{\bf k}}{2k_0(2\pi)^3} \bra{0} C_\mu ({\bf k})
C^{\mu\dagger}({\bf k}) \ket{0} \,,
\end{align}
using the definition of $C^{\mu\dagger}({\bf k})$ in
(\ref{Cmudefn}). This equation then compares with (\ref{rhoem1}) and
we shall require similar manipulations of ${\bf p}$ and ${\bf P}$.
We have
\begin{multline}
{\cal P}_{\rm em} = - \frac{1}{\hbar}\int \frac{d^3{\bf
k}}{2k(2\pi)^3}
 \frac{d^3{\bf p}}{\sqrt{2p_0}(2\pi\hbar)^3} \frac{d^3{\bf
 p}'}{\sqrt{2p_0'}(2\pi\hbar)^3}\\
   f^*({\bf p'}){\cal A}_\mu^*
({\bf p'},{\bf k}) f({\bf p}){\cal A}^\mu({\bf p},{\bf k})
 \langle 0|A({\bf P'})A^\dagger({\bf P})|0\rangle \,.
 \end{multline}
In order to perform the ${\bf p'}$ integration in the case of the
potential dependent on spatial direction $x^a$, we note that from
the commutation relations we have
\begin{align}
\langle 0|A({\bf P'})A^\dagger({\bf P})|0\rangle &= 2P_0
(2\pi\hbar)^3 \delta^3 ({\bf P}-{\bf P'}) \nol &= 2p_0 (2\pi\hbar)^3
\delta^3 ({\bf p}-{\bf p'}) \f{P_0}{p_0} \f{\partial p_a}{\partial
P_a} \,.
\end{align}
Thus the emission probability can be written
\begin{align}
{\cal P}_{\rm em} = -\f{1}{\hbar} \int \f{d^3{\bf p}}{(2\pi\hbar)^3}
|f({\bf p})|^2 \int \f{d^3{\bf k}}{2k_0(2\pi)^3} {\cal A}_\mu^*
({\bf p},{\bf k}){\cal A}^\mu({\bf p},{\bf k})\f{P_0}{p_0}
\f{\partial p_a}{\partial P_a} \,. \label{emprob1}
\end{align}
Given that (\ref{emprob1}) and (\ref{forprob1}) must sum to unity
for all $f({\bf p})$, and recalling the normalisation\footnote{The
expression for the normalisation $\langle i | i \rangle = 1$ in
(\ref{inorm}).} of $f({\bf p})$, we must have
\begin{align}
2\Im {\cal F}({\bf p}) = -\f{1}{\hbar} \int \f{d^3{\bf
k}}{2k_0(2\pi)^3} {\cal A}_\mu^* ({\bf p},{\bf k}){\cal A}^\mu({\bf
p},{\bf k})\f{P_0}{p_0}
\f{\partial p_a}{\partial P_a} \,. \label{unitarity1}
\end{align}

Via a similar argument presented in the derivation of $\langle x^i
\rangle_{\rm em}$ in (\ref{posexpem}) and (\ref{posexpem2}), we note
that for the $t$-dependent potential, the appropriate expressions
are
\begin{align}
{\cal P}_{\rm em} = -\f{1}{\hbar} \int \f{d^3{\bf p}}{(2\pi\hbar)^3}
|f({\bf p})|^2 \int \f{d^3{\bf k}}{2k_0(2\pi)^3} {\cal A}_\mu^*
({\bf p},{\bf k}){\cal A}^\mu({\bf p},{\bf k})\f{P_0}{p_0} \,,
\label{emprob2}
\end{align}
and
\begin{align} 2\Im {\cal F}({\bf p}) = -\f{1}{\hbar} \int
\f{d^3{\bf k}}{2k_0(2\pi)^3} {\cal A}_\mu^* ({\bf p},{\bf k}){\cal
A}^\mu({\bf p},{\bf k})\f{P_0}{p_0}
\,. \label{unitarity2}
\end{align}

\subsection{Position expectation value}

Combining the position expectation values from the two components of
the final state as given in (\ref{posexpfor}) and (\ref{posexpem})
we have (using the subscript $f$ to denote the full final state),
\begin{align}
\langle x^i (0) \rangle_{\rm f} &= \frac{i\hbar}{2}
\int\frac{d^3{\bf p}}{(2\pi\hbar)^3} \( f^*({\bf p})
\stackrel{\leftrightarrow}{\partial}_{p_i} f({\bf p})\) \[ 1-2\Im
{\cal F}({\bf p}) \] \nol & \quad -\hbar
\int\frac{d^3{\bf p}}{(2\pi\hbar)^3} |f({\bf p})|^2
\partial_{p_i} \Re {\cal F} \nol
& \quad -\frac{i}{2} \int \frac{d^3{\bf p}}{(2\pi\hbar)^3} f^*({\bf
p})\stackrel{\leftrightarrow}{\partial}_{p_i} f({\bf p})
 \int\frac{d^3{\bf k}}{2k(2\pi)^3} {\cal A}^{\mu
*}({\bf p},{\bf k}){\cal A}_\mu({\bf p},{\bf k})
 \f{P_0}{p_0}\f{\partial p_a}{\partial P_a}\f{\partial p_i}{\partial P_i}
\nol & \quad -
\frac{i}{2} \int \frac{d^3{\bf p}}{(2\pi\hbar)^3} | f({\bf
p})|^2
\int \frac{d^3{\bf k}}{2k(2\pi)^3}
{\cal A}^{\mu *}({\bf p},{\bf k})
\stackrel{\leftrightarrow}{\partial}_{p_i} {\cal A}_\mu({\bf p},{\bf
k}) \,,
\end{align}
for which we are using the space-dependent potential expressions.
The relation in (\ref{unitarity1}) can be used to eliminate the
imaginary part of the forward scattering. The result can be written
in three terms:
\begin{align}
& \langle x^i (0) \rangle_{\rm f} \nol &= \frac{i\hbar}{2}
\int\frac{d^3{\bf p}}{(2\pi\hbar)^3} \( f^*({\bf p})
\stackrel{\leftrightarrow}{\partial}_{p_i} f({\bf p})\) \nol
& \quad +  \int \frac{d^3{\bf p}}{(2\pi\hbar)^3} | f({\bf p})|^2 \[
-\hbar \partial_{p_i} \Re {\cal F} -
\frac{i}{2}
\int \frac{d^3{\bf k}}{2k(2\pi)^3}
{\cal A}^{\mu *}({\bf p},{\bf k})
\stackrel{\leftrightarrow}{\partial}_{p_i} {\cal A}_\mu({\bf p},{\bf
k}) \]
\nol
& \quad - \frac{i}{2}
\int\frac{d^3{\bf p}}{(2\pi\hbar)^3} \( f^*({\bf p})
\stackrel{\leftrightarrow}{\partial}_{p_i} f({\bf p})\)
\nol & \qquad \times \int\frac{d^3{\bf k}}{2k(2\pi)^3} {\cal A}^{\mu
*}({\bf p},{\bf k}){\cal A}_\mu({\bf p},{\bf k})
 \f{P_0}{p_0}\f{\partial p_a}{\partial P_a}\( \f{\partial p_i}{\partial
 P_i}-1\) \,.
 \end{align}
Now our task is to interpret these terms. The first term can be
recognized as the position expectation value of the non-interacting
state $\ket{i}$, which is written $\langle x^i(0)\rangle_{\rm i}$.
As this is our control particle, from the definition of the position
shift as the change in position due to radiation reaction effects,
we conclude that the position shift is the sum of the second and
third terms. The reader will recall that the function $f({\bf p})$
was defined to be sharply peaked about the final momentum. Calling
this final momentum ${\bf p}$ for simplicity\footnote{This is of
course equivalent to the momentum being peaked about $\tilde{\bf p}$
say, followed by a change of variables $\tilde{\bf p}\to{\bf p}$.},
we find that in the $\hbar\to 0$ limit, these two terms are given by
\begin{align}\label{deltaxQ1}
\delta x^i_{\rm Q1} &=  -\hbar \partial_{p_i} \Re {\cal F} -
\frac{i}{2}
\int \frac{d^3{\bf k}}{2k(2\pi)^3}
{\cal A}^{\mu *}({\bf p},{\bf k})
\stackrel{\leftrightarrow}{\partial}_{p_i} {\cal A}_\mu({\bf p},{\bf
k})  \,, \\
\delta x^i_{\rm Q2} &= -\frac{\langle x^i(0)\rangle_{\rm i}}{\hbar}
\int\frac{d^3{\bf k}}{2k(2\pi)^3} {\cal A}^{\mu
*}({\bf p},{\bf k}){\cal A}_\mu({\bf p},{\bf k})
\( \f{\partial p_i}{\partial
 P_i}-1\) \,. \label{deltaxQ2}
\end{align}
For $\delta x^i_{\rm Q2}$ we have again the fact that $\partial
p_i/\partial P_i - 1$ is of order $\hbar$ and consequently in this
$\hbar\to 0$ limit dropped the factor $\(P_0/p_0\)\(\partial
p_a/\partial P_a\)$. Dropping this factor also means that we may
continue with the results applying to both time and space dependent
potential cases. The quantum position shift $\delta x^i_{\rm Q}$ due
to radiation reaction of the scalar field can thus be written
\begin{align}
\delta x^i_{\rm Q} = \delta x^i_{\rm Q1} + \delta x^i_{\rm Q2} \,.
\end{align}

Now, in the choice of coordinate system that we have chosen we have
set $\langle x^i(0)\rangle_{\rm i}=0$, and thus $\delta x^i_{\rm
Q2}=0$ at this order. We have nevertheless kept this contribution up
to now in order to obtain a formula for the position expectation
value without making this assumption which would be given, in terms
of the above definitions, to order $\hbar$ by
\begin{align}
\langle x^i(0)\rangle_{\rm f} = \langle x^i(0)\rangle_{\rm i} +\delta x^i_{\rm Q1} + \delta x^i_{\rm
Q2} \,.
\end{align}
Indeed, we shall later show that $\delta x^i_{\rm Q2}$ gives the
correct correction to the position expectation value if we use our
freedom of coordinate system to choose $\langle x^i(0)\rangle_{\rm
i}\neq 0$.

Returning for now to our standard choice of coordinates, we find
that the quantum position shift is given by $\delta x^i_{\rm
Q}=\delta x^i_{\rm Q1}$ or
\begin{align}
\delta x^i_{\rm
Q} &=-\hbar \partial_{p_i} \Re {\cal F}-
\frac{i}{2}
\int \frac{d^3{\bf k}}{2k(2\pi)^3}
{\cal A}^{\mu *}({\bf p},{\bf k})
\stackrel{\leftrightarrow}{\partial}_{p_i} {\cal A}_\mu({\bf p},{\bf
k}) \,.
\end{align}
There are two contributions to this shift, coming naturally from the
forward scattering and emission contributions to the final state. In
order to analyse the quantum position shift and compare the result
with the classical position shift $\delta x^i_{\rm C}$ given in
(\ref{cshift}), we must calculate these contributions. For this
purpose firstly define
\begin{align}\label{deltaxfor}
 \delta x^i_{\rm for} &= -\hbar \partial_{p_i} \Re {\cal F}({\bf p}) \,,\\
 \delta x^i_{\rm em} &= -
\frac{i}{2}
\int \frac{d^3{\bf k}}{2k(2\pi)^3}
{\cal A}^{\mu *}({\bf p},{\bf k})
\stackrel{\leftrightarrow}{\partial}_{p_i} {\cal A}_\mu({\bf p},{\bf
k}) \,, \label{deltaxem}
\end{align}
as the quantum position shift due to forward scattering and emission
respectively.

\section{Emission Amplitude}\label{scalaremissioncalcsection}
We now come to the task of calculating the emission amplitude.
Recall that the emission amplitude was originally defined by its
presence in the one-photon sector of the final state which we repeat
here (with the original equation numbering)
\begin{align}
\ket{f}_{\rm em} &=\f{i}{\hbar}
\int \f{d^3{\bf p}}{(2\pi\hbar)^3 \sqrt{2p_0}} \int \f{d^3{\bf k}}{2k_0(2\pi)^3} {\cal A}^\mu({\bf p},{\bf k})
a^\dagger_\mu({\bf k})f({\bf p})A^\dagger({\bf P})\ket{0}
\,. \tag{\ref{sfem}}
\end{align}
Considering the form of the initial state, the above represents the
following state evolution
\begin{equation}
A^\dagger({\bf p})|0\rangle \to \cdots + \frac{i}{\hbar}
\int\frac{d^3{\bf k}}{2k(2\pi)^3} {\cal A}^{\mu}({\bf p},{\bf
k})a_\mu^\dagger({\bf k}) A^\dagger({\bf P})|0\rangle \,.
\label{evolA}
\end{equation}
As it stands this is merely a definition for ${\cal A}^\mu$. The
full first order evolution of this state in time-dependent
perturbation theory is
\begin{equation}\label{evolL}
A^\dagger({\bf p})|0\rangle \to \cdots -\frac{i}{\hbar}\int d^4 x
{\cal H}_I(x)A^\dagger({\bf p})|0\rangle\,,
\end{equation}
where ${\cal H}_I(x)$ is the interaction Hamiltonian density.
Comparing these two evolution expressions, we can write the emission
amplitude in terms of the interaction Hamiltonian density as
\begin{equation}
{\cal A}_\mu ({\bf p},{\bf k}) =
\frac{1}{\hbar}\int \frac{d^3{\bf p'}}{2p_0'(2\pi\hbar)^3} \int
d^4x\langle 0|a_\mu({\bf k})A({\bf p'}){\cal H}_I(x)A^\dagger({\bf
p})|0\rangle\,.
\end{equation}
The Hamiltonian density for scalar electrodynamics\footnote{A note
should be made at this point that we are using the normal ordering
from the free-field. The appropriate subtraction to be made is
technically the subtraction of the vacuum in the $V\to 0$ limit of
the potential. However, it can be shown that to at least order
$\hbar^2$ the difference between the methods is zero
\cite{HM6} and consequently we are justified in using the more
familiar normal ordering here.\label{normalfootnote1}} can be
obtained from the Lagrangian density by the standard method and was
given in (\ref{Hamiltonian}) and we repeat it here to aid the
reader:\footnote{Normal ordering on the $A_i A_i$ type term would
add an infinite constant term altering our definition of the mass
counter-term and would not affect the final result. The full
treatment of both the electromagnetic fields and scalar fields is
technically that noted in the previous footnote and gives the same
results as this treatment \cite{HM6}.}
\begin{equation}\tag{\ref{Hamiltonian}}
{\cal H}_I(x) = \frac{ie}{\hbar}A_\mu \normal\left[ \varphi^\dagger
D^\mu
\varphi - (D^\mu \varphi)^\dagger \varphi\right]\normal
 + \frac{e^2}{\hbar^2}\sum_{i=1}^3 A_iA_i
\normal\varphi^\dagger\varphi\normal \,.
\end{equation}
As was noted when this expression was originally introduced, it is
useful to observe the $i=1,2,3$ sum in the second term i.e. the
absence of the $A_0A_0$-type term. The first term is the contraction
between the electromagnetic field and the current of the scalar
field $J^\mu(x)$. It is this term with which we are presently
interested as it is the current-EM field coupling that produces the
photon emission process with which we may match the emission
amplitude expression. Thus, we have
\begin{align}
& \quad {\cal A}_\mu ({\bf p},{\bf k}) \nol & =
\frac{ie}{\hbar^2}\int
\frac{d^3{\bf p'}}{2p_0'(2\pi\hbar)^3} \int d^4x \langle
0|a_\mu({\bf k})A({\bf p'})\left\{ A_\nu \normal\left[
\varphi^\dagger D^\nu \varphi - (D^\nu\varphi)^\dagger
\varphi\right]\normal \right\}
A^\dagger({\bf p})|0\rangle \,.
\end{align}
By using the expansion of the fields $A_\mu$ and $\varphi$, and the
commutation relations for the annihilation and creation operators,
we readily find
\begin{align}\label{emisampPhi}
 & \quad {\cal A}_\mu ({\bf p},{\bf k}) \nol &= -ie\hbar \int \frac{d^3{\bf
p'}}{2p_0'(2\pi\hbar)^3} \int d^4x\,  e^{i k\cdot x}
 \left\{ \Phi^\dagger_{{\bf p}'}(x) D_\mu \Phi_{\bf p}(x) - \left[ D_\mu
\Phi_{{\bf p}'}(x)\right]^\dagger \Phi_{\bf p}(x) \right\} \,.
\end{align}
We may now proceed to substitute the appropriate expression for the
mode function into the above and calculate ${\cal A}^\mu$.

\subsection{Time dependent potential}
We begin by looking at the case of a time-dependent potential
$V^i(t)$ ($i=1,2,3$) with  $V^0(t) = 0$.~\footnote{If $V^0(t) \neq
0$, one is free to gauge away this component. Our choice here is for
consistency with $\partial_\mu A^\mu = 0$.} The system is
translationally invariant in the spatial directions and hence we can
let
\begin{equation}
\Phi_{{\bf p}}({\bf x},t) = \phi_{\bf p}(t)\exp\(i{\bf p}\cdot{\bf
x}/\hbar\) \,.
\label{Phiphi}
\end{equation}
The amplitude in a spatial direction for the $t$-dependent potential
is then
\begin{align}
{\cal A}^i({\bf p},{\bf k})
 &= -e\int\f{d^3{\bf p}'}{2p_0'(2\pi\hbar)^3}\int d^4x \,
 \phi^*_{\bf p'}(t)\phi_{\bf p}(t) \[p^i+p^{\prime i}
-2V^i(t)\]e^{i\[({\bf p}-{\bf p'})
\cdot {\bf x}/\hbar\]}\nol & \qquad \times e^{i(k t - {\bf k}\cdot{\bf x})}
\nol
&= -e\int dt \, e^{ikt} \phi^*_{\bf P}(t)\phi_{\bf p}(t)
\f{p^i-V^i(t)}{p_0} \,, \label{inter1}
\end{align}
with ${\bf P} = {\bf p}-\hbar {\bf k}$, where we have let $P^i + p^i
-2V^i(t) = 2[p^i-V(t)]$ as the difference $p^i - P^i$ is of order
$\hbar$.~\footnote{Later, when we consider the forward-scattering,
the semiclassical approximation for the emission probability
 is justified (\ref{imaginary}). This in turn shows the validity of the
physically reasonable assumption that a typical photon emitted has
energy of order $\hbar$.} However, it would be incorrect to equate
$\phi_{\bf P}(t)$ with $\phi_{\bf p}(t)$, because these functions
oscillate with periods of order $\hbar^{-1}$, as will be seen
shortly. For the time component we have $D_t=\d_t$ and simply obtain
\begin{equation}
{\cal A}^0({\bf p},{\bf k}) = -\f{ie\hbar}{2p_0}\int
dt\,\[\phi^*_{\bf P}\d_t\phi_{\bf p} -
 \(\d_t\phi^*_{\bf P}\)\phi_{\bf p}\] e^{ikt} \,. \label{inter2}
\end{equation}

To proceed further, we require the remaining undetermined factor
$\phi_{\bf p}(t)$ of the mode function to be approximated for the
field in the region ${\cal M}_I$ in a form suitable for taking the
$\hbar \to 0$ limit. This is of course precisely what we have in the
semiclassical approximation in (\ref{tdepwkb}). We repeat this
result here to aid the reader:
\begin{align}\label{wkbt}
\phi_{\bf p}(t) &=
 \sqrt{\f{p_0}{E_p(t)}}\exp\[-\f{i}{\hbar}\int^t_0E_p(\zeta)d\zeta
 \] \,,
\end{align}
where
\begin{equation}
E_p(t) \equiv \sqrt{|{\bf p}-{\bf V}(t)|^2+m^2} \,.
\end{equation}
We note that the local momentum and energy of the point particle
corresponding to the wave packet considered here are
\begin{align}
m\frac{d{\bf x}}{d\tau} & =  {\bf p} - {\bf V}(t) \,,\label{momentum1}\\
m\frac{dt}{d\tau} & =  E_p(t) \,.\label{energy1}
\end{align}
Now, the product of two wave functions in the emission amplitude
(\ref{inter1}) can be written
\begin{equation}
\phi^*_{\bf P}(t)\phi_{\bf p}(t) =
\f{p_0}{E_p(t)}
\exp\left\{-\f{i}{\hbar}\int^t_0
\[E_p(\zeta)-E_P(\zeta)\]d\zeta\right\} \,,
\label{phiproduct}
\end{equation}
where we have replaced $P_0$ and $E_P(t)$ in the pre-factor by $p_0$
and $E_p(t)$, respectively, due to the $\hbar\to 0$ limit.
\footnote{This is not contrary to our previous point regarding the
order of the product as the oscillation period, which is of ${\cal
O}(\hbar^{-1})$, is contained in the exponential. The pre-factor
replacements may thus be made analogously with those multiplicative
factors in (\ref{inter1}).} The integrand in the exponent can be
evaluated to lowest order in $\hbar$ by using (\ref{momentum1}) and
(\ref{energy1}) as
\begin{align}
E_p-E_P &= \f{\d E_p}{\d p^i}(P^i-p^i) \nol
 &= \f{dx^i}{dt}\hbar k^i \,, \label{wkbvel}
\end{align}
where the repeated indices $i$ are summed over. By substituting this
approximation in (\ref{phiproduct}) we find
\begin{align}
\phi^*_{\bf P}(t)\phi_{\bf p}(t) &=
\f{p_0}{E_p}\exp\(-i \int^t_0 dt\, \f{dx^i}{dt}
k^i\) \nol
 &= \f{p_0}{E_p}\exp\(-i{\bf k}\cdot {\bf x}\) \,,
\label{phipronice}
\end{align}
where we have used the fact that the particle passes through the
spacetime origin. By substituting
 this formula in (\ref{inter1}) and noting
(\ref{momentum1}) and (\ref{energy1}) we obtain
\begin{align}
{\cal A}^i({\bf p},{\bf k}) &= - e\int dt\, e^{ik\cdot
x}\f{dx^i}{dt}\nol
 &= - e \int d\xi\, \f{dx^i}{d\xi} e^{ik\xi} \,, \label{Aiai}
\end{align}
where we have defined $\xi\equiv t-{\bf n}\cdot {\bf x}$ with ${\bf
n} \equiv {\bf k}/k$\,. We emphasize that ${\bf x}$ and $\xi$ here
are functions of $t$ evaluated on the world line of the
corresponding classical particle passing through the spacetime
origin.

Let us now consider the time component ${\cal A}^0({\bf p},{\bf k})$
of the emission amplitude given by (\ref{inter2}). Note that from
the semiclassical expression (\ref{wkbt}) for $\phi_{\bf p}(t)$ we
have, to lowest order in $\hbar$,
\begin{equation}
\d_t\phi_{\bf p}(t) = -\f{i}{\hbar}E_p(t)\phi_{\bf p}(t) \,.
\end{equation}
By substituting this formula in (\ref{inter2}) we obtain
\begin{align}
{\cal A}^0({\bf p},{\bf k}) &= -\f{e}{2p_0}\int dt\, \left[E_p(t) +
E_P(t)\right]
\phi^*_{\bf P}(t)\phi_{\bf p}(t)e^{ikt} \nol
 &=-e\int dt\,e^{-i{\bf k}\cdot {\bf x}}e^{ikt} \nol
 &= -e \int d\xi\, \f{dt}{d\xi} e^{ik\xi} \,,
\end{align}
where we have let $E_P(t)=E_p(t)$ and used (\ref{phipronice}). By
combining this formula and (\ref{Aiai}) we obtain the following
concise expression for the $\hbar \to 0$ limit of the emission
amplitude:
\begin{equation}\label{amplitude}
{\cal A}^\mu({\bf p},{\bf k}) = -e \int d\xi\, \f{dx^\mu}{d\xi}
e^{ik\xi} \,.
\end{equation}

\subsection{Space-dependent potential}
Let us now consider the case where the potential is dependent on one
of the spatial coordinates, $z$ say, although the following will
apply to $x$ and $y$ equally by symmetry. The following calculations
are very similar to the previous $t$-dependent potential case,
albeit with subtle differences in the workings. We once again start
from the equation (\ref{emisampPhi})
\begin{align}
 & \quad {\cal A}_\mu ({\bf p},{\bf k}) \nol &= -ie\hbar \int \frac{d^3{\bf
p'}}{2p_0'(2\pi\hbar)^3} \int d^4x\,  e^{i k\cdot x}
 \left[ \Phi^\dagger_{\bf p'}(x) D_\mu \Phi_{\bf p}(x) - \left(D_\mu
\Phi_{\bf p'}(x)\right)^\dagger \Phi_{\bf p}(x) \right] \,.
\tag{\ref{emisampPhi}}
\end{align}
The potential is $V^\mu(z)$, with $V^3(z) = 0$. The translational
invariance in the $t$, $x$ and $y$ directions means that the mode
function can be decomposed by
\begin{equation}
\Phi_{\bf p} = \phi_{{\bf p}} (z) e^{-\f{i}{\hbar}\(p_0t-p_xx-p_yy\) }
\,.
\end{equation}
Here let us use the notation $\perp$ to represent the $x,y$
directions. The amplitude for the $x,y$ components is
\begin{align}
{\cal A}_\perp &= e\int\f{d^3p}{2p_0'(2\pi\hbar)^3}\int d^4x \,
 \phi^*_{\bf p'}(z)\phi_{\bf p}(z) \[p_\perp+p_\perp'-2V_\perp(z)\] \nol &
 \qquad \times
 e^{i\(({\bf p}_\perp-{\bf p'}_\perp) \cdot {\bf x}_\perp\)/\hbar}e^{-i(p_0-p_0')t}e^{i(k_0 t - {\bf k}\cdot{\bf x})}
\nol
&= e\int\f{d^3p}{2p_0'(2\pi\hbar)^3}\int dz
\,\phi^*_{\bf p'}(z)\phi_{\bf p}(z)
\[p_\perp+p_\perp'-2V_\perp(z)\] e^{-ik_zz} (2\pi\hbar)^3 \nol & \qquad \times \delta(p'_0+\hbar
k_0-p_0)\delta^2(p'_\perp+\hbar k_\perp -p_\perp)
\nol
&= e\int dz \, e^{-ik_zz} \phi^*_{{\bf P}}(z)\phi_{{\bf p}}(z)
\f{\[p_\perp-V_\perp(z)\]}{p_z} \,,
\end{align}
where in the last two lines we firstly integrated over $x,y,t$ then
used the fact that $dp_z/p_0=dp_0/p_z$ before integrating out the
delta functions. For the last line, $P$ is defined via the
conservation of transverse momentum and energy represented by the
delta functions, i.e. ${\bf P}_{\perp} = {\bf p}_{\perp}-\hbar{\bf
k}_{\perp}$ and $\sqrt{{\bf P}^2+m^2} = \sqrt{{\bf p}^2+m^2}-\hbar
k_0$. Consequently, following analogous reasoning set out in the
$t$-dependent potential evaluation, we take $P_\perp + p_\perp -
2V_\perp(z) = 2\[p_\perp-V_\perp(z)\]$ as the difference is of order
$\hbar$. Once again, $\phi^*_{{\bf P}}(z)\phi_{{\bf p}}(z)$ must be
treated carefully. For the $t$ component the only difference to the
above is the covariant derivative $D_t$ gives an extra minus sign.
This gives
\begin{equation}
{\cal A}_t = -e\int dz \, e^{-ik_zz} \phi^*_{{\bf p'}}(z)\phi_{{\bf
p}}(z)
\f{\[p_0-V_t(z)\]}{p_z} \,.
\end{equation}
For the $z$ component we have $D_z=\d_z$ thus simply obtain
\begin{align}
{\cal A}_z &= -ie\hbar \int\f{d^3p}{2p_0'(2\pi\hbar)^3}\int d^4x
\,
\[\phi^*_{{\bf p'}}\d_z\phi_{{\bf p}} - \(\d_z\phi^*_{{\bf p'}}\)\phi_{{\bf p}}\]
\nol & \qquad \times
e^{-\f{i}{\hbar}\((p_0-p'_0)t-(p_x-p_x')x-(p_y-p_y')y\) }e^{i(k_0 t
- {\bf k}\cdot{\bf x})} \nol
 &= -\f{ie\hbar}{2p_z}\int dz\,\[\phi^*_{{\bf p'}}\d_z\phi_{\bf p} -
 \(\d_z\phi^*_{\bf p'}\)\phi_{\bf p}\] e^{-ik_zz} \,,
\end{align}
where we have again changed the integration from $p_z$ to $p_0$.
This expression is similar to the case ${\cal A}_0$ for $V(t)$ in
(\ref{inter2}).

We now need the semiclassical expression for the remainder of the
mode function substitution. This was found in
(\ref{scalarsemispace}) and as before, we reproduce it here to aid
the reading of this calculation.
\begin{equation}
\phi_{\bf p}(t) =
 \sqrt{\f{p_z}{\kappa_{\bf p}(z)}}\exp\[\f{i}{\hbar}\int^z_0\kappa_{\bf p}(\zeta)d\zeta
 \] \,,
\end{equation}
where
\beq
\kappa_p(z) =
\sqrt{\(p_0-V_t(t)\)^2-\(p_x-V_x(t)\)^2-\(p_y-V_y(t)\)^2-m^2} \,.
\eeq
This time we note that $(p_\perp-V_\perp(z))$,$(p_0-V_t(z))$ are the
$i$,$t$ components ($\perp=x,y$) of the momentum of the
corresponding classical particle and $\kappa_p$ is the $z$
component, i.e. $dx^\perp/d\tau$,$dt/d\tau$ and $dz/d\tau$
respectively. The product expression is now written
\begin{equation}
\phi^*_{\bf P}(z)\phi_{\bf p}(z) =
\sqrt{\f{P_zp_z}{\kappa_{\bf P}(z)\kappa_{\bf p}(z)}}
\exp\[\f{i}{\hbar}\int^z_0
\(\kappa_{\bf p}(\zeta)-\kappa_{\bf P}(\zeta)\)d\zeta\] \,.
\end{equation}
Again, to lowest order in $\hbar$ we can change $P_z$ to $p_z$ and
$\kappa_{\bf P}(z)$ to $\kappa_{\bf p}(z)$ as the difference is of
order $\hbar$. The difference in the $\kappa$ terms to lowest order
in $\hbar$ is
\begin{align}
\kappa_{\bf p}-\kappa_{\bf P} &= \f{\d \kappa_{\bf p}}{\d p_0}(P_0-p_0)
+\f{\d \kappa_{\bf p}}{\d {\bf p}_\perp}\cdot({\bf P}_\perp-{\bf
p}_\perp) \nol
 &= \f{dt}{dz}\hbar k_0 - \f{d{\bf x}_\perp}{dz}\cdot\hbar {\bf
 k}_\perp \,.
\end{align}
Thus the product of $\phi$'s can be written
\begin{align}
\phi^*_{\bf p'}(z)\phi_{\bf p}(z) &=
\f{p_z}{\kappa_{\bf p}}\exp\[\f{i}{\hbar}\int^z_0 dz\, \f{dt}{dz}\hbar k_0-\f{dx_\perp}{dz} \hbar
k_\perp\] \nol
 &= \f{p_z}{\kappa_{\bf p}}\exp\(i(k_0t-k_\perp\cdot x_\perp)\)
 \,.
\end{align}

This gives the $i$ component of the emission amplitude to be
\begin{align}
{\cal A}_i &= e\int dz \, e^{-ik_zz}\f{p_z}{\kappa_{\bf
p}}\exp\(i(k_0t-k_\perp\cdot x_\perp)\)
\f{\[p_i-V_i(z)\]}{p_z} \nol
 &= e \int d\xi\, \f{dx_i}{d\xi} e^{ik\xi} \,,
\end{align}
using $\xi=t-{\bf n}\cdot{\bf x}$, ${\bf n}={\bf k}/k$ as defined
previously. Similarly, the $t$ component gives
\begin{equation}
{\cal A}_t = -e \int d\xi\, \f{dx_i}{d\xi} e^{ik\xi} \,.
\end{equation}

Consider now the $z$ component of the amplitude given previously as
\beq
{\cal A}_z
 = -\f{ie\hbar}{2p_z}\int dz\,\[\phi^*_{{\bf p'}}\d_z\phi_{\bf p} -
 \(\d_z\phi^*_{\bf p'}\)\phi_{\bf p}\] e^{-ik_zz} \,.
\eeq
Note here that from equation for $\phi_{\bf p}$ we have, for lowest
order in $\hbar$
\beq
\d_z\phi_{\bf p}(z) = \f{i}{\hbar}\kappa_{\bf p}(z)\phi_{\bf p}(z)
\,,
\eeq
thus the amplitude becomes
\begin{align}
{\cal A}_z
 &= \f{e}{2p_z}\int dz\, \(\kappa_{\bf p}+\kappa_{\bf p'}\)\phi^*_{\bf p'}\phi_{\bf p} e^{-ik_zz} \nol
 &= e\int dz\, e^{i(k_0t-k_xx-k_yy-k_zz)} \nol
 &= e \int d\xi\, \f{dz}{d\xi} e^{ik\xi} \,.
\end{align}
Raising the indices gives
\begin{equation} \label{emamp}
{\cal A}^\mu = -e \int d\xi\, \f{dx^\mu}{d\xi} e^{ik\xi} \,.
\end{equation}
This is the same expression as (\ref{amplitude}) which we obtained
for the $t$-dependent potential. Given the symmetry between the
spatial components we can thus use this amplitude to calculate the
position shift for a potential dependent on a single space-time
coordinate. In the expression for the amplitude $x^\mu$ is the
classical trajectory of a particle with final momentum $p$ that
passes through $(t,{\bf x})=(0,{\bf 0})$. This emission
amplitude.\footnote{which is already under the $\hbar\to 0$ limit,}
is identical with that for a classical point charge passing through
$(t,{\bf x})=(0,{\bf 0})$

\subsection{Cut-off}\label{cutoffsection}
The expression for the emission amplitude (\ref{emamp}) is currently
ill-defined because the integrand does not tend to zero as $\xi \to
\pm \infty$. To counter this pathology, we introduce a smooth
cut-off function $\chi(\xi)$ which has the properties:
\begin{itemize}
\item $\chi(\xi)$ takes the value $1$ whilst the acceleration is
non-zero.
\item $\lim_{\xi\to\pm\infty} \chi(\xi) = 0$.
\end{itemize}
Also, we take $\chi(\xi)$ to be a member of a family of such cut-off
functions such that we can take the limit $\chi\to 1$, with the
property
\begin{itemize}
\item $\int_{-\infty}^{+\infty}[\chi'(\xi)]^2d\xi \to 0$.
\end{itemize}
The cut-off version of the emission amplitude is thus written
\begin{equation}
{\cal A}^\mu({\bf p},{\bf k})  =  -e \int_{-\infty}^{+\infty} d\xi\,
\frac{dx^\mu}{d\xi}\chi(\xi)\,e^{ik\xi}\,,\label{cut-off}
\end{equation}
and is now well-defined. We shall make use of both this expression
for ${\cal A}^\mu$ along with the result of integrating
(\ref{cut-off}) by parts
\begin{equation}\label{cut-off2}
{\cal A}^\mu({\bf p},{\bf k}) =
-\frac{ie}{k}\int_{-\infty}^{+\infty}d\xi\, \left[ \frac{d^2
x^\mu}{d\xi^2} +
\frac{dx^\mu}{d\xi}\chi'(\xi)\right]e^{ik\xi}\,,
\end{equation}
where we have used the condition that $\chi(\xi)=1$ if
$d^2x^\mu/d\xi^2 \neq 0$.

\subsection{Larmor Formula}

The reader may recall that when calculating the contributions to the
position shift we included the term $\delta x_{\rm Q2}$, defined in
(\ref{deltaxQ2})
\begin{equation}
\delta x^i_{\rm Q2} = -\frac{\langle x^i(0)\rangle_{\rm i}}{\hbar}
\int\frac{d^3{\bf k}}{2k(2\pi)^3} {\cal A}^{\mu
*}({\bf p},{\bf k}){\cal A}_\mu({\bf p},{\bf k})
\( \f{\partial p_i}{\partial
 P_i}-1\) \,, \tag{\ref{deltaxQ2}}
\end{equation}
which was evaluated as zero due to the arrangement in the model that
the control particle passes through the spacetime origin. It was
also stated at the time that this contribution is in fact that which
describes the additional shift produced should the control particle
not be at spatial origin at $t=0$. Whilst the truth of this
statement does not, due to the model, affect the results we wish to
obtain, it is nonetheless worth briefly taking an aside to consider.
In the chapter on the classical position shift, we considered the
effect of such a change in the point of measurement whilst analysing
the linear acceleration due to a space-dependent potential. We refer
the reader to the results (\ref{extraaa}) and (\ref{Larmor}), where
we found that the extra contribution to the position shift was given
by
\begin{equation}
\delta z_{\rm extra}  =
-\frac{z_0}{m\gamma_0^3 v_0^2}E_{\rm em} \,, \tag{\ref{extraaa}}
\end{equation}
where $z_0\neq 0$ is the position of the particle at $t=0$ and $v_0$
is its final speed. This contribution is written in terms of the
energy $E_{\rm em}$ emitted as radiation
\begin{equation}
E_{\rm em} =  \frac{2\alpha_c}{3}\int_{-\infty}^0 (\gamma^3
\ddot{z})^2\,dt \,,
\tag{\ref{Larmor}}
\end{equation}
which we noted is the relativistic Larmor formula for
one-dimensional motion.

Let us use the newly derived emission amplitude to calculate $\delta
x_{\rm Q2}$ for this case of linear acceleration. Choosing the
direction $z$ as with the classical formulae above, we have $i=3$
and
\begin{equation}
\delta z_{\rm Q2}  =  - \frac{z_0}{\hbar}
\int\frac{d^3{\bf k}}{2k(2\pi)^3}  {\cal A}^{\mu *} ({\bf p},{\bf k}){\cal A}_\mu({\bf
p},{\bf k})
\left( \frac{dp}{dP} - 1\right) \,, \label{q2q2}
\end{equation}
where we use $p=p_z$ and $P=P_z$ for simplicity. Firstly, we need to
find an expression for $dp/dP$ in terms of $p$ and $k$ to order
$\hbar$. The energy conservation equation $p_0 - P_0 = \hbar k$
gives a one-to-one relation between $p$ and $P$ for a given $k$
after letting the transverse momenta ${\bf P}_\perp^2 = {\bf
p}_\perp^2 = 0$, because these are of order $\hbar^2$ here. We then
write
\begin{align}
P_0 = \(P^2+m^2\)^{1/2} = p_0-\hbar k\,,
\end{align}
and thus
\begin{align}
P^2&=p_0^2-2\hbar k p_0-m^2+{\cal O}(\hbar^2)
\nol &= p^2 - 2\hbar k p_0+{\cal O}(\hbar^2) \,.
\end{align}
Solving for $P$ and expanding the resultant square root gives to
order $\hbar$
\begin{equation}
P = p-\f{p_0}{p} \hbar k = p - \( 1+ \f{m^2}{p^2} \)^{1/2}\hbar k
\,.
\end{equation}
This leads us, to order $\hbar$, to
\begin{equation}
\frac{dP}{dp} = 1 + \frac{m^2}{p^2 p_0}\hbar k \,,
\end{equation}
and finally,
\begin{align}
\frac{dp}{dP} &= 1 - \frac{m^2}{p^2 p_0}\hbar k \,.\label{dpdp}
\end{align}
By using this formula in Eq.~(\ref{q2q2}) we obtain
\begin{equation}
\delta z_{\rm Q2}  =
- \frac{m^2z_0}{p^2 p_0}{\cal E}_{\rm em}  \,, \label{deltazq2}
\end{equation}
where
\begin{equation}\label{emprob}
{\cal E}_{\rm em} \equiv - \int\frac{d^3{\bf k}}{2k(2\pi)^3} k {\cal
A}^{\mu *}({\bf p},{\bf k}){\cal A}_\mu({\bf p},{\bf k}) \,,
\end{equation}
is the expectation value of the energy emitted as radiation. By
comparing (\ref{deltazq2}) with the classical expression
(\ref{extraaa}), it can be seen that the equality $\delta z_{\rm Q2}
=
\delta z_{\rm extra}$ will hold if we can show that
\begin{equation}\label{larmoreq}
{\cal E}_{\rm em} =\frac{2\alpha_c}{3} \int_{-\infty}^0 (\gamma^3
\ddot{z})^2\,dt \,,
\end{equation}
which is or course identical to the relativistic generalization of
the \emph{classical} Larmor formula. For this purpose, we now use
the emission amplitude. It is convenient to use the form
(\ref{cut-off2}) of the emission amplitude, which was produced by
integration by parts. Substituting this expression into the
expectation value (\ref{emprob}) we obtain
\begin{multline}
{\cal E}_{\rm em}   =  - e^2\int d\Omega \int_0^\infty
\frac{dk}{16\pi^3} \int_{-\infty}^{+\infty}
d\xi'\int_{-\infty}^{+\infty}d\xi \\  \times
\left[ \frac{d^2 x^\mu}{d{\xi'}^2} +
\frac{dx^\mu}{d\xi'}\chi'(\xi')\right]
\left[ \frac{d^2 x_\mu}{d\xi^2} +
\frac{dx_\mu}{d\xi}\chi'(\xi)\right]e^{ik(\xi-\xi')} \,,
\end{multline}
where $d\Omega$ is the solid angle in the ${\bf k}$-space and now
$\xi=t-z\cos\theta$. We may extend the integration range for $k$
from $[0,+\infty)$ to $(-\infty,+\infty)$ and divide by two.  We can
then integrate over $k$ to produce the delta function and integrate
out the variable $\xi'$ to find
\begin{align}
{\cal E}_{\rm em}  & =  - \frac{e^2}{16\pi^2}
\int d\Omega\int_{-\infty}^{+\infty}d\xi
\left\{ \frac{d^2x_\mu}{d\xi^2}\frac{d^2 x^\mu}{d\xi^2} +
\frac{dx_\mu}{d\xi}\frac{dx^\mu}{d\xi}
\left[\chi'(\xi)\right]^2\right\} \,.
\end{align}
Noting that $(dx_\mu/d\xi)(dx^\mu/d\xi)$ is bounded, the second term
tends to zero in the limit $\chi(\xi)\to 1$ due to the requirement
\begin{equation}
\int_{-\infty}^{+\infty}[\chi'(\xi)]^2d\xi \to 0 \,.
\end{equation}
Hence, we have in this limit
\begin{equation}\label{radia}
{\cal E}_{\rm em}  =
 - \frac{\alpha_c}{4\pi}
\int_{-\infty}^{+\infty}d\xi
\int d\Omega \,\frac{d^2x_\mu}{d\xi^2}\frac{d^2 x^\mu}{d\xi^2} \,,
\end{equation}
where $\alpha_c \equiv e^2/4\pi$ as before. Now, one can readily
show that
\begin{equation}
  \frac{d^2x^{\mu}}{d\xi^2} =
\left(\frac{dt}{d\xi}\right)^3\left[\frac{d\xi}{dt}\frac{d^2
  x^{\mu}}{dt^2}-
  \frac{d^2\xi}{dt^2}\frac{dx^{\mu}}{dt}\right] \,.
\end{equation}
By substituting $d\xi/dt = 1 - \dot{z}\cos\theta$ we find
\begin{eqnarray}
\frac{d^2z}{d\xi^2} &=&
\frac{\ddot{z}}{(1-\dot{z}\cos\theta)^3}\,, \label{d2zdxi2}\\
\frac{d^2 t}{d\xi^2} & = & \frac{d^2z}{d\xi^2}\,\cos\theta \,,
\label{d2tdxi2}
\end{eqnarray}
and hence
\begin{equation}
\frac{d^2x^\mu}{d\xi^2}\frac{d^2x_\mu}{d\xi^2}
= - \frac{\ddot{z}^2 \sin^2\theta}
{(1-\dot{z}\cos\theta)^5}\,\frac{dt}{d\xi} \,.
\end{equation}
By substituting this formula in (\ref{radia}) we obtain
\begin{align}
{\cal E}_{\rm em} & = \f{\alpha_c}{4\pi} \int^\infty_{-\infty} d\xi
\int d\Omega \frac{\ddot{z}^2 \sin^2\theta}
{(1-\dot{z}\cos\theta)^5}\,\frac{dt}{d\xi} \nol &=\f{\alpha_c}{2}
\int^\infty_{-\infty} dt
\int^\pi_0 d\theta \frac{\ddot{z}^2 \sin^3\theta}
{(1-\dot{z}\cos\theta)^5} \nol &=\f{\alpha_c}{2}
\int^\infty_{-\infty} dt
\ddot{z}^2\f{4}{3} \f{1}{(1-\dot{z}^2)^3}
\nol
&=\f{2\alpha_c}{3}\int^0_{-\infty} dt (\gamma^3 \ddot{z})^2 \,.
\end{align}
The limits for the last line were changed from $(-\infty,\infty)$ to
$(-\infty,0)$ by virtue of the fact that $\ddot{z}=0$ for $t
\geq 0$. Consequently, we have
\begin{equation}
\delta z_{\rm Q2} = \delta z_{\rm extra} \,,
\end{equation}
as required.

\subsection{Position shift}

We now use the emission amplitude expressions to find the emission
contribution to the quantum position shift. Recall that this was
given in (\ref{deltaxem}) as
\begin{align}
 \delta x^i_{\rm em} &= -
\frac{i}{2}
\int \frac{d^3{\bf k}}{2k(2\pi)^3}
{\cal A}^{\mu *}({\bf p},{\bf k})
\stackrel{\leftrightarrow}{\partial}_{p_i} {\cal A}_\mu({\bf p},{\bf
k}) \,. \tag{\ref{deltaxem}}
\end{align}
In the product ${\cal A}^{\mu *}({\bf p},{\bf k}) {\partial}_{p_i}
{\cal A}_\mu({\bf p},{\bf k})$, we shall use the expression
(\ref{cut-off2}) for ${\cal A}^{\mu *}$ and (\ref{cut-off}) for
${\cal A}_\mu$ (and vice-versa for the conjugate).\footnote{Equation
(\ref{cut-off}) was where we introduced the cut-off in ${\cal
A}^\mu$ and (\ref{cut-off2}) was the result of integrating by
parts.} This leads to
\begin{align}
 \delta x^i_{\rm em}&= - \frac{e^2}{2}\int d\Omega \int_0^\infty
\frac{k^2 dk}{16\pi^3 k^2} \int_{-\infty}^{+\infty}
d\xi'\int_{-\infty}^{+\infty}d\xi \nol & \quad \times
\left\{\left[ \frac{d^2 x^\mu}{d(\xi')^2} +
\frac{dx^\mu}{d(\xi')}\chi'(\xi')\right] \frac{\partial\
}{\partial p_{i}}\left( \frac{dx_\mu}{d\xi}\right)\chi(\xi) +
(\xi\leftrightarrow \xi')\right\} e^{ik(\xi-\xi')} \,,
\end{align}
where $d\Omega$ is in angular part of the ${\bf k}$ integration in
spherical polar coordinates. Due to the symmetry in the integration,
we make the swap back $\xi\leftrightarrow
\xi'$ to the second term. However this will change the
exponential, producing the same overall effect as the transformation
$k\to -k$ on the first term, viz
\begin{align}
\delta x^i_{\rm em}&= - \frac{e^2}{2}\int d\Omega \int_0^\infty
\frac{k^2 dk}{16\pi^3 k^2} \int_{-\infty}^{+\infty}
d\xi'\int_{-\infty}^{+\infty}d\xi \nol & \quad \times
\left[ \frac{d^2 x^\mu}{d(\xi')^2} +
\frac{dx^\mu}{d(\xi')}\chi'(\xi')\right] \frac{\partial\
}{\partial p_{i}}\left( \frac{dx_\mu}{d\xi}\right)\chi(\xi)
\[ e^{ik(\xi-\xi')} +  e^{-ik(\xi-\xi')} \] \,.
\end{align}
This expression makes it clearer that the second term is the
conjugate of the first, as is known from (\ref{deltaxem}). Making
the substitution of integration variables $k\to -k$ for the second
term, we see that it is integrand is identical to the first, but the
integration range is now $(-\infty,0)$. We may thus rewrite
\begin{align}
\delta x^i_{\rm em}&= - \frac{e^2}{2}\int d\Omega \int_{-\infty}^\infty
\frac{k^2 dk}{16\pi^3 k^2} \int_{-\infty}^{+\infty}
d\xi'\int_{-\infty}^{+\infty}d\xi \nol & \quad \times
\left[ \frac{d^2 x^\mu}{d(\xi')^2} +
\frac{dx^\mu}{d(\xi')}\chi'(\xi')\right] \frac{\partial\
}{\partial p_{i}}\left( \frac{dx_\mu}{d\xi}\right)\chi(\xi)
 e^{ik(\xi-\xi')} \,,
\end{align}
where the $k$ integration is over the full range $(-\infty,\infty)$.
We can now integrate over $k$ to produce the delta function
$2\pi\delta(\xi-\xi')$ and consequently integrate out $\xi'$ to give
\begin{align}
\delta x^i_{\rm em}&=  - \frac{e^2}{16\pi^2}\int d\Omega
\int_{-\infty}^{+\infty} d\xi \left[ \frac{d^2
x^\mu}{d\xi^2} + \frac{dx^\mu}{d\xi}\chi'(\xi)\right]
\frac{\partial\ }{\partial p_{i}}\left(
\frac{dx_\mu}{d\xi}\right)\chi(\xi) \nol
&=  - \frac{e^2}{16\pi^2}\int d\Omega \int d\xi\left[ \frac{d^2
x^\mu}{d\xi^2}\frac{\partial\ }{\partial p_{i}}
\left(\frac{dx^\mu}{d\xi}\right) + \frac{1}{4}\frac{\partial\
}{\partial p_{i}} \left(
\frac{dx^\mu}{d\xi}\frac{dx_\mu}{d\xi}\right)
\frac{d\ }{d\xi}(\chi(\xi))^2\right] \,, \label{cutoffcalc2}
\end{align}
where, whilst combining terms we have recalled the property
$\chi(\xi)=1$ if $d^2x^\mu/d\xi^2 \neq 0$. Noting that
\begin{equation}
\frac{dx^\mu}{d\xi}\frac{dx_\mu}{d\xi} =
\left(\frac{d\tau}{d\xi}\right)^2
\frac{dx^\mu}{d\tau}\frac{dx_\mu}{d\tau} = \left(
\frac{d\tau}{d\xi}\right)^2 \,,
\end{equation}
we find that the second term in (\ref{cutoffcalc2}) above is
proportional to the integral
\begin{equation}
I \equiv \int d\Omega \int_{-\infty}^{+\infty} d\xi \frac{d\ }{d\xi}
\left\{\frac{\partial\ }{\partial p_{i}}
\left(\frac{d\tau}{d\xi}\right)^2\right\}(\chi(\xi))^2 \,.
\end{equation}
We shall now show that this integral is in fact zero: Owing to the
fact that $\partial /
\partial p_i$ is taken with $\xi$ fixed, the $\xi$-derivative and
the $p_{i}$-derivative commute. Hence
\begin{equation}
I = \int d\Omega \int_{-\infty}^{+\infty} d\xi \frac{\partial\
}{\partial p_{i}} \left[\frac{d\
}{d\xi}\left(\frac{d\tau}{d\xi}\right)^2\right] (\chi(\xi))^2 \,.
\end{equation}
This expression is simplified by the observation that the quantity
inside the square brackets is nonzero only if the acceleration is
nonzero, or correspondingly when $\chi(\xi)=1$. Relocating the
$p_{i}$ differentiation outside the integration in $I$ we have
\begin{align}
I &= \frac{\partial\ }{\partial p_{i}} \int d\Omega
\int_{-\infty}^{+\infty}d\xi\, \frac{d\
}{d\xi}\left(\frac{d\tau}{d\xi}\right)^2 \nol &= \frac{\partial\
}{\partial p_{i}} \int d\Omega \left[ \left(
\frac{d\tau}{d\xi}\right)^2 \right]_{\xi=-\infty}^{\xi=+\infty}
\,.
\end{align}
Now, the quantity to be differentiated is
\begin{align}
\int d\Omega \left(\frac{d\tau}{d\xi}\right)^2 & =  2\pi
\int_{0}^\pi d\theta\,\sin\theta \left( \frac{dt}{d\tau} -
\frac{dz}{d\tau}\,\cos\theta \right)^{-2}\nol\
& = \frac{4\pi}{(dt/d\tau)^2 - (dz/d\tau)^2} = 4\pi \,,
\end{align}
i.e. a constant, ergo the integral $I=0$ as stated. If one recalls
the definition $\alpha_c=e^2/4\pi$, then the emission contribution
to the quantum position shift can now be written in a fairly compact
form
\begin{equation}
\delta x^i_{\rm em} = -\f{\alpha_c}{4\pi} \int d\Omega \int d\xi
\f{d^2 x^\mu}{d\xi^2} \f{\partial }{\partial p_{i}}
\(\f{dx_\mu}{d\xi}\) \,. \label{emshiftxi2}
\end{equation}
We additionally note that this expression is now independent of the
cut-off function.

Whilst (\ref{emshiftxi2}) is a fairly simple compact expression, it
is still one in a somewhat different form to that of the similarly
compact classical position shift as given in (\ref{cshift}), which
is written in terms of $t$ rather than $\xi$. We thus need to
eliminate the variable $\xi$ using its definition $\xi=t-{\bf
n}\cdot{\bf x}$.~\footnote{Recall that we defined ${\bf n}={\bf
k}/k$.} Before tackling the second derivative term, we may again
interchange the order of the $p_{i}$ and $\xi$ derivatives and
change integration variables from $\xi$ to $t$:
\begin{equation}
\delta x^i_{\rm em}  = -\frac{\alpha_c}{4\pi}\int d\Omega \int dt\,
\frac{d^2 x_{\mu}}{d\xi^2}
\frac{d\ }{dt}\left(\frac{\partial x^\mu}{\partial p_i}
\right)_{\xi} \,. \label{emshiftxi3}
\end{equation}
We remind the reader that in the above expression $d\Omega$ is the
solid angle in the wave-number space ${\bf k}$ of the emitted
photon. We have additionally placed the subscript $\xi$ on the final
partial derivative to emphasize that this variable is held fixed,
which will be important when it is evaluated. Furthermore we remind
the reader that the momentum $p_{i}$ is the \emph{final} momentum in
the measurement region ${\cal M}_+$. Consequently we would write the
velocity $dx^\mu/dt = \left(\partial x^\mu/\partial t\right)_{p_i}$.

Proceeding with the evaluation of (\ref{emshiftxi3}) via the
elimination of $\xi$, one can readily write $d^2 x^\mu/d\xi^2$ in
terms of $t$-derivatives by using $d/d\xi =
(1-n^i\dot{x}^i)^{-1}d/dt$ as follows:
\beq\label{d2xdxi2}
\f{d^2x^\mu}{d\xi^2} = \dot{\xi}^{-3}
\[\(1-n^i\dot{x}^i\)\ddot{x}^\mu +
n^i\ddot{x}^i\dot{x}^\mu\] \,,
\eeq
where $\dot{\xi} = 1 - n^i \dot{x}^i$. Here and in the rest of this
section, Latin indices, which we recall take the spatial values 1 to
3, are summed over when repeated. The time and space components of
(\ref{d2xdxi2}) can separately be given as
\begin{align}
\f{d^2t}{d\xi^2} &= \dot{\xi}^{-3}\,n^i\ddot{x}^i\,,
\label{secondder1}\\
\f{d^2x^j}{d\xi^2} &=\dot{\xi}^{-3}
\[\(1-n^i\dot{x}^i\)\ddot{x}^j +
n^i\ddot{x}^i\dot{x}^j\] \,.
\label{secondder2}
\end{align}
Next we express $(\partial x^\mu/\partial p_i)_\xi$ in
(\ref{emshiftxi3}) in the form involving $t$ rather than $\xi$ as
follows.  Note first
\begin{equation}
dx^\mu =
\frac{dx^\mu}{dt}dt + \left(\frac{\partial x^\mu}
{\partial p_i}\right)_t dp_i \,.
\end{equation}
The zeroth component of this equation is in fact trivial because
$(\partial t/\partial p_i)_t = 0$.  By substituting $dt=d\xi +
n^kdx^k$ in this equation with $\mu=j$ we have
\begin{align}
dx^i & =  \frac{dx^i}{dt} d\xi + \frac{dx^i}{dt} n^j dx^j + \left(
\frac{\partial x^i}{\partial p^j}
\right)_t dp^j  \,. \label{dxpart2}
\end{align}
We can solve (\ref{dxpart2}) for $dx^i$ by first observing that by
contracting both sides of (\ref{dxpart2}) with $n^j$ we can solve to
find an expression for $n^k dx^k$, viz
\begin{equation} n^k dx^k = \frac{n^j v^j}{1-n^jv^j}d\xi +
\frac{n^k}{1-n^j v^j}\left(\frac{\partial x^k}{\partial p^j}
\right)_t dp^j \,,
\end{equation}
where $v^i \equiv \dot{x}^i =dx^i/dt$. By substituting this back
into (\ref{dxpart2}), we solve for $dx^i$
\begin{equation}dx^i = \frac{v^i}{1-n^jv^j}d\xi +
\frac{[\delta^{ik}(1-n^lv^l)+n^kv^i]}{1-n^lv^l}
\left(\frac{\partial x^k}{\partial p^j}\right)_t dp^j \,.
\end{equation}
Hence
\begin{equation}
\left(\frac{\partial x^i}{\partial p_j}\right)_\xi
= \frac{[\delta^{ik}(1-n^lv^l)+n^kv^i]}{1-n^lv^l}
\left(\frac{\partial x^k}{\partial p_j}\right)_t \,.
\label{partial2}
\end{equation}
With $\xi$ fixed we have $dt - n^i dx^i = d\xi = 0$. Thus,
\begin{align}
\left(\frac{\partial t}{\partial p_j}\right)_\xi
& =  n^i \left(\frac{\partial x^i}{\partial p_j}\right)_\xi \nol & =
\frac{n^i}{1-n^l v^l}\left(\frac{\partial x^i}{\partial p_j}
\right)_t \,. \label{partial3}
\end{align}

By substituting the pairs (\ref{secondder1}), (\ref{secondder2}) and
(\ref{partial2}), (\ref{partial3}) in the position shift
(\ref{emshiftxi3}) we find, after a straightforward albeit lengthy
amount of rearranging,
\begin{multline}
\delta x^i_Q
= -\frac{\alpha_c}{4\pi}\int dt\left\{
\[ I_2^{kj} \gamma^{-2}a^j
- I_0 a^k - I_1^j a^j v^k - I_1^k ({\bf a}\cdot {\bf v})
\] \frac{d\ }{dt}\left(\frac{\partial x^k}{\partial p^i}\right)_t
\right. \\
\left. + \[
I_3^{kjl}\gamma^{-2}a^ja^l - 2I_2^{kj}a^j({\bf a}\cdot {\bf v})
 - I_1^k {\bf a}^2  \] \left(\frac{\partial x^k}{\partial p^i}
\right)_t\right\} \,,  \label{almostthere}
\end{multline}
where
\begin{align}
I_0 &\equiv \displaystyle\int d\Omega \df{1}{\dot{\xi}^2}
= 4\pi\gamma^2 \,, \label{integral1} \\
I_1^i &\equiv  \displaystyle\int d\Omega \df{n^i}{\dot{\xi}^3}
= 4\pi\gamma^4 v^i \,, \\
I_2^{ij} &\equiv  \displaystyle\int d\Omega
\df{n^i n^j}{\dot{\xi}^4}
=  \f{16}{3}\pi\gamma^6 v^i v^j   +
\f{4}{3}\pi\gamma^4 \delta^{ij} \,, \\
I_3^{ijk} &\equiv  \displaystyle\int d\Omega
\df{n^in^jn^k}{\dot{\xi}^5} =  8\pi\gamma^8 v^iv^jv^k +\f{4}{3}\pi\gamma^6\(
v^i\delta^{jk}
 +v^j\delta^{ik}+v^k\delta^{ij} \) \,. \label{integral4}
\end{align}
Evaluation of these solid-angle integrals is facilitated by noting
that the last three integrals are proportional to partial
derivatives of $I_0$ with respect to $v^i$, explicitly
\begin{equation}
I_n^{i_1\ldots i_n} = \f{1}{(n+1)!}\f{\partial}{\partial v^{i_n}}
I_{n-1} \quad n=1,2,3 \,.
\end{equation}

Substitution of (\ref{integral1})--(\ref{integral4}) in
(\ref{almostthere}) yields
\begin{multline}
\delta x^i_{\rm em}  = \f{2\alpha_c}{3}\int dt \left\{
\[\gamma^4 ({\bf a}\cdot {\bf v}) v^k + \gamma^2 a^k \] \f{d\ }{dt} \(
\f{\partial x^k}{\partial p^i}\)_t  \right. \\  \left. +
\[\gamma^6 ({\bf a}\cdot {\bf v})^2 v^k +\gamma^4{\bf a}^2 v^k  \] \(
\f{\partial x^k}{\partial p^i}\)_t \right\} \,, \label{xiQOK}
\end{multline}
which, by using the fact that ${\bf a}(t) \neq 0$ only for $t < 0$
and for a finite interval of time to integrate the first term by
parts, becomes
\begin{multline}
\delta x^i_{\rm em} = -\f{2\alpha_c}{3}\int dt \left\{
\f{d}{dt}\[ \gamma^4 ({\bf a}\cdot {\bf v}) v^k + \gamma^2 a^k\]
\right. \\ \left. -\gamma^6 ({\bf a}\cdot {\bf v})^2 v^k -\gamma^4{\bf a}^2 v^k
\right\}\(
\f{\partial x^k}{\partial p^i}\)_t \,.
\end{multline}
Comparison with the expression for the Lorentz-Dirac force in
(\ref{3dLD}) demonstrates that we have
\begin{equation} \label{qemshift}
\delta x^i_{\rm em} = - \int^0_{-\infty}
dt\, {\bf {\cal F}}^{j}_{\rm LD} \pd{x^j}{p^i}{t} \,.
\end{equation}
We recognize this as equal to the classical position shift
(\ref{cshift}).

\section{Forward Scattering}

The forward scattering contribution to the position shift was shown
in (\ref{deltaxfor}) to be equal to
\begin{align}\tag{\ref{deltaxfor}}
 \delta x^i_{\rm for} &= -\hbar \partial_{p_i} \Re {\cal F}({\bf
 p}) \,.
\end{align}
Considering this expression there are two points to bear in mind
when calculating the forward scattering amplitude. Firstly, we note
that we only require the real part of the amplitude. Recall that the
imaginary part was canceled via its relation to the emission
probability. Secondly, we note that we have an additional $\hbar$
factor multiplying ${\cal F}({\bf p})$. This is not contrary to the
fact that (\ref{deltaxfor}) is in the $\hbar\to 0$ limit, as we
shall shortly see that ${\cal F}$ is of order $\hbar^{-2}$. We shall
thus actually need the first
\emph{two} orders, $\hbar^{-2}$ and $\hbar^{-1}$, in ${\cal F}$.
With the extra $\hbar$ factor, these orders will contribute at
orders $\hbar^{-1}$ and $\hbar^0$. In taking the $\hbar\to 0 $
limit, one would wish the former to be zero or canceled.
Additionally, should the latter be non-zero, then we would have an
additional position shift contribution in the $\hbar\to 0$ limit. As
the emission position shift has already been shown to give the
classical position shift expression, such an additional contribution
would present a quantum correction to the classical theory.

In this section we shall in fact show that the overall contribution
from the forward scattering towards the position shift is zero when
we take the renormalisation of the mass into account. This is in
fact in keeping with the classical description of radiation
reaction, although with some subtle differences, and we shall return
to this later. We proceed with the calculation of the forward
scattering amplitude and shall then continue to calculate the
contribution to the renormalised forward scattering made by the mass
counterterm. The forward scattering amplitude was first introduced,
and consequently given its definition, by its presence as part of
the amplitude of the zero-photon sector of the final state, viz
\begin{align}
\ket{f}_{\rm for} &=  \int \f{d^3{\bf p}}{(2\pi\hbar)^3 \sqrt{2p_0}}
\[ 1+i{\cal F}({\bf p}) \] f({\bf p}) A^\dagger({\bf p}) \ket{0}
\,.
\tag{\ref{sffor}}
\end{align}
If we equate this expression with the state evolution in
time-dependent perturbation theory, then to ${\cal O}(e^2)$ we have
the following zero photon sectors:
\begin{align}\label{forham1}
i{\cal F}({\bf p})A^\dagger({\bf p})\ket{0} &= \left. -\f{i}{\hbar}
\int d^4x {\cal H}_{I} A^\dagger({\bf p})\ket{0} \right|_{\rm
zero-photon}
\nol &\left. \quad + \f{1}{2}\(\f{-i}{\hbar}\)^2 \int d^4x d^4x' T \[ {\cal H}_I(x') {\cal
H}_I(x) \] A^\dagger({\bf p}) \ket{0}  \right|_{\rm zero-photon}\,,
\end{align}
where $T[\ldots]$ represents the usual time ordering and ${\rm
zero-photon}$ indicates that we require only the zero photon terms.
We note that we need both the first and second orders in the
interaction Hamiltonian for this order of $e^2$. Recall that ${\cal
H}_I$ was given by
\begin{equation}\tag{\ref{Hamiltonian}}
{\cal H}_I(x) = \frac{ie}{\hbar}A_\mu \normal\left[ \varphi^\dagger
D^\mu
\varphi - (D^\mu \varphi)^\dagger \varphi\right]\normal
 + \frac{e^2}{\hbar^2}\sum_{i=1}^3 A_iA_i
\normal\varphi^\dagger\varphi\normal \,.
\end{equation}
We see here that the first term in (\ref{Hamiltonian}) contributes
at second order in ${\cal H}_I$ for (\ref{forham1}), whilst the
second term in (\ref{Hamiltonian}) contributes at first order in
${\cal H}_I$. Operating on both sides of (\ref{forham1}) with the
bra-state $\bra{0} A({\bf p'})$, which invokes the zero photon
condition, and using the commutation relations for the resulting
inner product, we may rearrange the result to produce the forward
scattering:
\begin{align}
{\cal F}({\bf p}) &= -\f{1}{\hbar} \int \f{d^3{\bf
p'}}{2p_0'(2\pi\hbar)^3} d^4x \bra{0} A({\bf p'}) {\cal H}_I(x)
A^\dagger({\bf p}) \ket{0} \nol & \quad + \f{i}{2\hbar^2} \int
\f{d^3{\bf p'}}{2p_0'(2\pi\hbar)^3} d^4x d^4x' \bra{0} A({\bf p'})
T \[ {\cal H}_I(x') {\cal H}_I(x) \] A^\dagger({\bf p}) \ket{0}
\,.
\label{FHamexp}
\end{align}
The portion of the above expression at first order in ${\cal H}_I$
we label ${\cal F}_1$ and correspondingly name the remainder ${\cal
F}_2$.~\footnote{In \cite{HM4} this notation was the reverse of that
given here.}

\subsection{${\cal F}_1$}
The calculation of this first order ${\cal F}_1$ part is fairly
straightforward and we present it here. Substituting the interaction
Hamiltonian into the definition of ${\cal F}_1$, we have
\begin{align}\label{for1}
{\cal F}_1({\bf p}) &= -\f{1}{\hbar}\f{e^2}{\hbar^2} \int \f{d^3{\bf
p'}}{2p_0'(2\pi\hbar)^3} d^4x \sum_{i=1}^3 \bra{0} A({\bf p'})
A_i(x) A_i(x) \normal \varphi^\dagger (x) \varphi (x) \normal
A^\dagger({\bf p}) \ket{0} \nol &= -\f{e^2}{\hbar^3} \int \f{d^3{\bf
p'}}{2p_0'(2\pi\hbar)^3} d^4x \sum_{i=1}^3 \bra{0} A_i A_i
\ket{0} \bra{0} A({\bf p'}) \normal \varphi^\dagger \varphi \normal A^\dagger({\bf p}) \ket{0}
\,.
\end{align}
Let us analyse the two inner products in turn, using the appropriate
commutation relations. For the first product, involving the
electromagnetic field (which is not normal ordered) only the
annihilation operator from the first field and the creation operator
from the second field will give non-zero results. Thus
\begin{align}
\bra{0} A_i A_i \ket{0} &= \int \f{d^3{\bf k}}{(2\pi)^32k_0}\f{d^3{\bf k'}}{(2\pi)^32k_0'}
\sum_{i=1}^3 \bra{0} a_i({\bf k}) a_i^\dagger({\bf k'}) \ket{0}
e^{-i({\bf k}-{\bf k'})\cdot{\bf x}}
\nol
&= \int \f{d^3{\bf k}}{(2\pi)^32k_0}\f{d^3{\bf k'}}{(2\pi)^32k_0'}
\sum_{i=1}^3 \( -\hbar g_{ii} 2k_0'(2\pi)^3\delta^3({\bf k}-{\bf
k'}) \) e^{-i({\bf k}-{\bf k'})\cdot{\bf x}} \nol &= 3\hbar \int
\f{d^3{\bf k}}{(2\pi)^32k_0} \,, \label{F1aprod}
\end{align}
where we recall that we are using the time-like sign convention
$(+---)$. The second product, involving the scalar particle fields
is
\begin{align}
\bra{0} A({\bf p'}) \normal \varphi^\dagger \varphi \normal A^\dagger({\bf p}) \ket{0}
&= \hbar^2 \int \f{d^3{\bf p^{(2)}}}{(2\pi\hbar)^32p_0^{(2)}} \int
\f{d^3{\bf p^{(3)}}}{(2\pi\hbar)^32p_0^{(3)}}
\nol & \quad \times \bra{0} A({\bf p'}) A^\dagger({\bf p^{(2)}}) A({\bf p^{(3)}})  A^\dagger({\bf p})
\ket{0} \Phi^*_{{\bf p^{(2)}}}(x) \Phi_{{\bf p^{(3)}}}(x) \,.
\end{align}
The remaining operators from the normal ordered fields are
annihilated by the vacuum (after normal ordering). We also recall
the overall prefactor of $\hbar$ in the expression for $\varphi$
(see (\ref{scalar_expansion})) thus leading to the $\hbar^2$ above.
Using the commutation relations we have
\begin{align}
\bra{0} A({\bf p'}) \normal \varphi^\dagger \varphi \normal A^\dagger({\bf p}) \ket{0}
&= \hbar^2 \Phi^*_{{\bf p'}}(x) \Phi_{{\bf p}}(x) \,,
\label{doublephiprod}
\end{align}
and consequently, combining with (\ref{F1aprod}),
\begin{align}
{\cal F}_1({\bf p})&= -3e^2 \int \f{d^3{\bf p'}}{2p_0'(2\pi\hbar)^3}
d^4x  \Phi^*_{{\bf p'}}(x) \Phi_{{\bf p}}(x)  \int \f{d^3{\bf
k}}{(2\pi)^32k_0} \nol &= -3e^2 \int \f{d^3{\bf
p'}}{2p_0'(2\pi\hbar)^3} d^4x  \phi^*_{{\bf p'}}(t) \phi_{{\bf
p}}(t) e^{-i({\bf p'}-{\bf p})\cdot{\bf x}/\hbar} \int \f{d^3{\bf
k}}{(2\pi)^32k_0}
\nol
&= -\f{3e^2}{2p_0}\int dt |\phi_{\bf p}(t)|^2 \int \f{d^3{\bf
k}}{(2\pi)^32k_0} \,.
\end{align}
The use of $k$ in the integration hides the $\hbar$ dependence of
this term. Changing the integration variable to the photon momentum
$K=\hbar k$, we use $d^3{\bf k}/k = \hbar^{-2}d^3{\bf
K}/K$.~\footnote{Where $K=|{\bf K}|$ analogously to $k$.} Thus
\begin{align}\label{F1endk}
{\cal F}_1({\bf p})&= -\f{3e^2}{2\hbar^2p_0}\int dt |\phi_{\bf
p}(t)|^2 \int \f{d^3{\bf K}}{(2\pi)^32K} \,.
\end{align}

\subsection{${\cal F}_2$}

The portion of the forward scattering at second order in the
Hamiltonian forms the bulk of the calculation and is more
complicated than the above first order term. We start from
\begin{align}\label{for2}
{\cal F}_2({\bf p}) &= + \f{i}{2\hbar^2} \int \f{d^3{\bf
p'}}{2p_0'(2\pi\hbar)^3} d^4x d^4x' \bra{0} A({\bf p'}) T \[ {\cal
H}_I(x') {\cal H}_I(x) \] A^\dagger({\bf p}) \ket{0} \,.
\end{align}
That part of ${\cal H}_I$ which is relevant\footnote{Recall that we
only require terms up to ${\cal O}(e^2)$.} is
\begin{equation}
\frac{ie}{\hbar}A_\mu \normal\left[ \varphi^\dagger
D^\mu
\varphi - (D^\mu \varphi)^\dagger \varphi\right]\normal\,
= \frac{ie}{\hbar}A_\mu \normal\left[ \varphi^\dagger
\stackrel{\leftrightarrow}{D^\mu}
\varphi \right]\normal \,,
\end{equation}
where we have used the more compact notation
$\stackrel{\leftrightarrow}{D^\mu} = \stackrel{\rightarrow}{D^\mu}
-\stackrel{\leftarrow}{D^{\dagger\mu}}$. The inner product in
(\ref{for2}) can be separated into the electromagnetic and scalar
field parts:
\begin{align}
 & \bra{0} A({\bf p'}) T \[ {\cal H}_I(x') {\cal
H}_I(x) \] A^\dagger({\bf p}) \ket{0} \nol &= \frac{-e^2}{\hbar^2}
\bra{0} A({\bf p'}) T \[ A_{\mu'} (x') \normal
\varphi^\dagger(x')
\stackrel{\leftrightarrow}{D^{\mu'}}
\varphi (x')  \normal
A_\nu (x) \normal  \varphi^\dagger(x)
\stackrel{\leftrightarrow}{D^\nu}
\varphi (x)  \normal \] A^\dagger({\bf p}) \ket{0} \nol
&= \frac{-e^2}{\hbar^2} \bra{0} A({\bf p'}) T \[  \normal
\varphi^\dagger(x')
\stackrel{\leftrightarrow}{D^{\mu'}}
\varphi (x') \normal
 \normal \varphi^\dagger(x)
\stackrel{\leftrightarrow}{D^\nu}
\varphi (x) \normal \] A^\dagger({\bf p}) \ket{0} \nol & \quad \times
\bra{0}T \[ A_{\mu'} (x') A_\nu (x) \] \ket{0} \,.
\end{align}
The second inner product is simply the photon propagator, the
expression for which is
\begin{align}
& \quad \bra{0}T \[ A_{\mu'} (x') A_\nu (x) \] \ket{0}
\nol &= -\hbar g_{\mu'\nu} \int \f{d^3{\bf k}}{(2\pi)^3 2k_0} \[
\theta(t-t') e^{-ik\cdot(x-x')} +\theta(t'-t)e^{ik\cdot(x-x')} \]
\,.
\end{align}
The scalar field product is more complicated and the full notation
becomes somewhat cumbersome; In full, with the expansions of the
fields, we have
\begin{align}
& \quad \bra{0} A({\bf p'}) T \[ \, \normal \varphi^\dagger(x')
\stackrel{\leftrightarrow}{D^{\mu'}} \varphi (x') \normal \quad
 \normal \varphi^\dagger(x) \stackrel{\leftrightarrow}{D^\nu}
\varphi (x) \normal \, \] A^\dagger({\bf p}) \ket{0} \nol
&= \hbar^4 \int \f{d^3{\bf p^{(2)}}}{(2\pi\hbar)^3 2p_0^{(2)}}
\f{d^3{\bf p^{(3)}}}{(2\pi\hbar)^3 2p_0^{(3)}}
\f{d^3{\bf p^{(4)}}}{(2\pi\hbar)^3 2p_0^{(4)}} \f{d^3{\bf
p^{(5)}}}{(2\pi\hbar)^3 2p_0^{(5)}} \nol & \quad T \bra{0} A({\bf
p'}) \normal \(A^\dagger_2 \Phi^*_2(x') + B_2
\bar {\Phi}_2(x') \)\stackrel{\leftrightarrow}{D^{\mu'}} \(
A_3\Phi_3(x') + B^\dagger_3 \bar{\Phi}^*_3 (x') \) \normal \nol &
\quad
\normal \(A^\dagger_4 \Phi^*_4(x) + B_4
\bar {\Phi}_4(x) \)\stackrel{\leftrightarrow}{D^{\nu}} \(
A_5\Phi_5(x) + B^\dagger_5 \bar{\Phi}^*_5 (x) \) \normal
A^\dagger({\bf p})\ket{0} \,,
\end{align}
where the subscript gives the momentum of the operator or mode
function, eg $A_j \equiv A({\bf p^{(j)}})$. Considering only the
operators momentarily, we have for each interaction Hamiltonian term
\begin{align}
\normal \(A^\dagger_i  + B_i\) \( A_j + B^\dagger_j  \) \normal &=
A^\dagger_i A_j +A^\dagger_i B^\dagger_j +B_iA_j +B^\dagger_j B_i
\,.
\end{align}
Operating on this combination on the left by $\bra{0} A({\bf p'})$,
the second and fourth terms are annihilated by the vacuum.
Similarly, when operated on the right by $A^\dagger({\bf p})\ket{0}$
the third and fourth are annihilated. Overall, in terms of just the
operators, we then have
\begin{align}
\bra{0} A({\bf p'})A^\dagger_2 A_3A^\dagger_4 A_5A^\dagger({\bf
p})\ket{0} +
\bra{0} A({\bf p'})B_2 A_3A^\dagger_4 B^\dagger_5A^\dagger({\bf
p})\ket{0} \,,
\end{align}
from which we obtain the delta functions
\begin{align}
& (2\pi\hbar)^9 2^3p_0^{(2)}p_0^{(4)}p_0^{(5)}
\delta^3 ({\bf p'}-{\bf p^{(2)}}) \delta^3({\bf p^{(3)}}-{\bf
p^{(4)}}) \delta^3 ({\bf p}-{\bf p^{(5)}})  \nol +& (2\pi\hbar)^9
2^3p_0^{(2)}p_0^{(3)}p_0^{(4)}
\delta^3 ({\bf p'}-{\bf p^{(4)}}) \delta^3({\bf p^{(2)}}-{\bf
p^{(5)}}) \delta^3 ({\bf p}-{\bf p^{(3)}}) \,,
\end{align}
and an additional term representing the vacuum pair creation and
annihilation event which we can ignore.\footnote{The same event was
ignored for the initial state calculations in this chapter and
explained in the first paragraph at the beginning of section
\ref{scalarinitialsection}} Using the appropriate mode functions
and integrating out the delta functions, what remains is the
following
\begin{align}
& \hbar^4 \int \f{d^3{\bf q}}{(2\pi\hbar)^3 2q_0} \nol & T \[
\Phi^*_{{\bf p'}}(x') \stackrel{\leftrightarrow}{D^{\mu'}}
\Phi_{{\bf q}}(x') \Phi^*_{{\bf q}}(x)
\stackrel{\leftrightarrow}{D^{\nu}} \Phi_{{\bf p}}(x) +
\bar{\Phi}_{{\bf q}}(x') \stackrel{\leftrightarrow}{D^{\mu'}}
\Phi_{{\bf p}}(x') \Phi^*_{{\bf p'}}(x)
\stackrel{\leftrightarrow}{D^{\nu}} \bar{\Phi}_{{\bf q}}(x)
\] \,, \nol
\end{align}
where we have used ${\bf q}$ for the internal momentum (previously
${\bf p^{(3)}}$ in one case and ${\bf p^{(5)}}$ in the other).

We can now combine this result with that for the electromagnetic
fields.
\begin{align}
& \quad {\cal F}_2({\bf p}) \nol &= \f{ie^2\hbar}{2}
\int d^4x d^4x' \f{d^3{\bf p'}}{2p_0'(2\pi\hbar)^3}  \f{d^3{\bf q}}{(2\pi\hbar)^3 2q_0}
 \f{d^3{\bf k}}{(2\pi)^3 2k_0}
 \nol
 & \left\{ \theta(t'-t) e^{ik\cdot(x-x')} \right. \nol &  \(\Phi^*_{{\bf p'}}(x') \stackrel{\leftrightarrow}{D'_{\mu}}
\Phi_{{\bf q}}(x') \Phi^*_{{\bf q}}(x)
\stackrel{\leftrightarrow}{D^{\mu}} \Phi_{{\bf p}}(x) +
\Phi^*_{{\bf p'}}(x) \stackrel{\leftrightarrow}{D^{\mu}} \bar{\Phi}_{{\bf q}}(x)
\bar{\Phi}_{{\bf q}}(x') \stackrel{\leftrightarrow}{D'_{\mu}} \Phi_{{\bf p}}(x')
\) \nol &
 +\theta (t-t')e^{-ik\cdot(x-x')} \nol & \left.
\(\Phi^*_{{\bf p'}}(x) \stackrel{\leftrightarrow}{D_{\mu}} \Phi_{{\bf q}}(x)
\Phi^*_{{\bf q}}(x') \stackrel{\leftrightarrow}{{D'}^{\mu}} \Phi_{{\bf p}}(x') +
\Phi^*_{{\bf p'}}(x') \stackrel{\leftrightarrow}{{D'}^{\mu}} \bar{\Phi}_{{\bf q}}(x')
\bar{\Phi}_{{\bf q}}(x) \stackrel{\leftrightarrow}{D_{\mu}} \Phi_{{\bf p}}(x)
\) \right\} \,,
\end{align}
where we have rearranged the last term in each curved bracket so
that it matches the first. The two time ordered terms are then the
same under the $(x,x')$ integration symmetry. We choose to combine
them as follows:
\begin{align}
 {\cal F}_2({\bf p})  &= ie^2\hbar
\int d^4x d^4x' \f{d^3{\bf p'}}{2p_0'(2\pi\hbar)^3}  \f{d^3{\bf q}}{(2\pi\hbar)^3 2q_0}
 \f{d^3{\bf k}}{(2\pi)^3 2k_0}
 \nol
 & \quad \left\{ \theta(t-t') \Phi^*_{{\bf p'}}(x) \stackrel{\leftrightarrow}{D_{\mu}} \Phi_{{\bf q}}(x)
\Phi^*_{{\bf q}}(x') \stackrel{\leftrightarrow}{{D'}^{\mu}} \Phi_{{\bf p}}(x')e^{-ik\cdot(x-x')}
\right. \nol & \quad \left. + \theta(t'-t) \Phi^*_{{\bf p'}}(x) \stackrel{\leftrightarrow}{D^{\mu}} \bar{\Phi}_{{\bf q}}(x)
\bar{\Phi}_{{\bf q}}(x') \stackrel{\leftrightarrow}{D'_{\mu}} \Phi_{{\bf p}}(x')e^{ik\cdot(x-x')}
 \right\} \,.
\end{align}
These two terms represent the particle and antiparticle loops
respectively. We denote them as follows:
\begin{multline}
{\cal F}_{2a} ({\bf p})  = ie^2\hbar
\int d^4x d^4x' \f{d^3{\bf p'}}{(2\pi\hbar)^32p_0'}  \f{d^3{\bf q}}{(2\pi\hbar)^3 2q_0}
 \f{d^3{\bf k}}{(2\pi)^3 2k_0}
 \\ \left\{ \theta(t-t') \Phi^*_{{\bf p'}}(x) \stackrel{\leftrightarrow}{D_{\mu}} \Phi_{{\bf q}}(x)
\Phi^*_{{\bf q}}(x') \stackrel{\leftrightarrow}{{D'}^{\mu}} \Phi_{{\bf p}}(x')e^{-ik\cdot(x-x')}
\right\} \,,
\end{multline}
and
\begin{multline}
{\cal F}_{2b}({\bf p})  = ie^2\hbar
\int d^4x d^4x' \f{d^3{\bf p'}}{(2\pi\hbar)^32p_0'}  \f{d^3{\bf q}}{(2\pi\hbar)^3 2q_0}
 \f{d^3{\bf k}}{(2\pi)^3 2k_0}
 \\ \left\{ \theta(t'-t) \Phi^*_{{\bf p'}}(x) \stackrel{\leftrightarrow}{D^{\mu}} \bar{\Phi}_{{\bf q}}(x)
\bar{\Phi}_{{\bf q}}(x') \stackrel{\leftrightarrow}{D'_{\mu}} \Phi_{{\bf
p}}(x')e^{ik\cdot(x-x')}
 \right\} \,.
\end{multline}
From the integrand of the particle loop ${\cal F}_{2a}$, we have
\begin{align}
& \quad \(\phi^*_{{\bf p'}}(t)e^{-i{\bf p'}\cdot{\bf
x}/\hbar}\stackrel{\leftrightarrow}{D_{\mu}}
 \phi_{{\bf q}}(t)e^{i{\bf q}\cdot{\bf x}/\hbar}
 \phi^*_{{\bf q}}(t')e^{-i{\bf q}\cdot{\bf x'}/\hbar}
 \stackrel{\leftrightarrow}{{D'}^{\mu}}\phi_{{\bf p}}(t')e^{i{\bf
 p}\cdot{\bf x'}/\hbar}\)e^{-ik\cdot(x-x')} \nol
&= \phi^*_{{\bf p'}}(t)\phi_{{\bf p}}(t') \[
-\hbar^2\stackrel{\leftrightarrow}{\partial}_t
\stackrel{\leftrightarrow}{\partial}_{t'} +\( {\bf p}+{\bf q}-2{\bf
V}(t) \) \cdot \( {\bf p}+{\bf q}-2{\bf V}(t') \) \]\phi_{{\bf
q}}(t)\phi^*_{{\bf q}}(t') \nol & \quad \times \f{1}{\hbar^2}
e^{-i\({\bf p'}-{\bf q}-{\bf K}\)\cdot{\bf x}/\hbar} e^{i\({\bf
p}-{\bf q}-{\bf K}\)\cdot{\bf x'}/\hbar} e^{-iK(t-t')/\hbar} \,,
\end{align}
where we have used the antisymmetry of
$\stackrel{\leftrightarrow}{\partial}$ and ${\bf K}=\hbar{\bf k}$.
Also we recall that for a spatial component $j$ we have $D_{j} =
\partial_j - iV^j/\hbar$. Integration over ${\bf x}$ and ${\bf x'}$
gives the delta functions $(2\pi\hbar)^6\delta^3({\bf p}-{\bf
q}-{\bf K})\delta^3({\bf p'}-{\bf p})$. Integrating out these delta
functions using the ${\bf p'}$ and ${\bf k}$
integrals\footnote{Recall that $d^3 {\bf k}/k_0 =
\hbar^{-2}d^3{\bf K}/K$.} we obtain, with $K=|{\bf p}-{\bf q}|$,
\begin{multline}
{\cal F}_{2a}({\bf p})  = \f{ie^2}{2p_0}
\int   \f{dt dt'd^3{\bf q}}{(2\pi\hbar)^3 2q_02K}
\theta(t-t')e^{-iK(t-t')/\hbar} \\ \times \phi^*_{{\bf p}}(t)\phi_{{\bf p}}(t')
\[ -\hbar^2\stackrel{\leftrightarrow}{\partial}_t
\stackrel{\leftrightarrow}{\partial}_{t'} +\( {\bf p}+{\bf q}-2{\bf
V}(t) \) \cdot \( {\bf p}+{\bf q}-2{\bf V}(t') \) \] \phi_{{\bf
q}}(t)\phi^*_{{\bf q}}(t') \,,
\end{multline}
where the time differentiations only apply to the $\phi$ terms on
that line. Similarly, from the integrand of ${\cal F}_{2b}$, we have
\begin{align}
& \quad \(\phi^*_{{\bf p'}}(t)e^{-i{\bf p'}\cdot{\bf
x}/\hbar}\stackrel{\leftrightarrow}{D_{\mu}}
 \bar{\phi}^*_{{\bf q}}(t)e^{-i{\bf q}\cdot{\bf x}/\hbar}
 \bar{\phi}_{{\bf q}}(t')e^{i{\bf q}\cdot{\bf x'}/\hbar}
 \stackrel{\leftrightarrow}{{D'}^{\mu}}\phi_{{\bf p}}(t')e^{i{\bf
 p}\cdot{\bf x'}/\hbar}\)e^{-ik\cdot(x-x')} \nol
&= \phi^*_{{\bf p'}}(t)\phi_{{\bf p}}(t') \[
-\hbar^2\stackrel{\leftrightarrow}{\partial}_t
\stackrel{\leftrightarrow}{\partial}_{t'} +\( {\bf p}-{\bf q}-2{\bf
V}(t) \) \cdot \( {\bf p}-{\bf q}-2{\bf V}(t') \)
\]\bar{\phi}^*_{{\bf q}}(t)\bar{\phi}_{{\bf q}}(t') \nol & \quad
\times \f{1}{\hbar^2} e^{-i\({\bf p'}+{\bf q}+{\bf K}\)\cdot{\bf
x}/\hbar} e^{i\({\bf p}+{\bf q}+{\bf K}\)\cdot{\bf x'}/\hbar}
e^{iK(t-t')/\hbar} \,,
\end{align}
and consequently
\begin{multline}
{\cal F}_{2b}({\bf p})  = \f{ie^2}{2p_0}
\int   \f{dt dt'd^3{\bf q}}{(2\pi\hbar)^3 2q_02K}
\theta(t'-t)e^{iK(t-t')/\hbar} \\ \times \phi^*_{{\bf p}}(t)\phi_{{\bf p}}(t')
\[ -\hbar^2\stackrel{\leftrightarrow}{\partial}_t
\stackrel{\leftrightarrow}{\partial}_{t'} +\( {\bf p}-{\bf q}-2{\bf
V}(t) \) \cdot \( {\bf p}-{\bf q}-2{\bf V}(t') \) \]
\bar{\phi}^*_{{\bf q}}(t)\bar{\phi}_{{\bf q}}(t') \,.
\end{multline}
Using the symmetry of the ${\bf q}$ integration, we can take ${\bf
q}\to-{\bf q}$. Recalling that from the semiclassical expansion,
$\bar{\phi}_{-{\bf q}} = \phi_{{\bf q}}$ we thus obtain
\begin{multline}
{\cal F}_{2b}({\bf p})  = \f{ie^2}{2p_0}
\int   \f{dt dt'd^3{\bf q}}{(2\pi\hbar)^3 2q_02K}
\theta(t'-t)e^{iK(t-t')/\hbar} \\ \times \phi^*_{{\bf p}}(t)\phi_{{\bf p}}(t')
\[ -\hbar^2\stackrel{\leftrightarrow}{\partial}_t
\stackrel{\leftrightarrow}{\partial}_{t'} +\( {\bf p}+{\bf q}-2{\bf
V}(t) \) \cdot \( {\bf p}+{\bf q}-2{\bf V}(t') \) \] \phi^*_{{\bf
q}}(t)\phi_{{\bf q}}(t') \,.
\end{multline}

To briefly summarize the current situation, the second part of the
forward scattering amplitude is given by the sum of the amplitudes
representing the particle and antiparticle loops and can be written
\begin{align}
{\cal F}_2({\bf p})  &  =  \frac{ie^2}{2p_0}
\int \frac{d^3{\bf q}}{2q_0(2\pi\hbar)^3}\frac{1}{2K}\int dt
dt' \nol & \quad \left\{ \theta(t-t')
\left[\phi_{\bf p}^*(t)\phi_{\bf p}(t')
\stackrel{\leftrightarrow}{\cal D}_1\hspace{-1mm}
(t,t',{\bf p},{\bf q})
\phi_{\bf q}(t)\phi_{\bf q}^*(t')\right]e^{-iK(t-t')/\hbar}
\right. \nol
& \quad + \left. \theta(t'-t)
\left[\phi_{\bf p}^*(t)\phi_{\bf p}(t')
\stackrel{\leftrightarrow}{\cal D}_1\hspace{-1mm}
(t,t',{\bf p},{\bf q})
\phi_{\bf q}^*(t)\phi_{\bf q}(t')\right]
e^{iK(t-t')/\hbar}\right\} \,,\label{firstF1}
\end{align}
where
\begin{equation}
\stackrel{\leftrightarrow}{\cal D}_1\hspace{-1mm}
(t,t',{\bf p},{\bf q}) \equiv
 - \hbar^2\stackrel{\leftrightarrow}{\partial}_{t}
\stackrel{\leftrightarrow}{\partial}_{t'}
+ \left[{\bf p}+{\bf q}-2{\bf V}(t)\right]\cdot
\left[{\bf p}+{\bf q}-2{\bf V}(t')\right]
\,, \label{defD1}
\end{equation}
is the differential operator found in both loops.\footnote{This was
the main reason for the earlier manipulations of the ${\bf q}$
integration; to show that the operator can in fact be written the
same in both loops.} In order to proceed further, it is convenient
to define the new time variables $\bar{t}$ and $\eta$ as follows:
\begin{itemize}
\item $t=\bar{t}-\df{\hbar\eta}{2}$\,,
\item $t'=\bar{t}+\df{\hbar\eta}{2}$\,.
\end{itemize}
The Jacobian is straightforward: $dtdt'=\hbar \, d\bar{t}d\eta$. For
the differential operator $\stackrel{\leftrightarrow}{\cal D}_1$, we
note that
\beq
\left[{\bf p}+{\bf q}-{\bf V}(t)\right]\cdot
\left[{\bf p}+{\bf q}-{\bf V}(t')\right]
= \left[{\bf p}+ {\bf q}-{\bf V}(\bar{t})\right]^2 + {\cal
O}(\hbar^2) \,.
\eeq
Including the Heaviside functions in our current considerations, we
find that the amplitude ${\cal F}_2({\bf p})$ can be rewritten as
follows:
\beq  \label{momint}
{\cal F}_2({\bf p}) =  \frac{ie^2}{2\hbar^2p_0}
\int \frac{d^3{\bf q}}{2q_0(2\pi)^3}\frac{1}{2K}
\int d\bar{t} \left[ G_{-}({\bf p},{\bf q},\bar{t}) + G_{+}
({\bf p},{\bf q},\bar{t})\right] \,,
\eeq
with
\begin{align}
G_{-}({\bf p},{\bf q},\bar{t}) & =  \int_{-\infty}^0 d\eta
\left[ \phi_{\bf p}^*(t)\phi_{\bf p}(t')
\stackrel{\leftrightarrow}{\cal D}_2\hspace{-1mm}
(\bar{t},\eta,{\bf p},{\bf q})
\phi_{\bf q}(t)\phi_{\bf q}^*(t')\right] e^{iK\eta}\,, \\
G_{+}({\bf p},{\bf q},\bar{t}) & = \int_0^{\infty}d\eta
\left[ \phi_{\bf p}^*(t)\phi_{\bf p}(t')
\stackrel{\leftrightarrow}{\cal D}_2\hspace{-1mm}
(\bar{t},\eta,{\bf p},{\bf q})
\phi_{\bf
q}^*(t)\phi_{\bf q}(t')\right] e^{-iK\eta}\,,
\end{align}
where
\beq\label{defD2}
\stackrel{\leftrightarrow}{\cal D}_2\hspace{-1mm}
(\bar{t},\eta,{\bf p},{\bf q})
\equiv - \frac{\hbar^2}{4}\stackrel{\leftrightarrow}{\partial}_{\bar{t}}^2
-\stackrel{\leftrightarrow}{\partial}_{\eta}^2 + \left[{\bf p}+{\bf
q}-2{\bf V}(\bar{t})\right]^2 + {\cal O}(\hbar^2)
\,.
\eeq
The $G_-$ term results from the particle loop contribution and the
$G_+$ term from the antiparticle loop. The reader will notice that
in the definitions for $G_\pm$ we have yet to convert the $\phi$'s
to the new time variables. This task is the more complicated,
involving the semiclassical expansion, and the evaluation of $G_\pm$
forms the bulk of our work in finding the forward scattering
amplitude. Thus we have presented the above definitions first for
aid of presentation. Apropos the mode function, the semiclassical
expansion of the time-dependent factor $\phi_{\bf q}(t)$, can be
written
\beq \label{scalartmodeho}
\phi_{\bf q}(t) = \sqrt{\frac{q_0}{ E_q(t)}}\varphi_{\bf q}(t)
\exp \left[ - \frac{i}{\hbar}\int_0^t  E_q(\zeta)\,d\zeta\right]\,,
\eeq
where the higher order $\hbar$ corrections are contained within
$\varphi_{\bf q}(t)$ and thus we require $\varphi_{\bf q}(t)
\to 1$ as $\hbar \to 0$. Substitution into the field equation
(\ref{phitKG}) for $\phi(t)$ gives the $\hbar$ expansion of
$\varphi_{\bf q}(t)$ as
\beq
\varphi_{\bf q}(t) = 1 + i\hbar \varphi^{(1)}_{\bf q}(t) + {\cal
O}(\hbar^2)\,.
\eeq
The explicit form of $\varphi^{(1)}_{\bf q}(t)$ will actually be
unnecessary for our calculations, though it can easily be found.
Note also that
\beq
\phi_{\bf q}^*(t)\phi_{\bf q}(t')
= \frac{q_0\varphi_{\bf q}^*(t)\varphi_{\bf q}(t')} {\sqrt{ E_q(t)
E_q(t')}}
\exp\left[ -i\int_{-\eta/2}^{\eta/2}
 E_q(\bar{t}+\hbar\zeta)\,d\zeta\right]\,.
\eeq
We note the following: Converting to $(\bar{t},\eta)$ variables we
have $\varphi_{\bf q}^*(t)\varphi_{\bf q}(t')
= 1 + {\cal O}(\hbar^2)$ and $\sqrt{ E_q(t) E_q(t')} =
 E_q(\bar{t}) + {\cal O}(\hbar^2)$. We also note the lack of an
order $\hbar$ term in (\ref{defD2}) because
\beq
\left[{\bf p}+{\bf q}-{\bf V}(t)\right]\cdot
\left[{\bf p}+{\bf q}-{\bf V}(t')\right]
= \left[{\bf p}+ {\bf q}-{\bf V}(\bar{t})\right]^2 + {\cal
O}(\hbar^2)\,.
\eeq
With these relations in mind, it can readily be shown that the
functions $G_{\pm}({\bf p},{\bf q},t)$ are of the form
\begin{multline}
G_{\pm} ({\bf p},{\bf q},t) = \pm\int_{0}^{\pm\infty}d\eta
\left[ f_{\pm}({\bf p},{\bf q},t) + {\cal O}(\hbar^2)\right]
\\ \times \exp\left\{ \mp i\int_{-\eta/2}^{+\eta/2}
d\zeta \left[\pm  E_p(t+\hbar\zeta) +  E_q(t+\hbar\zeta) +
K\right]\right\} \,, \label{Gpm}
\end{multline}
where after performing the appropriate differentiations, the
function $f_{\pm}({\bf p},{\bf q},t)$ can be found as
\beq\label{fplus}
f_{\pm}({\bf p},{\bf q},t)  = \left\{- \[
 E_p(t)\mp E_q(t)\]^2
+ \[{\bf p}+{\bf q}-2{\bf V}(t)\]^2\right\}|\phi_{\bf p}(t)|^2
|\phi_{\bf q}(t)|^2 \,.
\eeq
The points noted above conspire to produce the important fact that
there are no terms of order $\hbar$ in the pre-factor of
(\ref{Gpm}), inside the first square brackets. Having now removed
the last trace of the original time variables, we may clean up the
notation by changing $\bar{t}\to t$ and consider the evaluation of
the above $G_\pm$ integrals.

Let us first consider the integral $G_{+}({\bf p},{\bf q},t)$. We
change the integration variable from $\eta$ to $\beta$ defined by
the following relation:
\beq \label{defbeta}
[ E_p(t) +  E_q(t) + K]\beta\equiv
\int_{-\eta/2}^{\eta/2}
\left[ E_p(t+\hbar\zeta)
+  E_q(t+\hbar\zeta) + K\right]\,d\zeta \,.
\eeq
Expanding the integrand and integrating the right hand side, we
obtain
\begin{align}
\[ E_p(t) +  E_q(t) + K \]\beta &=
\[  E_p(t) +  E_q(t) + K \] \eta
+ \hbar^2\f{\eta^3}{3.2^3} \[  \ddot{E}_p(t) +
 \ddot{E}_q(t)\] \nol & \quad +{\cal O}(\hbar^4\eta^5) \,.
\end{align}
From the above it is evident that $\eta=\beta+{\cal O}(\hbar^2)$.
Hence we solve this equation for $\eta$ as a function of $\beta$ for
small $\hbar$ and find
\beq
\eta = \left[ 1 - \frac{1}{24}\frac{ \ddot{E}_p(t)+
 \ddot{E}_q(t)}
{ E_p(t)+ E_q(t)+K}\hbar^2\beta^2 + {\cal
O}(\hbar^4\beta^4)\right]\beta \,, \label{etabetaexp}
\eeq
and
\beq \label{detabetaexp}
d\eta = \left[ 1 - \frac{1}{8}\frac{ \ddot{E}_p(t) +
\ddot{E}_q(t)} { E_p(t)+ E_q(t)+K}\hbar^2\beta^2 + {\cal O}
(\hbar^4\beta^4)\right] d\beta \,.
\eeq
We label the Jacobian $\hbar$ expansion contained in the square
brackets by \newline $J({\bf p},{\bf q},t,\hbar\beta)$. The integral
we are concerned with is then
\beq\label{Gplus}
G_{+}({\bf p},{\bf q},t) = \int_{0}^{\infty}d\beta
\left[ f_{+}({\bf p},{\bf q},t) + {\cal O}(\hbar^2)\right]\,
\exp\left\{ -i\left[ E_p(t)
+  E_q(t) + K\right]\beta\right\} \,.
\eeq
The integration can be completed if we introduce a convergence
factor by replacing $K$ with $K-i\epsilon$. Accordingly, we obtain
\beq
G_{+}({\bf p},{\bf q},t) = -\frac{if_{+}({\bf p},{\bf q},t)} {
E_q(t)+ E_p(t)+K} + {\cal O}(\hbar^2) \,.
\eeq
The corresponding contribution to the forward-scattering amplitude
can be seen with reference to (\ref{momint}) as\footnote{Recall the
change of notation $\bar{t}\to t$ since that reference.}
\beq
{\cal F}_{2+}({\bf p}) =  \frac{e^2}{2\hbar^2 p_0}\int dt
\int \frac{d^3{\bf q}}{2q_0(2\pi)^3}\frac{1}{2K}
\frac{f_{+}({\bf p},{\bf q},t)}{ E_p(t)
+  E_q(t) + K} + {\cal O}(\hbar^0) \,,  \label{div1}
\eeq
The amplitude ${\cal F}_{2+}({\bf p})$ can readily be seen to be
ultra-violet divergent. This will not however be a cause of
difficulty, as we fully expect the results to be divergent.

Next we analyse the contribution from $G_{-}({\bf p},{\bf q},t)$,
with which we find ourselves additional difficulties. One cannot
proceed as above because of the infrared divergence in the ${\bf
q}$-integration as we shall see shortly. We start as in the previous
case and define the variable $\tilde{\beta}$ in analogy with the
variable $\beta$ in (\ref{defbeta}) as follows
\beq \label{deftildebeta}
[- E_p(t) +  E_q(t) + K]\tilde{\beta} \equiv
\int_{-\eta/2}^{\eta/2}
\left[- E_p(t+\hbar\zeta)
+  E_q(t+\hbar\zeta) + K\right]\,d\zeta \,.
\eeq
With foresight knowledge of the new divergence as $K\to
0$~\footnote{Recall $d^3{\bf q}=d^3{\bf K}$.} we should check the
validity of using the variable $\tilde{\beta}$ in such
circumstances. For small $K = \|{\bf p} - {\bf q}\|$, we have
\beq\label{smallK}
- E_p(t)+ E_q(t) + K \approx
 K - {\bf v}(t)\cdot {\bf K}\,,
\eeq
where ${\bf v}(t) = [{\bf p}-{\bf V}(t)]/ E_p(t)$ is the velocity of
the classical particle with final momentum ${\bf p}$.~\footnote{See
(\ref{wkbvel}) for this result.} Hence, in the limit $K\to 0$ one
finds
\beq
\tilde{\beta} = \frac{1}{1-{\bf v}(t)\cdot{\bf n}}
\int_{-\eta/2}^{\eta/2} \[1-{\bf v}(t+\hbar\zeta)\cdot{\bf n}
\]d\zeta\,,
\eeq
where ${\bf n} \equiv {\bf K}/K$.  Thus, if we write $d\eta =
\tilde{J}({\bf p},{\bf q},t,\hbar\tilde{\beta})d\tilde{\beta}$, then
the function $\tilde{J}({\bf p},{\bf q},t,\hbar\tilde{\beta})$ is
finite as $K\to 0$ and we can now safely use the $\tilde{\beta}$
definition (\ref{deftildebeta}). The expression corresponding to
(\ref{Gplus}) can be given in the following form:
\begin{align}
G_{-}({\bf p},{\bf q},t) & =  \int_{-\infty}^{0} d\tilde{\beta}
\left[ f_{-}({\bf p},{\bf q},t)
+ \sum_{n,d}\hbar^n\tilde{\beta}^d f_{nd-}({\bf p},{\bf q},t)\right]
\nol & \qquad\times \exp\left\{ i\left[ - E_p(t) +  E_q(t) + K\right]
\tilde{\beta}\right\}\nol
& =  \frac{-if_{-}({\bf p},{\bf q},t)} {- E_p(t)+  E_q(t)+K} +
\sum_{n,d}\frac{(-i)^d d!\hbar^n f_{nd-}({\bf p},{\bf q},t)}
{\left[- E_p(t)+ E_q(t)+K\right]^{d+1}}
 \,, \label{hexp}
\end{align}
with $n\geq 2$ and $n\geq d\geq 0$. The higher order pre-factors
$f_{nd-}({\bf p},{\bf q},t)$, which are finite as $K \to 0$, can not
be removed in the $\hbar\to 0$ limit as for $G_{+}$ due to the
infrared divergence. This can be seen clearly if we substitute
(\ref{hexp}) into the amplitude expression (\ref{momint}):
\begin{align}
{\cal F}_{2-}({\bf p}) &=  \frac{ie^2}{2\hbar^2p_0}
\int \frac{d^3{\bf q}}{(2\pi)^32q_0}\frac{1}{2K}
\int dt \nol &  \qquad\[ \frac{-if_{-}({\bf p},{\bf q},t)}
{- E_p(t)+  E_q(t)+K} + \sum_{n,d}\frac{(-i)^d d!\hbar^n
f_{nd-}({\bf p},{\bf q},t)} {\left[- E_p(t)+ E_q(t)+K\right]^{d+1}}
\] \,. \label{div2}
\end{align}
Here we see that the terms with $d \geq 1$ are infrared divergent in
the ${\bf q}$-integration because of the limit $\lim_{K\to 0}[-
E_p(t)+ E_q(t)+K] \to 0$ as can be seen from (\ref{smallK}).

We can approach this difficulty by separating the infrared divergent
section of the integral via the addition of a cut-off in the ${\bf
q}$ (or equivalently ${\bf K}$) integral. We may then consider the
situation above the cut-off and return to the problematic
sub-cut-off area later. Let us thus cut-off the integral by
requiring
\begin{equation}
K \geq K_0 = \hbar^\alpha \lambda \,,
\end{equation}
with $\lambda$ a positive constant and where $\f{3}{4}<\alpha<1$.
The reasoning for choosing these precise limits on the choice of
$\alpha$ will become apparent in later stages of the calculations.
Above the cut-off $K_0$, we find that the contributions of the terms
of ${\cal F}_{2-}$ to the ${\bf q}$ integration have the small-$K$
behaviour
\begin{equation}
\begin{cases}
\hbar^{n}K_0^{1-d} = \hbar^{n + (1-d)\alpha}\lambda^{1-d} & \text{if $d \geq
2$}
\\ \hbar^n \log(\hbar^\alpha \lambda) & \text{if $d=1$}
\end{cases} \,.
\end{equation}
Since $1-\alpha > 0$, $n \geq 2$ and $n\geq d$, from their
appropriate definitions, we have
\begin{equation}
n + (1-d)\alpha \geq  2-\alpha \,,
\end{equation}
for the $\hbar$ power in the $d \geq 2$ case above.\footnote{ $n +
(1-d)\alpha \geq n +(1-n)\alpha = n(1-\alpha)+\alpha \geq
2(1-\alpha)-\alpha = 2-\alpha$\,.} We thus see that in the $\hbar\to
0$ limit, the higher order $f_{nd-}$ terms will not contribute above
the cut-off.\footnote{This explains our choice of the upper limit on
$\alpha$.} In this arena we are thus left only with the leading
order term containing $f_-$, in a situation analogous to ${\cal
F}_{2+}$. We give this leading order contribution, over the full
range both above and below the cut-off, the label ${\cal F}_{2-}^0$,
\begin{align}
{\cal F}_{2-}^0 =  \frac{e^2}{2\hbar^2 p_0}\int dt
\int \frac{d^3{\bf q}}{2q_0(2\pi)^3}\frac{1}{2K}
\frac{f_{-}({\bf p},{\bf q},t)}{\(- E_p(t)
+  E_q(t) + K\)} \,. \label{F20sub}
\end{align}
For ${\cal F}_{2+}$, we effectively have ${\cal F}_{2+}={\cal
F}_{2+}^0$ to order $\hbar^{-1}$ (which is the highest order we
require).

At this point we pause to take stock of the various contributions to
the forward scattering. Firstly we have the leading order terms. We
combine ${\cal F}_{1}$, ${\cal F}_{2+}^0$ and ${\cal F}_{2-}^0$ and
let
\begin{equation}
{\cal F}^0 ={\cal F}_{1}+{\cal F}_{2+}^0+{\cal F}_{2-}^0 \,.
\end{equation}
What remains is the contribution of the higher order terms from
${\cal F}_{2-}$, from below the cut-off. We shall attack this in a
round-about way. Labelling the full ${\cal F}_{2-}$ term below the
cut-off by ${\cal F}_{2-}^<$, and the leading order term in the same
range by ${\cal F}_{2-}^{<,0}$, the desired contribution can be
calculated as
\begin{equation}
{\cal F}^{<,{\rm ho}}={\cal F}_{2-}^<-{\cal F}_{2-}^{<,0} \,.
\end{equation}
Thus the forward scattering amplitude\footnote{non-renormalised so
far,} is given by
\begin{align}
{\cal F}({\bf p}) = {\cal F}^0({\bf p})+{\cal F}^{<,{\rm ho}}({\bf
p}) \,.
\end{align}
With this aside complete, we now return to finish the calculation of
these two terms.

Firstly we turn to the higher order contributions ${\cal F}^{<,{\rm
ho}}$ which we approach as described above. For these terms, we are
interested in their behaviour at small-$k$. For the full-term ${\cal
F}_{2-}^<$, we consequently back-track somewhat to the expression in
(\ref{firstF1}), the appropriate part of which gives
\begin{align}
{\cal F}_2^<({\bf p})  &=  \frac{ie^2}{2p_0}
\int_< \frac{d^3{\bf q}}{2q_0(2\pi\hbar)^3}\frac{1}{2K}\int dt
dt' \theta(t-t') \nol & \qquad \times
\left[\phi_{\bf p}^*(t)\phi_{\bf p}(t')
\stackrel{\leftrightarrow}{\cal D}_1\hspace{-1mm}
(t,t',{\bf p},{\bf q})
\phi_{\bf q}(t)\phi_{\bf q}^*(t')\right]e^{-iK(t-t')/\hbar} \,.
\label{firstF1p}
\end{align}
The subscript $<$ on the ${\bf q}$ integral indicates that we
integrate below the cut-off. We change the integration variable,
firstly to the photon momentum ${\bf K}={\bf p}-{\bf q}$ and then to
the wave number ${\bf k}={\bf K}/\hbar$. Ergo $d^3{\bf q}/(\hbar^3K)
= d^3{\bf k}/(\hbar k)$ and write
\begin{align}
{\cal F}_2^<({\bf p}) &=  \frac{ie^2}{2\hbar p_0}
\int dt dt'\int_{k \leq \hbar^{\alpha-1}\lambda}
\frac{d^3{\bf k}}{2q_0(2\pi)^3}\frac{1}{2k}\theta(t-t')\nol
& \qquad \times
\left[\phi_{\bf p}^*(t)\phi_{\bf p}(t')
\stackrel{\leftrightarrow}{\cal D}_1\hspace{-1mm}
(t,t',{\bf p},{\bf q})
\phi_{\bf q}(t)\phi_{\bf q}^*(t')\right]e^{-ik(t-t')} \,.
\end{align}
 In the $\hbar\to 0$ limit
we note that the upper limit $\hbar^{\alpha-1}\lambda$ of
integration for $k$ becomes infinite. Now we have ${\bf q}= {\bf p}
- \hbar {\bf k}$. Hence, we have ${\bf q} \to {\bf p}$ for all ${\bf
k}$ as $\hbar\to 0$ because $\hbar\cdot\hbar^{\alpha-1}\lambda \to
0$. Using these limits, and with reference to our previous
calculations for ${\cal F}_{2-}$, the exponential factor can take
the form
\begin{multline}\label{truncate1}
\exp\left\{ i\int_{t}^{t'} d\zeta\left[K +  E_q(\zeta)- E_p (\zeta)\right]/\hbar\right\} \\ =
\exp\left\{ i\int_{t}^{t'} \left[ k - \frac{\partial E_p(\zeta)}{\partial {\bf p}}\cdot{\bf k} +
\frac{1}{2}\frac{\partial^2 E_p(\zeta)}{\partial
p^i\partial p^j}\hbar k^ik^j + \cdots\right]d\zeta\right\} \,.
\end{multline}
In order to truncate the series in the exponent at the second term
for all ${\bf k}$ in the integration range, one would require $\hbar
(\hbar^{\alpha-1})^2 \to 0$ as $\hbar\to 0$ i.e. $\alpha >
\frac{1}{2}$.
 This is naturally satisfied due to the earlier choice of $\alpha
 >\frac{3}{4}$. Thus for the ${\bf q}\to {\bf p}$ limit we write
\begin{align}
{\cal F}_2^<({\bf p}) &=  \frac{ie^2}{\hbar 2 p_0}
\int dt dt'\int_{k \leq \hbar^{\alpha-1}\lambda}
\frac{d^3{\bf k}}{2p_0(2\pi)^3}\frac{1}{2k}\theta(t-t')\nol
& \quad \times
\left[ -2 E_p(t) 2 E_p(t') + 4 \({\bf p}-{\bf V}(t)\)\cdot\({\bf p}-{\bf V}(t')\)
\right]|\phi_{\bf p}(t)|^2|\phi_{\bf p}(t')|^2 \nol
& \qquad \exp \left[ i k(t'-t) - i\int_{t}^{t'}
\frac{{\bf p}-{\bf V}(\zeta)}{ E_p(\zeta)}\cdot {\bf k}
d\zeta\right] \(1+{\cal O}(\hbar k^2)\) \nol &= \frac{ie^2}{\hbar}
\int dt dt'\int_{k \leq \hbar^{\alpha-1}\lambda}
\frac{d^3{\bf k}}{(2\pi)^32k}\theta(t-t')\nol
& \quad \times
\left[ -1 +  \f{\({\bf p}-{\bf V}(t)\)\cdot\({\bf p}-{\bf V}(t')\)}{ E_p(t)E_p(t')}
\right] \nol
& \qquad \exp \left[ i k(t'-t) - i\int_{t}^{t'}
\frac{{\bf p}-{\bf V}(\zeta)}{ E_p(\zeta)}\cdot {\bf k}
d\zeta \right] + {\cal O}(\hbar^{4\alpha-4}) \,,
\end{align}
where we have used $|\phi_{\bf p}(t)|^2=p_0/ E_p(t) + {\cal
O}(\hbar^2)$ and recall from (\ref{momentum1}) that the local
momentum is $md{\bf x}/d\tau={\bf p}-{\bf V}$. The extra ${\cal
O}(\hbar^{4\alpha-4})$ is the result of the combination of $\hbar
k^2$ from the higher order dependence with the $k/\hbar$ dependence
multiplying the entire integrand. These non-leading terms do not
contribute to the position shift provided the overall order is
greater than $\hbar^{-1}$, which is the case since $4\alpha-4>-1$
because of our earlier requirement that $\alpha>\frac{3}{4}$. We
consequently drop this contribution from now on. Recalling that
$[{\bf p}-{\bf V}(t)]/ E_p(t)$ is the velocity of the corresponding
classical particle, $d{\bf x}/dt$, we obtain at leading order
\begin{align}
{\cal F}_2^{<}({\bf p}) &= \frac{ie^2}{\hbar}\int_{k \leq
\hbar^{\alpha-1}\lambda}
\frac{d^3{\bf k}}{(2\pi)^32k} dtdt' \theta(t-t') \[
-\f{d x^\mu}{dt}\f{d x_\mu}{dt'}\] e^{i\(k(t'-t)-{\bf k}\cdot({\bf
x}(t)-{\bf x}(t'))\)} \nol & = -\frac{ie^2}{\hbar}\int_{k \leq
\hbar^{\alpha-1}\lambda}
\frac{d^3{\bf k}}{(2\pi)^32k}\int_{-\infty}^{+\infty}d\xi
\int_{-\infty}^{+\infty}d\xi'
\theta(\xi-\xi')\frac{dx^\mu}{d\xi}\frac{dx_\mu}{d\xi'}
e^{ik(\xi'-\xi)} \,, \label{Fsub2}
\end{align}
where in analogy with the emission amplitude, we have defined $\xi
\equiv t_1 - {\bf n}\cdot {\bf x}(t_1)$ and $\xi' \equiv t_2 - {\bf
n}\cdot{\bf x}(t_2)$ with ${\bf n} \equiv {\bf k}/k$.

The Heaviside function can be rewritten in the form
\begin{equation} \label{Heavisign}
\theta(\xi-\xi')=\f{1}{2} + \f{1}{2}\epsilon(\xi-\xi') \,,
\end{equation}
where\footnote{The function $\epsilon(x)$ is the sign function and
sometimes written $\mbox{sgn}(x)$.}
\begin{equation}\label{signfunc}
\epsilon(\xi-\xi') \equiv
\begin{cases} 1 & \text{if $\xi>\xi'$} \\ -1 & \text{if $\xi<\xi'$}
\end{cases} \,.
\end{equation}
The use of (\ref{Heavisign}) in place of the step function in ${\cal
F}_2^<$ in (\ref{Fsub2}) has the advantage that it splits the
 real and imaginary parts of the expression.
The reader will recall that only the real part of the forward
scattering amplitude affects the position shift. Also worthy of
recall is the equivalence demonstrated between the emission
probability and the imaginary part of ${\cal F}({\bf p})$ as given
by (\ref{unitarity1}). Using the above, we find twice the imaginary
part as
\begin{equation}
2\Im\,{\cal F}^{<}({\bf p}) = \frac{e^2}{\hbar}\int_{k \leq
\hbar^{\alpha-1}\lambda}
\frac{d^3{\bf k}}{(2\pi)^32k}\int_{-\infty}^{+\infty}d\xi
\int_{-\infty}^{+\infty}d\xi'
\frac{dx^\mu}{d\xi}\frac{dx_\mu}{d\xi'}
e^{ik(\xi'-\xi)} \,. \label{imaginary}
\end{equation}
In the limit $\hbar^{\alpha-1}\lambda \to \infty$ this expression
coincides with the emission probability using our expression for the
emission amplitude, as required by unitarity.\footnote{This
demonstrates the semiclassical approximation for the emission
probability and thus validates the previous physically reasonable
assumption that a typical photon energy emitted has energy of order
$\hbar$.}
 The real part can similarly be written
\begin{align}
& \quad \Re \,{\cal F}_2^<({\bf p})
\nol & =  \frac{ie^2}{2\hbar}\int_{k \leq \hbar^{\alpha-1}\lambda}
\frac{kdkd\Omega}{2(2\pi)^3}
\int_{-\infty}^{+\infty}
d\xi \int_{-\infty}^{+\infty} d\xi'
\epsilon(\xi'-\xi)
\frac{dx^\mu}{d\xi}\chi(\xi)\frac{dx_\mu}{d\xi'}\chi(\xi')
e^{ik(\xi'-\xi)} \,.
\end{align}
Here we have again introduced the cut-off function $\chi(\xi)$ as
defined in section \ref{cutoffsection}. In addition, we have used
the antisymmetry of the sign function
$-\epsilon(\xi-\xi')=\epsilon(\xi'-\xi)$ and expanded the $d^3{\bf
k}$ in spherical polar coordinates. Integration by parts with
respect to $\xi'$ gives
\begin{multline}
\Re \,{\cal F}_2^<({\bf p})
 =  -\frac{ie^2}{2\hbar}\int_{k \leq \hbar^{\alpha-1}\lambda}
\frac{kdkd\Omega}{2(2\pi)^3}
\int_{-\infty}^{+\infty}
d\xi \int_{-\infty}^{+\infty} d\xi' \\
\f{d}{d\xi'}\[\epsilon(\xi'-\xi)\frac{dx_\mu}{d\xi'}\chi(\xi')\]
\frac{dx^\mu}{d\xi}\chi(\xi)
\f{e^{ik(\xi'-\xi)}}{ik} \,.
\end{multline}
Alternatively, we can integrate by parts with respect to $\xi$ to
obtain a similar result. Adding the two expressions and dividing by
two we obtain the symmetrized version
\begin{multline}
\Re\,{\cal F}_2^{<}({\bf p})  =
 - \frac{e^2}{4\hbar}\int_{k\leq \hbar^{\alpha-1}\lambda}
\frac{dkd\Omega}{2(2\pi)^3}
\int_{-\infty}^{+\infty}
d\xi \int_{-\infty}^{+\infty} d\xi'
\left\{ 4\delta(\xi'-\xi) \frac{dx^\mu}{d\xi'}\frac{dx_\mu}{d\xi}\right.
\\ \left.
+ \epsilon(\xi'-\xi)\left[ \(\frac{d}{d\xi'}-\frac{d}{d\xi}\)
\frac{dx^\mu}{d\xi'}
\chi(\xi')\frac{dx_\mu}{d\xi}\chi(\xi)\right]
\right\}e^{ik(\xi'-\xi)}  \,, \label{realF2}
\end{multline}
where we have used the result $d \epsilon(x)/dx = 2 \delta (x)$ and
we have taken the limit $\chi(\xi)\to 1$ for the first term.
Consider for a moment the second term of (\ref{realF2}): Due to the
$\xi\leftrightarrow\xi'$ symmetry of the factors inside the curly
brackets, in the $\hbar\to 0$ limit we can extend the integration
range of $k$ from $(0,\infty)$ to $(-\infty,+\infty)$ and divide by
two once more. Consequently the $k$ integration produces the delta
function $\delta(\xi'-\xi)$. The second term of (\ref{realF2}) is
zero when $\xi=\xi'$ and consequently we can say that the
contribution from this term to $\Re {\cal F}_2^<$ is of order higher
than $\hbar^{-1}$. The first term thus remains which we can rewrite
as
\begin{align}
\Re\,{\cal F}_2^{<}({\bf p})  &=
 - \frac{e^2}{2(2\pi)^3\hbar}\int_{k\leq \hbar^{\alpha-1}\lambda}
dkd\Omega  d\xi
 \frac{dx^\mu}{d\xi}\frac{dx_\mu}{d\xi} \nol
 &=  - \frac{e^2}{16\pi^3\hbar}\int_{k\leq \hbar^{\alpha-1}\lambda}
dkd\Omega  dt
 \frac{dx^\mu}{dt}\frac{dx_\mu}{dt}\f{dt}{d\xi} \nol
 &= - \frac{e^2\lambda}{16\pi^3\hbar^{2-\alpha}}\int_{-\infty}^{+\infty}
dt \int d\Omega \frac{1-{\bf v}^2}{1-{\bf n}\cdot {\bf v}}
\,, \label{F2subfin}
\end{align}
to order $\hbar^{-1}$. We have integrated over $k$ and noted that
$\dot{\xi} = 1-{\bf n}\cdot{\bf v}$ and
$(dx^\mu/dt)(dx_\mu/dt)=1-{\bf v}^2$.

We now turn our attention to the leading order term of ${\cal
F}_{2-}$ below the cut-off, which by expanding $f_-$ in
(\ref{F20sub}) is given by
\begin{multline}
{\cal F}_{2}^{<,0} =  \frac{e^2}{2\hbar^2 p_0}\int dt
\int_{K \leq \hbar^\alpha \lambda} \frac{d^3{\bf
q}}{2q_0(2\pi)^3}\frac{1}{2K} \\
\frac{-\( E_p((t)+ E_q(t)\)^2 + \({\bf p}+{\bf q}-2{\bf V}(t)\)^2}{\(- E_p(t)
+  E_q(t) + K\)} |\phi_{\bf p}(t)|^2|\phi_{\bf q}(t)|^2 \,.
\label{F20sub2}
\end{multline}
Using the small-$K$ approximation (\ref{smallK}) and noting the
following equations
\begin{align}
|\phi_{\bf p}(t)|^2 &= \f{p_0}{ E_p(t)}+{\cal O}(\hbar^2) \,,
\\
\f{{\bf p}-{\bf V}(t)}{ E_p(t)} &= {\bf v} \,,
\end{align}
we find
\begin{align}
{\cal F}_{2}^{<,0} &=  \frac{e^2}{2(2\pi)^3\hbar^2 }\int dt
\int_{K \leq \hbar^\alpha \lambda} \frac{d^3{\bf
q}}{K}\f{1}{4 E_p^2(t)}
\frac{-4 E_p^2 + 4\({\bf p}-{\bf V}(t)\)^2}{K\(1-{\bf n}\cdot{\bf
v}\)} \nol &= -\frac{e^2}{16\pi^3\hbar^2 }\int dt
\int_{K \leq \hbar^\alpha \lambda}dKd\Omega
\frac{1 -{\bf v}^2}{\(1-{\bf n}\cdot{\bf
v}\)} \nol &= -
\frac{e^2\lambda}{16\pi^3\hbar^{2-\alpha}}\int_{-\infty}^{+\infty}
dt \int d\Omega \frac{1-{\bf v}^2}{1-{\bf n}\cdot {\bf v}} \,.
\end{align}
We recognize the same expression arrived at in this limit for ${\cal
F}_2^<$ in (\ref{F2subfin}).

From the above results we thus conclude that ${\cal F}_{2}^{<,0}$ is
equal to the leading term of ${\cal F}_2^<$. Hence ${\cal F}^{<,{\rm
ho}}={\cal F}_2^<-{\cal F}_{2}^{<,0}$ is of order $\hbar^{-1}$, but
is purely imaginary at this order. Due to the fact that only the
real part of the forward scattering affects the position shift, the
only remaining contributions are those grouped under ${\cal F}^0$
and it is these terms to which we now draw our attention.

The leading order part of the forward scattering amplitude was
earlier defined by ${\cal F}^0 = {\cal F}_1 +{\cal F}_{2+}^0 + {\cal
F}_{2-}^0$ where we have so far found that
\begin{align}
{\cal F}_1({\bf p})&= -\f{3e^2}{2\hbar^2p_0}\int dt |\phi_{\bf
p}(t)|^2 \int \f{d^3{\bf
K}}{(2\pi)^32K} \tag{\ref{F1endk}} \\
{\cal F}_{2+}({\bf p}) &=  \frac{e^2}{2\hbar^2 p_0}\int dt
\int \frac{d^3{\bf q}}{2q_0(2\pi)^3}\frac{1}{2K}
\frac{f_{+}({\bf p},{\bf q},t)}{ E_p(t)
+  E_q(t) + K} + {\cal O}(\hbar^0)  \tag{\ref{div1}} \\
{\cal F}_{2-}^0 ({\bf p}) &=  \frac{e^2}{2\hbar^2 p_0}\int dt
\int \frac{d^3{\bf q}}{2q_0(2\pi)^3}\frac{1}{2K}
\frac{f_{-}({\bf p},{\bf q},t)}{\(- E_p(t)
+  E_q(t) + K\)} \,. \tag{\ref{F20sub}}
\end{align}
Substituting in the expressions for $f_\pm$ the ${\cal F}_{2\pm}$
terms are\footnote{We have removed the $(t)$ from the $E$'s and
${\bf V}$'s for ease of presentation as there is no risk of
confusion here.}
\begin{align}
& \quad {\cal F}_{2\pm} \nol &= \frac{e^2}{2\hbar^2 p_0}\int dt
\int \frac{d^3{\bf q}}{2q_0(2\pi)^3}\frac{1}{2K} \nol & \qquad
\frac{\[-\( E_p\mp E_q\)^2 + \({\bf p} + {\bf q} - 2{\bf V}\)^2\]}
{\(\pm E_p +  E_q + K\)} |\phi_{\bf p}(t)|^2 |\phi_{\bf q}(t)|^2
\nol &= \frac{e^2}{2\hbar^2 p_0}\int dt |\phi_{\bf p}(t)|^2
\int \frac{d^3{\bf q}}{(2\pi)^32 E_q(t)}\frac{1}{2K}
\frac{\[-\( E_p\mp E_q\)^2 + \({\bf p} + {\bf q} - 2{\bf V}\)^2\]}
{\(\pm E_p +  E_q + K\)} \,.
\end{align}
We slightly modify the expression for ${\cal F}_1$, using the
variable of integration ${\bf q}={\bf p}-{\bf K}$ defined and used
in the ${\cal F}_2$'s, to bring it in line with the others in form.
\begin{align}
{\cal F}_1 ({\bf p})&= \f{e^2}{2\hbar^2p_0}\int dt |\phi_{\bf
p}(t)|^2 \int
\frac{d^3{\bf q}}{(2\pi)^32 E_q(t)}\frac{1}{2K} \[ -6 E_q(t) \]
\,.
\end{align}
Combining the terms of ${\cal F}^0({\bf p})$, we therefore write
\begin{align}
& \quad {\cal F}^0({\bf p})\nol &= \f{e^2}{2\hbar^2p_0}\int dt
|\phi_{\bf p}(t)|^2 \int
\frac{d^3{\bf q}}{(2\pi)^32 E_q}\frac{1}{2K}
\Big[ -6 E_q \nol & \quad
 \left.+ \frac{\[-\( E_p+ E_q\)^2 + \({\bf p} + {\bf q} - 2{\bf V}\)^2\]}
{\(- E_p +  E_q + K\)}+ \frac{\[-\( E_p- E_q\)^2 + \({\bf p} + {\bf
q} - 2{\bf V}\)^2\]} {\( E_p +  E_q + K\)} \] \,.
\end{align}
We define the momenta $\tilde{\bf p}={\bf p}-{\bf V}(t)$ and
$\tilde{\bf q}={\bf q}-{\bf V}(t)$, so that we have $
E_p=\sqrt{\tilde{\bf p}^2+m^2}$ and similarly for $ E_q$. After
these substitutions, we are free to change the integration variable
from ${\bf q}$ to $\tilde{\bf q}$. This gives
\begin{align}
{\cal F}^0({\bf p}) &= -\f{e^2}{2\hbar^2p_0}\int dt |\phi_{\bf
p}(t)|^2 \int
\frac{d^3\tilde{\bf q}}{(2\pi)^32E_q}\frac{1}{2K} \nol &  \quad
\[ 6E_q
+ \frac{\[\(E_p+E_q\)^2 - \(\tilde{\bf p} + \tilde{\bf q}\)^2\]}
{\(-E_p + E_q + K\)}+ \frac{\[\(E_p-E_q\)^2 - \(\tilde{\bf p} +
\tilde{\bf q}\)^2\]} {\(E_p + E_q + K\)} \] \,. \label{F0final}
\end{align}
This term is then our remaining contribution to the forward
scattering amplitude. It is real and of order $\hbar^{-2}$, thus
would contribute at order $\hbar^{-1}$ to the position shift. It is
also divergent. We shall now show that this divergent contribution
is exactly canceled by the contribution from the divergent mass
counterterm when we renormalise the mass.

\subsection{Renormalisation}
We achieve renormalisation of the mass by the counterterm addition
to the Lagrangian
\begin{align}
\delta{\cal L} = \f{\delta m^2}{\hbar^2} \varphi^\dagger \varphi
\,.
\end{align}
This in turn provides an additional contribution to the interaction
Hamiltonian that is included in the forward scattering, viz
\begin{align}
\delta{\cal H}_{I} = -\f{\delta m^2}{\hbar^2} \varphi^\dagger
\varphi \,.
\end{align}
This term contributes at first order in ${\cal
H}_I$\footnote{$\delta m^2$ is of order $e^2$ as will be seen
shortly.} as in (\ref{FHamexp}), thus
\begin{align}
\delta{\cal F}({\bf p}) &= \f{1}{\hbar} \int \f{d^3{\bf
p'}}{2p_0'(2\pi\hbar)^3} d^4x \f{\delta m^2}{\hbar^2}\bra{0}  A({\bf
p'})
\normal \varphi^\dagger(x) \varphi(x) \normal
A^\dagger({\bf p}) \nol &= \f{1}{\hbar} \int \f{d^3{\bf
p'}}{2p_0'(2\pi\hbar)^3} d^4x \delta m^2  \Phi^*_{{\bf p'}}(x)
\Phi_{{\bf p}}(x) \nol &= \f{1}{2\hbar p_0} \int dt |\phi_{\bf
p}(t)|^2 \delta m^2 \,,\label{delFwithdelm}
\end{align}
where we used (\ref{doublephiprod}) for the inner product. To
compute the counterterm we first find the self-energy $\Sigma(p)$.
Using the Feynman rules for the standard covariant perturbation
theory of scalar electrodynamics we obtain
\begin{align}
\Sigma(p)  =
\frac{e^2}{\hbar}\int\frac{d^4 q}{(2\pi)^4 i}\left\{
\frac{(p+q)^2}{\left[q^2 - m^2+i\epsilon\right]
\left[(p-q)^2 + i\epsilon\right]} -
\frac{4}{\left[(p-q)^2+i\epsilon\right]}\right\} \,,
\end{align}
where we use $q=p-K$. The convergence factors $i\epsilon$ are added,
along with $\epsilon>0$ and the usual assumption that the limit
$\epsilon\to 0$ is to be taken at the end. We integrate over the
$q_0$ component in order to compare with our previous results. In
the denominators we have the terms
\begin{align}
\left[q^2 - m^2+i\epsilon\right] &= \[ q_0 +\omega -i\delta
\]\[q_0-\omega +i\delta\] \\
\left[(p-q)^2+i\epsilon\right] &= \[ q_0-p_0+K -i\delta \]\[
q_0-p_0-K +i\delta \] \,,
\end{align}
where we define ${\bf K}={\bf p}-{\bf q}$ with $K=|{\bf K}|$,
$\omega = \sqrt{{\bf q}^2+m^2}$ and $\delta>0$ with the limit
$\delta\to 0$ assumed. For contour integration, the poles in the
upper half plane are clearly
\begin{align}
q_0 = \begin{cases} -\omega+i\delta \\ p_0-K+i\delta \end{cases}
\,.
\end{align}
Thus integrating, we obtain
\begin{multline} \label{counter3}
\Sigma (p) =
\f{e^2}{\hbar}\int\f{d^3 q}{(2\pi)^3} \\ \left\{
\f{\(p_0-\omega\)^2-\({\bf p}+{\bf q}\)^2}{\(-2\omega\)\(\(p_0+\omega\)^2-K^2\)} +
\f{\(2p_0-K\)^2-\({\bf p}+{\bf q}\)^2}{\(\(p_0-K\)^2-\omega^2\)\(-2K\)} - \f{4}{\(-2K\)}
\right\} \,.
\end{multline}
We note that
\begin{align}
& \quad \f{1}{2\omega\(\(p_0+\omega\)^2-K^2\)} +
\f{1}{2K\(\(p_0-K\)^2-\omega^2\))} \nol &=
\f{1}{2\omega 2K} \[ \f{1}{p_0-\omega-K}-\f{1}{p_0+\omega-K} \] +
\f{1}{2\omega 2K} \[ \f{1}{p_0+\omega-K}-\f{1}{p_0+\omega+K} \] \nol
&= -\f{1}{2\omega 2K}
\[\f{1}{\(p_0+\omega+K\)}+\f{1}{\(-p_0+\omega+K\)}-\] \,,
\end{align}
and
\begin{align}
& \quad \f{\(p_0-\omega\)^2}{2\omega\(\(p_0+\omega\)^2-K^2\)} +
\f{\(2p_0-K\)^2}{2K\(\(p_0-K\)^2-\omega^2\))}
\nol &= \f{\omega^2K + K^2\omega - p_0^2K-4p_0^2\omega}{2\omega K
\(p_0+\omega+K\)\(-p_0+\omega+K\)} \nol
&=
\f{1}{2K}-\f{1}{2\omega2K}\[\f{\(p_0-\omega\)^2}{\(p_0+\omega+K\)} +
\f{\(p_0+\omega\)^2}{\(-p_0+\omega+K\)}\] \,.
\end{align}
Comparing these results with the expression (\ref{counter3}) we can
write the counterterm
\begin{multline}
\Sigma (p)  =
\f{e^2}{\hbar}\int\f{d^3 q}{(2\pi)^32\omega}\f{1}{2K}  \\ \[ 6\omega
+ \f{\(p_0+\omega\)^2-\({\bf p}+{\bf q}\)^2}{\(-p_0+\omega+K\)} +
\f{\(p_0-\omega\)^2-\({\bf p}+{\bf q}\)^2}{\(p_0+\omega+K\)} \]
\,. \label{counter4}
\end{multline}
Clearly, we may change the variable of integration from ${\bf q}$ to
$\tilde{\bf q}={\bf q}-{\bf V}(t)$. We then have $\omega \to
E_q=\sqrt{\tilde{\bf q}^2+m^2}$. The counterterm $\delta m^2$ is
obtained by evaluating the self-energy $\Sigma(p)$ on the mass-shell
i.e. with $p_0=E_p$ and with ${\bf p}=\tilde{\bf p}$. It does not
matter which point on the mass-shell is invoked because it is well
known that the mass counterterm does not in fact explicitly depend
on ${\bf p}$, but only on $p_0^2-{\bf p}^2=m^2=E_p^2-\tilde{\bf
p}^2$. Technically, the mass-shell involves the physical mass $m_P$,
however the counterterm $\delta m^2$ is itself is of order ${\cal
O}(e^2)$ and for overall calculations at that order we may use $m$.
Therefore we may see that the inner integral in (\ref{F0final}) is
nothing but $e^{-2}\delta m^2$, and thus independent of $p$, which
shows that $\delta {\cal F} = -{\cal F}^0$ on comparison with
(\ref{delFwithdelm}) as stated. Consequently, the renormalised
forward scattering amplitude does not contribute to the position
shift.

\chapter{Quantum Green's Function Decomposition} \label{greenschapter}
\begin{quote}
In this chapter we present an alternative derivation of some of the
results for the position shift of the quantum scalar field based on
the Green's function decomposition description of classical
radiation reaction.
\end{quote}

In the previous chapter we established that the classical position
shift was reproduced in the $\hbar\to 0$ limit for the ${\cal
O}(e^2)$ perturbation theory of quantum scalar electrodynamics. In
fact, we showed that the position shift was entirely due to the
emission process. Whilst giving equality between the two results our
previous working does not however make clear any reasoning as for
\emph{why} this should come about. Given that the position shifts
are equal, we may wish to know if the treatment of radiation
reaction is the same in both theories. We may similarly ask what
connections and differences there are between the two approaches
with regards to the position shift. These questions are the subject
of this chapter which we present as a short aside to the body of the
work. That is not, however, to say that it is unimportant. On the
contrary, here we present the clues gleaned mathematically from the
results as to the interpretation of the quantum position shift
contributions and the interpretation of the connections between
classical and quantum theory with which we may view the body of work
presented so far.

In order to attain these goals, we shall return to the earlier
results for the scalar field and rework them to find expressions
involving the Green's functions that were used in the classical
derivation of the radiation reaction force. As such, we shall be
using some of the model and results from the previous chapter and
the appropriate descriptions and results shall be introduced again
here when required.

We use the model of a wave packet of the scalar particle passing
through a time-dependent potential for a finite period in the past
of the measurements. Let the state of the wave packet of a scalar
particle with momentum peaked about ${\bf p}$ be given by $\ket{{\bf
p}}$. We recall that the final state for a particle undergoing
radiation reaction is given to ${\cal O}(e^2)$ in our notation by
\begin{align}
 [1 + i{\cal F}({\bf p})]\ket{{\bf p}} + \frac{i}{\hbar} \int
\frac{d^3{\bf k}}{(2\pi)^3 2k} {\cal A}^\mu({\bf p},{\bf k})
\hat{a}_\mu^\dagger({\bf k})\ket{{\bf P}}\,, \label{evolp}
\end{align}
with $k\equiv \|{\bf k}\|$ and ${\bf P} = {\bf p}-\hbar{\bf k}$. We
further recall that we found the position shift in the $\hbar\to 0$
limit to be given by
\begin{align}
 \delta x^i_{\rm Q} = & - \frac{i}{2} \int \frac{d^3{\bf
k}}{2k(2\pi)^3} {\cal A}^{\mu *}({\bf p},{\bf k})
\stackrel{\leftrightarrow}{\partial}_{{p}^i} {\cal A}_\mu( {\bf
p},{\bf k})  - \hbar \partial_{p^i} {\rm Re}\,{\cal F}({\bf p})\,.
\label{xiQ}
\end{align}
We approach the quantum emission and forward scattering processes
and results in turn, starting with the former.

\section{Emission decomposition}\label{greensemissiondecomp}
The emission contribution to the quantum position shift (\ref{xiQ})
is
\begin{equation}
\label{deltaem}
\delta x_{\rm em}^i = -\f{i}{2}\int \f{d^3{\bf k}}{2k(2\pi)^3}
{\cal A}^{\mu
*}({\bf p},{\bf k})
\stackrel{\leftrightarrow}{\partial}_{p^i} {\cal A}_\mu ({\bf p},{\bf
k})\,.
\end{equation}
We used the semiclassical approximation to find that the emission
amplitude can be written in the $\hbar\to 0$ limit as
\begin{equation}
{\cal A}^\mu({\bf p},{\bf k})  =  -e \int_{-\infty}^{+\infty} d\xi\,
\frac{dx^\mu}{d\xi}\chi(\xi)\,e^{ik\xi} \,, \tag{\ref{cut-off}}
\end{equation}
where we have included the cut-off function $\chi(\xi)$ that takes
the value $1$ when the external force is nonzero and smoothly
becomes zero for large $|t|$~\footnote{This expression was arrived
at for both the time and space-dependent potentials.}. We now note
that this expression for the emission amplitude coincides with that
from a classical point charge to order $\hbar^0$.
\begin{equation}
{\cal A}^\mu_C({\bf p},{\bf k}) = - \int d^4x \, e^{ik\cdot x}
j^\mu(x)\,,
\end{equation}
with the current $j^\mu(x)$ given by
\begin{equation}
j^\mu(x) = e \frac{dx^\mu}{dt} \delta^3({\bf x}-{\bf X}_{\bf
p}(t))\chi(t)
\,,  \label{fakecurrent}
\end{equation}
where ${\bf X}_{\bf p}(t)$ is the path of the classical particle
which passes through the origin with momentum ${\bf p}$. The
classical field emitted from the current (\ref{fakecurrent}) is
\begin{equation}
A^\mu_{-}(x) = \int d^4x' G^{\ph\mu}_{-\ph\nu'}(x-x') j^{\nu'}(x')
\,, \label{retarded}
\end{equation}
where $G_-$ is the retarded Green's function. Now it is well known,
and fairly straightforward to show, that the retarded Green's
function can be written as~\cite{Itzykson}
\begin{equation}
G_{-\mu\nu'}(x-x') = i g_{\mu\nu'}\theta(t-t')\int \frac{d^3{\bf
k}}{2k(2\pi)^3}\left[ e^{-ik\cdot (x-x')} - e^{ik\cdot(x-x')}
\right]\,.
\end{equation}
These equations together imply that we can rewrite the classical
(retarded) field in terms of the classical, and thus the quantum,
emission amplitude i.e.
\begin{equation}
A^\mu_{-}(x) = -i \int\f{d^3{\bf k}}{2k(2\pi)^3}
\[ {\cal A}^\mu ({\bf p},{\bf k})e^{-ik\cdot x} -
{\cal A}^{\mu *}({\bf p},{\bf k}) e^{ik\cdot x} \]\,,
\end{equation}
for large enough $t$ such that $\chi(t)=0$. This gives us a fourier
expansion of the classical field, written in terms of the quantum
emission amplitude. We can consequently reverse this to rewrite the
quantum emission amplitude in terms of the classical field. Firstly,
we define the positive and negative frequency parts of the classical
field as follows:
\begin{align}
A^{(+) \mu} &= -i\int \f{d^3{\bf k}}{2k(2\pi)^3} {\cal A}^\mu
e^{-ik\cdot x} \,, \label{Aplus}\\
A^{(-) \mu} &= +i\int \f{d^3{\bf k}}{2k(2\pi)^3} {\cal A}^{\mu *}
e^{ik\cdot x} \,. \label{Aneg}
\end{align}
Note also that they are complex conjugates; $A^{(+) \mu *} = A^{(-)
\mu}$. Inverting the fourier expansion of the field, we have the
amplitude in terms of ${\cal A}^{(+) \mu}$
\beq
{\cal A}^\mu = 2k\int d^3{\bf x} A^{(+) \mu} e^{ik\cdot x} \,.
\eeq
The position shift can thus be rewritten as follows
\begin{align}
\delta x_{em}^i &=\f{i}{2}\int \f{d^3{\bf k}}{2k(2\pi)^3}\( 2k\int d^3{\bf
x'} A^{(-) \mu} e^{-ik\cdot x'} \)
\stackrel{\leftrightarrow}{\partial}_{p^i} \(2k\int d^3{\bf x} A^{(+)}_\mu
e^{ik\cdot x}\) \nol &= \f{i}{2}\int \f{d^3{\bf k}}{(2\pi)^3}\( \int
d^3{\bf x'} A^{(-) \mu} e^{-ik\cdot x'} \)
\stackrel{\leftrightarrow}{\partial}_{p^i} \(\int d^3{\bf x} A^{(+)}_\mu
e^{ik\cdot x}\) \nol & \quad \times \left\{ ik-(-ik)\right\} (-i)
\,,
\nonumber
\end{align}
(the last factor is just the remaining $2k$).
\begin{align}
&= \f{i}{2}\int \f{d^3{\bf k}}{(2\pi)^3}\( \int d^3{\bf x'} A^{(-)
\mu} e^{-ik\cdot x'} \)
\(-i\stackrel{\leftrightarrow}{\partial}_t\)\stackrel{\leftrightarrow}{\partial}_{p^i}
\(\int d^3{\bf x} A^{(+)}_\mu e^{ik\cdot x}\) \nol &= \f{1}{2}\int
\f{d^3{\bf k}}{(2\pi)^3}\( \int d^3{\bf x'} A^{(-) \mu} e^{-ik\cdot x'} \)
\stackrel{\leftrightarrow}{\partial}_t\stackrel{\leftrightarrow}{\partial}_{p^i}
\(\int d^3{\bf x} A^{(+)}_\mu e^{ik\cdot x}\) \nol &= \f{1}{2} \int d^3{\bf
x} A^{(-) \mu}
\stackrel{\leftrightarrow}{\partial}_t\stackrel{\leftrightarrow}{\partial}_{p^i}
A^{(+)}_\mu \,.
\end{align}
Now the position shift is real $\delta x ^*=\delta x$ and also, due
to the time derivative acting on the exponential in (\ref{Aplus})
and (\ref{Aneg}), we have
\begin{equation}
\int d^3{\bf x} A^{(+) \mu}
\stackrel{\leftrightarrow}{\partial}_t\stackrel{\leftrightarrow}{\partial}_{p^i}
A^{(+)}_\mu= \int d^3{\bf x} A^{(-) \mu}
\stackrel{\leftrightarrow}{\partial}_t\stackrel{\leftrightarrow}{\partial}_{p^i}
A^{(-)}_\mu =0 \,.
\end{equation}
Hence
\begin{align}
 \int d^3{\bf x} A^\mu_-
\stackrel{\leftrightarrow}{\partial}_t\stackrel{\leftrightarrow}{\partial}_{p^i}A_{-\mu}
&= \int d^3{\bf x} \[ A^{(+) \mu}
\stackrel{\leftrightarrow}{\partial}_t\stackrel{\leftrightarrow}{\partial}_{p^i}A^{(-)}_\mu
+ A^{(-) \mu}
\stackrel{\leftrightarrow}{\partial}_t\stackrel{\leftrightarrow}{\partial}_{p^i}A^{(+)}_\mu
\] \nol
&= 2\int d^3{\bf x} A^{(-) \mu}
\stackrel{\leftrightarrow}{\partial}_t\stackrel{\leftrightarrow}{\partial}_{p^i}A^{(+)}_\mu
\,.
\end{align}
Thus
\begin{equation}
\delta x_{em}^i = \f{1}{4} \int d^3{\bf x} A^\mu_-
\stackrel{\leftrightarrow}{\partial}_t\stackrel{\leftrightarrow}{\partial}_{p^i}
A_{-\mu} \,.
\end{equation}
Furthermore
\begin{align}
A^\mu_-
\stackrel{\leftrightarrow}{\partial}_t\stackrel{\leftrightarrow}{\partial}_{p^i}
A_{-\mu} &= A^\mu_-
\stackrel{\leftrightarrow}{\partial}_t\(\partial_{p^i} A_{-\mu}\) -
\(\partial_{p^i} A^\mu_-\)\stackrel{\leftrightarrow}{\partial}_t
A_{-\mu}
\nol
&= - \(\partial_{p^i}  A_{-\mu}\)
\stackrel{\leftrightarrow}{\partial}_t A^\mu_--
\(\partial_{p^i}A^\mu_-
\)\stackrel{\leftrightarrow}{\partial}_t A_{-\mu} \nol
&= -2
\(\partial_{p^i}A^\mu_-\)\stackrel{\leftrightarrow}{\partial}_t
A_{-\mu} \,.
\end{align}
Thus
\begin{equation}
\label{fieldposshift} \delta x_{em}^i = -\f{1}{2} \int d^3{\bf x}
\(\partial_{p^i}A^\mu_-\)\stackrel{\leftrightarrow}{\partial}_t
A_{-\mu} \,.
\end{equation}
We now have an expression for the quantum position shift in terms of
the retarded classical field.

\subsection{Green's function substitution}
We can use the Green's function decomposition of the classical field
to rewrite the quantum position shift in terms of Green's functions
instead of fields. Substituting the Green's function expression for
the fields in the position shift, we obtain\footnote{We have left
off the arguments of the functions for brevity as they are obvious
from the indices}
\begin{align}
\delta x_{em}^i &= -\f{1}{2} \int d^3{\bf x} \partial_{p^i} \[ \int d^4 x'
G_-^{\mu\nu'}j_{\nu'} \] \stackrel{\leftrightarrow}{\partial}_t
\[ \int d^4 x'' G_{-\mu\rho''}j^{\rho''} \] \nol
&= -\f{1}{2} \int d^4x' d^4x'' \[ \int d^3{\bf x} G_-^{\mu\nu'}
\stackrel{\leftrightarrow}{\partial}_t G_{-\mu\rho''} \] j^{\rho''}
\partial_{p^i} j_{\nu'} \,. \label{greenem2}
\end{align}
We note that using the Kirchhoff representation, the regular Green's
function can be written in terms of the retarded one, viz
\begin{align}
G_{{\rm R}\alpha''\beta'}(x''-x') = -\f{1}{2} \int_{t=T} d^3{\bf x}
\,
{{G_{-}}^\mu}_{\alpha''}(x-x'')\stackrel{\leftrightarrow}{\partial}_t
G_{-\mu\beta'}(x-x') \,,
\end{align}
for $x_0 > {\rm max}(x^{\prime}_0,x^{\prime\prime}_0)$. Substituting
into (\ref{greenem2}) we obtain
\begin{align}
\delta x_{\rm em} & = \int d^4x' d^4x''\,
\partial_{p^i} j_{\rho''}(x'') G_R^{\rho''\nu'}(x''-x')j_{\nu'}(x')\nonumber \\
  & = \int
d^4x\, \partial_{p^i}j^\mu(x)A_{{\rm R}\mu}(x)
\,, \label{standard0}
\end{align}
where we have changed the notation slightly and identified the
regular field generated by $G_R$.

The partial derivative acts on the current $j^\mu(x)$. Let $j_p(x)$
be the current following the path with final momentum $p$ and let
$j_{p+\Delta p}$ be the current following the path with final
momentum $p+\Delta p$. This second path will be shifted from the
original path, ${\bf X}$, by ${\bf
\Delta X}$. Explicitly these currents can be written
\begin{align}
j^\mu_p(x) &= e\f{dx^\mu}{dt}\delta^3({\bf x}-{\bf X}) \label{jpexpl}\\
j^\mu_{p+\Delta p}(x) &= e\f{dx^\mu}{dt}\delta^3({\bf x}-{\bf X
-\Delta X}) \,. \label{jpdexpl}
\end{align}
The derivative can then be written in limit form as
\beq
\partial_{p^i}j^\mu(t,{\bf x}) = \lim_{\Delta p^i\to 0}
\frac{j^\mu_{p+\Delta p}(x) - j^\mu_p(x)}{\Delta p^i}\,.
\eeq
Now, defining the four-vector $\Delta X^\alpha = (0,{\bf \Delta
X})$, we write
\begin{align}
\delta x^i_{\rm em} &= \lim_{\Delta p^i\to 0} \f{1}{\Delta p^i} \int
d^4x \[ j^\mu_{p+\Delta p}(x)-j^\mu_p(x) \] A_{{\rm R}\mu}(x) \,.
\end{align}
Substituting the explicit expressions for the currents,
(\ref{jpexpl}) and (\ref{jpdexpl}), we find the position shift as
\begin{align}
\delta x^i_{\rm em} &= \lim_{\Delta p^i\to 0} \f{1}{\Delta p^i} \int d^4x
\, \[ e\f{dx^\mu}{dt}\delta^3({\bf x}-{\bf X -\Delta
X})-e\f{dx^\mu}{dt}\delta^3({\bf x}-{\bf X})
\]A_{{\rm R}\mu}(x)
\nol
&= \lim_{\Delta p^i\to 0}
\f{e}{\Delta p^i} \int dt \, \( \(\f{dX^\mu}{dt}+\f{d\Delta
X^\mu}{dt}\)A_{{\rm R}\mu}((X+\Delta X)-\f{dX^\mu}{dt} A_{{\rm
R}\mu}(X)\)
\nol
&=
\lim_{\Delta p^i\to 0} \f{e}{\Delta p^i} \int dt \, \(
\(\f{dX^\mu}{dt}+\f{d\Delta X^\mu}{dt}\)
\[A_{{\rm R}\mu}(X)+\Delta X^\alpha \nabla_\alpha A_{{\rm R}\mu}(X) \] \right. \nol & \qquad
\qquad \qquad \qquad \quad \left. -\f{dX^\mu}{dt}
A_{{\rm R}\mu}(X)\) \nol &= \lim_{\Delta p^i\to 0} \f{e}{\Delta p^i}
\int dt \,
\(\f{dX^\mu}{dt}\Delta X^\alpha \nabla_\alpha
A_{{\rm R}\mu}(X)+\f{d\Delta X^\mu}{dt}A_{{\rm R}\mu}(X)\) \,,
\end{align}
where due to the limit (recall $\Delta X \to 0$ as $\Delta p
\to 0$) we need keep only the terms up to first order in $\Delta X$.
Integrating the second term of the integrand by
parts\footnote{Recall that $A_{{\rm R}\mu}$ vanishes at the position
of the charge when there is no acceleration.} we obtain
\begin{align}
\delta x^i_{\rm em} &= \lim_{\Delta p^i\to 0} \f{e}{\Delta p^i} \int dt \,
\(\f{dX^\mu}{dt}\Delta X^\alpha \nabla_\alpha
A_{{\rm R}\mu}(X)-\Delta X^\mu \f{d}{dt}A_{{\rm R}\mu}(X)\) \nol &=
\lim_{\Delta p^i\to 0} \f{e}{\Delta p^i} \int dt \,
\(\f{dX^\mu}{dt}\Delta X^\alpha \nabla_\alpha
A_{{\rm R}\mu}(X)-\Delta X^\mu \f{dX^\alpha}{dt}\nabla_\alpha
A_{{\rm R}\mu}(X)\) \,.
\end{align}
Swapping the spacetime indices in the two sums in the second term,
the position shift can be rewritten as
\begin{align}
\delta x^i_{\rm em} &= \lim_{\Delta p^i\to 0} e \int dt \,
\f{dX^\mu}{dt}\f{\Delta X^\alpha}{\Delta p^i} \[ \nabla_\alpha A_{{\rm
R}\mu}(X) - \nabla_\mu A_{{\rm R}\alpha}(X) \] \nol &= e \int dt \,
F^R_{\mu\alpha} \f{dX^\mu}{dt} \( \f{\partial X^\alpha}{\partial
p^i} \)_t \nol & =  -\int_{-\infty}^0 dt\,{\cal F_{\rm LD}}^j
\,\left(
\frac{\partial X^j}{\partial p^i}\right)_t \,.
\end{align}
where we recall from the introduction  that the radiative
electromagnetic field tensor is defined in terms of the
regular/radiative field analogously to the standard field tensor and
that the Lorentz-Dirac force is given as the Lorentz-force generated
by this field (see (\ref{FmunuR}) and (\ref{LDeqR})). We recognise
this result as the classical position shift given by (\ref{cshift}).
This calculation is analogous to the derivation of the Lorentz force
from the standard Lagrangian for a point charge in an external
electromagnetic field (see, e.g. Ref.~\cite{Jackson}). We have made
the upper bound of the $t$-integration to $t=0$ in the last line
because ${\cal F_{\rm LD}}^j =0$ for $t > 0$. Thus, we have shown
that the contribution from the emission of a photon to the position
shift agrees with the classical counterpart using the Green's
function method.

\section{Forward-Scattering decomposition}

We now turn to the forward-scattering contribution to the position
shift given by
\beq
\delta x_{\rm for}^i =  - \hbar\partial_{p^i} {\rm
Re} \,{\cal F}({\bf p}) \,.
\eeq
We have shown that this contribution vanishes in the end. More
precisely, the leading order terms of the real part of the
forward-scattering amplitude are exactly canceled by the
contribution from the mass counter-term, i.e.\ it is eliminated to
order $\hbar^0$ by the mass renormalisation. Here we shall see that
the field generated by the singular Green's function appears in the
calculation of $\delta x^i_{\rm for}$. We recall that in the
classical theory, this contribution to the field is, as the name
implies, singular and is subsequently subtracted from the field in a
process akin to the mass renormalisation.

\begin{figure}
\begin{center}
\begin{pspicture}(-4,-1)(4,2.5)
\psline[linewidth=0.5mm,linestyle=dashed]{->}(-3.5,0)(-3,0)
\psline[linewidth=0.5mm,linestyle=dashed]{->}(-3,0)(0,0)
\psline[linewidth=0.5mm,linestyle=dashed]{->}(0,0)(3,0)
\psline[linewidth=0.5mm,linestyle=dashed](3,0)(3.5,0)
\pscurve[linewidth=0.5mm]{*-*}(-2,0)(-2,.5)(-1.5,.5)%
(-1.4,1.2)(-.9,1.1)(-.8,1.6)(-.3,1.5)(0,2)(.3,1.5)(.8,1.6)(.9,1.1)%
(1.4,1.2)(1.5,.5)(2,.5)(2,0)
\rput(-3,-.5){${\bf p}$}
\rput(0,-.5){${\bf p}-\hbar{\bf k}$}
\rput(3,-.5){${\bf p}$}
\rput(0,2.4){$\hbar{\bf k}$}
\end{pspicture}
\caption{The one-loop diagram contributing to the
forward-scattering amplitude: the dashed and wavy lines represent
the scalar and photon propagators, respectively.}
\label{feynman}
\end{center}
\end{figure}
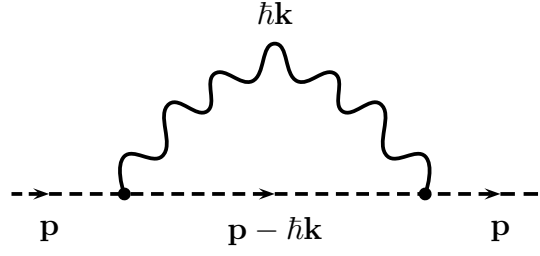

The forward-scattering amplitude comes from the one-loop diagram
shown in Fig.~\ref{feynman} and the additional loop diagram from the
seagull vertex. For the contribution from the intermediate {\em
particle} state (as opposed to {\em anti-particle} state), we
divided the momentum integral for the virtual photon in this loop
diagram into two parts; one with momentum $\hbar\|{\bf k}\|$ less
than $\hbar^{\alpha}\lambda$ and the other with momentum larger than
$\hbar^{\alpha}\lambda$, where $\alpha$ and $\lambda$ are constants.
We chose $\alpha$ to satisfy $\frac{3}{4} < \alpha < 1$. However the
condition $\alpha <1$ will suffice for our current purpose. Denoting
the first part with the virtual-photon momentum below the cut-off by
${\cal F}^{<}({\bf p})$, we found that to lowest order in $\hbar$
\begin{align}
\hbar {\cal F}^{<}({\bf p}) = & - ie^2 \int_{k \leq
\hbar^{\alpha-1}\lambda}
\frac{d^3{\bf k}}{(2\pi)^32k}\int_{-\infty}^{+\infty}dt
\int_{-\infty}^{+\infty}dt'
\nol &\times \theta(t-t')\frac{dX_{\bf p}^\mu}{dt}
\frac{dX_{{\bf p}\mu}}{dt'}
e^{ik(t'-t) - i{\bf k}\cdot({\bf X}_{\bf p}(t')-{\bf X}_{\bf p}(t))}
\,.
\end{align}
In the classical limit $\hbar\to 0$, the ${\bf k}$-integration will
have no restriction because $\hbar^{\alpha-1}\lambda \to
\infty$. As we did for the emission process, we can replace part of this
expression with one containing Green's functions. Firstly, we note
the presence of the classical currents and take advantage of the
symmetry in the integrations to write
\begin{align}
\hbar {\cal F}^{<}({\bf p}) &=  - \f{1}{\hbar} \int d^4x d^4x'
 j^\mu(x) j^{\nu'}(x')i\hbar g_{\mu\nu'}\int
\frac{d^3{\bf k}}{(2\pi)^32k}\theta(t-t')
e^{-ik\cdot(x-x')} \nol &= - \f{1}{2\hbar} \int d^4x d^4x'
 j^\mu(x) j^{\nu'}(x')
 \nol & \quad \times i\hbar g_{\mu\nu'} \int
\frac{d^3{\bf k}}{(2\pi)^32k} \[ \theta(t-t')
e^{-ik\cdot(x-x')}+\theta(t'-t) e^{-ik\cdot(x'-x)} \] \,.
\end{align}
Within this expression we recognize the form of the Feynman
propagator, which can be given by
\begin{align}
G_F^{\mu\nu'}(x-x') = & -i\hbar g^{\mu\nu'}
\int \f{d^3k}{(2\pi)^3 2k} \[ \theta(t-t')e^{-ik\cdot(x-x')} +
\theta(t'-t)e^{-ik\cdot(x'-x)} \] \,.
\end{align}
We consequently find that in this limit
\begin{equation}
\hbar{\cal F}^{<}({\bf p})
= \frac{1}{2\hbar}\int d^4 x d^4
x'\,j_\mu(x)j_{\nu'}(x')G_F^{\mu\nu'}(x-x') \,.
\label{greenforfey}
\end{equation}
The contraction of the photon propagator with the external particle
currents in the above expression is reminiscent of the one-loop
diagram. However, it should be stressed that the above expression
contracts this propagator with the
\emph{classical} currents and the validity is limited by the
presence of both the $\hbar\to 0$ limit and the low-energy photon
sector. To further our manipulation of the Green's functions, let us
write the Feynman propagator as the sum of the real and imaginary
parts:
\begin{align}
G_F^{\mu\nu'}(x-x') &= -\hbar G_S^{\mu\nu'}(x-x') - \f{i\hbar}{2}
G^{(1)\mu\nu'}(x-x') \,, \label{feysplit}
\end{align}
where we have Hadamard's elementary form, given by
\begin{equation}\label{hadamardG1}
\hbar G^{(1)\mu\nu'}(x-x') =
\bra{0} \left\{ \hat{A}^\mu(x),\hat{A}^{\nu'}(x') \right\} \ket{0} \,,
\end{equation}
with $\hat{A}^\mu(x)$ being the quantum electromagnetic potential.
We thus spot the presence of the singular Green's function in our
calculation. Before returning to this point, we briefly look at the
imaginary part. Above the cut-off, there is no imaginary
contribution in the $\hbar\to 0$ limit. By unitarity, ${\rm
Im}\,{\cal F}^{<}({\bf p})$ is required to equal half the emission
probability. This has previously been shown to be the case by direct
computation. It can also easily be shown from the above expression
(\ref{greenforfey}) using (\ref{feysplit}), (\ref{hadamardG1}) and
the symmetry of the integration and anticommutator:
\begin{align}
{\rm Im} \, {\cal F}^{<}({\bf p}) &= -\f{1}{2\hbar^2} \int d^4x
d^4x' j_\mu(x) \bra{0} \hat{A}^\mu(x)
\hat{A}^{\nu'}(x')
\ket{0} j_{\nu'}(x')  \nonumber \\
&=-\f{1}{2\hbar^2} \int d^4x d^4x' j_\mu(x) j_{\nu'}(x')
e^{-ik\cdot(x-x')} \nol
 &=
 -\f{1}{2\hbar} \int \f{d^3 {\bf k}}
{2k(2\pi)^3} {\cal A}_\mu^*({\bf p},{\bf k}){\cal A}^\mu({\bf
p},{\bf k}) \,.
\end{align}
Returning now to the real part we have
\begin{align}
\hbar {\rm Re} \, {\cal F}^{<}({\bf p})
& = -\f{1}{2}\int d^4x d^4x' \, j_\mu(x) j_\nu (y) G_S^{\mu\nu}(x-y)
\,.
\end{align}
Using the symmetry of $G_S^{\mu\nu'}(x-x')$, we obtain
\begin{equation}
-\hbar \partial_{p^i} {\rm Re} \, {\cal F}^{<}({\bf p}) = \int
d^4x\,
 \partial_{p^i} j^\mu (x) A_{S\mu}(x) \,, \label{singcont}
\end{equation}
where $A_S^\mu(x)$ is the singular part of the self-field given by
\begin{equation}
A_S^\mu(x) = \int d^4 x'\,G_S^{\mu\nu'}(x-x')j_{\nu'}(x') \,.
\end{equation}
Equation (\ref{singcont}) is analogous to the expression for the
emission contribution to the position shift (\ref{standard0}), which
was in terms of the regular field. If we add (\ref{singcont}) to the
emission contribution (\ref{standard0}), then
\begin{equation}
\delta x_{\rm em} - \hbar\partial_{p^i}{\rm Re}\,{\cal F}^{<}({\bf p})
= \int d^4x\,
 \partial_{p^i} j_\mu (x)A_{-}^\mu(x) \,,
\end{equation}
where the self-field, $A_{-}^\mu(x)$, is given by (\ref{retarded}).
Thus, the one-photon emission process and the low-energy part of the
forward-scattering process are incorporated in the classical
self-field $A_{-}^\mu(x)$ if one sees this field from the viewpoint
of quantum derivation of the self-force. The regular part,
$A_R^\mu(x)$, of the self-field in classical electrodynamics
corresponds to the emission process in QED and the singular part,
$A_S^\mu(x)$, to the low-energy forward-scattering process.  The
remaining high-energy and intermediate anti-particle state
contributions to the forward-scattering amplitude in QED have no
classical counterpart. The forward-scattering contribution as a
whole vanishes if one includes the quantum mass counter-term, as was
shown in Chapter \ref{scalarshiftchapter}.

\chapter{Spinor Quantum Position Shift}\label{spinorshiftchapter}
\begin{quote}
In this chapter we repeat our derivations and calculations for the
quantum position shift using the canonical theory of quantum
electrodynamics based on the model of the Dirac spinor field. We
again combine the effects of the photon emission, forward scattering
and mass renormalisation in the $\hbar\to 0$ limit in order to
compare the result with the classical theory.
\end{quote}

In this chapter we shall replace the scalar quantum field model with
the more realistic spinor field of quantum electrodynamics in order
to calculate the quantum position shift. We shall thus start our
quantum position shift calculations from scratch using the spinor
field definitions and the spinor semiclassical expansions from
Chapter \ref{introchapter} section \ref{spinordefnsection} and
Chapter \ref{Semiclassical} section
\ref{semiclassspinorsection} respectively. Much of the path that
we shall tread here will be familiar from the scalar work and some
of the expressions derived from this source will be the same as
before. Naturally, the classical position shift is unchanged, but we
note that the classical theory does not include the concept of spin.
Despite the similarities with our previous scalar work, there are
however differences due to the construction of the fields, not least
the addition of spin to consider in the interactions and evolutions.
Whilst using the same approach as before, we shall nonetheless tread
carefully and repeat most of the calculations from the new spinor
particle definitions. For the potential, we shall look at the
time-dependent (and spatially independent) case $V(t)$ throughout
and thus take advantage of the conservation of momentum.

\section{Initial control state}
We again start with the expressions describing the initial control
state, but this time for a spinor particle. As per the introduction
to the spinor field definitions in Chapter
\ref{introchapter} section \ref{spinordefnsection}, we define the
initial incoming wave packet of the spinor field with spin labeled
by $\alpha$ as
\begin{equation}\tag{\ref{spinistate}}
| i \rangle = \int\f{d^3 {\bf p}}{(2\pi\hbar)^3}\sqrt{\f{m}{p_0}}
f({\bf p}) b^\dagger_\alpha({\bf p}) |0\rangle \,,
\end{equation}
where we recall that $f$ is sharply peaked about the initial
momentum in the region ${\cal M}_-$ and normalised via $\langle
i|i\rangle=1$, viz
\begin{equation}\tag{\ref{normalisef}}
\int \f{d^3 {\bf p}}{(2\pi\hbar)^3} f^*({\bf p})f({\bf p}) = 1 \,.
\end{equation}
The spinor initial state differs from the scalar state only in the
presence of the spinor field creation operator, with spin index, and
also the factor multiplying the basic Lorentz-invariant measure:
$1/(2p_0)\to m/p_0$, which is due to the canonical convention chosen
for these fields. The outgoing wave packet of our control particle,
which we recall does not undergo radiation reaction in ${\cal M}_I$,
is given by the same expression, albeit with $f$ now sharply peaked
about the final momentum $\bar{\bf p}$ in the region ${\cal M}_+$
and $\alpha$ now represents the spin of the outgoing state. As with
the scalar field, we let the potential satisfy $|V_0|<2m$, thus
precluding the possibility of particle pair creation. The associated
vacuum effects can then be safely ignored and the charge density can
be considered equivalent to the probability density for a
one-particle state. The expectation value of the density of the
state $\ket{i}$, $\langle
\rho(x) \rangle = \langle i |:\psi^\dagger\psi :|i\rangle$ is
given as follows
\begin{align}
& \quad \langle \rho(x) \rangle
\nol
&= \langle 0 |  \int\f{d^3 {\bf
p}'}{(2\pi\hbar)^3}\sqrt{\f{m}{p_0'}} f^*({\bf p}') b_\alpha({\bf
p}') \nol & \quad : \[ \int \f{d^3 {\bf p}''}{(2\pi\hbar)^3}
\f{m}{p_0''} \sum_\beta b^\dagger_\beta({\bf p}'')\Phi^{\beta}(p'') \int \f{d^3 {\bf p}'''}{(2\pi\hbar)^3}
\f{m}{p_0'''}\sum_\gamma b_\gamma({\bf p}''')\Phi^{\gamma\dagger}(p''') \] : \nol & \quad \times
 \int\f{d^3 {\bf p}}{(2\pi\hbar)^3}\sqrt{\f{m}{p_0}} f({\bf p})
b^\dagger_\alpha({\bf p}) |0\rangle \nol
 &= \int\f{d^3 {\bf p}}{(2\pi\hbar)^3}\f{d^3 {\bf p}'}{(2\pi\hbar)^3}
\f{d^3 {\bf p}''}{(2\pi\hbar)^3}\f{d^3 {\bf p}'''}{(2\pi\hbar)^3}
\f{m}{\sqrt{p_0p_0'}}\f{m^2}{p_0''p_0'''}
 f^*({\bf p}')f({\bf p})\sum_{\beta , \gamma}
\Phi^{\beta\dagger}(p'')\Phi^{\gamma}(p''')  \nol
& \quad \times \langle 0 | b_\alpha({\bf p}') b^\dagger_\beta({\bf
p}'') b_\gamma({\bf p}''')b^\dagger_\alpha({\bf p}) | 0 \rangle
\nol
 &=  \int\f{d^3 {\bf p}}{(2\pi\hbar)^3}\f{d^3 {\bf p}'}{(2\pi\hbar)^3}
\f{d^3 {\bf p}''}{(2\pi\hbar)^3}\f{d^3 {\bf p}'''}{(2\pi\hbar)^3}
\f{m}{\sqrt{p_0p_0'}}\f{m^2}{p_0''p_0'''}
 f^*({\bf p}')f({\bf p})\sum_{\beta
, \gamma} \Phi^{\beta\dagger}(p'')\Phi^{\gamma}(p''')  \nol &
\quad \times\f{{\bf p}_0''}{m}(2\pi\hbar)^3\delta^3({\bf p}'-{\bf p}'')\delta_{\alpha\beta}
\f{{\bf p}_0'''}{m}(2\pi\hbar)^3\delta^3({\bf p}'''-{\bf p})\delta_{\alpha\gamma}  \nol
 &= \int\f{d^3 {\bf p}}{(2\pi\hbar)^3}\f{d^3 {\bf p}'}{(2\pi\hbar)^3}\f{m}{\sqrt{p_0
 p_0'}}f^*({\bf p}')f({\bf
 p})\Phi_\alpha^{\dagger}({\bf p'})\Phi_\alpha({\bf p}) \,.
\end{align}
We wish to measure the position expectation value at time $t=0$. As
this lies, by definition, far into the region ${\cal M}_+$ we may
use the mode functions for the free field i.e. $\Phi_\alpha({\bf
p})=u_\alpha(p)e^{ip\cdot x/\hbar}$. Hence
\begin{align}
\langle x^i (0) \rangle=& \int d^4 x \, x^i \langle i |:\psi^\dagger\psi
:|i\rangle \nol =&
\int d^3x \int\f{d^3 {\bf p}}{(2\pi\hbar)^3}\f{d^3 {\bf p}'}{(2\pi\hbar)^3}\f{m}{\sqrt{p_0
 p_0'}}f^*({\bf p}')f({\bf p}) u_\alpha^\dagger(p')u_\alpha(p)
 e^{-i({\bf p}-{\bf p'})\cdot x/\hbar} x^i \nol
 =&  \int\f{d^3 {\bf p}}{(2\pi\hbar)^3}\f{d^3 {\bf p}'}{(2\pi\hbar)^3}\f{m}{\sqrt{p_0
 p_0'}}f^*({\bf p}')f({\bf p}) u_\alpha^\dagger(p')u_\alpha(p) \nol & \times \int
 d^3x (-i\hbar)\partial_{p_i} e^{-i({\bf p}-{\bf p'})\cdot x} \,. \nol
 \intertext{Integrating by parts, and integrating out the resultant
 delta function, we obtain}
 \langle x^i (0) \rangle =& i\hbar\int\f{d^3 {\bf p}}{(2\pi\hbar)^3}
 m\( \f{f^*({\bf p})}{\sqrt{p_0}} u_\alpha^\dagger(p) \) \partial_{p_i} \(
\f{f({\bf p})}{\sqrt{p_0}}u_\alpha(p) \) \,.
\end{align}
This expression will in fact be the one we need to recall when
comparing with the final interacting state. However, if we complete
the $p^i$ differentiation, the resulting terms turn out to be of
different orders in $\hbar$. Although we are only dealing with the
$\hbar\to 0$ limit, for completeness we shall differentiate and
analyse these terms further: The term differentiated with respect to
the momentum in the $x^i$ direction is
\begin{align}
\partial_{p_i} \(\f{f({\bf p})}{\sqrt{p_0}}u_\alpha(p) \) =& \f{\partial_{p_i}
f({\bf p})}{\sqrt{p_0}}u_\alpha(p)-\f{f({\bf
p})p_i}{2p_0^{5/2}}u_\alpha(p) +\f{f({\bf
p})}{\sqrt{p_0}}\partial_{p_i}u_\alpha(p) \,,
\end{align}
thus giving the position expectation value as
\begin{align}
& i\hbar\int\f{d^3 {\bf p}}{(2\pi\hbar)^3}\f{m}{p_0} \nol &  \Big\{
\,
 \[f^*({\bf p})\partial_{p_i}
f({\bf p})\] u_\alpha^\dagger(p) u_\alpha(p) -|f({\bf
p})|^2\f{p_i}{2p_0^{2}}u_\alpha^\dagger(p)u_\alpha(p) +|f({\bf
p})|^2 u_\alpha^\dagger(p)\partial_{p_i}u_\alpha(p)
\Big\} \,.
\end{align}
From the definition of the free spinor in (\ref{freeu})
\begin{equation}
u_\alpha(p) = \sqrt{\df{p_0+m}{2m}} \begin{pmatrix} s_\alpha \\
\df{\itbd{\sigma}\cdot{\bf p}}{p_0+m} s_\alpha \end{pmatrix} \,,
\tag{\ref{freeu}}
\end{equation}
 we have the normalisation $u_\alpha^\dagger(p)u_\alpha(p)=p_0/m$. We now
calculate the product
$u_\alpha^\dagger(p)\partial_{p_i}u_\alpha(p)$. The momentum
derivatives of the two factors in the spinor (\ref{freeu}) are
\begin{align}
\partial_{p_i} \(\sqrt{\f{p_0+m}{2m}}\) &= \f{1}{2}
\f{1}{\sqrt{2m\(p_0+m\)}} \f{p_i}{p_0} \\
\partial_{p_i} \(\f{\itbd{\sigma}\cdot{\bf p}}{\sqrt{2m\(p_0+m\)}} \) &=
\f{\itbd{\sigma}\cdot{\bf n_i}}{\sqrt{2m\(p_0+m\)}} -
\f{1}{2}\f{1}{\sqrt{2m}}\f{\itbd{\sigma}\cdot{\bf
p}}{\(p_0+m\)^{3/2}}\f{p_i}{p_0} \,,
\end{align}
where ${\bf n_i}$ is the unit vector is the $i$th direction.

The general expression $u_\alpha^{\dagger} \partial_{p_i} u_\beta$
is thus given by
\begin{align}
& \quad u_\alpha^{\dagger}(p) \partial_{p_i} u_\beta(p) \nol &= \(
\sqrt{\f{p_0+m}{2m}}
\,
s_\alpha^\dagger \)\(
 \f{1}{\sqrt{2m\(p_0+m\)}} \f{p_i}{2p_0} s_\beta\) \nonumber \\
& \quad + \(\sqrt{\f{p_0+m}{2m}} \, s_\alpha^\dagger \,
\f{\itbd{\sigma}
\cdot {\bf p} }{p_0+m}\) \(\f{\itbd{\sigma}\cdot{\bf
n_i}}{\sqrt{2m\(p_0+m\)}}\, s_\beta -
\f{\itbd{\sigma}\cdot{\bf p}}{\(p_0+m\)^{3/2}}\f{p_i}{2p_0\sqrt{2m}} \, s_\beta\)
\nonumber
\\
&= \f{1}{2}\f{p_i}{2mp_0} s_\alpha^\dagger s_\beta +
\f{s_\alpha^\dagger \( \itbd{\sigma}\cdot{\bf p}\,
\itbd{\sigma}\cdot{\bf n_i} \) s_\beta}{2m(p_0+m)}  - \f{1}{2}
\f{p_i}{2mp_0}\f{s_\alpha^\dagger
\( \itbd{\sigma}\cdot{\bf p}\itbd{\sigma}\cdot{\bf p} \) s_\beta}{(p_0+m)^2}  \nonumber \\
&= \f{1}{2}\f{p_i}{2mp_0} s_\alpha^\dagger s_\beta +
\f{s_\alpha^\dagger \( p_i + i ({\bf
p}\times {\bf n_i})\cdot \itbd{\sigma} \) s_\beta}{2m(p_0+m)}  -
\f{1}{2}
\f{p_i}{2mp_0}\f{s_\alpha^\dagger {\bf p}^2 s_\beta}{(p_0+m)^2}  \nonumber \\
&= \f{1}{2}\f{p_i}{2mp_0} s_\alpha^\dagger s_\beta +
\f{p_i}{2m(p_0+m)}s_\alpha^\dagger s_\beta + \f{is_\alpha^\dagger
{\bf n_i}\cdot\(\itbd{\sigma}\times{\bf p} \)  s_\beta}{2m(p_0+m)}
\nol & \quad -
\f{1}{2}
\f{p_i}{2mp_0}\f{(p_0^2-m^2)}{(p_0+m)^2} s_\alpha^\dagger s_\beta \nonumber \\
&= \f{p_i}{2mp_0} s_\alpha^\dagger s_\beta + \f{i}{2m(p_0+m)}
s_\alpha^\dagger {\bf n_i}\cdot\(\itbd{\sigma}\times{\bf p} \)
s_\beta \,,
\end{align}
where we have used $ \itbd{\sigma}\cdot{\bf
a}\,\itbd{\sigma}\cdot{\bf b} = {\bf a}\cdot{\bf b}I + i ({\bf
a}\times{\bf b})\cdot \itbd{\sigma}$ and ${\bf p}\cdot{\bf p} =
p_0^2-m^2$. We consequently obtain
\beq
u_\alpha^\dagger \partial_{p_i} u_\beta =
\f{p_i}{2mp_0}\delta_{\alpha\beta}
+
\f{i}{2m(p_0+m)} s_\alpha^\dagger
{\bf n_i}\cdot\(\itbd{\sigma}\times{\bf p} \) s_\beta
\label{udiffuterm}\,.
\eeq

As a short aside, we can look at an interpretation of the second
term in (\ref{udiffuterm}). In the case where $\alpha=\beta$, as we
require, then $s_\alpha^\dagger {\bf
n_i}\cdot\(\itbd{\sigma}\times{\bf p}
\) s_\alpha = {\bf n_i}\cdot\itbd{\xi}\times{\bf p}$ where
$\itbd{\xi}$ is the unit vector in the direction of the spin
(positive for spin up, negative for spin down). This term is an
example of the effects of the addition of spin to the quantum model,
in this case on the measurement of the position in the $i$
direction. Now, the expression measures the component in the $i$
direction of the vector $\itbd{\xi}\times{\bf p}$, perpendicular to
the spin and the momentum (and is zero when these
coincide)\footnote{The term is also zero in the direction of either
the spin or the momentum.} and could be described as providing a
change, in the momentum and consequently the position, due to the
interaction between the momentum and the spin. This type of effect
can be seen by analysing the Poincar\'{e} algebra of the generators
for a boost and a rotation. Note that the same algebra is obeyed by
the spin and boost operators for a spinor. If $K_i$ represents the
generator of a boost in the $i$ direction and $J_j$ represents the
generator of a rotation in the $j$ direction then we have
\begin{align}
\[K_i,K_j\] &= -i \epsilon_{ijk} J_k \,, \\
\[J_i,K_j\] &= i \epsilon_{ijk} K_k \,,
\end{align}
where $\epsilon_{ijk}$ is the Levi-Civita symbol. The first relation
leads to the Thomas precession correction to the spin-orbit
interaction. The second relation is related to the current effect.
This term is naturally not present for calculations using the scalar
field. Our extra term is thus a mathematical consequence of the fact
that the spin and boost operators do not commute.

Returning to our main calculation, the position expectation value
(at $t=0$) of the initial state can be given as
\begin{align}
& \quad i\hbar \int \f{d^3{\bf p}}{(2\pi\hbar)^3} m
\(\f{f^*({\bf p})u_\alpha^\dagger(p)}{\sqrt{p_0}}\)\partial_{p_i}\(\f{f({\bf p})u_\alpha(p)}{\sqrt{p_0}}\)
\nol
&= i\hbar \int \f{d^3{\bf p}}{(2\pi\hbar)^3} \Big\{ m f^*({\bf
p})\partial_{p_i} f({\bf p})\f{1}{p_0}\f{p_0}{m}
\nol &  \quad + m|f({\bf p})|^2
\(-\f{p_i}{2p_0^3}\f{p_0}{m}
+\f{1}{p_0}\[ \f{p_i}{2mp_0}  +
\f{i{\bf n_i}\cdot\itbd{\xi}\times{\bf p}}{2m(p_0+m)}\]\)\Big\} \nol
 &= i\hbar
\int \f{d^3{\bf p}}{(2\pi\hbar)^3} \left\{ f^*({\bf p})\partial_{p_i} f({\bf p}) +
|f({\bf p})|^2\f{i{\bf n_i}\cdot\itbd{\xi}\times{\bf
p}}{2p_0(p_0+m)}\right\} \,. \label{spinorcontrolexpand}
\end{align}
Because, after a $(2\pi\hbar)^3$ pre-multiple, $f^*({\bf
p})\partial_p f({\bf p})$ is of order $\hbar^{-1}$, whereas $|f({\bf
p})|^2$ is ${\cal O}(\hbar^0)$, the second term in
(\ref{spinorcontrolexpand}) is of order $\hbar$ and in the classical
limit the spin-related effect given above does not contribute. In
this limit we have
\begin{equation}
\langle x^i \rangle |_{t=0} = i\hbar \int \f{d^3{\bf p}}{(2\pi\hbar)^3} f^*({\bf p})\partial_{p_i}
f({\bf p}) \,.
\end{equation}
We recognize this expression as the same as that was reached for the
scalar field in (\ref{nonradpev}). Once again, as the position shift
is real, we may write
\begin{align} \label{nonradpev2}
 \langle x^i \rangle |_{t=0} = \frac{i\hbar}{2}
\int\frac{d^3{\bf p}}{(2\pi\hbar)^3} f^*({\bf p})
\stackrel{\leftrightarrow}{\partial}_{p_i} f({\bf p}) \,.
\end{align}

\section{Final interacting state}
For the final state of a particle undergoing radiation reaction, we
again start with an incoming wave packet of the form $\ket{i}$. The
interactions to order $e^2$ for the spinor field are, as with the
scalar field, composed of the photon emission sector and a one-loop
forward scattering sector along of course with the null interaction.
We shall use notation similar to the scalar field for the amplitudes
of these processes. These amplitudes we shall of course later
calculate (in sections \ref{spinoremissioncalcsection} and
\ref{spinorforwardscattcalcsection} of this chapter) using the semiclassical spinor
expansions for the interacting region ${\cal M}_I$, whilst the
measurement of the position shift takes place at $t=0$ inside ${\cal
M}_+$. Let us start with the final state giving the definitions of
the amplitudes:
\beq
[1 + i{\cal F}({\bf p})]b^\dagger_\alpha({\bf p})|0\rangle  +
\frac{i}{\hbar}
\int \frac{d^3{\bf k}}{2k(2\pi)^3} {\cal A}^{(\beta)\mu}_{(\alpha)}({\bf p},{\bf k})
a_\mu^\dagger({\bf k})b^\dagger_\beta({\bf p}')|0\rangle \,,
\eeq
where ${\cal F}$ represents the forward scattering amplitude from
the one loop self-interaction and ${\cal A}$ represents the
amplitude from the one-photon emission. In the case of the forward
scattering and non-interacting processes, the spin and momentum of
the final states are the same. Note however, that for the one-photon
emission this is no longer the case. The final momentum is labelled
${\bf p'}$ here and the final spin $\beta$. The emission amplitude
thus contains two spin indices. Nevertheless, to lowest order in
$\hbar$ the spin does not change. In addition, by momentum
conservation, the final momentum is equal to ${\bf p}-\hbar{\bf k}$
which we shall label ${\bf P}$. These relations will be proved later
when we explicitly calculate the emission amplitude, but we shall
utilize them now in order to simplify the following calculations and
drop the spin indices on the emission amplitude. Let us thus define
the following parts of the final state
\begin{align}\label{ffor}
|f\rangle_{\rm for} &= \int
\f{d^3{\bf p}}{(2\pi\hbar)^3}\sqrt{\f{m}{p_0}}
\[1+i{\cal F}({\bf p})\]f({\bf p}) b^\dagger_\alpha ({\bf p}) |0\rangle \\ \label{fem}
|f\rangle_{\rm em} &= \f{i}{\hbar}\int
\f{d^3{\bf p}}{(2\pi\hbar)^3}\f{d^3 {\bf k}}{(2\pi)^3 2k_0}\sqrt{\f{m}{p_0}}
{\cal A}^\mu({\bf k},{\bf p}) f({\bf p}) a_\mu^\dagger({\bf k})
b^\dagger_\alpha ({\bf P}) |0\rangle \,.
\end{align}
As there will be no cross-term between these two states, the final
state density is the sum of the densities of the above two state. We
now proceed to calculate these densities and consequently obtain an
expression for the position expectation value of the final state in
terms of the two amplitudes. These calculations follow those from
the scalar field very closely and the reader may wish to refer to
them.

\subsection{Zero photon sector}

The zero photon sector density is given by
\beq
{}_{\rm for}\langle f | :\psi^\dagger \psi: |f\rangle_{\rm for}
\,.
\eeq
The state $|f\rangle_{\rm for}$ is nearly identical to that used for
the scalar case and again note that the calculation of the density
is identical to that for the non-interacting state with the
substitution $f({\bf p})
\to
\[1+i{\cal F}({\bf p})\]f({\bf p})$. The density is thus
\beq
\int\f{d^3 {\bf p}}{(2\pi\hbar)^3}\f{d^3 {\bf p}'}{(2\pi\hbar)^3}\f{m}{\sqrt{p_0
 p_0'}}\[1-i{\cal F}^*({\bf p}')\]f^*({\bf p}')
 \[1+i{\cal F}({\bf p})\]f({\bf p})
 \Phi^{\alpha\dagger}({\bf p'})\Phi^{\alpha}({\bf p}) \,.
\eeq
The $t=0$ position expectation value of the $|f\rangle_{for}$ state
is hence
\begin{align}
\langle x^i(0) \rangle_{\rm for}  &= \f{i\hbar}{2} \int \f{d^3{\bf p}}{(2\pi\hbar)^3}m
\nol & \quad \times \(\[1-i{\cal F}^*({\bf p})\]\f{f^*({\bf p})}{\sqrt{p_0}}u^\dagger(p)\)
\stackrel{\leftrightarrow}{\partial}_{p_i} \(\[1+i{\cal F}({\bf p})\]
\f{f({\bf p})}{\sqrt{p_0}}u(p)\) \,.
\end{align}
Expanding out the terms to order $e^2$ (i.e. ignoring the ${\cal
F}^*{\cal F}$ type terms) we have
\begin{align}\label{fspos}
\langle x^i(0) \rangle_{\rm for}
&= \f{i\hbar}{2} \int \f{d^3{\bf p}}{(2\pi\hbar)^3} m
\(\f{f^*({\bf p})u^\dagger(p)}{\sqrt{p_0}}\)\stackrel{\leftrightarrow}{\partial}_{p_i}\(\f{f({\bf p})u(p)}{\sqrt{p_0}}\)
\[1-2\Im{\cal F}\] \nol
& \quad  -\hbar\int \f{d^3{\bf p}}{(2\pi\hbar)^3} |f({\bf p})|^2
\partial_{p_i}
 \Re {\cal F}({\bf p}) \,.
\end{align}
The expression obtained is analogous to the scalar case in that we
find the appropriate form of the non-interacting position combined
with $\[1-2\Im{\cal F}\]$ which we shall deal with later, and a
further term dependent on the real part of the forward scattering
amplitude.\footnote{In fact, careful analysis of the calculation
would show that the momentum derivative of the imaginary part of
${\cal F}$ is necessarily zero.}

\subsection{One photon sector}
The density and position expectation value for the one photon sector
is more complicated due to the fact that the final state electron is
now moving with momentum ${\bf P}$ rather than ${\bf p}$. We dealt
with this problem before with the scalar field, and as we are
dealing with a time-dependent potential, we may make use of the
conservation of momentum.\footnote{The reader may recall that the
time-dependent case is slightly more straight forward in this
respect than that for the space-dependent case.} The density is
given by
\begin{align}
& \quad {}_{\rm em}\langle f|:\psi^\dagger \psi:|f\rangle_{\rm em}
\nol &=  \f{-i}{\hbar}\int
\f{d^3{\bf p}'}{(2\pi\hbar)^3}\f{d^3 {\bf k}'}{(2\pi)^3 2k_0'}\sqrt{\f{m}{p_0'}}
{\cal A}^{\nu\dagger}({\bf k}',{\bf p}') f({\bf p}')
 \nol & \quad \times \f{i}{\hbar}\int
\f{d^3{\bf p}}{(2\pi\hbar)^3}\f{d^3 {\bf k}}{(2\pi)^32k_0}\sqrt{\f{m}{p_0}}
{\cal A}^\mu({\bf k},{\bf p}) f({\bf p}) \nol & \quad \times \langle
0 | a_\nu({\bf k}') b_\alpha ({\bf P}') :
\psi^\dagger \psi :a_\mu^\dagger({\bf k}) b^\dagger_\alpha ({\bf P}) |0\rangle
\nol
&= -\f{1}{\hbar} \int
\f{d^3{\bf p}'d^3{\bf p}}{(2\pi\hbar)^6 \sqrt{p_0p_0'}}\f{d^3 {\bf k}}{(2\pi)^3 2k_0}m
\Big(f^*({\bf p}'){\cal A}_\mu^* ({\bf p}',{\bf k})\Phi^\dagger_\alpha ({\bf P'})\Big)\Big(f({\bf p}){\cal
A}^\mu({\bf p},{\bf k})\Phi_\alpha ({\bf P}) \Big) \,,
\end{align}
where ${\bf P'}={\bf p'}-\hbar{\bf k'}$ and we have used the
anticommutation relations for spinor field (\ref{anticomms}), as
used for the initial control particle calculations, and the
commutation relations for the electromagnetic field
(\ref{EMcommutation}). Note that the mode functions present in this
expression are those of the free field for the density in the ${\cal
M}_+$ region. The position expectation value (at $t=0$) is therefore
\begin{align}
& \quad \langle x^i(0) \rangle_{\rm em} \nol &= -\f{1}{\hbar} \int
\f{d^3{\bf p}'d^3{\bf p} m}{(2\pi\hbar)^6 \sqrt{p_0p_0'}}\f{d^3 {\bf k}}{(2\pi)^3 2k_0}
\Big(f^*({\bf p}'){\cal A}_\mu^* ({\bf p}',{\bf k})u^\dagger_\alpha(P')\Big)\Big(f({\bf p}){\cal
A}^\mu({\bf p},{\bf k})u_\alpha(P)\Big) \nol & \quad \times \int
d^3x
\partial_{P_i}
\(i\hbar e^{-i\({\bf P}-{\bf P}'\cdot {\bf x}\)/\hbar}\) \nol &= -i
\int
\f{d^3{\bf p}'d^3{\bf p}}{(2\pi\hbar)^6}\f{d^3 {\bf k}}{(2\pi)^3 2k_0}m \nol & \quad \times
\Big(\f{f^*({\bf p}')}{\sqrt{p_0'}}{\cal A}_\mu^* ({\bf p}',{\bf k})
u^\dagger_\alpha(P')\Big)\f{\partial}{\partial P_i}\Big(\f{f({\bf
p})}{\sqrt{p_0}}{\cal A}^\mu({\bf p},{\bf
k})u_\alpha(P)\Big)(2\pi\hbar)^3
\delta^3({\bf P}-{\bf P}') \,.
\end{align}
Given the definitions of ${\bf P}$ and ${\bf P'}$, we have
$\delta^3({\bf P}-{\bf P'})=\delta^3({\bf p}-{\bf p'})$. The
position shift due to emission is therefore given by the expression
\begin{multline}
\langle x^i (0) \rangle_{\rm em} = -\f{i}{2} \int
\f{d^3{\bf p}}{(2\pi\hbar)^3}\f{d^3 {\bf k}}{(2\pi)^3 2k_0}m \\
\Big(\f{f^*({\bf p})}{\sqrt{p_0}}{\cal A}_\mu^* ({\bf p},{\bf k})u^\dagger_\alpha(P)\Big)\stackrel{\leftrightarrow}{\partial}_{P_i}\Big(\f{f({\bf p})}{\sqrt{p_0}}{\cal
A}^\mu({\bf p},{\bf k})u_\alpha(P)\Big) \,.
\end{multline}
This can be split into two parts:
\begin{align}
& \quad \langle x^i (0) \rangle_{\rm em} \nol &= -\f{i}{2} \int
\f{d^3{\bf p}}{(2\pi\hbar)^3p_0}\f{d^3 {\bf k}}{(2\pi)^3 2k_0}m
\( {\cal A}_\mu^* ({\bf p},{\bf k}) \stackrel{\leftrightarrow}{\partial}_{P_i} {\cal A}^\mu({\bf p},{\bf k})
\)
|f({\bf p})|^2  u^\dagger_\alpha(P)u_\alpha(P) \nol & \quad
-\f{i}{2}  \int \f{d^3{\bf p}}{(2\pi\hbar)^3} m
\(\f{f^*({\bf p})}{\sqrt{p_0}}u^\dagger_\alpha(P)\) \stackrel{\leftrightarrow}{\partial}_{P_i}
\(\f{f({\bf p})}{\sqrt{p_0}}u_\alpha(P)\) \nol & \qquad \times \int \f{d^3
{\bf k}}{(2\pi)^3 2k_0} {\cal A}_\mu^* ({\bf p},{\bf k}) {\cal
A}^\mu({\bf p},{\bf k}) \,.
\end{align}
The first integral gives to lowest order
\begin{align}
&\quad -\f{i}{2} \int \f{d^3{\bf p}}{(2\pi\hbar)^3}|f({\bf p})|^2
\int \f{d^3 {\bf k}}{(2\pi)^3 2k_0}
\( {\cal A}_\mu^* ({\bf p},{\bf k}) \stackrel{\leftrightarrow}{\partial}_{p_i} {\cal A}^\mu({\bf p},{\bf k})
\)
 \f{P_0}{p_0}\f{\partial p_i}{\partial P_i}
 \nol
&= -\f{i}{2} \int \f{d^3{\bf p}}{(2\pi\hbar)^3}|f({\bf p})|^2 \int
\f{d^3 {\bf k}}{(2\pi)^3 2k_0}
\( {\cal A}_\mu^* ({\bf p},{\bf k}) \stackrel{\leftrightarrow}{\partial}_{p_i} {\cal A}^\mu({\bf p},{\bf k})
\) \,,
\end{align}
where in the last line we have made use of the fact that to order
$\hbar^0$, $P_0=p_0$ and $\partial p_i/\partial P_i = 1$.

\subsection{Unitarity}
In the scalar calculation, we gained a further relation between the
imaginary part of the forward scattering and the emission
probability using the normalisation of the final state, viz
\beq\label{funitarity}
\langle f | f \rangle = 1 \,.
\eeq
We can complete the same calculation again and find that we do in
fact find the same relation. The final state is given by the sum of
$\ket{f}_{\rm for}$ and $\ket{f}_{\rm em}$, in (\ref{ffor}) and
(\ref{fem}) respectively. The left hand side of (\ref{funitarity})
above is thus,
\begin{align}
& \quad \langle f | f \rangle \nol &= \int \f{d^3{\bf
p}}{(2\pi\hbar)^3}
\f{d^3{\bf p}'}{(2\pi\hbar)^3}\sqrt{\f{m}{p'_0}}\sqrt{\f{m}{p_0}}
 f^*({\bf p})\[ 1-i{\cal F}^*({\bf p})\]\[ 1+i{\cal F}({\bf p}')\]f({\bf p}')
 \langle 0 | b_\alpha({\bf p})b^\dagger_\alpha({\bf p}')|0\rangle
\nol
& \quad + \f{1}{\hbar^2}\int \f{d^3{\bf p}}{(2\pi\hbar)^3}
\f{d^3{\bf p}'}{(2\pi\hbar)^3}\f{d^3{\bf k}}{(2\pi)^32k_0}\f{d^3{\bf k}'}{(2\pi)^32k_0'}
\sqrt{\f{m}{p'_0}}\sqrt{\f{m}{p_0}} \nol & \quad \times
f^*({\bf p}) {\cal A}^{*\mu}(k,{\bf p})f({\bf p}'){\cal
A}^\nu(k',{\bf p}')
\langle 0 | b_\alpha({\bf P}) a_\mu({\bf k})a_\nu^\dagger({\bf k}')b^\dagger_\alpha({\bf P}') |0
\rangle \,.
\end{align}
We again make use of the conservation of momentum, with
\begin{equation}
\f{P_0}{m}\delta^3({\bf P}-{\bf P'})=\f{p_0}{m}\delta^3({\bf p}-{\bf
p'})\f{P_0}{p_0} \,.
\end{equation}
Hence to order $e^2$ (i.e. only up to first order in ${\cal F}$) we
have
\begin{align}
\langle f | f \rangle &=
\int \f{d^3{\bf p}}{(2\pi\hbar)^3} |f({\bf p})|^2 \(1-2\Im {\cal F}({\bf p})\) \nol
& \quad -\f{1}{\hbar} \int \f{d^3{\bf p}}{(2\pi\hbar)^3}\f{d^3{\bf
k}}{(2\pi)^32k_0} |f({\bf p})|^2 {\cal A}_\mu^*({\bf p},{\bf k})
{\cal A}^\mu({\bf p},{\bf k})\f{P_0}{p_0} \,.
\end{align}
As this is equal to $1$ (by (\ref{funitarity})) and $f({\bf p})$ is
normalised by (\ref{normalisef}), we obtain
\beq
\int \f{d^3{\bf p}}{(2\pi\hbar)^3} |f({\bf p})|^22\Im {\cal F}({\bf p}) =
-\f{1}{\hbar}\int
\f{d^3{\bf p}}{(2\pi\hbar)^3}|f({\bf p})|^2 \int \f{d^3{\bf k}}{(2\pi)^32k_0}
 {\cal A}_\mu^*({\bf p},{\bf k}) {\cal A}^\mu({\bf p},{\bf
 k})\f{P_0}{p_0} \,,
\eeq
as before. Consequently, using the delta function limit for $|f({\bf
p})|^2$,
\beq\label{unitarity}
2\Im {\cal F}({\bf p}) = -\f{1}{\hbar}\int \f{d^3{\bf
k}}{(2\pi)^32k_0} {\cal A}_\mu^*({\bf p},{\bf k}) {\cal A}^\mu({\bf
p},{\bf k})\f{P_0}{p_0} \,,
\eeq
where we have relabeled the final peak momentum $\bar{\bf p}\to {\bf
p}$.

\subsection{Position of the final state}

If we add the contributions to the position expectation value from
the forward scattering and emission sectors we obtain
\begin{align}
& \quad \langle x^i (0) \rangle \nol &= \f{i\hbar}{2} \int
\f{d^3{\bf p}}{(2\pi\hbar)^3} m
\(\f{f^*({\bf p})u^\dagger(p)}{\sqrt{p_0}}\)\stackrel{\leftrightarrow}{\partial}_{p_i}\(\f{f({\bf p})u(p)}{\sqrt{p_0}}\)
\[1-2\Im{\cal F}({\bf p})\] \nol
&  \quad -\hbar\int \f{d^3{\bf p}}{(2\pi\hbar)^3} |f({\bf p})|^2
\partial_{p_i} \Re {\cal F}({\bf p})
\nol
& \quad -\f{i}{2} \int \f{d^3{\bf p}}{(2\pi\hbar)^3}|f({\bf p})|^2
\int \f{d^3 {\bf k}}{(2\pi)^3 2k_0}
\({\cal A}_\mu^*({\bf p},{\bf {\bf k}})
\stackrel{\leftrightarrow}{\partial}_{p_i}{\cal A}^\mu({\bf p},{\bf k})\)
 \nol
&
 \quad -\f{i}{2} \int
\f{d^3{\bf p}}{(2\pi\hbar)^3}\f{d^3 {\bf k}}{(2\pi)^3 2k_0} m
\(\f{f^*({\bf p})u^\dagger_\alpha(P)}{\sqrt{p_0}}\)\stackrel{\leftrightarrow}{\partial}_{P_i}\(\f{f({\bf p})u_\alpha(P)}{\sqrt{p_0}}\)
{\cal A}_\mu^*({\bf p},{\bf k}){\cal A}^\mu({\bf p},{\bf k}) \,.
\end{align}
Using the unitarity condition (\ref{unitarity}) we remove the
imaginary part of ${\cal F}$ to produce
\begin{align}
& \quad \langle x^i (0) \rangle  \nol &= i\hbar \int \f{d^3{\bf
p}}{(2\pi\hbar)^3} m
\(\f{f^*({\bf p})u^\dagger_\alpha(p)}{\sqrt{p_0}}\)\f{\partial}{\partial p_i}\(\f{f({\bf p})u_\alpha(p)}{\sqrt{p_0}}\)
\nol
& \quad -\hbar\int \f{d^3{\bf p}}{(2\pi\hbar)^3} |f({\bf p})|^2
\d_p \Re {\cal F}({\bf p})
\nol
& \quad -\f{i}{2}  \int
\f{d^3{\bf p}}{(2\pi\hbar)^3}|f({\bf p})|^2 \int \f{d^3 {\bf k}}{(2\pi)^3 2k_0}
{\cal A}_\mu^*({\bf p},{\bf k})
\stackrel{\leftrightarrow}{\partial}_{p_i}{\cal A}^\mu({\bf p},{\bf k})
\nol
& \quad + \f{i}{2}\int \f{d^3{\bf p}}{(2\pi\hbar)^3}\f{d^3 {\bf
k}}{(2\pi)^3 2k_0} m {\cal A}_\mu^*({\bf p},{\bf k}) {\cal
A}^\mu({\bf p},{\bf k}) \nol &
\[ \f{P_0}{p_0}\(\f{f^*({\bf p})u^\dagger_\alpha(p)}{\sqrt{p_0}}\)\stackrel{\leftrightarrow}{\partial}_{p_i}
\(\f{f({\bf p})u_\alpha(p)}{\sqrt{p_0}}\)
-\(\f{f^*({\bf p})u^\dagger_\alpha(P)}{\sqrt{p_0}}\)
\stackrel{\leftrightarrow}{\partial}_{P_i}\(\f{f({\bf
p})u_\alpha(P)}{\sqrt{p_0}}\)\] \,.
\end{align}
To ${\cal O}(\hbar^0)$ and using the sharply peaked property of
$f({\bf p})$ we thus obtain
\begin{align}
 \langle x^i(0) \rangle
& = \f{i\hbar}{2} \int \f{d^3{\bf p}}{(2\pi\hbar)^3} f^*({\bf
p})\stackrel{\leftrightarrow}{\partial}_{p_i}f({\bf p})
\nol &  \quad -\hbar\partial_{p_i} \Re {\cal F}({\bf p})
 -\f{i}{2} \int \f{d^3 {\bf k}}{(2\pi)^3 2k_0}
{\cal A}_\mu^*({\bf p},{\bf k})
\stackrel{\leftrightarrow}{\partial}_{p_i}{\cal A}^\mu({\bf p},{\bf
k}) \,.
\end{align}
The first term is the position of the non-radiating particle which
we recall is at the origin. There are thus two contributions to the
position shift, the emission shift $\delta x^i_{\rm em}$ and forward
scattering shift $\delta x^i_{\rm for}$ defined as follows:
\begin{align}\label{emshift}
 \delta x^i_{\rm em} =& -\f{i}{2} \int \f{d^3 {\bf k}}{(2\pi)^3 2k_0}
{\cal A}_\mu^*({\bf p},{\bf k})\stackrel{\leftrightarrow}{\partial}_{p_i}{\cal A}^\mu({\bf p},{\bf k}) \\
\delta x^i_{\rm for} =& -\hbar\partial_{p_i}\Re{\cal F}({\bf p})
\,. \label{forshift}
\end{align}
In this limit the expression for the position expectation value, in
terms of the amplitudes, is the same as that obtained for the scalar
field. It is now our task to evaluate these two expressions.

\section{Emission Amplitude}\label{spinoremissioncalcsection}

The emission process, resulting from the first order interaction
term, is given by
\begin{align}
b_\alpha^\dagger({\bf p})|0\rangle \to ... -\f{i}{\hbar}\int
d^4x\,{\cal H}_I (x) b_\alpha^\dagger({\bf p})|0\rangle \,.
\end{align}
The QED interaction Hamiltonian for the coupling of the spinor and
electromagnetic fields is the negative of the interaction
Lagrangian.\footnote{As in the scalar case, we note that we are
using the free-field normal ordering operators in the interaction
Hamiltonian (see footnote \ref{normalfootnote1} at the beginning of
the scalar Emission Amplitude calculation). Again, however, it can
be shown that this is justified to order $\hbar^2$
\cite{HM6}.} Unlike the more complicated situation we had to deal
with for the scalar field, we have just the one coupling term to
consider. Substituting the concrete expression for ${\cal H}_I$, and
we obtain
\begin{align}
&-\f{i}{\hbar} \int d^4x {\cal H}_I(x) b^\dagger_\alpha ({\bf p})
\ket{0}
\nol
=& -\f{ie}{\hbar} \int d^4x \, : A_\mu \bar{\psi}\gamma^\mu\psi :
b^\dagger_\alpha ({\bf p}) \ket{0} \nol =& -\f{ie}{\hbar} \int d^4x
\f{d^3{\bf k}}{(2\pi)^3 2k_0}\f{d^3{\bf p}'}{(2\pi\hbar)^3
p_0'/m}\f{d^3{\bf p}''}{(2\pi\hbar)^3 p_0''/m} \nol &
\times a^\dagger_\mu ({\bf k}) e^{ik\cdot x}  \bar{\Phi}^\beta ({\bf p}')
\gamma^\mu \Phi^\gamma ({\bf p''}) b^\dagger_\beta({\bf p}') b_\gamma ({\bf p}'') b^\dagger_\alpha({\bf p})
\ket{0} \nol
 =& -\f{ie}{\hbar} \int d^4x
\f{d^3{\bf k}}{(2\pi)^3 2k_0}\f{d^3{\bf p}'}{(2\pi\hbar)^3 p_0'/m}
 e^{ik\cdot x}  \bar{\Phi}^\beta ({\bf p'})
\gamma^\mu \Phi_\alpha ({\bf p})  a^\dagger_\mu ({\bf k})
b^\dagger_\beta({\bf p}') \ket{0} \,,
\end{align}
where we have ignored the separate particle creation vacuum process
which is not part of the evolution of the state. The emission
amplitude is thus given by the expression
\begin{equation}
{\cal A}^{(\beta)\mu}_{(\alpha)} ({\bf p},{\bf k}) = -e \int d^4x
\f{d^3 {\bf p}'}{(2\pi\hbar)^3p_0'/m} \bar{\Phi}^\beta ({\bf p'}) \gamma^\mu
\Phi_\alpha ({\bf p}) e^{ik\cdot x} \,.
\end{equation}
We have indices on the amplitude to represent the initial and final
spins. The fields involved in the interaction are the non-free
fields from the region ${\cal M}_I$. The mode functions in the
emission amplitude are therefore those for the non-free field. We
proceed substituting the semiclassical expansion of these mode
functions from (\ref{wkbspinormodep}). As we wish to take the
$\hbar\to 0$ limit, we shall only need the ${\cal O}(\hbar^0)$ terms
in the expansion.
\begin{align}
{\cal A}^{(\beta)\mu}_{(\alpha)} ({\bf p},{\bf k}) =& -e\int
\f{d^4x\,d^3p'}{(2\pi\hbar)^3p_0'/m}
 \phi_{p'}^*(t)\phi_p(t) e^{-i{\bf p'}\cdot{\bf x}/\hbar}e^{i{\bf p}\cdot{\bf
 x}/\hbar} e^{ik\cdot x}
 \bar{u}^\beta(p',t)\gamma^\mu u_\alpha(p,t) \nol
 =& -e\int
\f{d^4x\,d^3p'}{(2\pi\hbar)^3p_0'/m} \phi_{p'}^*(t)\phi_p(t) e^{ikt}
\bar{u}^\beta(p',t)\gamma^\mu u_\alpha(p,t) e^{i({\bf p}-{\bf p'}-\hbar{\bf
k})/\hbar} \,.
\end{align}
The spatial integration gives the delta function corresponding to
the conservation of momentum ${\bf p}={\bf p'}+\hbar{\bf k}$. In our
previous working, we stated that we had conservation of momentum and
defined the final momentum as ${\bf P}={\bf p}-\hbar{\bf k}$. The
above calculation demonstrates this conservation (with the
substitution ${\bf p'}={\bf P}$).
\begin{align}
{\cal A}^{(\beta)\mu}_{(\alpha)} ({\bf p},{\bf k}) &= -e\int
\f{dt\,d^3{\bf p}'}{p_0'/m}
\sqrt{\f{p_0'p_0}{E_{p'} E_p}}
\bar{u}^\beta(p',t)\gamma^\mu u_\alpha(p,t) \nol & \quad \times \exp
\(-\f{i}{\hbar}\int^t_0 \(E_p(\zeta)-E_{p'}(\zeta)\)d\zeta\)
e^{ikt} \delta^3({\bf p}-{\bf p'}-\hbar{\bf k}) \,.
\end{align}
The exponential can be written (using the delta function) as in the
scalar case:
\begin{align}
\exp\(-\f{i}{\hbar}\int_0^t \( E_p(\zeta)-E_{p'}(\zeta)\) d\zeta\) =&
\exp\(-\f{i}{\hbar} \[ \partial_{\bf p} \int_0^t
E_p(\zeta)d\zeta\]\cdot\[{\bf p}-{\bf p'} \] \) \nol =&
\exp\(-\f{i}{\hbar} \[  \int_0^t
\f{d {\bf x}}{d \zeta} d\zeta\]\cdot\[\hbar {\bf k}\] \) \nol
 =& \exp \(-i{\bf k}\cdot{\bf x}\) \,.
\end{align}
We now look at the spinor factor. The component with $\gamma^0$ is
to order $\hbar^0$, using the zeroth order spinor given in
(\ref{zerothorderspinoru}),
\begin{align}
& \bar{u}^\beta(p',t)\gamma^0 u_\alpha(p,t) \nol =&
\f{\sqrt{(E_{p'}+m)(E_p+m)}}{2m} s^{\beta\dagger} U_{p'}^\dagger(t)
\[ 1+
\f{\itbd{\sigma}\cdot{\bf p'}(t)\,\itbd{\sigma}\cdot{\bf p}(t)}{(E_{p'}+m)(E_p+m)} \] U_p(t)
s_\alpha \,,
\end{align}
with $U_p(t)$ defined in (\ref{Udefn}). We can take ${\bf p'}$ to
${\bf p}$ in all the terms to lowest $\hbar$ order, including the
unitary matrix $U_{p'}^\dagger(t)$. This component thus simplifies
to
\begin{align}
\bar{u}^\beta(p',t)\gamma^0 u^\alpha(p,t) = \f{E_p}{m}\delta^\beta_\alpha + {\cal
O}(\hbar) \,.
\end{align}
Similarly,
\begin{align}
& \bar{u}^\beta(p',t)\gamma^i u_\alpha(p,t) \nol =&
\f{\sqrt{(E_{p'}+m)(E_p+m)}}{2m} s^{\beta\dagger} U_{p'}^\dagger(t)
\[ \f{\itbd{\sigma}\cdot{\bf p'}(t)\,\sigma^i}{E_{p'}+m} +
\f{\sigma^i\,\itbd{\sigma}\cdot{\bf p}(t)}{E_p+m} \] U_p(t) s_\alpha
+{\cal O}(\hbar) \,.
\end{align}
As before, we change ${\bf p'}\to{\bf p}$ to the lowest order and
note that
\begin{equation}
\f{\itbd{\sigma}\cdot{\bf p}(t)\,\sigma^i}{E_{p'}+m} +
\f{\sigma^i\,\itbd{\sigma}\cdot{\bf p}(t)}{E_p+m} =
2\f{p^i(t)}{E_p+m}\delta^\beta_\alpha \,.
\end{equation}
Overall, we thus obtain
\begin{align}
\bar{u}^\beta(p',t)\gamma^\mu u_\alpha(p,t) = \f{\tilde{p}^\mu}{m}\delta^\beta_\alpha +
{\cal O}(\hbar) \,.
\end{align}
Consequently, the emission amplitude can be written to lowest order
as
\begin{align}
{\cal A}^{(\beta)\mu}_{(\alpha)} (p,k) =& -e \int dt
\f{\tilde{p}^\mu}{E_p} e^{ik\cdot x} \delta^\beta_\alpha \nol =& -e
\int dt \f{dx^\mu}{dt} e^{ik\cdot x} \delta^\beta_\alpha \nol =& -e
\int d\xi \f{d x^\mu}{d\xi} e^{ik\xi} \delta^\beta_\alpha \,,
\end{align}
where we define $\xi:= t-{\bf k}\cdot{\bf x}/k_0$. We see that as
stated previously, the spin does not change in the lowest $\hbar$
order. Therefore
\begin{align}
{\cal A}^\mu (p,k) =& -e \int d\xi \f{d x^\mu}{d\xi} e^{ik\xi} \,.
\end{align}
This is the same expression as obtained for the lowest order
emission amplitude for the scalar field (\ref{amplitude}) and is
written in terms of the classical trajectory. We additionally note
that the amplitude is equal to the classical amplitude.
Consequently, using either of the methods from the scalar
calculations (Chapter \ref{scalarshiftchapter} section
\ref{scalaremissioncalcsection} and Chapter
\ref{greenschapter} section
\ref{greensemissiondecomp}) we find that the position shift due to
emission $\delta x^i_{\rm em}$ is equal to the classical position
shift $\delta x^i_C$. From Chapter
\ref{greenschapter} we can rewrite the shift as follows:
\begin{align}
 \delta x^i_{\rm em} & = -\f{i}{2} \int \f{d^3 {\bf k}}{(2\pi)^3 2k_0}
\({\cal A}_\mu^*({\bf p},{\bf k})\partial_{p_i}{\cal A}^\mu({\bf p},{\bf k})\)
\nol
& = \int d^4x\, \partial_{p^i}j^\mu(x)A_{{\rm R}\mu}(x) \nol & =
-\int_{-\infty}^0 dt\,{\cal F}^j_{\rm LD}\,\left(
\frac{\partial x^j}{\partial p^i}\right)_t \,,
\end{align}
where $A_{{\rm R}\mu}(x)$ is the regular field, constructed from the
regular Green's function $1/2(G_{-}-G_{+})$, which acts on the
classical particle to produce the radiation reaction force
\cite{Poisson}. This was the same situation we had for the scalar
field, whereby the position shift due to the emission could be
equated with that due to the regular (or radiative) field and
consequently the full classical position shift. As with the previous
case though, we still have another quantum contribution to the
position shift from the forward scattering which we must take into
account and calculate.

\section{Forward Scattering}\label{spinorforwardscattcalcsection}

We now consider the forward scattering amplitude. As before, we
calculate the amplitude using the semiclassical mode functions for
the interacting region ${\cal M}_I$ and expect this result to also
be divergent. To this we can add the amplitude due to the QED mass
counterterm, thereby renormalising the forward scattering. It is
well known that the counter term is also divergent. We shall see
however, that the situation is not as straightforward as the scalar
renormalisation. Before continuing with the amplitude calculation,
we briefly recall that the position shift term is
\begin{equation}
\delta x^i_{\rm for} = -\hbar\partial_{p_i}\Re{\cal F}({\bf p})
\,, \tag{\ref{forshift}}
\end{equation}
containing a multiplying factor of $\hbar$. Given that the forward
scattering is of order $\hbar^{-2}$, we are interested in the
$\hbar^{-2}$ and $\hbar^{-1}$ terms of the real part of ${\cal F}$
leading to position shift contributions at order $\hbar^{-1}$ and
$\hbar^0$ respectively. For the scalar field, the former
contribution canceled upon renormalisation, whereas the latter
contribution was zero due to ${\cal F}$ being imaginary at that
order. With these comments and the previous method and calculation
in mind, we proceed with the spinor amplitude.

The relative simplicity of the interaction Hamiltonian for the
spinor field when compared with the previous scalar case means that
the forward scattering process is simply the one-loop process and is
the zero-photon sector of the second order interaction term:
\begin{equation}
i{\cal F}({\bf p}) b_\alpha^\dagger({\bf p}) \ket{0} = \left.
\f{1}{2}\(\f{-i}{\hbar}\)^2
\int d^4xd^4x' T \[ {\cal H}_I(x'){\cal H}_I(x) \] b_\alpha^\dagger({\bf p})
\ket{0} \right|_{\rm
zero-photon}\,,
\end{equation}
with ${\cal H}_I=e\bar{\psi}\slashed{A}\psi$. Operating on both
sides with $\bra{0}b_\alpha({\bf p'})$ we have
\begin{multline}
\bra{0} b_\alpha({\bf p'}) i{\cal F}({\bf p}) b_\alpha^\dagger({\bf p}) \ket{0}\\ =   -\f{1}{2}\f{1}{\hbar^2}
\int d^4xd^4x'\bra{0} b_\alpha({\bf p'}) T \[ :e\bar{\psi}(x')\slashed{A}(x')\psi(x')::e\bar{\psi}(x)\slashed{A}(x)\psi(x): \]
 b_\alpha^\dagger({\bf p}) \ket{0} \,, \nonumber
\end{multline}
thus
\begin{multline}
{\cal F}({\bf p})  = \f{i}{2}\f{e^2m}{\hbar^2p_0}
\int \f{d^3{\bf p'}}{(2\pi\hbar)^3} d^4xd^4x' \bra{0}T \[A_\mu(x')A_\nu(x)\]\ket{0} \\
\times \bra{0} b_\alpha({\bf p'}) T \[ :\bar{\psi}(x')\gamma^\mu\psi(x')::\bar{\psi}(x)\gamma^\nu\psi(x): \] b_\alpha^\dagger({\bf p})
\ket{0} \,. \label{Fprods2}
\end{multline}
Here we have, as one would expect from a one-loop diagram, the
photon propagator
\begin{multline}
\bra{0}T \[A_\mu(x')A_\nu(x)\]\ket{0} \\ =  -\hbar g_{\mu\nu} \int \f{d^3{\bf k}}{2k(2\pi)^3}
\[ \theta(t'-t)
e^{-ik\cdot(x'-x)}+\theta(t-t')e^{-ik\cdot(x-x')}\] \,,
\end{multline}
and a time-ordered combination of the spinor fields which we shall
denote
\begin{equation}
T_S(x,x') = \bra{0} b_\alpha({\bf p'}) T \[
:\bar{\psi}(x')\gamma^\mu\psi(x')::\bar{\psi}(x)\gamma^\nu\psi(x):
\] b_\alpha^\dagger({\bf p})
\ket{0} \,.
\end{equation}
Writing $T_S$ with the field expansions the term for each normal
ordered product, in which we have only one space-time variable, is
of the form
\begin{multline}
:\bar{\psi}\gamma^\nu \psi: \, =  \int \f{d^3 {\bf
p}}{(2\pi\hbar)^3}
\f{m}{p_0} \f{d^3 {\bf p}'}{(2\pi\hbar)^3} \f{m}{p_0'} \\ :
\[ b_\alpha^\dagger({\bf p})\bar{\Phi}^{\alpha}({\bf p}) +
d_\alpha({\bf p})\bar{\Psi}^{\alpha}({\bf p}) \] \gamma^\nu
\[ b_\beta({\bf p'})\Phi^{\beta}({\bf p'}) +
d_\beta^\dagger({\bf p'})\Psi^{\beta}({\bf p'}) \] : \,.
\end{multline}
The creation and annihilation operators form the following
combinations
\begin{multline}
: b_\alpha^\dagger({\bf p}) b_\beta({\bf p'}) + d_\alpha({\bf p})
b_\beta({\bf p'}) + b_\alpha^\dagger({\bf p}) d_\beta^\dagger({\bf
p'}) + d_\alpha({\bf p}) d_\beta^\dagger({\bf p'}) :\, \\ =
b_\alpha^\dagger({\bf p}) b_\beta({\bf p'}) + d_\alpha({\bf p})
b_\beta({\bf p'}) + b_\alpha^\dagger({\bf p}) d_\beta^\dagger({\bf
p'}) - d_\beta^\dagger({\bf p'}) d_\alpha({\bf p}) \,.
\end{multline}
When operated on the left by $\bra{0} b_\gamma({\bf p''})$ the third
and fourth terms vanish and when operated on the right by
$b_\gamma^\dagger({\bf p''})\ket{0}$, the second and fourth terms
drop out. Hence, in terms of the operators only, the form of $T_S$
is
\begin{align}
& \bra{0} b_\alpha({\bf p'}) \( b_\beta^\dagger b_\gamma + d_\beta
b_\gamma
\)\(b_\delta^\dagger b_\epsilon + b_\delta^\dagger d_\epsilon^\dagger \)
b_\alpha^\dagger({\bf p}) \ket{0} \nol =&\, \bra{0} b_\alpha({\bf
p'}) b_\beta^\dagger b_\gamma b_\delta^\dagger b_\epsilon
b_\alpha^\dagger({\bf p}) \ket{0} - \bra{0} b_\alpha({\bf p'})
b_\delta^\dagger d_\beta d_\epsilon^\dagger b_\gamma
b_\alpha^\dagger({\bf p})
\ket{0} + V \,,
\end{align}
where V is the process with a vacuum graph (creation of
electron-positron pair and photon and subsequent annihilation
separate from the original particle and an unaffected particle)
which as before, we can ignore. In both the above calculations, we
have the familiar minus signs due to the anticommutation relations
used for spinor fields. Reintroducing the mode functions is
straightforward by matching their spins indices to those on the
creation/annihilation operators in the expansion above:
\begin{align}
T_S(x,x')  & = \int
\f{d^3{\bf p^{(2)}}m}{(2\pi\hbar)^3p^{(2)}_0} \f{d^3{\bf p^{(3)}}m}{(2\pi\hbar)^3p^{(3)}_0}
\f{d^3{\bf p^{(4)}}m}{(2\pi\hbar)^3p^{(4)}_0} \f{d^3{\bf p^{(5)}}m}{(2\pi\hbar)^3p^{(5)}_0} \nol
& \qquad T \Big[ \bar{\Phi}^\beta({\bf p^{(2)}},x)\gamma^\mu
\Phi^\gamma({\bf p^{(3)}},x)
\bar{\Phi}^\delta({\bf p^{(4)}},x') \gamma^\nu \Phi^\epsilon({\bf p^{(5)}},x')
 \nol  & \qquad  \qquad\times \bra{0} b_\alpha({\bf p}')
b_\beta^\dagger({\bf p^{(2)}}) b_\gamma({\bf p^{(3)}})
b_\delta^\dagger({\bf p^{(4)}}) b_\epsilon ({\bf
p^{(5)}})b_\alpha^\dagger({\bf p}) \ket{0}\nol & \qquad -
\bar{\Psi}^\beta({\bf p^{(2)}},x)\gamma^\mu \Phi^\gamma({\bf p^{(3)}},x)
\bar{\Phi}^\delta({\bf p^{(4)}},x') \gamma^\nu \Psi^\epsilon({\bf p^{(5)}},x')
\nol & \qquad  \qquad\times \bra{0} b_\alpha({\bf p}') b_\delta^\dagger({\bf p^{(4)}})
d_\beta ({\bf p^{(2)}})d_\epsilon^\dagger ({\bf p^{(5)}})
b_\gamma({\bf p^{(3)}}) b_\alpha^\dagger({\bf p})\ket{0} \Big] \,,
\end{align}
which simplifies when we use the anticommutation relations, to
produce
\begin{multline}
 T_S(x,x')   = \int \f{d^3{\bf q}m}{(2\pi\hbar)^3q_0}
 T
 \[ \bar{\Phi}_\alpha({\bf p'},x)\gamma^\mu \Phi^\gamma({\bf q},x)
\bar{\Phi}_\gamma({\bf q},x') \gamma^\nu \Phi_\alpha({\bf p},x') \right. \\ \left. -
\bar{\Psi}^\beta({\bf q},x)\gamma^\mu \Phi_\alpha({\bf p},x)
\bar{\Phi}_\alpha({\bf p'},x') \gamma^\nu \Psi_\beta({\bf q},x') \] \,,
\end{multline}
where we have changed the remaining integration variable to ${\bf
q}$ in both cases. The reader will recall that in the forward
scattering (\ref{Fprods2}), the product $T_S$ is integrated over $x$
and $x'$ and contracted with the metric $g_{\mu\nu}$. As a
consequence of the resulting symmetry, both orders of the time
variables give the same result and therefore we take one and
multiply by two. The forward scattering (not forgetting the photon
propagator term) can then be written
\begin{multline}
{\cal F}({\bf p})  = -\f{ie^2m}{\hbar p_0}
\int \f{d^3{\bf p'}}{(2\pi\hbar)^3} \f{d^3{\bf k}}{2k(2\pi)^3}
\f{d^3{\bf q}m}{(2\pi\hbar)^3q_0} d^4xd^4x' \theta(t-t')
e^{-ik\cdot(x-x')} \\
 \[ \bar{\Phi}_\alpha({\bf p'},x)\gamma^\mu \Phi^\gamma({\bf q},x)
\bar{\Phi}_\gamma({\bf q},x') \gamma_\mu \Phi_\alpha({\bf p},x') \right. \\ \left. -
\bar{\Psi}^\beta({\bf q},x)\gamma^\mu \Phi_\alpha({\bf p},x)
\bar{\Phi}_\alpha({\bf p'},x') \gamma_\mu \Psi_\beta({\bf q},x') \] \,.
\end{multline}
As with the one-loop part of the scalar field, we split this up into
the `particle loop' and `anti-particle loop', defined respectively
by\footnote{The plus and minus designation will become clear later.
It is however the same as that used in the scalar ${\cal F}_2$
cases.}
\begin{multline} \label{Fmin}
{\cal F}_-({\bf p})  = -\f{ie^2m}{\hbar p_0}
\int \f{d^3{\bf p'}}{(2\pi\hbar)^3} \f{d^3{\bf k}}{2k(2\pi)^3}
\f{d^3{\bf q}m}{(2\pi\hbar)^3q_0} d^4xd^4x' \theta(t-t')
 \\
 \times \bar{\Phi}_\alpha({\bf p'},x)\gamma^\mu \Phi^\gamma({\bf q},x)
\bar{\Phi}_\gamma({\bf q},x') \gamma_\mu \Phi_\alpha({\bf p},x')
e^{-ik\cdot(x-x')} \,,
\end{multline}
\begin{multline} \label{Fplus}
{\cal F}_+({\bf p})  = \f{ie^2m}{\hbar p_0}
\int \f{d^3{\bf p'}}{(2\pi\hbar)^3} \f{d^3{\bf k}}{2k(2\pi)^3}
\f{d^3{\bf q}m}{(2\pi\hbar)^3q_0} d^4xd^4x' \theta(t-t')
 \\
 \times \bar{\Phi}_\alpha({\bf p'},x') \gamma^\mu \Psi_\beta({\bf q},x') \bar{\Psi}^\beta({\bf q},x)\gamma_\mu \Phi_\alpha({\bf p},x)
 e^{-ik\cdot(x-x')} \,,
\end{multline}
where we have rearranged the spinor mode functions in the last line.

\subsection{Particle Loop}
The ${\cal F}_-({\bf p})$ half of the forward scattering amplitude
is the result of the particle loop process. Using the semiclassical
expansion, the spinor mode functions in ${\cal F}_-({\bf p})$, along
with the $k$ exponential,\footnote{which is really the free
electromagnetic field mode function,} can be written
\begin{align}
&\quad  \bar{\Phi}_\alpha({\bf p'},x)\gamma^\mu \Phi^\gamma({\bf
q},x)
\bar{\Phi}_\gamma({\bf q},x') \gamma_\mu \Phi_\alpha({\bf p},x')
e^{-ik\cdot(x-x')} \nol
 &= \[ \bar{u}_\alpha
(p',t) \gamma^\mu u^\gamma (q,t) \bar{u}_\gamma (q,t') \gamma_\mu
u_\alpha (p,t') \] \nol &\qquad \times
 \phi^*_{p'}(t) \phi_q(t) \phi^*_q(t')\phi_p(t') e^{-i{\bf p'}\cdot{\bf x}/\hbar}e^{i{\bf q}\cdot{\bf x}/\hbar}e^{-i{\bf
q}\cdot{\bf x'}/\hbar}e^{i{\bf p}\cdot{\bf x'}/\hbar}
e^{-ik\cdot(x-x')} \,,
\end{align}
where we have the time-dependent spinor expansions inside the square
brackets and the scalar semiclassical terms and the exponentials
outside. Completing the spatial integrals produces the delta
functions $\delta^3({\bf p}-{\bf q}-{\bf K})\delta^3({\bf p'}-{\bf
q}-{\bf K})$ with ${\bf K}:=\hbar {\bf k}$. Using $\delta^3({\bf
p}-{\bf q}-{\bf K})\delta^3({\bf p}-{\bf p'})$ and integrating over
${\bf p'}$, the particle loop contribution can be written
\begin{align}
{\cal F}_-({\bf p}) &= -\f{ie^2m}{\hbar p_0}
 \int \f{d^3{\bf k}}{2k(2\pi)^3}
\f{d^3{\bf q}m}{(2\pi\hbar)^3q_0} dt dt' \theta(t-t') (2\pi\hbar)^3 \delta^3({\bf p}-{\bf q}-{\bf K})
 \nol & \qquad
 \times \[{\cal S}_-\] \phi^*_{p}(t) \phi_q(t) \phi^*_q(t')\phi_p(t')
 e^{-iK(t-t')/\hbar} \,,
 \end{align}
where we have defined the semiclassical particle loop spinor
combination
\begin{align}
{\cal S}_- = \bar{u}_\alpha (p,t) \gamma^\mu u^\gamma (q,t)
\bar{u}_\gamma (q,t') \gamma_\mu u_\alpha (p,t') \,.
\end{align}
We shall at times refer to the spinors in between the $\gamma$
matrices as the \emph{inner} spinors, and those on the outside as
the \emph{outer} spinors. Following our previous method, the next
step is to change the time variables of integration in order to
expand the difference in terms of $\hbar$, i.e.
\begin{align}
t &= \bar{t}-\f{\hbar}{2}\eta \\
t' &= \bar{t} + \f{\hbar}{2}\eta \,,
\end{align}
with $\theta(t-t') = \theta(-\eta)$ and
\begin{align}
dtdt'=\hbar dtd\eta \,.
\end{align}
The semiclassical scalar component and the action of this variable
change is familiar from the scalar field calculations and we present
a summary here. We have
\beq
\phi_{\bf p}(t) = \sqrt{\frac{p_0}{E_p(t)}}
\exp \left[ - \frac{i}{\hbar}\int_0^t E_p(t')\,dt'\right]\,.
\eeq
The term $\varphi_{\bf p}(t)$, which in (\ref{scalartmodeho})
represented the higher order expansion terms, is not used here as
these terms are contained within the expansion of the spinors. The
product can be written (mixing both variable sets) by
\beq
 \phi^*_p(t)\phi_p(t') = \frac{p_0}
{\sqrt{E_p(t)E_p(t')}}
\exp\left[ -i\int_{-\eta/2}^{\eta/2}
E_p(\bar{t}+\hbar\zeta)\,d\zeta\right]\,.
\eeq
Changing the variables in the prefactor to the exponential, the
product is equal to $|\phi_p(\bar{t})|^2+{\cal O}(\hbar^2)$, which
leads us overall to produce
\begin{align}
& \quad \phi^*_{p}(t) \phi_q(t) \phi^*_q(t')\phi_p(t')
e^{iK(t-t')/\hbar} \nol & = |\phi_p(\bar{t})|^2|\phi_q(\bar{t})|^2
\exp\left[ i\int_{-\eta/2}^{\eta/2} \(-
E_p(\bar{t}+\hbar\zeta)+E_q(\bar{t}+\hbar\zeta)+K\)\,d\zeta\right]\,.
\end{align}
We recall that we shall need to be careful when integrating over
$\eta$ due to the infrared divergence. As before we can use the
variable $\tilde{\beta}$ to replace $\eta$ and aid the integration,
where
\beq\label{defbeta2}
[-E_p(t) + E_q(t) + K]\tilde{\beta}\equiv
\int_{-\eta/2}^{\eta/2}
\left[-E_p(t+\hbar\zeta)
+ E_q(t+\hbar\zeta) + K\right]\,d\zeta\,.
\eeq
We have already studied and confirmed the validity of this
transformation when dealing with the scalar field, and the same
argument can be repeated here briefly to aid recall. For low $K =
\|{\bf p} - {\bf q}\| \to 0$, we have the same equation as before:
\beq\tag{\ref{smallK}}
-E_p(t)+E_q(t) + K \approx
 K - {\bf v}(t)\cdot {\bf K}\,,
\eeq
where ${\bf v}(t) = [{\bf p}-{\bf V}(t)]/E_p(t)$ is the velocity of
the classical particle with final momentum ${\bf p}$
 and thus with ${\bf n} \equiv {\bf K}/K$,
\beq
\tilde{\beta} = \frac{1}{1-{\bf v}(t)\cdot{\bf n}}
\int_{-\eta/2}^{\eta/2} \[1-{\bf v}(t+\hbar\zeta)\cdot{\bf n}
\]d\zeta\,.
\eeq
Therefore, writing $d\eta = J({\bf p},{\bf
q},t,\hbar\tilde{\beta})d\tilde{\beta}$, the function $J({\bf
p},{\bf q},t,\hbar\tilde{\beta})$ is finite as $K\to 0$ and we can
use the variable $\tilde{\beta}$ to replace $\eta$. The expansions
of $\eta$ and $d\eta$ are
\beq
\eta = \left[ 1 - \frac{1}{24}\frac{\(-\ddot{E}_{\bf p}(t)+
\ddot{E}_{\bf q}(t)\)}
{\(-E_p(t)+E_q(t)+K\)}\hbar^2\tilde{\beta}^2 + {\cal
O}(\hbar^4\tilde{\beta}^4)\right]\tilde{\beta} \,,
\label{etatbetaexp}
\eeq
and
\beq
d\eta = \left[ 1 - \frac{1}{8}\frac{\(-\ddot{E}_{\bf p}(t)
+\ddot{E}_{\bf q}(t)\)} {\(-E_p(t)+E_q(t)+K\)}\hbar^2\tilde{\beta}^2
+ {\cal O} (\hbar^4\tilde{\beta}^4)\right] d\tilde{\beta}\,.
\eeq
Integration of $\tilde{\beta}$ will produce further powers of
$\(-E_p(t)+E_q(t)+K\)$ in the denominator, thus producing an
infrared divergences as $\lim_{K\to 0}[-E_p(t)+E_q(t)+K]
\to 0$ as before. Before completing this integration, we must
return to the spinor combination ${\cal S}_-$.

The particle loop spinor combination, we recall, was defined as
\begin{align}
{\cal S}_- = \bar{u}_\alpha (p,t) \gamma^\mu u_\beta (q,t)
\bar{u}^\beta (q,t') \gamma_\mu u_\alpha (p,t') \,.
\end{align}
The spinors here are the time-dependent semiclassical expansion
spinors derived previously Chapter \ref{Semiclassical}, section
\ref{semiclassspinorsection}. The relevant identities and $\hbar$ expansions, including the
change of time variables to $(\bar{t},\eta)$, to enable the
calculation of ${\cal S}_-$ are detailed in the Appendix
\ref{spinoridappendix}, with a summary in section \ref{spinoridsummary}. We quote the necessary results as they are needed
here. Building up the ${\cal S}_-$ combination from the inner
spinors, we have\footnote{Recall that $\tilde{\bf q}={\bf q}-{\bf
V}(t)$ and similarly for $\tilde{\bf p}$.} to ${\cal O}(\hbar)$
\begin{align}
& \quad u_\alpha (q,t) \bar{u}^\alpha (q,t') \nol &= \f{\gamma\cdot
\tilde{q} +m}{2m} +
\( \f{\hbar}{m(2E_q)^2} -
\f{i\hbar\eta}{2m2E_q} \)\gamma^5
\itbd{\gamma}\cdot\tilde{\bf q}\times\dot{\tilde{{\bf q}}} -\( \f{i\hbar}{(2E_q)^2}
+ \f{\hbar\eta}{2E_q.2} \)
\gamma^0 \itbd{\gamma}\cdot\dot{\tilde{{\bf q}}} \,.
\end{align}
Here and later all energies and momenta without explicit time
arguments are evaluated at $\bar{t}$. Sandwiching the expression
immediately above between the contracted gamma matrices, we find
\begin{align}
\gamma^\mu u_\alpha (q,t) \bar{u}^\alpha (q,t')\gamma_\mu &= 2-\f{\gamma\cdot \tilde{q}}{m} +
\( \f{\hbar}{2mE_q^2} -
\f{i\hbar\eta}{2mE_q} \)\gamma^5
\itbd{\gamma}\cdot\tilde{\bf q}\times\dot{\tilde{{\bf q}}} \,.
\end{align}
The particle loop combination can then be written
\begin{align}
{\cal S}_- &= \bar{u}_\alpha (p,t) \gamma^\mu u_\beta (q,t)
\bar{u}^\beta (q,t') \gamma_\mu u_\alpha (p,t') \nol &=
2\[\bar{u}_\alpha (p,t)u_\alpha (p,t') \]
-\f{E_q}{m}\[\bar{u}_\alpha (p,t)\gamma^0 u_\alpha (p,t') \]
+\f{\tilde{\bf q}}{m} \cdot
\[\bar{u}_\alpha (p,t)\itbd{\gamma} u_\alpha (p,t') \] \nol & \quad +
\( \f{\hbar}{2mE_q^2} - \f{i\hbar\eta}{2mE_q}\)
\[ \bar{u}_\alpha (p,t)\gamma^5 \itbd{\gamma}\cdot\tilde{\bf q}\times\dot{\tilde{{\bf q}}}u_\alpha (p,t')
\] \,.
\end{align}
Substituting into this expression the relevant spinor identities, we
have
\begin{align}
{\cal S}_- &= 2\[1-\(
\f{\hbar}{2mE_p}+\f{i\hbar\eta}{2m}\)\f{\itbd{\xi}\cdot\tilde{\bf
p}\times\dot{\tilde{{\bf p}}}}{E_p} \]
 -\f{E_q}{m}\[ \f{E_p}{m} \]
\nol & \quad +\f{\tilde{\bf q}}{m} \cdot
\[\f{\tilde{\bf p}}{m} - \( \f{\hbar}{2mE_p}+\f{i\hbar\eta}{2m}\)
\( \f{\itbd{\xi}\cdot\tilde{\bf p}\times\dot{\tilde{{\bf p}}}\,\tilde{\bf p}}{E_p(E_p+m)} +\itbd{\xi}\times
\(\dot{\tilde{{\bf p}}}-\f{\dot{E}_p\tilde{\bf p}}{E_p+m} \) \)  \]
\nol & \quad -
\( \f{\hbar}{2mE_q^2} - \f{i\hbar\eta}{2mE_q}\)
\[ \itbd{\xi}\cdot\tilde{\bf q}\times\dot{\tilde{{\bf q}}}
+ \f{\tilde{\bf p}\cdot\tilde{\bf q}\times\dot{\tilde{{\bf
q}}}\,\itbd{\xi}\cdot\tilde{\bf p}}{m(E_p+m)} \] \nol &= 2 -
\f{\tilde{q}\cdot \tilde{p}}{m^2} -\(
\f{\hbar}{mE_p}+\f{i\hbar\eta}{m}\)\f{\itbd{\xi}\cdot\tilde{\bf
p}\times\dot{\tilde{{\bf p}}}}{E_p}
\nol & \quad - \( \f{\hbar}{2m^2E_p}+\f{i\hbar\eta}{2m^2}\)
\( \f{\itbd{\xi}\cdot\tilde{\bf p}\times\dot{\tilde{{\bf p}}}\,\tilde{\bf p}\cdot\tilde{\bf q}}{E_p(E_p+m)} +\tilde{\bf q}\cdot\itbd{\xi}\times
\(\dot{\tilde{{\bf p}}}-\f{\dot{E}_p\tilde{\bf p}}{E_p+m} \) \)
\nol & \quad -
\( \f{\hbar}{2mE_q^2} - \f{i\hbar\eta}{2mE_q}\)
\[ \itbd{\xi}\cdot\tilde{\bf q}\times\dot{\tilde{{\bf q}}}
+ \f{\tilde{\bf p}\cdot\tilde{\bf q}\times\dot{\tilde{{\bf
q}}}\,\itbd{\xi}\cdot\tilde{\bf p}}{m(E_p+m)}
\] \,.
\end{align}
Changing the variables\footnote{$(t,t')\to(\bar{t},\eta)$ variables
and also changing to ${\bf K}=\hbar{\bf k}$,} and adding the scalar
part of the expansion, we can write the contribution to the forward
scattering as
\begin{align}
{\cal F}_-({\bf p}) &= -\f{ie^2m}{\hbar^2 p_0}
 \int \f{d^3{\bf K}}{2K(2\pi)^3}
\f{d^3{\bf q}m}{(2\pi\hbar)^3q_0} d\bar{t} d\eta \theta(-\eta) (2\pi\hbar)^3 \delta^3({\bf p}-{\bf q}-{\bf K})
 \nol &
 \times \[{\cal S}_-\] |\phi_p(\bar{t})|^2|\phi_q(\bar{t})|^2
\exp\left[ i\int_{-\eta/2}^{\eta/2} \(-
E_p(\bar{t}+\hbar\zeta)+E_q(\bar{t}+\hbar\zeta)+K\)\,d\zeta\right]
\,.
\end{align}

The factor ${\cal S}_-$ is present in the integrand subject to the
delta function $\delta({\bf p}-{\bf q}-{\bf K})$ and the integrals
over $d^3{\bf K}$ and $d^3{\bf q}$. Integrating out this delta
function using the ${\bf q}$ integral, we can replace $\tilde{\bf
q}$ by $\tilde{\bf q} \to \tilde{\bf p}-{\bf K}$. We also have
$\dot{\tilde{{\bf q}}}=\dot{\tilde{{\bf p}}}$, as both are in fact
equal to $-\dot{\bf V}$. The energy $E_q$ can then be regarded as
defined by $E_q=\sqrt{(\tilde{\bf p}-{\bf K})^2+m^2}$.  Within the
resulting $K$ integral, the only preferred direction about which to
choose an axis is that given by $\tilde{\bf p}$. We may thus replace
${\bf K}$ by ${\bf K}
\to \({\bf K}\cdot\tilde{\bf p}/\tilde{\bf p}^2\)\tilde{\bf p}$. Combining all of the above statements, we may
produce the `effective version'\footnote{The effective version under
the delta function.} of ${\cal S}_-$ under the integrations via the
transformation
\begin{align}
\tilde{\bf q} \to \tilde{\bf p} \(1-\f{{\bf K}\cdot\tilde{\bf p}}{\tilde{\bf p}^2}
\) \,.
\end{align}
This produces
\begin{align}
{\cal S}_-^{\rm eff} &= 2 - \f{E_q E_p}{m^2} + \f{\tilde{\bf
p}\cdot\tilde{\bf p}}{m^2}
\(1-\f{{\bf K}\cdot\tilde{\bf p}}{\tilde{\bf p}^2} \) -\(
\f{\hbar}{mE_p}+\f{i\hbar\eta}{m}\)\f{\itbd{\xi}\cdot\tilde{\bf p}\times\dot{\tilde{{\bf p}}}}{E_p}
 \nol & \quad - \( \f{\hbar}{2m^2E_p}+\f{i\hbar\eta}{2m^2}\) \( \f{\itbd{\xi}\cdot\tilde{\bf p}\times\dot{\tilde{{\bf p}}}\,\tilde{\bf p}^2}{E_p(E_p+m)} +\tilde{\bf p}\cdot\itbd{\xi}\times
\(\dot{\tilde{{\bf p}}}-\f{\dot{E}_p\tilde{\bf p}}{E_p+m} \) \)\(1-\f{{\bf K}\cdot\tilde{\bf p}}{\tilde{\bf p}^2} \)
\nol & \quad -
\( \f{\hbar}{2mE_q^2} - \f{i\hbar\eta}{2mE_q}\)
\[ \itbd{\xi}\cdot\tilde{\bf p}\times\dot{\tilde{{\bf p}}} + \f{\tilde{\bf p}\cdot\tilde{\bf p}\times\dot{\tilde{{\bf q}}}\,\itbd{\xi}\cdot\tilde{\bf p}}{m(E_p+m)} \]\(1-\f{{\bf
K}\cdot\tilde{\bf p}}{\tilde{\bf p}^2} \) \nol &= 2 - \f{E_q
E_p}{m^2} + \f{\tilde{\bf p}\cdot\tilde{\bf p}}{m^2}
\(1-\f{{\bf K}\cdot\tilde{\bf p}}{\tilde{\bf p}^2} \) -\(
\f{\hbar}{mE_p}+\f{i\hbar\eta}{m}\)\f{\itbd{\xi}\cdot\tilde{\bf p}\times\dot{\tilde{{\bf p}}}}{E_p}
 \nol & \quad - \( \f{\hbar}{2m^2E_p}+\f{i\hbar\eta}{2m^2}\)  \itbd{\xi}\cdot\tilde{\bf p}\times\dot{\tilde{{\bf p}}}\(\f{-m}{E_p}\)
 \(1-\f{{\bf K}\cdot\tilde{\bf p}}{\tilde{\bf p}^2} \)
\nol & \quad -
\( \f{\hbar}{2mE_q^2} - \f{i\hbar\eta}{2mE_q}\)
\itbd{\xi}\cdot\tilde{\bf p}\times\dot{\tilde{{\bf p}}}\(1-\f{{\bf K}\cdot\tilde{\bf p}}{\tilde{\bf p}^2}
\) \,.
\end{align}
Hence
\begin{align}
{\cal S}_-^{\rm eff} &= 2 - \f{E_q E_p}{m^2} + \f{\tilde{\bf
p}\cdot\tilde{\bf p}}{m^2}
 -\f{\hbar}{2m}\itbd{\xi}\cdot\tilde{\bf p}\times\dot{\tilde{{\bf p}}} \(
\f{1}{E_p^2}+\f{1}{E_q^2} \)
\nol & \quad -\f{i\hbar\eta}{2m} \itbd{\xi}\cdot\tilde{\bf
p}\times\dot{\tilde{{\bf p}}} \(
\f{1}{E_p}-\f{1}{E_q}\) \nol & \quad - \f{{\bf K}\cdot\tilde{\bf p}}{\tilde{\bf p}^2} \[ \f{\tilde{\bf p}^2}{m^2} + \itbd{\xi}\cdot\tilde{\bf p}\times\dot{\tilde{{\bf p}}} \( \f{\hbar}{2m} \( \f{1}{E_p^2}-\f{1}{E_q^2} \) +
\f{i\hbar\eta}{2m} \( \f{1}{E_p} + \f{1}{E_q} \) \) \] \,.
\end{align}
We change variables to $\tilde{\beta}$ instead of $\eta$ and rewrite
the notation $\bar{t}$ as $t$. Writing ${\cal S}_-^{\rm
eff}(t,\tilde{\beta})$ to indicate the new time variable and
notation, we now find the forward scattering contribution as
\begin{align}
{\cal F}_-({\bf p}) &= -\f{ie^2m^2}{\hbar^2 p_0}
 \int \f{d^3{\bf K}}{(2\pi)^32Kq_0}
 dt d\tilde{\beta} J({\bf p},{\bf
q},t,\hbar\tilde{\beta})
\theta(-\tilde{\beta})|\phi_p(t)|^2|\phi_q(t)|^2
 \nol & \quad
 \times {\cal S}_-^{\rm eff}(t,\tilde{\beta})
\exp\left[ i\(-E_p+E_q+K\)\tilde{\beta} \right] \,.
\end{align}
Within the integrand we have terms of order $\hbar^0$, $\hbar$ and
higher order terms resulting from the $\hbar$-expansion of $\eta\to
\tilde{\beta}$ in ${\cal S}_-^{\rm eff}(t,\tilde{\beta})$ and in
$J({\bf p},{\bf q},t,\hbar\tilde{\beta})$. Here we again note that
the higher order terms in $\hbar$ ($\hbar^2$ and above) can no
longer be assumed to be zero in the classical limit due to the
infrared divergence. Now, note that every occurrence of
$\tilde{\beta}$ is of the form $\hbar\tilde{\beta}$. Therefore all
higher order terms are of the form $\hbar^n\tilde{\beta}^d f_{nd}$
where $f_{nd}$ is some function (for each choice of $n$ and $d$) and
$n\geq 1$, $n\geq d
\geq 0$. To complete the integration over $\tilde{\beta}$ we must use a Wick
rotation i.e. replace $K$ by $K-i\epsilon$. We shall then use the
integral
\begin{equation}\label{betaint}
\lim_{\epsilon\to 0} \int^0_{-\infty} \tilde{\beta}^d
e^{iX\tilde{\beta}}e^{\epsilon\tilde{\beta}}d\tilde{\beta} = \f{d!
(i)^{d-1}}{X^{d+1}} \,,
\end{equation}
to integrate over $\tilde{\beta}$. Further noting that we can write
$|\phi_q(t)|^2=q_0/E_q + {\cal O}(\hbar^2)$, the result can be
written as
\begin{multline} \label{Fwithfs}
{\cal F}_-({\bf p}) = -\f{ie^2m^2}{\hbar^2 p_0}
 \int \f{d^3{\bf K}}{(2\pi)^32KE_q}
 dt |\phi_p(t)|^2
 \\ \times \left[
 \f{(-i)\(f_{00}+\hbar f_{10}\)}{\(-E_p+E_q+K\)} + \f{\hbar f_{11}}{\(-E_p+E_q+K\)^2}
 +\sum_{\stackrel{n\geq 2}{n\geq d \geq 0}} \f{\hbar^n f_{nd} d! i^{d-1}}{\(-E_p(t)+E_q(t)+K\)^{d+1}}
 \right] \,,
\end{multline}
where we have written the first three terms explicitly, outside of
the summation sign $\sum$. From ${\cal S}_-^{\rm eff}$ we can give
these three terms as
\begin{align} \label{f00}
f_{00} &= 2 - \f{E_qE_p}{m^2} + \f{\tilde{\bf p}\cdot\tilde{\bf
p}}{m^2}\(1 - \f{{\bf K}\cdot\tilde{\bf p}}{\tilde{\bf p}^2}\)\,, \\
f_{10} &= \f{1}{2m}\itbd{\xi}\cdot\tilde{\bf p}
\times\dot{\tilde{{\bf p}}} \[ -\( \f{1}{E_p^2}+\f{1}{E_q^2} \)
- \f{{\bf K}\cdot\tilde{\bf p}}{\tilde{\bf p}^2} \(
\f{1}{E_p^2}-\f{1}{E_q^2} \) \] \,,\label{f01}\\
f_{11} &= \f{i}{2m}
\itbd{\xi}\cdot\tilde{\bf p} \times\dot{\tilde{{\bf p}}}
\[ -\( \f{1}{E_p}-\f{1}{E_q} \)- \f{{\bf K}\cdot\tilde{\bf p}}{\tilde{\bf p}^2} \(
\f{1}{E_p}+\f{1}{E_q} \) \] \,. \label{f11}
\end{align}
Here we can see the infrared divergences in the higher order terms.
The remaining terms of the summation, i.e. with $n\geq 2$, have the
same form as the higher order terms dealt with in the scalar field
calculations. Thus we introduce a cut-off and integrate above
$K_0=\hbar^\alpha\lambda$ with $3/4<\alpha<1$ (for later reasons)
and $\lambda$ a positive constant. The small-$K$ contributions
behave like
\begin{align}
\begin{cases} \hbar^nK_0^{1-d} = \hbar^{n+(1-d)\alpha}\lambda^{1-d}
& \text{for $d\geq 2$} \\ \hbar \log(\hbar\alpha) & \text{for the
$d=1$ term}
\end{cases} \,.
\end{align}
Given the limits on $n$, $d$ and $\alpha$, we again have
$n+(1-d)\alpha\geq 2-\alpha$. Considering the additional
$\hbar^{-2}$ multiplying the integral (after converting $k$ to $K$)
and the $\hbar$ multiplying the whole forward scattering, we find
that these higher order terms do not contribute in the $\hbar\to 0$
limit to the position shift. Above the cut-off this leaves the
lowest order term and the first order correction to be considered.
We must still analyse the contributions below the cut-off, however.
Fortunately, this calculation is again analogous to what we
previously encountered. We can in fact show that the real
contribution to ${\cal F}_-$ below that cut-off comes entirely from
the leading order term to order $\hbar^{-1}$~\footnote{That is order
$\hbar^{-1}$ for ${\cal F}_-$ which one recalls has a prefactor of
$1/\hbar^2$ multiplying the integrand and is additionally multiplied
by a further $\hbar$ when producing the position shift
contribution.}, which in our current notation is the $f_{00}$ term
in (\ref{Fwithfs}). The remaining contribution to ${\cal F}_-$ at
this order is again imaginary. We demonstrate this as previously, by
calculating the low-$K$ contribution of the full ${\cal F}_-$ and
comparing it with the leading order contribution. We additionally
recall that it is in the classical limits with which all our results
are phrased.

Firstly, below the cut-off ${\cal F}_-$ can be written
\begin{align}
{\cal F}_-^< =& -\f{ie^2}{\hbar}  \int dtdt'
\int_{k<\hbar^{\alpha-1}\lambda}\f{d^3k}{2k(2\pi)^3}
\f{d^3q}{(2\pi\hbar)^3}\f{m^2}{q_0p_0} \theta(t-t')(2\pi\hbar)^3 \delta^3 ({\bf p}-{\bf q}-{\bf K})
 \nol &
\times \[
\bar{u}_\alpha(p,t)\gamma^\mu u^\gamma(q,t) \bar{u}_\gamma (q,t')\gamma_\mu
u_\alpha(p,t') \] \phi^*_p(t)\phi_q(t)\phi^*_q(t')\phi_p(t')
e^{-K(t-t')/\hbar} \,.
\end{align}
For small $k$ we have
\begin{align} \label{trunc2}
& \exp\(i\int^t_0 \(K+E_q(\zeta)-E_p(\zeta)\)d\zeta/\hbar\) \nol =&
\exp \( i\int^t_0 \( k - \f{\partial E_p(\zeta)}{\partial{\bf
p}}\cdot{\bf k} + \f{1}{2} \f{\partial^2 E_p(\zeta)}{\partial p^i
\partial p^j}
\hbar k^i k^j +\ldots \) d\zeta \) \nol =& \exp \(ikt-\int^t_0 \f{d{\bf
x}}{d\zeta}\cdot{\bf k} d\zeta \) \,,
\end{align}
where the series in the exponential has been truncated at the second
term due to the condition $\alpha>3/4$ (see the equivalent section
of the scalar calculations, (\ref{truncate1})). Thus ${\cal F}_-^<$
becomes for small $k$ (with $q\to p$)
\begin{align}
{\cal F}_-^< &= -\f{ie^2}{\hbar}  \int dtdt'
\int_{k<\hbar^{\alpha-1}\lambda}\f{d^3k}{2k(2\pi)^3}
\f{m^2}{p_0^2} \theta(t-t') \f{p_0}{E_p(t)}\f{p_0}{E_p(t')}
 \nol &
\qquad \times \[
\bar{u}_\alpha(p,t)\gamma^\mu u^\gamma(p,t) \bar{u}_\gamma (p,t')\gamma_\mu
u_\alpha(p,t') \]
\exp\(ik(t'-t)-{\bf k}\cdot\({\bf x'}-{\bf x}\)\) \,.
\end{align}
Let us denote the spinor combination in the square brackets above by
\begin{align}
{\cal S}^< (t,t')= \bar{u}_\alpha(p,t)\gamma^\mu u^\gamma(p,t)
\bar{u}_\gamma (p,t')\gamma_\mu u_\alpha(p,t') \,.
\end{align}
In line with (\ref{trunc2}), we consider only the lowest order
terms, i.e. $\hbar^0$ terms, for this spinor combination. At the end
of the calculation, we shall note that higher order terms, of order
$\hbar$ and above, do not contribute here. Firstly, we have the
factor $\bar{u}_\alpha(p,t)\gamma^\mu u_\gamma(p,t)$. From the
derivations for the equal time spinor identities in Appendix
\ref{spinoridappendix}\footnote{The particle equal time spinor
identities are in section
\ref{partequaltimesection} of Appendix
\ref{spinoridappendix}.}, we
write
\begin{align}
\bar{u}_\alpha (p)\gamma^0 u_\gamma (p)
&= \f{E_p}{m}\delta_{\alpha\gamma} +{\cal O}(\hbar)\,,
\end{align}
and
\begin{align}
\bar{u}_\alpha (p)\itbd{\gamma} u_\alpha (p)
&= \f{\tilde{\bf p}}{m}\delta_{\alpha\gamma}+{\cal O}(\hbar) \,.
\end{align}
The components of the other factor, $\bar{u}_\gamma (p,t')\gamma_\mu
u_\alpha(p,t')$, can then be obtained by reversing the spin indices
and using the notation $\tilde{\bf p}'={\bf p}-{\bf V}(t')$,
$E_p'=\sqrt{\tilde{\bf p}'^2+m^2}$. Combining the identities and
summing over the index $\gamma$, we obtain
\begin{align}
{\cal S}^< &= \f{E_pE_p'}{m^2} -
 \f{\tilde{\bf p}\cdot\tilde{\bf p}'}{m^2} +{\cal O}(\hbar) \,.
 \end{align}

We note straight away that the order $\hbar^0$ term here gives the
same contribution as the scalar case: With
$\tilde{p}(t)=(E_q(t),\tilde{\bf q}(t))$, we have
\begin{align}
& \quad \Re {\cal F}_-^<({\bf p})\nol  &=-\f{ie^2}{\hbar}  \int
dtdt'
\int_{k<\hbar^{\alpha-1}\lambda}\f{d^3k}{2k(2\pi)^3}
\f{m^2}{E_p(t)E_p(t')} \theta(t-t')
\[
\f{\tilde{p}^\mu(t)}{m}\cdot\f{\tilde{p}_\mu(t')}{m}  \] e^{ik\cdot(x'-x)}
\nol
&= -\f{ie^2}{\hbar}  \int dtdt'
\int_{k<\hbar^{\alpha-1}\lambda}\f{d^3k}{2k(2\pi)^3}
 \theta(t-t')
\f{dx^\mu}{dt}\f{dx_\mu}{dt'} e^{ik\cdot(x'-x)}
\nol
&=  -\f{ie^2}{\hbar}  \int d\xi d\xi'
\int_{k<\hbar^{\alpha-1}\lambda}\f{d^3k}{2k(2\pi)^3}
 \theta(\xi-\xi')
\f{dx^\mu}{d\xi}\f{dx_\mu}{d\xi'}e^{ik(\xi'-\xi)} \,.
\end{align}
and thus
\begin{align}
\Re {\cal F}_-^<({\bf p}) &=
-\f{e^2\lambda}{16\pi^3\hbar^{2-\alpha}}
\int  d\Omega dt
\f{1-{\bf v}^2}{1-{\bf n}\cdot{\bf v}}  \,,
\end{align}
We can thus see that any terms of order $\hbar$ from the spinor
combination would produce a contribution to the position shift of
order $\hbar^\alpha$ and thus would not contribute here, in the
$\hbar\to 0$ limit, as stated before.

We now show that the above expression can be arrived at from the
leading order term of ${\cal F}_-$ in (\ref{Fwithfs}). Consequently,
the higher order $f_{nd}$ terms (which we recall do not contribute
above the cut-off) do not contribute to the real part of ${\cal
F}_-$ to order $\hbar^{-1}$. The leading order term, below the
cut-off, we denote ${\cal F}_-^{<,0}$. From (\ref{Fwithfs}) we can
write
\begin{equation} \label{Fwithfs3}
{\cal F}_-^{<,0,1}({\bf p}) = -\f{e^2}{\hbar^2}
 \int_{K\leq\hbar^\alpha \lambda} \f{d^3{\bf K}}{(2\pi)^32K}
 dt \f{m^2}{E_pE_q}
 \f{f_{00}}{\(-E_p+E_q+K\)}  \,.
\end{equation}
In the $K\to 0$ limit, with $q\to p$, we have
\begin{align} \label{f00lowk}
f_{00} &\to 2 - \f{E_p^2}{m^2} + \f{\tilde{\bf p}\cdot\tilde{\bf
p}}{m^2} \,.
\end{align}
thus
\begin{multline} \label{Fwithfs4}
{\cal F}_-^{<,0}({\bf p}) = -\f{e^2}{\hbar^2}
 \int_{K\leq\hbar^\alpha \lambda} \f{d^3{\bf K}}{(2\pi)^32K}
 dt \f{m^2}{E_p^2}
 \\ \times \f{1}{\(-E_p+E_q+K\)}\[ 2 -
 \f{\tilde{p}\cdot\tilde{p}}{m^2}  \] \,.
\end{multline}
In the small $K$ limit we have
\beq
-E_p+E_q+K\approx K-{\bf v}(t)\cdot {\bf K} \,,
\eeq
where ${\bf v}(t) = \tilde{\bf p}/E_p = \[{\bf p}-{\bf V}(t)\]/E_p$
is the velocity of a particle with final momentum $p$. Thus
splitting the ${\bf K}$ integral into spherical polars and
integrating $K$ gives
\begin{align}
{\cal F}_-^{<,0}({\bf p}) &= -\f{e^2}{2(2\pi)^3\hbar^2}
\int_{K\leq\hbar^\alpha \lambda}
 \f{dKd\Omega K^2}{K} dt
\f{1-{\bf v}^2}{K\(1-{\bf n}\cdot{\bf v}\)}\nol &=
-\f{e^2\lambda}{16\pi^3\hbar^{2-\alpha}} \int dt d\Omega
\f{1-{\bf v}^2}{\(1-{\bf n}\cdot{\bf v}\)} \,,
\end{align}
as required.

\subsection{Antiparticle Loop}
The antiparticle loop contribution to the forward scattering is
represented by ${\cal F}_+$ in (\ref{Fplus}):
\begin{multline} \tag{\ref{Fplus}}
{\cal F}_+({\bf p})  = \f{ie^2m}{\hbar p_0}
\int \f{d^3{\bf p'}}{(2\pi\hbar)^3} \f{d^3{\bf k}}{2k(2\pi)^3}
\f{d^3{\bf q}m}{(2\pi\hbar)^3q_0} d^4xd^4x' \theta(t-t')
 \\
 \times \bar{\Phi}_\alpha({\bf p'},x') \gamma^\nu \Psi_\beta({\bf q},x') \bar{\Psi}^\beta({\bf q},x)\gamma^\mu \Phi_\alpha({\bf p},x)
 e^{-ik\cdot(x-x')} \,.
\end{multline}
The spinor and electromagnetic mode functions can be expanded to
give
\begin{align}
& \quad \bar{\Phi}_\alpha({\bf p'},x') \gamma^\mu \Psi_\beta({\bf
q},x') \bar{\Psi}^\beta({\bf q},x)\gamma_\mu \Phi_\alpha({\bf p},x)
 e^{-ik\cdot(x-x')} \nol
 &= \[ \bar{u}_\alpha(p',t') \gamma^\mu v_\beta(q,t')
 \bar{v}^\beta(q,t) \gamma_\mu u_\alpha(p,t) \]  \nol & \qquad \times \phi^*_{p'}(t')
 \bar{\phi}^*_q (t') \bar{\phi}_q(t) \phi_p(t)
e^{-i{\bf p'}\cdot{\bf x'}}e^{-i{\bf q}\cdot{\bf x'}} e^{i{\bf
q}\cdot{\bf x}}e^{i{\bf p}\cdot{\bf x}} e^{-ik\cdot(x-x')} \,.
\end{align}
The spatial integrals produce the delta functions
$(2\pi\hbar)^6\delta^3({\bf p}+{\bf q}+{\bf K})\delta^3({\bf p}-{\bf
p'})$. Integrating over ${\bf p'}$ we write the antiparticle loop
contribution in analogy to the particle loop as
\begin{align}
{\cal F}_+({\bf p})  &= \f{ie^2m}{\hbar p_0}
\int \f{d^3{\bf p'}}{(2\pi\hbar)^3} \f{d^3{\bf k}}{2k(2\pi)^3}
\f{d^3{\bf q}m}{(2\pi\hbar)^3q_0} d^4xd^4x' \theta(t-t')
 \nol & \qquad
 \times \[{\cal S}_+ \] \phi^*_{p'}(t')
 \bar{\phi}^*_q (t') \bar{\phi}_q(t) \phi_p(t)
 e^{-ik(t-t')} \,,
\end{align}
where the antiparticle loop semiclassical spinor combination is
defined as\footnote{We also use analogous terminology to that for
${\cal S}_-$ to refer to the inner and outer spinors. In this case,
the inner spinors are the anti-particle spinors.}
\begin{equation}
{\cal S}_+ = \bar{u}_\alpha(p',t') \gamma^\mu v_\beta(q,t')
 \bar{v}^\beta(q,t) \gamma_\mu u_\alpha(p,t) \,.
 \end{equation}
As we did for the particle loop, we use the spinor identities from
the appendix to obtain an expression for ${\cal S}_+$. We shall also
use the same time variable transformation
\begin{align}
t &= \bar{t}-\f{\hbar}{2}\eta \\
t' &= \bar{t} + \f{\hbar}{2}\eta \,,
\end{align}
and expand in terms of $\hbar$. It is important to note the two
differences between ${\cal S}_+$ and ${\cal S}_-$. Firstly, and most
obviously, is the presence of the antiparticle (negative energy)
spinors as the inner spinors forming the loop part. Further to this
we note that the order of the time variables is reversed. The time
order reversal is, in terms of the $(\bar{t},\eta)$ variables,
simply the transformation $\eta\to -\eta$. Consequently, from the
inner spinors we obtain, to order $\hbar$
\begin{align}
v_\alpha (q,t') \bar{v}^\alpha (q,t) &= \f{\gamma\cdot \tilde{q}_+
-m}{2m} -
\(\f{\hbar}{4mE_{q_+}^2}-\f{i\hbar\eta}{4mE_{q_+}}\) \gamma^5
\itbd{\gamma}\cdot\tilde{\bf q}_+\times\dot{\tilde{{\bf q}}}_+ \nol & \quad -\(\f{i\hbar}{4E_{q_+}^2}
 + \f{\hbar\eta}{4E_{q_+}} \)
\gamma^0 \itbd{\gamma}\cdot\dot{\tilde{{\bf q}}}_+ \,,
\end{align}
where we recall the definitions $\tilde{\bf q}_+={\bf q}+{\bf V}(t)$
and $E_{q_+}=\sqrt{\tilde{\bf q}_+^2+m^2}$. Thus with the contracted
gamma matrices,
\begin{align}
\gamma^\mu v_\alpha (q,t') \bar{v}^\alpha (q,t) \gamma_\mu &= -2 -\f{\gamma\cdot \tilde{q}_+}{m} -
\(\f{\hbar}{2mE_{q_+}^2}-\f{i\hbar\eta}{2mE_{q_+}}\) \gamma^5
\itbd{\gamma}\cdot\tilde{\bf q}_+\times\dot{\tilde{{\bf q}}}_+ \,.
\end{align}
The antiparticle loop contribution is thus written
\begin{align}
{\cal S}_+ &= \bar{u}_\alpha (p,t') \gamma^\mu v_\beta (q,t')
\bar{v}^\beta (q,t) \gamma_\mu u_\alpha (p,t) \nol &=
-2\[\bar{u}_\alpha (p,t')u_\alpha (p,t) \]
-\f{E_{q_+}}{m}\[\bar{u}_\alpha (p,t')\gamma^0 u_\alpha (p,t) \]
+\f{\tilde{\bf q}_+}{m} \cdot
\[\bar{u}_\alpha (p,t')\itbd{\gamma} u_\alpha (p,t) \] \nol & \quad -
\( \f{\hbar}{2mE_{q_+}^2} - \f{i\hbar\eta}{2mE_{q_+}}\)
\[ \bar{u}_\alpha (p,t')\gamma^5 \itbd{\gamma}\cdot\tilde{\bf q}_+\times\dot{\tilde{{\bf q}}}_+u_\alpha (p,t)
\] \,.
\end{align}
Substituting in the outer spinor identities (being careful with the
$\eta$ signs), we have
\begin{align}
{\cal S}_+ &= -2\[1-\( \f{\hbar}{2mE_p}-\f{i\hbar\eta}{2m}\)
\f{\itbd{\xi}\cdot\tilde{\bf p}\times\dot{\tilde{{\bf p}}}}{E_p} \]
 -\f{E_{q_+}}{m}\[ \f{E_p}{m} \]
\nol & \quad +\f{\tilde{\bf q}_+}{m} \cdot
\[\f{\tilde{\bf p}}{m} - \( \f{\hbar}{2mE_p}-\f{i\hbar\eta}{2m}\)
\( \f{\itbd{\xi}\cdot\tilde{\bf p}\times\dot{\tilde{{\bf p}}}\,\tilde{\bf p}}{E_p(E_p+m)} +\itbd{\xi}\times
\(\dot{\tilde{{\bf p}}}-\f{\dot{E}_p\tilde{\bf p}}{E_p+m} \) \)  \]
\nol & \quad +
\( \f{\hbar}{2mE_{q_+}^2} - \f{i\hbar\eta}{2mE_{q_+}}\)
\[ \itbd{\xi}\cdot\tilde{\bf q}_+\times\dot{\tilde{{\bf q}}}_+
+ \f{\tilde{\bf p}\cdot\tilde{\bf q}_+\times\dot{\tilde{{\bf
q}}}_+\,\itbd{\xi}\cdot\tilde{\bf p}}{m(E_p+m)} \] \nol &= -2 -
\f{\tilde{q}_+\cdot \tilde{p}}{m^2} +\(
\f{\hbar}{mE_p}+\f{i\hbar\eta}{m}\)\f{\itbd{\xi}\cdot\tilde{\bf
p}\times\dot{\tilde{{\bf p}}}}{E_p}
 \nol & \quad- \( \f{\hbar}{2m^2E_p}-\f{i\hbar\eta}{2m^2}\)
 \( \f{\itbd{\xi}\cdot\tilde{\bf p}\times\dot{\tilde{{\bf p}}}\,
 \tilde{\bf p}\cdot\tilde{\bf q}_+}{E_p(E_p+m)} +\tilde{\bf q}_+\cdot\itbd{\xi}\times
\(\dot{\tilde{{\bf p}}}-\f{\dot{E}_p\tilde{\bf p}}{E_p+m} \) \)
\nol & \quad +
\( \f{\hbar}{2mE_{q_+}^2} - \f{i\hbar\eta}{2mE_{q_+}}\)
\[ \itbd{\xi}\cdot\tilde{\bf q}_+\times\dot{\tilde{{\bf q}}}_+
+ \f{\tilde{\bf p}\cdot\tilde{\bf q}_+\times\dot{\tilde{{\bf
q}}}_+\,\itbd{\xi}\cdot\tilde{\bf p}}{m(E_p+m)}
\] \,.
\end{align}

The integrand of ${\cal F}_+$ is under both $K$ and $q$ integrals,
but this time the delta function present is $\delta({\bf p}+{\bf
q}+{\bf K})$. We thus change the variables of integration via
\begin{align}
{\bf q}&\to -{\bf q} \\ {\bf K}& \to-{\bf K} \,.
\end{align}
We then have the following relations:
\begin{align}
d^3{\bf q}&\to d^3{\bf q} & d^3{\bf K} &\to d^3{\bf K} \nol
\tilde{\bf q}_+ &\to -\tilde{\bf q} & \dot{\tilde{{\bf q}}}_+ &=
-\dot{\tilde{{\bf q}}} \nol E_{q_+} &\to E_q & K &\to K \,,
\end{align}
along with the new delta function $\delta^3({\bf p}-{\bf q}-{\bf
K})$, which is now the same as for ${\cal S}_-$. Labelling ${\cal
S}_+$ under this transformation by ${\cal S}_+'$, we have
\begin{align}
{\cal S}_+' &= -2 - \f{E_qE_p}{m^2}-\f{\tilde{\bf q}\cdot\tilde{\bf
p}}{m^2} +\( \f{\hbar}{mE_p}+\f{i\hbar\eta}{m}\)
\f{\itbd{\xi}\cdot\tilde{\bf p}\times\dot{\tilde{{\bf p}}}}{E_p}
\nol & \quad + \( \f{\hbar}{2m^2E_p}-\f{i\hbar\eta}{2m^2}\)
 \( \f{\itbd{\xi}\cdot\tilde{\bf p}\times\dot{\tilde{{\bf p}}}\,\tilde{\bf p}\cdot\tilde{\bf q}}{E_p(E_p+m)}
  +\tilde{\bf q}\cdot\itbd{\xi}\times
\(\dot{\tilde{{\bf p}}}-\f{\dot{E}_p\tilde{\bf p}}{E_p+m} \) \)
\nol & \quad +
\( \f{\hbar}{2mE_{q}^2} - \f{i\hbar\eta}{2mE_{q}}\)
\[ \itbd{\xi}\cdot\tilde{\bf q}\times\dot{\tilde{{\bf q}}} +
\f{\tilde{\bf p}\cdot\tilde{\bf q}\times\dot{\tilde{{\bf q}}}\,\itbd{\xi}\cdot\tilde{\bf p}}{m(E_p+m)}
\] \,.
\end{align}
For scalar semiclassical terms, we note
$\bar{\phi}_{-q}(t)=\phi_q(t)$. Hence
\begin{align}
\phi^*_{p'}(t')
 \bar{\phi}^*_q (t') \bar{\phi}_q(t) \phi_p(t) \to
 \phi^*_{p'}(t')
 \phi^*_q (t') \phi_q(t) \phi_p(t) \,.
\end{align}
For this scalar combination, we recall that
\beq
\phi_{\bf p}(t) = \sqrt{\frac{p_0}{E_p(t)}}
\exp \left[ - \frac{i}{\hbar}\int_0^t E_p(t')\,dt'\right]\,,
\eeq
and so we can write
\begin{multline}
\phi^*_{p'}(t')  \phi^*_q (t') \phi_q(t) \phi_p(t)
e^{iK(t-t')/\hbar} \\ = |\phi_p(\bar{t})|^2|\phi_q(\bar{t})|^2
\exp\left[ i\int_{-\eta/2}^{\eta/2} \(
E_p(\bar{t}+\hbar\zeta)+E_q(\bar{t}+\hbar\zeta)+K\)\,d\zeta\right]
\,.
\end{multline}
Again, we can change the variable to help with the integration over
$\eta$ and define
\beq
\(
E_p(t)+E_q(t)+K\)
\beta =\int_{-\eta/2}^{\eta/2} \(
E_p(t+\hbar\zeta)+E_q(t+\hbar\zeta)+K\)\,d\zeta \,.
\eeq
As we recall from the scalar antiparticle loop, which used the same
variable change at this point, there is no infrared divergence
problem resulting from the $\beta$ integration, so the previous
manipulations, including consideration of the higher order terms,
are not needed here. The equations for $\eta$ and $d\eta$ were given
in the scalar work as
\beq
\eta = \left[ 1 - \frac{1}{24}\frac{ \ddot{E}_p(t)+
 \ddot{E}_q(t)}
{ E_p(t)+ E_q(t)+K}\hbar^2\beta^2 + {\cal
O}(\hbar^4\beta^4)\right]\beta  \,, \tag{\ref{etabetaexp}}
\eeq
and
\beq \tag{\ref{detabetaexp}}
d\eta = \left[ 1 - \frac{1}{8}\frac{ \ddot{E}_p(t) +
\ddot{E}_q(t)} { E_p(t)+ E_q(t)+K}\hbar^2\beta^2 + {\cal O}
(\hbar^4\beta^4)\right] d\beta \,.
\eeq
In this case we can ignore the higher order $\hbar^2$ terms.

Using the above, and changing notation $\bar{t}\to t$, we write
${\cal F}_+$ as
\begin{align}
{\cal F}_+({\bf p})  &= \f{ie^2m}{p_0}
\int \f{d^3{\bf k}}{2k(2\pi)^3}
\f{d^3{\bf q}m}{(2\pi\hbar)^3q_0} dt d\beta \theta(-\beta)
(2\pi\hbar)^3 \delta^3({\bf p}-{\bf q}-{\bf K})
 \nol & \qquad
 \times \[{\cal S}_+' \]|\phi_p(\bar{t})|^2|\phi_q(\bar{t})|^2
\exp\left[ i\(
E_p+E_q+K\)
\beta \right] \,.
\end{align}
As with the particle spinor combination, we produce an `effective'
version of ${\cal S}_+'$ under the delta function.  The previous
argument relating to the preferred direction of the axis for the
${\bf K}$ integration is still valid and thus we use the
transformation
\begin{align}
\tilde{\bf q} \to \tilde{\bf p}\(1-\f{{\bf K}\cdot\tilde{\bf
p}}{\tilde{\bf p}^2} \) \,,
\end{align}
along with $\dot{\tilde{{\bf q}}}=\dot{\tilde{{\bf p}}}$. Proceeding
as such, and recalling that we now have $\eta \to \beta$ we have
\begin{align}
{\cal S}_+^{'\rm eff} &= -2 - \f{E_q E_p}{m^2} - \f{\tilde{\bf
p}\cdot\tilde{\bf p}}{m^2}
\(1-\f{{\bf K}\cdot\tilde{\bf p}}{\tilde{\bf p}^2} \) +\(
\f{\hbar}{mE_p}-\f{i\hbar\beta}{m}\)\f{\itbd{\xi}\cdot\tilde{\bf p}\times\dot{\tilde{{\bf p}}}}{E_p}
 \nol & \quad - \( \f{\hbar}{2m^2E_p}-\f{i\hbar\beta}{2m^2}\) \( -\f{\itbd{\xi}\cdot\tilde{\bf p}\times\dot{\tilde{{\bf p}}}\,\tilde{\bf p}^2}{E_p(E_p+m)} -\tilde{\bf p}\cdot\itbd{\xi}\times
\(\dot{\tilde{{\bf p}}}-\f{\dot{E}_p\tilde{\bf p}}{E_p+m} \) \) \nol & \qquad \times \(1-\f{{\bf K}\cdot\tilde{\bf p}}{\tilde{\bf p}^2} \)
\nol & \quad +
\( \f{\hbar}{2mE_q^2} - \f{i\hbar\beta}{2mE_q}\)
\[ \itbd{\xi}\cdot\tilde{\bf p}\times\dot{\tilde{{\bf p}}}
+ \f{\tilde{\bf p}\cdot\tilde{\bf p}\times\dot{\tilde{{\bf
q}}}_+\,\itbd{\xi}\cdot\tilde{\bf p}}{m(E_p+m)} \]
\(1-\f{{\bf K}\cdot\tilde{\bf p}}{\tilde{\bf p}^2} \) \nol
&= -2 - \f{E_q E_p}{m^2} - \f{\tilde{\bf p}\cdot\tilde{\bf p}}{m^2}
\(1-\f{{\bf K}\cdot\tilde{\bf p}}{\tilde{\bf p}^2} \) +\(
\f{\hbar}{mE_p}-\f{i\hbar\beta}{m}\)\f{\itbd{\xi}\cdot\tilde{\bf p}\times\dot{\tilde{{\bf p}}}}{E_p}
 \nol & \quad - \( \f{\hbar}{2m^2E_p}-\f{i\hbar\beta}{2m^2}\)  \itbd{\xi}\cdot\tilde{\bf p}\times\dot{\tilde{{\bf p}}}\f{m}{E_p} \(1-\f{{\bf K}\cdot\tilde{\bf p}}{\tilde{\bf p}^2}
\)
\nol & \quad +
\( \f{\hbar}{2mE_q^2} - \f{i\hbar\beta}{2mE_q}\)
\itbd{\xi}\cdot\tilde{\bf p}\times\dot{\tilde{{\bf p}}}\(1-\f{{\bf K}\cdot\tilde{\bf p}}{\tilde{\bf p}^2}
\) \,.
\end{align}
Hence
\begin{align}
{\cal S}_+^{'\rm eff}(t,\beta) &= -2 - \f{E_q E_p}{m^2} -
\f{\tilde{\bf p}\cdot\tilde{\bf p}}{m^2}
 +\f{\hbar}{2m}\itbd{\xi}\cdot\tilde{\bf p}\times\dot{\tilde{{\bf p}}} \(
\f{1}{E_p^2}+\f{1}{E_q^2} \)
\nol & -\f{i\hbar\beta}{2m} \itbd{\xi}\cdot\tilde{\bf
p}\times\dot{\tilde{{\bf p}}} \(
\f{1}{E_p}+\f{1}{E_q}\) \nol & \quad + \f{{\bf K}\cdot\tilde{\bf p}}{\tilde{\bf p}^2}
\[ \f{\tilde{\bf p}^2}{m^2} + \itbd{\xi}\cdot\tilde{\bf p}\times\dot{\tilde{{\bf p}}}
\( \f{\hbar}{2m} \( \f{1}{E_p^2}-\f{1}{E_q^2} \)
-
\f{i\hbar\beta}{2m} \( \f{1}{E_p} - \f{1}{E_q} \) \) \] \,.
\end{align}

Returning to the expression for ${\cal F}_+$, we recall that
$|\phi_q(t)|^2 = q_0/E_q +{\cal O}(\hbar^2)$ and integrate out the
delta function to produce
\begin{multline}
{\cal F}_+({\bf p})  = \f{ie^2m^2}{\hbar^2 p_0}
\int dt \f{d^3{\bf K}}{(2\pi)^32KE_q}
 d\beta |\phi_p(\bar{t})|^2 \theta(-\beta)
\\ \times \[{\cal S}_+^{'\rm eff}(t,\beta)\]
\exp\left[ i\(
E_p+E_q+K\)
\beta \right] \,,
\end{multline}
where we have changed the variable from ${\bf k}$ to ${\bf K}$ in
the measure\footnote{thus acquiring an $1/\hbar^2$ prefactor,} and
we can now consider the definition of $E_q$ to be
$E_q=\sqrt{\(\tilde{\bf p}-{\bf K}\)^2+m^2}$. Integrating over
$\beta$ using the convergence factor, as in the particle loop case
in (\ref{betaint}), we find
\begin{align}
{\cal F}_+({\bf p})  &= \f{e^2m^2}{\hbar^2 p_0}
\int dt \f{d^3{\bf K}}{(2\pi)^32KE_q}
 |\phi_p(\bar{t})|^2
 \[ \f{\(f^+_{00}+\hbar f^+_{01}\)}{\(E_p+E_q+K\)} + \f{i\hbar f^+_{11}}{\(E_p+E_q+K\)^2} \]
 \,, \label{Fpluswithfs}
\end{align}
with
\begin{align}
f^+_{00} &= -2 - \f{E_q E_p}{m^2} - \f{\tilde{\bf p}\cdot\tilde{\bf
p}}{m^2}\(1-\f{{\bf K}\cdot\tilde{\bf p}}{\tilde{\bf p}^2}\) \\
f^+_{10} &= \f{1}{2m}\itbd{\xi}\cdot\tilde{\bf
p}\times\dot{\tilde{{\bf p}}}\[  \(
\f{1}{E_p^2}+\f{1}{E_q^2} \) + \f{{\bf K}\cdot\tilde{\bf p}}{\tilde{\bf
p}^2} \(\f{1}{E_p^2}-\f{1}{E_q^2} \) \] \\
f^+_{11} &= \f{i}{2m} \itbd{\xi}\cdot\tilde{\bf
p}\times\dot{\tilde{{\bf p}}} \[ -\(
\f{1}{E_p}+\f{1}{E_q}\) - \f{{\bf K}\cdot\tilde{\bf p}}{\tilde{\bf p}^2}
\(\f{1}{E_p}-\f{1}{E_q}\) \] \,.
\end{align}

As a recap, the contributions to the real part of the forward
scattering are the leading and first order $\hbar$ terms from both
the particle and antiparticle loops as given in (\ref{Fwithfs}) and
(\ref{Fpluswithfs}). The higher order contributions from
(\ref{Fwithfs}) do not contribute to the real part to order
$\hbar^{-1}$. As with the scalar case, we now renormalise the mass
and calculate the contribution from the mass counterterm towards the
renormalised forward scattering.

\subsection{Mass renormalisation}

The contribution to the forward scattering due to the
renormalisation with the mass counterterm $\delta m$ is given by
\begin{align}
\delta {\cal F} = \f{m}{\hbar p_0}\int dt |\phi_p(t)|^2 \bar{u}_\alpha (p)\delta m u_\alpha
(p) \,,
\end{align}
where the mass counter term itself is given by
\begin{align}
\delta m =  \Sigma (p) |_{\slashed{p}=m_P} \,,
\end{align}
and we define
\begin{align}
\delta m_t = \bar{u}_\alpha (p) \Sigma (p) |_{\slashed{p}=m_P} u_\alpha
(p) \,.
\end{align}
The self energy $\Sigma (p)$ (not to be confused with the term
$\Sigma (t)$ used for the spinors), is calculated for the fermions
in the absence of the potential and represents the loop given by the
fermion and photon. The outer spinors found in $\delta {\cal F}$
however are the semiclassical spinors used in the presence of the
classical non-perturbative potential. To obtain the mass counterterm
the self-energy is evaluated on the mass shell, i.e. with $p=p(t)$
and $\slashed{p}=m_P$. We recall that the counterterm is thus
independent of the momentum. Hence $\delta m$ is calculated using
the standard QED without the external potential, but $\delta m_t$
depends on ${\bf V}(t)$ through the time-dependent momenta of the
spinors $u_\alpha$. Using the Feynman rules for the free field, we
have
\begin{align}
\Sigma (p) &= -\f{ie^2}{\hbar} \int \f{d^4 K}{(2\pi)^4}
\f{g_{\mu\nu}}{K^2+i\epsilon} \gamma^\mu
\f{1}{\slashed{p}-\slashed{K}-m+i\epsilon} \gamma^\nu \nol
&= -\f{ie^2}{\hbar}\int \f{d^4 K}{(2\pi)^4}
 \gamma^\mu
\f{\slashed{p}-\slashed{K}-m}{\(K^2+i\epsilon\)\((p-K)^2-m^2+i\epsilon\)}
\gamma_\mu \nol
&=-\f{ie^2}{\hbar}\int \f{d^4 K}{(2\pi)^4}
\f{-2(\slashed{p}-\slashed{K})+4m}{\(K^2+i\epsilon\)\((p-K)^2-m^2+i\epsilon\)}
\,, \nol
\end{align}
which is taken to be evaluated in the limit $\epsilon \to 0^+$. We
shall complete the $k_0$ portion of the integration by complex
contour integration. Let us define $\omega=\sqrt{|{\bf p}-{\bf
K}|^2+m^2}$ and rewrite the denominator as
\begin{align}
\( K_0 - |{\bf K}| + i\delta \) \( K_0 + |{\bf K}| - i\delta \)
\( K_0 - p_0 + \omega - i\delta \) \( K_0 - p_0 - \omega - i\delta
\) \,,
\end{align}
with the limit $\delta\to 0^+$. The poles in the upper half plane
are
\begin{align}
K_0 = -|{\bf K}| +i\delta\, ; \, K_0 = p_0-\omega + i \delta \,.
\end{align}
Enclosing these poles by the contour (anticlockwise) the residue
theorem gives us
\begin{align}
& \quad \Sigma (p) \nol &= -\f{ie^2}{\hbar} \(2\pi i\)
\int \f{d^3{\bf K}}{(2\pi)^4} \[ \left. \f{-2(\slashed{p}-\slashed{K})+4m}{\( K_0 - |{\bf K}| + i\delta
\) \( \(p_0-K_0\)^2 - \omega^2 +i\epsilon\)} \right|_{K_0=-|{\bf
K}| + i\delta} \right. \nol & \qquad \left. +
\left. \f{-2(\slashed{p}-\slashed{K})+4m}{\(K_0^2-|{\bf K}|^2
+i\epsilon \)\(K_0-p_0-\omega -i\delta\)}
\right|_{K_0=p_0-\omega+i\delta} \] \,.
\end{align}
Making the substitutions for $K_0$, defining $K=|{\bf K}|$, and
taking the appropriate limits, we obtain
\begin{align}
\Sigma(p) &= \f{e^2}{\hbar} \int \f{d^3{\bf K}}{(2\pi)^3} \nol & \quad  \[ \f{4m - 2\gamma^0 \(p_0+K\) +
2\itbd{\gamma}\cdot \({\bf p}-{\bf K}\)}{(-2K)
\(\(p_0+K\)^2-\omega^2\)} +
\f{4m - 2\gamma^0 \omega + 2\itbd{\gamma}\cdot \({\bf p}-{\bf K}\)}{
\(\(p_0-\omega\)^2-K^2\) \(-2\omega\)} \] \nol
&= -\f{e^2}{\hbar} \int \f{d^3{\bf K}}{(2\pi)^3} \[ 4m \[ \f{1}{2K
\(\(p_0+K\)^2-\omega^2\)} + \f{1}{2\omega\(\(p_0-\omega\)^2-K^2\)}
\] \right. \nol & \qquad \left. - 2\gamma^0 \[\f{p_0+K}{2K
\(\(p_0+K\)^2-\omega^2\)} + \f{\omega}{2\omega\(\(p_0-\omega\)^2-K^2\)}
\] \right. \nol & \qquad \left. + 2\itbd{\gamma}\cdot \({\bf p}-{\bf K}\)
\[ \f{1}{2K
\(\(p_0+K\)^2-\omega^2\)} + \f{1}{2\omega\(\(p_0-\omega\)^2-K^2\)}
\]
\] \,.
\end{align}
At this point, we make a short aside to note the following
\begin{align}
\f{1}{\(p_0+K\)^2-\omega^2} &= \f{1}{2\omega} \( \f{1}{p_0-\omega+K}
- \f{1}{p_0+\omega+K} \) \\
\f{1}{\(p_0-\omega\)^2-K^2} &= \f{1}{2K} \( \f{1}{p_0-\omega-K} -
\f{1}{p_0-\omega+K} \) \\
\f{p_0+K}{(\(p_0+K\)^2-\omega^2} &= \f{1}{2} \(\f{1}{p_0-\omega+K} +
\f{1}{p_0+\omega+K} \) \,.
\end{align}
Using these expansions, we can now write
\begin{align}
\Sigma (p) &=
- \f{e^2}{\hbar} \int \f{d^3{\bf K}m}{(2\pi)^32K\omega} \nol &
\quad \[ 2
\[\f{1}{p_0-\omega-K}- \f{1}{p_0+\omega+K} \] -\f{\gamma^0
\omega}{m} \[\f{1}{p_0-\omega-K}+ \f{1}{p_0+\omega+K} \]
\right. \nol & \qquad  \left. +\f{\itbd{\gamma}\cdot\({\bf p}-{\bf K}\)}{m} \[
\f{1}{p_0-\omega-K}- \f{1}{p_0+\omega+K} \] \] \,.
\end{align}
Evaluating the self-energy on the mass shell using the particle (as
opposed to antiparticle) momentum $\tilde{\bf p}={\bf p}-{\bf
V}(t)$, we have $p_0\to E_p(t)$ and $\omega
\to E_q(t)$. As with the scalar counterterm, the mass shell uses the physical mass $m_P$, however given that the
counterterm is again of order $e^2$, for our calculations we may use
$m$ and there is no difference in the result at this order. Thus,
with some rearrangement,
\begin{align}
\Sigma (\tilde{p}) &=
- \f{e^2}{\hbar} \int \f{d^3{\bf K}m}{(2\pi)^32KE_q}
\[ \f{1}{E_p-E_q-K} \[ 2 - \f{E_q\gamma^0}{m} +\f{\(\tilde{\bf p}-{\bf
K}\)\cdot\itbd{\gamma}}{m} \] \right. \nol & \quad \left. +
\f{1}{E_p+E_q+K} \[ -2 - \f{E_q\gamma^0}{m} -\f{\(\tilde{\bf p}-{\bf
K}\)\cdot\itbd{\gamma}}{m} \] \] \,.
\end{align}

Adding the semiclassical outer spinors, with the momentum
$\tilde{p}$, the (time-dependent) mass counter term is
\begin{align}
& \quad \delta m_t \nol &= \bar{u}_\alpha(\tilde{p})
\Sigma(\tilde{p}) u_\alpha (\tilde{p}) \nol &= - \f{e^2}{\hbar} \int
\f{d^3{\bf K}m}{(2\pi)^32KE_q} \nol &
\[ \f{1}{E_p-E_q-K} \[ 2 [\bar{u}_\alpha(p) u_\alpha (p)]
- \f{E_q}{m}[\bar{u}_\alpha(p)\gamma^0 u_\alpha (p)]
+\f{\(\tilde{\bf p}-{\bf K}\)}{m}\cdot
[\bar{u}_\alpha(p)\itbd{\gamma} u_\alpha (p)]\]
\right.
\nol & \quad
\left. +
\f{1}{E_p+E_q+K} \[ -2[\bar{u}_\alpha(p) u_\alpha (p)]
- \f{E_q}{m}[\bar{u}_\alpha(p)\gamma^0 u_\alpha (p)]
-\f{\(\tilde{\bf p}-{\bf
K}\)}{m}\cdot[\bar{u}_\alpha(p)\itbd{\gamma} u_\alpha (p)]
\] \] \,.
\end{align}
Using the first order identities we have
\begin{align}
\delta m_t  &= - \f{e^2}{\hbar} \int \f{d^3{\bf K}m}{(2\pi)^32KE_q} \nol
& \left\{ \f{1}{E_p-E_q-K} \[ 2 \( 1-\f{\hbar}{2m}
\f{\itbd{\xi}\cdot\tilde{\bf p}\times\dot{\tilde{{\bf p}}}}{E_p^2}
\) - \f{E_q}{m}\f{E_p}{m}  \right. \right. \nol & \qquad \left.
\left. +
\f{\(\tilde{\bf p}-{\bf K}\)}{m}\cdot
\( \f{\tilde{\bf p}}{m} -\f{\hbar}{2mE_p} \( \f{\itbd{\xi}\cdot\tilde{\bf p}\times\dot{\tilde{{\bf p}}}\, \tilde{\bf p}}{E_p \(E_p+m\)} + \itbd{\xi}\times
\(\dot{\tilde{{\bf p}}} - \f{\dot{E}_p\tilde{\bf p}}{E_p+m}\)\) \)
\]\right.  \nol
&  + \f{1}{E_p+E_q+K} \[ -2\( 1-\f{\hbar}{2m}
\f{\itbd{\xi}\cdot\tilde{\bf p}\times\dot{\tilde{{\bf p}}}}{E_p^2}
\) - \f{E_q}{m}\f{E_p}{m}  \right. \nol & \qquad \left.\left. -
\f{\(\tilde{\bf p}-{\bf K}\)}{m}\cdot\(
\f{\tilde{\bf p}}{m} -\f{\hbar}{2mE_p} \( \f{\itbd{\xi}\cdot\tilde{\bf p}\times\dot{\tilde{{\bf p}}}\, \tilde{\bf p}}{E_p \(E_p+m\)} + \itbd{\xi}\times
\(\dot{\tilde{{\bf p}}} - \f{\dot{E}_p\tilde{\bf p}}{E_p+m}\)\) \)
\]  \right\} \,.
\end{align}
We again make use of the symmetry present and choose the axis for
integration along the direction $\tilde{\bf p}$ thus transforming
${\bf K}\to \({\bf K}\cdot\tilde{\bf p}/\tilde{\bf p}^2\)\tilde{\bf
p}$. Simplifying, we obtain
\begin{align}
\delta m_t &= - \f{e^2}{\hbar} \int \f{d^3{\bf K}m}{(2\pi)^32KE_q}
\nol & \quad \left\{ \[ 2-\f{E_qE_p}{m^2} +\f{\tilde{\bf p}^2}{m^2} - \f{\hbar}{2m}
\f{\itbd{\xi}\cdot\tilde{\bf p}\times\dot{\tilde{{\bf p}}}}{E_p^2} - \f{{\bf
K}\cdot\tilde{\bf p}}{\tilde{\bf p}^2} \[ \f{\tilde{\bf p}^2}{m^2} +
\f{\hbar}{2m}\f{\itbd{\xi}\cdot\tilde{\bf p}\times\dot{\tilde{{\bf p}}}}{E_p^2} \]
\]  \right. \nol & \qquad \times \f{1}{E_p-E_q-K} \nol &  +
\[ -2-\f{E_qE_p}{m^2} -\f{\tilde{\bf p}^2}{m^2} + \f{\hbar}{2m}
\f{\itbd{\xi}\cdot\tilde{\bf p}\times\dot{\tilde{{\bf p}}}}{E_p^2} +\f{{\bf
K}\cdot\tilde{\bf p}}{\tilde{\bf p}^2} \[ \f{\tilde{\bf p}^2}{m^2} +
\f{\hbar}{2m}\f{\itbd{\xi}\cdot\tilde{\bf p}\times\dot{\tilde{{\bf p}}}}{E_p^2} \]
\] \nol & \left. \qquad \times \f{1}{E_p+E_q+K} \right\} \,.
\end{align}
The mass counter term contribution to the forward scattering is
therefore
\begin{align}
{\cal F}^{\delta m} &=- \f{e^2m}{p_0\hbar^2} \int dt
\f{d^3{\bf K}m}{(2\pi)^32KE_q} |\phi_p(t)|^2 \nol & \times
\left\{ \[ 2-\f{E_qE_p}{m^2} +\f{\tilde{\bf p}^2}{m^2} - \f{\hbar}{2m}
\f{\itbd{\xi}\cdot\tilde{\bf p}\times\dot{\tilde{{\bf p}}}}{E_p^2} - \f{{\bf
K}\cdot\tilde{\bf p}}{\tilde{\bf p}^2} \[ \f{\tilde{\bf p}^2}{m^2} +
\f{\hbar}{2m}\f{\itbd{\xi}\cdot\tilde{\bf p}\times\dot{\tilde{{\bf p}}}}{E_p^2} \]
\]  \right. \nol & \qquad \f{1}{E_p-E_q-K} \nol &  +
\[ -2-\f{E_qE_p}{m^2} -\f{\tilde{\bf p}^2}{m^2} + \f{\hbar}{2m}
\f{\itbd{\xi}\cdot\tilde{\bf p}\times\dot{\tilde{{\bf p}}}}{E_p^2} +\f{{\bf
K}\cdot\tilde{\bf p}}{\tilde{\bf p}^2} \[ \f{\tilde{\bf p}^2}{m^2} +
\f{\hbar}{2m}\f{\itbd{\xi}\cdot\tilde{\bf p}\times\dot{\tilde{{\bf p}}}}{E_p^2} \]
\] \nol & \qquad \left.\f{1}{E_p+E_q+K} \right\} \,.
\end{align}

This counter term contribution is to be added to the leading and
first order loop contributions to the forward scattering from
(\ref{Fwithfs}) and (\ref{Fpluswithfs}). These two contributions can
be written
\begin{align}
{\cal F}_- &= \f{e^2m}{p_0\hbar^2} \int dt
\f{d^3{\bf K}m}{(2\pi)^22KE_q} |\phi_p(t)|^2 \nol & \times
  \left\{ \[2 - \f{E_q E_p}{m^2} + \f{\tilde{\bf p}\cdot\tilde{\bf p}}{m^2}
 -\f{\hbar}{2m}\itbd{\xi}\cdot\tilde{\bf p}\times\dot{\tilde{{\bf p}}} \(
\f{1}{E_p^2}+\f{1}{E_q^2} \) \right. \right. \nol & \qquad  \left. - \f{{\bf K}\cdot\tilde{\bf p}}{\tilde{\bf p}^2}
\( \f{\tilde{\bf p}^2}{m^2}
+ \f{\hbar}{2m}\itbd{\xi}\cdot\tilde{\bf p}\times\dot{\tilde{{\bf
p}}}
\(
\f{1}{E_p^2}-\f{1}{E_q^2} \)\) \]
\f{1}{(E_p-E_q-K)} \nol & \left.
-\f{\hbar}{2m}
\itbd{\xi}\cdot\tilde{\bf p} \times\dot{\tilde{{\bf p}}}
\[ \( \f{1}{E_p}-\f{1}{E_q} \)+ \f{{\bf K}\cdot\tilde{\bf p}}{\tilde{\bf p}^2} \(
\f{1}{E_p}+\f{1}{E_q} \) \]\f{1}{(E_p-E_q-K)^2} \right\} \,,
\end{align}
and
\begin{align}
{\cal F}_+ &= \f{e^2m}{p_0\hbar^2} \int dt
\f{d^3{\bf K}m}{(2\pi)^22KE_q}|\phi_p(t)|^2\nol & \times
  \left\{ \[-2 - \f{E_q E_p}{m^2} - \f{\tilde{\bf p}\cdot\tilde{\bf p}}{m^2}
 +\f{\hbar}{2m}\itbd{\xi}\cdot\tilde{\bf p}\times\dot{\tilde{{\bf p}}} \(
\f{1}{E_p^2}+\f{1}{E_q^2} \) \right. \right. \nol & \qquad  \left. +\f{{\bf K}\cdot\tilde{\bf p}}{\tilde{\bf p}^2}
\( \f{\tilde{\bf p}^2}{m^2}
+ \f{\hbar}{2m}\itbd{\xi}\cdot\tilde{\bf p}\times\dot{\tilde{{\bf
p}}}
\(
\f{1}{E_p^2}-\f{1}{E_q^2} \)\) \]
\f{1}{(E_p+E_q+K)} \nol & \left.
+\f{\hbar}{2m}
\itbd{\xi}\cdot\tilde{\bf p} \times\dot{\tilde{{\bf p}}}
\[ \( \f{1}{E_p}+\f{1}{E_q} \)+ \f{{\bf K}\cdot\tilde{\bf p}}{\tilde{\bf p}^2} \(
\f{1}{E_p}-\f{1}{E_q} \) \]\f{1}{(E_p+E_q+K)^2} \right\} \,,
\end{align}
where we have reordered the terms and for ${\cal F}_- $ brought the
overall minus sign inside the integrand. Comparison between the mass
counter term and the loops terms shows that the counter term cancels
some, but not all of the loop contributions. From the derivation of
the mass counter term it is notable that those terms coming from the
time split are not present (as the self-energy is calculated at a
particular time - note for the `free' field, i.e. without the
potential, the momentum at different times is the same in the
absence of a further interaction). Additionally, the first order
corrections to the inner spinors of the loop are also not present -
the first order correction to the spinor is due to the presence of
the time-dependent potential. Adding ${\cal F}_-+{\cal F}_++{\cal
F}+{\cal F}^{\delta m}$ we find the renormalised forward scattering
contribution is
\begin{align}
{\cal F}^R({\bf p}) &=  \f{e^2m}{p_0\hbar^2} \int \f{dt\,d^3{\bf
K}m}{(2\pi)^32KE_q} |\phi_p(t)|^2 \f{\hbar}{2m}
\itbd{\xi}\cdot{\bf p}\times\dot{\bf p} \nol
& \times \left\{ \f{1}{E_q^2} \(
\f{1}{E_p+E_q+K}-\f{1}{E_p-E_q-K}\)\(1-\f{{\bf K}\cdot{\bf p}}{{\bf
p}^2}\)  \right. \nol &-
\f{1}{\(E_p-E_q-K\)^2}\(\f{1}{E_p}-\f{1}{E_q} \)
 + \f{1}{\(E_p+E_q+K\)^2}\(\f{1}{E_p}+\f{1}{E_q}\)
 \nol
& \left. +\f{{\bf K}\cdot{\bf p}}{{\bf p}^2} \[
-\f{1}{\(E_p-E_q-K\)^2}\(\f{1}{E_p}+\f{1}{E_q} \)  +
\f{1}{\(E_p+E_q+K\)^2}
\(\f{1}{E_p}-\f{1}{E_q}\) \] \right\} \,.
\end{align}
where we have dropped the tilde notation on the ${\bf p}$ due to the
fact that all the energy-momenta in the integrand are now the
time-dependent elements of $(E_p,{\bf p}-{\bf V}(t))$. This
contribution was not present in the scalar quantum position shift.
We can regard this term as a `correction' term leading to an
additional contribution to the position shift when compared with
either the scalar case or the classical case. As such, the term is
in need of interpretation. Before doing so, however, further
calculation and simplification of ${\cal F}^R$ will be useful.

\subsubsection{Integration and simplification}

Let us define ${\cal I}_K$ as the $K$ integral part of the
correction, viz
\begin{align}
{\cal F}^R({\bf p}) &= \f{e^2m^2}{p_0\hbar^2} \int dt
   |\phi_p(t)|^2 \f{\hbar}{2m} \itbd{\xi}\cdot{\bf p}\times\dot{\bf
   p} \, {\cal I}_K \,,
\end{align}
with
\begin{align}
{\cal I}_K &= \int \f{d^3{\bf K}}{(2\pi)^3 2|{\bf K}|E_q} \left\{
\f{1}{E_q^2} \( \f{1}{E_p+E_q+|{\bf K}|}-\f{1}{E_p-E_q-|{\bf K}|}\)\(1-\f{{\bf
K}\cdot{\bf p}}{{\bf p}^2}\)  \right. \nol &
    \quad -
   \f{1}{\(E_p-E_q-|{\bf K}|\)^2}\(\f{1}{E_p}-\f{1}{E_q} \)
   + \f{1}{\(E_p+E_q+|{\bf K}|\)^2}\(\f{1}{E_p}+\f{1}{E_q}\)
   \nol
&  \left. \quad +\f{{\bf K}\cdot{\bf p}}{{\bf p}^2} \[
   -\f{1}{\(E_p-E_q-|{\bf K}|\)^2}\(\f{1}{E_p}+\f{1}{E_q} \)  +
   \f{1}{\(E_p+E_q+|{\bf K}|\)^2}
   \(\f{1}{E_p}-\f{1}{E_q}\) \] \right\} \,.
\end{align}
The reader may notice that we have returned to the original notation
$|{\bf K}|$ for the modulus of the $3$-vector. The reason for this
is to prevent confusion in the following calculation in which we
shall need the $4$-vector $K$. In order to help with the calculation
${\cal I}_K$, let us look at the result of the $K_0$ integration of
the following two $4$-dimensional integrals over $K$:
\begin{align}
{\cal I}_A &= \int \f{d^4K}{(2\pi)^4 i}
   \f{K_0}{(K^2+i\varepsilon)\((K-p)^2-m^2+i\varepsilon\)^2} \,,
   \nol
{\cal I}_B &= \int \f{d^4K}{(2\pi)^4 i}
   \f{K_0}{(K^2+i\varepsilon)^2\((K-p)^2-m^2+i\varepsilon\)} \,,
\end{align}
with the limit $\varepsilon\to 0^+$. To clarify the previous comment
on the notation, we confirm that $K^2=K^\mu K_\mu$ here and for the
remainder of this section.

Firstly, for ${\cal I}_A$, we note that the denominator is
\begin{align}
& \qquad (K_0^2-|{\bf K}|^2+i\varepsilon)\((K_0-E_p)^2-|{\bf K}-{\bf
p}|^2-m^2+i\varepsilon\)^2 \nol &= (K_0^2-|{\bf K}|^2+i\varepsilon)
\((K_0-E_p)^2-E_q^2+i\varepsilon\)^2
\nol
&= (K_0+|{\bf K}|-i\delta)(K_0-|{\bf K}|+i\delta)
(K_0-E_p-E_q+i\delta)^2(K_0-E_p+E_q-i\delta)^2
\,,
\end{align}
with the limit $\delta \to 0^+$. The poles in the upper half plane
are
\begin{align}
K_0 &= -|{\bf K}| +i\delta \,, \nol K_0 &= E_p-E_q+i\delta \,,
\end{align}
where the second pole is second order. The residue for the
singularity at $K_0 = -|{\bf K}| +i\delta$, in the $\delta \to 0^+$
limit, is
\begin{align}
\f{1}{2} \f{1}{(E_p+E_q+|{\bf K}|)^2(E_p-E_q+|{\bf K}|)^2}\,,
\end{align}
and the residue at $K_0 = E_p-E_q+i\delta$ is
\begin{align}
\f{1}{4}\f{E_p^3-4E_p^2E_q+5E_pE_q^2-2E_q^3-|{\bf K}|^2E_p}{E_q^3(E_p-E_q+|{\bf K}|)^2(E_p-E_q-|{\bf K}|)^2}
\,.
\end{align}
The sum of the residues, after some rearranging, gives
\begin{multline}
-\f{1}{8|{\bf K}|} \[ \f{E_p+E_q}{(E_q+E_p+|{\bf K}|)^2E_q^2} +
\f{E_p-E_q}{(|{\bf K}|+E_q-E_p)^2E_q^2} \right. \\ \left. +
\f{E_p}{(|{\bf K}|+E_q-E_p)E_q^3} + \f{E_p}{(|{\bf K}|+E_q+E_p)E_q^3} \] \,.
\end{multline}
Consequently, the $K_0$ integration of ${\cal I}_A$ gives
\begin{align}
{\cal I}_A &= -\int\f{d^3{\bf K}}{(2\pi)^38|{\bf K}|}
\[ \f{E_p+E_q}{(E_q+E_p+|{\bf K}|)^2E_q^2}
+ \f{E_p-E_q}{(|{\bf K}|+E_q-E_p)^2E_q^2} \right. \nol & \qquad
\qquad \qquad \qquad \left. + \f{E_p}{(|{\bf K}|+E_q-E_p)E_q^3} +
\f{E_p}{(|{\bf K}|+E_q+E_p)E_q^3} \] \,.
\end{align}

Secondly, we analogously consider ${\cal I}_B$. The denominator
gives
\begin{align}
(K_0+|{\bf K}|-i\delta)^2(K_0-|{\bf K}|+i\delta)^2
(K_0-E_p-E_q+i\delta)(K_0-E_p+E_q-i\delta)
\,.
\end{align}
The poles in the upper half plane are again
\begin{align}
K_0 &= -|{\bf K}| +i\delta \,, \nol K_0 &= E_p-E_q+i\delta \,,
\end{align}
where this time the first pole is second order. For this integrand,
the residue at $K_0 = -|{\bf K}| +i\delta$, in the $\delta \to 0^+$
limit, is
\begin{align}
-\f{1}{2} \f{E_p+|{\bf K}|}{|{\bf K}|(E_p+E_q+|{\bf
K}|)^2(E_p-E_q+|{\bf K}|)^2} \,,
\end{align}
and the residue at $K_0 = E_p-E_q+i\delta$ is
\begin{align}
-\f{1}{2}\f{E_p-E_q}{E_q(E_p-E_q+|{\bf K}|)^2(E_p-E_q-|{\bf K}|)^2}
\,.
\end{align}
The sum of the residues rearranges to produce
\begin{align}
\f{1}{8|{\bf K}|E_q} \[ \f{1}{(E_p+E_q+|{\bf K}|)^2} - \f{1}{(E_p-E_q-|{\bf K}|)^2} \] \,.
\end{align}
The $K_0$ integration of ${\cal I}_B$ thus gives
\begin{align}
{\cal I}_B &=
\int\f{d^3{\bf K}}{(2\pi)^38|{\bf K}|E_q}
\[ \f{1}{(E_p+E_q+|{\bf K}|)^2} - \f{1}{(E_p-E_q-|{\bf K}|)^2} \] \,.
\end{align}

We now return to the correction integral. Expanding and rearranging
${\cal I}_K$ leads to the following
\begin{align}
{\cal I}_K &= \int \f{d^3{\bf K}}{(2\pi)^3 2|{\bf K}|} \left\{
\f{1}{E_pE_q^3} \( \f{E_p}{E_p+E_q+|{\bf K}|}+\f{E_p}{|{\bf K}|+E_q-E_p}\)
   \right. \nol & \qquad \qquad \qquad \qquad  \left. +\f{E_p-E_q}{\(E_p-E_q-|{\bf K}|\)^2E_q^2E_p}
   + \f{E_p+E_q}{\(E_p+E_q+|{\bf K}|\)^2E_q^2E_p}
   \right. \nol
&  \left. \quad -\f{{\bf K}\cdot{\bf p}}{{\bf p}^2} \[
\f{1}{E_pE_q^3} \( \f{E_p}{E_p+E_q+|{\bf K}|}+\f{E_p}{|{\bf K}|+E_q-E_p}\)
   \right. \right. \nol & \qquad \qquad \qquad \left. \left. +\f{E_p+E_q}{\(E_p-E_q-|{\bf K}|\)^2E_q^2E_p}  +
   \f{E_p-E_q}{\(E_p+E_q+|{\bf K}|\)^2E_q^2E_p}
    \] \right\} \nol
&= \f{4}{E_p}\int \f{d^3{\bf K}}{(2\pi)^3 8|{\bf K}|} \left\{
\f{1}{E_q^3}
\(
\f{E_p}{E_p+E_q+|{\bf K}|}+\f{E_p}{|{\bf K}|+E_q-E_p}\)
   \right. \nol & \qquad \qquad \qquad \qquad  \left. +\f{E_p-E_q}{\(E_p-E_q-|{\bf K}|\)^2E_q^2}
   + \f{E_p+E_q}{\(E_p+E_q+|{\bf K}|\)^2E_q^2}
   \right. \nol
&   \quad -\f{{\bf K}\cdot{\bf p}}{{\bf p}^2} \[ \f{1}{E_q^3} \(
\f{E_p}{E_p+E_q+|{\bf K}|}+\f{E_p}{|{\bf K}|+E_q-E_p}\)
   \right. \nol & \qquad \qquad \qquad  \left.+\f{E_p-E_q}{\(E_p-E_q-|{\bf K}|\)^2E_q^2}  +
   \f{E_p+E_q}{\(E_p+E_q+|{\bf K}|\)^2E_q^2} \right.  \nol & \qquad \qquad \qquad \left. \left.
   +\f{2E_q}{\(E_p-E_q-|{\bf K}|\)^2E_q^2} - \f{2E_q}{\(E_p+E_q+|{\bf K}|\)^2E_q^2}
    \] \right\} \,.
\end{align}

Using the above results for ${\cal I}_A$ and ${\cal I}_B$, we see
that we can rewrite ${\cal I}_K$ in terms of the $4$ dimensional
integrals:
\begin{align}
{\cal I}_K &= -\f{4}{E_p} \int \f{d^4K}{(2\pi)^4 i}
   \f{K_0}{(K^2+i\varepsilon)\((K-p)^2-m^2+i\varepsilon\)^2} \nol
   &
+\sum_i\f{4p_i}{E_p{\bf p}^2}\int \f{d^4K}{(2\pi)^4
i}\f{K_0K_i}{(K^2+i\varepsilon)\((K-p)^2-m^2+i\varepsilon\)^2}
\nol &
+\sum_i\f{8p_i}{E_p{\bf p}^2} \int \f{d^4K}{(2\pi)^4 i}
   \f{K_0K_i}{(K^2+i\varepsilon)^2\((K-p)^2-m^2+i\varepsilon\)}
   \,.
\end{align}

Let us now define the following $4$-dimensional integrals (in
Minkowski space):
\begin{align}
{\cal I}^{(1)}_\mu &= \int \f{d^4 K}{(2\pi)^4 i}
\f{K_\mu}{(K^2+i\varepsilon)((K-p)^2-m^2+i\varepsilon)^2} \,, \\
{\cal I}^{(2)}_{\mu\nu} &= \int \f{d^4 K}{(2\pi)^4 i}
\f{K_\mu K_\nu}{(K^2+i\varepsilon)((K-p)^2-m^2+i\varepsilon)^2} \,, \\
{\cal I}^{(3)}_{\mu\nu} &= \int \f{d^4 K}{(2\pi)^4 i}
\f{K_\mu K_\nu}{(K^2+i\varepsilon)^2((K-p)^2-m^2+i\varepsilon)}
\,.
\end{align}
We then obtain
\begin{align}
{\cal I}_K &= \f{4}{E_p} \[ -{\cal I}^{(1)}_0 + \sum_i
\f{p_i}{{\bf p}^2}\( {\cal I}^{(2)}_{0i} +2 {\cal I}^{(3)}_{0i} \)
\] \,.
\end{align}

We now proceed to calculate these $4$-dimensional integrals. We
start by noting the following identities, which we shall use in the
calculation:
\begin{align}
\f{1}{ab^2} &= \int^1_0 dy \f{2y}{((1-y)a+yb)^3} \,, \label{Feyntrick}\\
\int^\infty_0 \f{x^{\mu-1}}{(x+\alpha)^\nu} dx &= \alpha^{\mu-\nu}
\f{\Gamma(\nu-\mu)\Gamma(\mu)}{\Gamma(\nu)}\, \qquad \qquad \text{for positive $\alpha$}
\,. \label{intidalpha}
\end{align}
We shall also make use of the following time-coordinate rotation
from integration in Minkowski $K$ space to Euclidean $K_E$ space:
\begin{align}
K_4 &= -iK_0 \,,\\
K^2 &= -(K_1^2+K_2^2+K_3^2+K_4^2) = -K_E^2 \,, \\
d^4K_E &= d^3K\,dK_4 = -id^4K \,.
\end{align}
For $D$-dimensional Euclidean coordinates, the integration of a
function which is only dependent on the radial coordinate of the
hyperspherical polar coordinates can be written
\begin{align}
  \int d^D x f(r)
= \int dr \f{2\pi^{D/2}}{\Gamma(D/2)}r^{D-1} f(r) = \int d(r^2)
\f{\pi^{D/2}}{\Gamma(D/2)}(r^2)^{D/2-1} f(r)\,.
\end{align}
In $4$ dimensions, this becomes
\begin{align}
  \int d^4 x f(r)
= \int dr \f{2\pi^2}{\Gamma(2)}r^3 f(r) = \int d(r^2) \pi^2 r^2 f(r)
\,.
\end{align}

\subsubsection{First integral}
\begin{align}
{\cal I}^{(1)}_\mu &= \int \f{d^4 K}{(2\pi)^4 i}
   \f{K_\mu}{(K^2+i\varepsilon)((K-p)^2-m^2+i\varepsilon)^2} \nol
&= \int \f{d^4 K}{(2\pi)^4 i} \int^1_0 dy
   \f{2yK_\mu}{\[(1-y)(K^2+i\varepsilon)
   +y((K-p)^2-m^2+i\varepsilon)\]^3} \nol
&= \int^1_0 dy \int \f{d^4 K}{(2\pi)^4 i}
   \nol & \quad  \times \f{2yK_\mu}{\[ K^2-yK^2+(1-y)i\varepsilon +yK^2+yp^2-2yK\cdot p
   -ym^2 +yi\varepsilon \]^3} \nol
&= \int^1_0 dy \int \f{d^4 K}{(2\pi)^4 i}
   \f{2yK_\mu}{\[K^2-2yK\cdot p +y (p^2-m^2) +i\varepsilon\]^3} \nol
&= \int^1_0 dy \int \f{d^4 K}{(2\pi)^4 i}
   \f{2yK_\mu}{\[K^2-2yK\cdot p +i\varepsilon\]^3} \,.
\end{align}
Let
\begin{align}
\kappa_\mu=K_\mu-yp_\mu \,.
\end{align}
Then we have
\begin{align}
d^4K=d^4\kappa \,,
\end{align}
and
\begin{align}
\kappa^2 -y^2p^2
&= K^2+y^2p^2-2yK\cdot p -y^2p^2 \nol &=K^2-2yK\cdot p \,.
\end{align}
Thus we can change variables to produce
\begin{align}
{\cal I}^{(1)}_\mu &= \int^1_0 dy 2y\int \f{d^4 \kappa}{(2\pi)^4 i}
   \f{\kappa_\mu+yp_\mu}{\[\kappa^2-y^2p^2 +i\varepsilon\]^3} \,.
\end{align}
Using $p^2=m^2$, changing notation from $\kappa \to K$, we have
\begin{align}
{\cal I}^{(1)}_\mu &= \int^1_0 dy 2y\int \f{d^4 K}{(2\pi)^4 i}
   \f{K_\mu+yp_\mu}{\[K^2-y^2m^2 +i\varepsilon\]^3} \,.
\end{align}
The term proportional to $K_\mu$ in the numerator is odd and
integrates to zero, we thus have
\begin{align}
{\cal I}^{(1)}_\mu &= \int^1_0 dy 2y^2 p_\mu\int \f{d^4 K}{(2\pi)^4
i}
   \f{1}{\[K^2-y^2m^2 +i\varepsilon\]^3} \,. \label{4dI1}
\end{align}
Rotating the time coordinate to produce $4$-d Euclidean space, we
find:
\begin{align}
{\cal I}^{(1)}_\mu &= -\int^1_0 dy 2y^2 p_\mu\int \f{d^4
K_E}{(2\pi)^4}
   \f{1}{\[K_E^2+y^2m^2\]^3} \,,
\end{align}
where we have taken the limit $\varepsilon\to 0^+$ as the integral
converges. Note the overall minus sign and consequent rearrangement
of the denominator of the integrand. Using $4$-dimensional
hyperspherical polar coordinates, we have:
\begin{align}
{\cal I}^{(1)}_\mu &= -\int^1_0 dy 2y^2 p_\mu\int_0^\infty
\f{d(K_E^2)}{(2\pi)^4}
   \f{\pi^2 K_E^2}{\[K_E^2+y^2m^2\]^3} \,.
\end{align}
Using (\ref{intidalpha}), with $\mu=2$ and $\nu=3$, to perform the
$K_E^2$ integral, we obtain
\begin{align}
{\cal I}^{(1)}_\mu &= -\int^1_0 dy 2y^2
p_\mu\f{\pi^2}{(2\pi)^4y^2m^2}
     \f{\Gamma(1)\Gamma(2)}{\Gamma(3)} \nol
&= -\f{\pi^2 p_\mu}{(2\pi)^4m^2} \int^1_0 dy \f{y^2}{y^2} \nol &=
-\f{\pi^2 p_\mu}{(2\pi)^4m^2}
   \,.
\end{align}

\subsubsection{Second Integral}
The second integral follows the same method as the first:
\begin{align}
{\cal I}^{(2)}_{\mu\nu} &= \int \f{d^4 K}{(2\pi)^4 i}
   \f{K_\mu K_\nu}{(K^2+i\varepsilon)((K-p)^2-m^2+i\varepsilon)^2} \nol
&= \int \f{d^4 K}{(2\pi)^4 i} \int^1_0 dy
   \f{2yK_\mu  K_\nu}{\[(1-y)(K^2+i\varepsilon)
   +y((K-p)^2-m^2+i\varepsilon)\]^3} \nol
&= \int^1_0 dy \int \f{d^4 K}{(2\pi)^4 i}
   \f{2yK_\mu  K_\nu}{\[K^2-2yK\cdot p +y (p^2-m^2) +i\varepsilon\]^3} \nol
&= \int^1_0 dy \int \f{d^4 K}{(2\pi)^4 i}
   \f{2yK_\mu  K_\nu}{\[K^2-2yK\cdot p +i\varepsilon\]^3} \,.
\end{align}
We again use the change of variables given by
\begin{align}
\kappa_\mu=K_\mu-yp_\mu \,,\nol
d^4K=d^4\kappa \,,
\end{align}
and
\begin{align}
\kappa^2 -y^2p^2
&=K^2-2yK\cdot p \,,
\end{align}
to obtain
\begin{align}
{\cal I}^{(2)}_{\mu\nu} &= \int^1_0 dy 2y\int \f{d^4
\kappa}{(2\pi)^4 i}
   \f{\(\kappa_\mu+yp_\mu\)\(\kappa_\mu+yp_\mu\)}{\[\kappa^2-y^2m^2 +i\varepsilon\]^3} \,.
\end{align}
We change the notation as $\kappa \to K$ and expand the numerator.
Those terms proportional to $K_\mu$ or $K_\nu$ odd functions of
$K_\mu$ and $K_\nu$, and integrate to zero. The term proportional to
$K_\mu K_\nu$ is an odd function when $\mu\neq\nu$, which is the
case that we require. Thus for the $\mu=0,\,\nu=i$ elements, we have
\begin{align}
{\cal I}^{(2)}_{0i} &= \int^1_0 dy 2y^3 E_p p_i\int \f{d^4
K}{(2\pi)^4 i}
   \f{1}{\[K^2-y^2m^2 +i\varepsilon\]^3} \,.
\end{align}
The $K$ integral can be recognised as the same as that in ${\cal
I}^{(1)}_\mu$ in (\ref{4dI1}), thus we have
\begin{align}
{\cal I}^{(2)}_{0i} &= -\int^1_0 dy 2y^3 E_p p_i
\f{\pi^2}{(2\pi)^4y^2m^2}
     \f{\Gamma(1)\Gamma(2)}{\Gamma(3)} \nol
&= -\f{\pi^2 E_p p_i}{(2\pi)^4m^2} \int^1_0 dy \f{y^3}{y^2} \nol &=
-\f{1}{2}\f{\pi^2 E_p p_i}{(2\pi)^4m^2}
   \,.
\end{align}

\subsubsection{Third Integral}
For the third integral we proceed using the same method as before,
but note the change in the denominator (and thus the $K$ integral)
from the previous cases. We have
\begin{align}
{\cal I}^{(3)}_{\mu\nu} &= \int \f{d^4 K}{(2\pi)^4 i}
   \f{K_\mu K_\nu}{(K^2+i\varepsilon)^2((K-p)^2-m^2+i\varepsilon)} \nol
&= \int \f{d^4 K}{(2\pi)^4 i}\int^1_0 dy
   \f{2y K_\mu K_\nu}{\[ y(K^2+i\varepsilon) + (1-y)((K-p)^2-m^2+i\varepsilon)\]^3} \nol
&= \int \f{d^4 K}{(2\pi)^4 i}\int^1_0 dy
   \f{2y K_\mu K_\nu}{\[K^2 - (1-y)2K\cdot p +i\varepsilon \]^3}
   \,.
\end{align}
This time we change variables using
\begin{align}
\kappa'_\mu = K_\mu - (1-y)p_\mu \,,
\end{align}
and
\begin{align}
{\kappa'}^2-(1-y)^2p^2 = K^2 - (1-y)2K\cdot p \,.
\end{align}
Thus we have
\begin{align}
{\cal I}^{(3)}_{\mu\nu} &= \int^1_0 dy 2y \int \f{d^4
\kappa'}{(2\pi)^4 i}
   \f{\(\kappa'_\mu+(1-y)p_\mu\)\(\kappa'_\nu+(1-y)p_\nu\)}{\[{\kappa'}^2-(1-y)^2p^2+i\varepsilon
   \]^3} \,.
\end{align}
The current situation is analogous to the calculation for ${\cal
I}^{(2)}_{\mu\nu}$. After changing the notation as $\kappa'\to K$,
we can again remove the terms proportional to $K_\mu$ and $K_\nu$.
Similarly, we again require only the $\mu=0,\,\nu=i$ terms, and thus
can also remove the $K_0K_i$ term. The remainder gives
\begin{align}
{\cal I}^{(3)}_{0i} &= \int^1_0 dy 2y(1-y)^2 E_p p_i\int \f{d^4
K}{(2\pi)^4 i}
   \f{1}{\[K^2-(1-y)^2m^2 +i\varepsilon\]^3} \,.
\end{align}
Rotating the $K_0$ coordinate, we produce
\begin{align}
{\cal I}^{(3)}_{0i} &= -\int^1_0 dy 2y(1-y)^2 E_p p_i\int \f{d^4
K_E}{(2\pi)^4 i}
   \f{1}{\[K_E^2+(1-y)^2m^2\]^3} \nol
&= -\int^1_0 dy 2y(1-y)^2 E_p p_i\int \f{d (K_E)^2}{(2\pi)^4}
   \f{\pi^2 K_E^2}{\[K_E^2+(1-y)^2m^2\]^3} \nol
&= -\int^1_0 dy 2y(1-y)^2 E_p p_i
   \f{\pi^2}{(2\pi)^4(1-y)^2m^2}\f{\Gamma(1)\Gamma(2)}{\Gamma(3)} \nol
&= -\f{\pi^2E_p p_i}{(2\pi)^4m^2}\int^1_0 dy y
   \nol
&= -\f{1}{2}\f{\pi^2E_p p_i}{(2\pi)^4m^2} \,.
\end{align}

\subsubsection{Evaluation}
Collecting together the results, we have
\begin{align}
{\cal I}^{(1)}_0 &= -\f{\pi^2 E_p}{(2\pi)^4m^2}
   \,, \nol
{\cal I}^{(2)}_{0i} &= -\f{1}{2}\f{\pi^2 E_p p_i}{(2\pi)^4m^2} \,,
\nol {\cal I}^{(3)}_{0i} &= -\f{1}{2}\f{\pi^2E_p p_i}{(2\pi)^4m^2}
\,.
\end{align}
We recall that
\begin{align}
{\cal I}_K &= \f{4}{E_p} \[ -{\cal I}^{(1)}_0 + \sum_i
\f{p_i}{{\bf p}^2}\( {\cal I}^{(2)}_{0i} +2 {\cal I}^{(3)}_{0i} \)
\] \,.
\end{align}
Thus
\begin{align}
{\cal I}_K &= -\f{2\pi^2 }{(2\pi)^4m^2}
 \,.
\end{align}
The correction term for the forward scattering is then given by
\begin{align}
{\cal F}^R({\bf p}) &= \f{e^2m^2}{p_0\hbar^2} \int dt
   |\phi_p(t)|^2 \f{\hbar}{2m}\, \itbd{\xi}\cdot{\bf p}\times\dot{\bf
   p} \,{\cal I}_K \nol
&= -\f{e^2}{16\pi^2p_0m\hbar} \int dt
   |\phi_p(t)|^2 \, \itbd{\xi}\cdot{\bf p}\times\dot{\bf
   p} \,. \label{FRcorrection}
\end{align}
With this simplified expression, we can now proceed to consider the
interpretation of the correction.

\subsection{Vertex correction}
The external potential is regarded as classical, but is coupled to
the spinor field via the interaction term $\gamma^\mu V_\mu$ and
this vertex has an associated one-loop correction. The one-loop
process thus not only alters the propagator, but also the
interaction with the external field. This correction is well known
and responsible for the anomalous magnetic moment.\footnote{The
anomalous magnetic moment at the one-loop level for QED was first
derived by Schwinger \cite{Schwinger}. It is currently the most
accurately tested and confirmed prediction in the history of
physics.} Our external potential, although considered
non-perturbatively, acts like a minimally substituted
electromagnetic external field. As such, whilst the one-loop
corrections to the vertices for the interaction between the
electromagnetic field and the spinor field in the emission and
one-loop forward scattering diagrams are of higher order in $e^2$
than those with which we are concerned and thus ignored, the
external potential $V$ is still coupled to the spinor field via the
term $\gamma^\mu V_\mu$. We stress that this effect is simply one
that the current one-loop forward scattering process has on the
external field coupling, and not an additional process which we are
now adding. We shall show that the one-loop correction to the vertex
is entirely responsible for the `correction' term ${\cal F}^R$ which
we have found.

As we are simply interpreting ${\cal F}^R$, and not deriving the
vertex correction from scratch, it will be sufficient to quote some
of the relevant theory. The renormalised vertex is given
by\footnote{See for example ($7-54$) on p340 of Itzykson and Zuber's
Quantum Field Theory
\cite{Itzykson}. The theory discussed briefly here is given in
more detail in
\cite{Itzykson} in p340-341 and p347 in particular.}
\begin{equation}
\Gamma_\rho^R(p',p) = \gamma_\rho F_1(q^2) + \frac{i}{2m}\sigma_{\rho
\nu} q^\nu F_2(q^2)\,,
\end{equation}
where $F_i$ are the form factors, the evaluation of which for our
circumstances is given shortly; $p,p'$ are the momenta before and
after the vertex and $q$ is the momentum transfer $p'-p$. We also
have $\sigma^{\rho \nu} = \frac{i}{2}[\gamma^\rho,\gamma^\nu]$. For
the coupling of the spinor field to the external potential,
\begin{equation}
V^\rho(x) = \int \frac{d^4q}{(2\pi)^4}V^\mu(q)e^{-iq\cdot x}\,,
\end{equation}
we have
\begin{equation}
\partial^\nu V^\rho(x) = \int \frac{d^4q}{(2\pi)^4}
(-iq^\nu) V^\mu(q)e^{-iq\cdot x}\,.
\end{equation}
The interaction of the classical potential can be regarded as taking
place is the so-called quasi-static limit, $q\to 0$ and from the
above, the momentum transfer is replaced by the derivative operator.
Consequently, the coupling term $\overline{\psi}\gamma_\rho
\psi\,V^\rho$ changes to
\begin{equation}
\overline{\psi}\gamma_\rho \psi\,F_1(-\hbar^2\partial^2)V^\rho
-
\frac{\hbar}{2m}\overline{\psi}\sigma_{\rho\nu}\psi
F_2(-\hbar^2\partial^2)
\partial^\nu V^\rho\,.
\end{equation}
We are only interested in the lowest $\hbar$ order (renormalised)
vertex corrections and it can be shown\footnote{This calculation is
performed in the referenced pages in \cite{Itzykson} in the previous
footnote.} that in our limit we thus obtain
\begin{equation}
\overline{\psi}\gamma_\rho \psi\,F_1(0)V^\rho
-
\frac{\hbar}{2m}\overline{\psi}\sigma_{\rho\nu}\psi
F_2(0)
\partial^\nu V^\rho
= \overline{\psi}\gamma_\rho \psi\,V^\rho -
\frac{\alpha_c}{4\pi m}\overline{\psi}\sigma_{\rho\nu}\psi\,
\partial^\nu V^\rho\,. \label{vertexterms2}
\end{equation}
We recall that $\alpha_c = e^2/4\pi$ and consequently, the second
term of (\ref{vertexterms2}) gives our correction term as a result
of including the renormalised vertex:
\begin{equation}
- \frac{e^2}{16\pi^2 m}\overline{\psi}\sigma_{\rho\nu}\psi\,
\partial^\nu V^\rho\,. \label{vertexcorrection}
\end{equation}
It is this correction term in which we are interested. As in the
case of the mass counter term, we could regard this correction as an
interaction in the Lagrangian producing a Feynman diagram
contribution. We can then calculate the contribution of this term
towards the forward scattering.\footnote{This interaction term is
\emph{part} of the forward-scattering already considered. We aim
here to show that it is this part which is solely responsible for
${\cal F}^R$.}

Now, our external potential has only space components and they are
only dependent on time $t$. The term (\ref{vertexcorrection}) can
thus be rewritten as
\begin{equation}
 \frac{e^2}{16\pi^2 m}\overline{\psi}\sigma_{k 0}\psi\,\dot{V}_k
=-\frac{e^2}{16\pi^2 m}\overline{\psi}\sigma_{k
0}\psi\,\dot{\tilde{p}}_k\,,
\end{equation}
Using the definition $\sigma^{\rho \nu} =
\frac{i}{2}[\gamma^\rho,\gamma^\nu]$, we find that this term is
equal to
\begin{align}
\f{ie^2}{16\pi^2 m} \(\psi^\dagger \itbd{\gamma} \psi
\) \cdot \dot{\tilde{\bf p}}
\end{align}
If we continue analogously to the calculation of the contribution of
the counter term, then the contribution from this vertex correction
term towards the forward scattering amplitude, ${\cal F}_{\rm Vtx}$,
can be written
\begin{align}
{\cal F}_{\rm Vtx}({\bf p}) & =  - \frac{1}{\hbar}
\int \frac{d^3{\bf p}'}{(2\pi\hbar)^3}\frac{m}{p_0}
\int d^4 x \langle 0|b_\alpha({\bf p}')
\left(\frac{ie^2}{16\pi^2 m} \( \psi^\dagger(x)\itbd{\gamma} \psi(x)\,\)\cdot \dot{\tilde{\bf p}}
\right) b_\alpha^\dagger({\bf p})|0\rangle\nonumber \\
& = - \frac{ie^2}{16\pi^2 \hbar p_0}\int dt\, |\phi_p(t)|^2
\[ u^\dagger_\alpha (p)\itbd{\gamma} u_\alpha(p)\,\]\cdot
\dot{\tilde{\bf p}}\,. \label{Fvertex1}
\end{align}
All momenta in the integrand are the time dependent
$\tilde{p}=(E_p,{\bf p}-{\bf V}(t))$ and so we drop the tilde
notation.\footnote{This change of notation, performed for the
purpose of simplicity and legibility, was also enacted in the main
forward scattering calculation at a similar point and is thus also
needed here for the purposes of comparison.} The spinor combination
in the integrand of this equation can be straightforwardly
calculated as follows:
\begin{align}
& \[u^\dagger_\alpha(p)\itbd{\gamma} u_\alpha(p)\,\]
\cdot\dot{{\bf p}} \nol
& =
\frac{E_p+m}{2m}
\left(s_\alpha^\dagger\,\,s_\alpha^\dagger
\dfrac{\itbd{\sigma}\cdot {\bf p}}{E_p+m}\right)
\left( \begin{array}{cc} 0  & \sigma^k \\ -\sigma^k & 0
\end{array}\right)
\left( \begin{array}{c} s_\alpha \\ \dfrac{\itbd{\sigma}\cdot
{\bf p}}{E_p+m}s_\alpha\end{array}
\right)\dot{p}^k \nonumber \\
& =   \frac{1}{2m} s_\alpha^\dagger \left[ \sigma^k, \sigma^j\right]
s_\alpha p^j
\dot{p}^k \nonumber \\
& =  \frac{i}{m}
\epsilon^{nkj}s_\alpha^\dagger \sigma^n s_\alpha\,
p^j \dot{p}^k\nonumber \\
& =  \frac{i}{m}
\epsilon^{nkj}\xi^n p^j \dot{p}^k \nonumber \\
& =  - \frac{i}{m}
\itbd{\xi}\cdot({\bf p}\times \dot{\bf p})\,. \label{vertexcombo}
\end{align}
where $\epsilon^{nkj}$ is the usual Levi-Civita antisymmetric
symbol. Substituting (\ref{vertexcombo}) in to (\ref{Fvertex1}) we
find
\begin{equation}
{\cal F}_{\rm Vtx}({\bf p}) = - \frac{e^2}{16\pi^2 \hbar mp_0}\int
dt\,|\phi_p(t)|^2\,
\itbd{\xi}\cdot({\bf p}\times \dot{\bf p})\,.
\end{equation}
Comparison with (\ref{FRcorrection}) shows that
\begin{equation}
 {\cal F}^R({\bf p}) = {\cal F}_{\rm Vtx}({\bf p}) \,.
\end{equation}
We can consequently deduce that the correction to the forward
scattering and thus the subsequent correction to the position shift
are due to the renormalised one-loop vertex correction. Finally, we
conclude that the quantum position shift for the spinor field in the
$\hbar\to 0$ limit is given by
\begin{align}
\delta x = -\int_{-\infty}^0 dt\,{\cal F}^j_{\rm LD}\,\left(
\frac{\partial x^j}{\partial p^i}\right)_t
+\frac{e^2}{16\pi^2mp_0}\partial_{p^i} \int dt\,|\phi_p(t)|^2\,
\itbd{\xi}\cdot({\bf p}\times \dot{\bf p}) \,. \label{spinorshift}
\end{align}

\chapter{Summary and Conclusion}\label{summarychapter}
\begin{quote}
In this chapter we summarize the work which has been presented and
discuss the results of our investigations. We also discuss possible
avenues for future research on this topic.
\end{quote}

In this work we have investigated the effects of radiation reaction
in classical and quantum electrodynamics on the position of a
particle. We defined the position shift to be the change in position
due to the effects of radiation reaction and calculated this
quantity for the theories with which we were interested. The
equations of motion are a fundamental part of any theory, and the
observation of dynamics is likewise fundamental to our ability to
discern between rival theories and question our understanding. The
reader may recall from the introduction, that the phenomenon of
radiation reaction alters the usual equations of motion and
consequently an understanding of radiation reaction and its effects
lies at the heart of accurately understanding dynamics. One could
regard the predicted position, or predicted position expectation
value in order to fully include quantum theories, as one of the most
important predictions of a theory. The change in this prediction
after the addition of a new phenomenon, is consequently a sensible
choice of measure to use in order to help understand our theoretical
models.

The classical theory of radiation reaction is not without its
problems, both in implementation and especially in interpretation.
As we previously explained, most of these problems are related to
the third order nature of the resulting equations of motion. It has
been these difficulties, along with the recent renewed interest and
progress on radiation reaction in curved space, that have motivated
this work. Our aim has been to look at the effects of radiation
reaction in classical electrodynamics and to compare the results
with the predictions of the so-called classical limit of the more
fundamental quantum field theory. A knowledge of the similarities
and differences between the two approaches to radiation reaction,
and in turn the similarities and differences in the results of our
investigations will hopefully aid a fuller understanding of how this
phenomenon can be interpreted within our theoretical models. Given
the debate about the classical theory and its interpretation, the
natural question is whether the predictions of the quantum theory,
in the classical limit, are the same as those of classical
electrodynamics. This is one of the main questions that we sought to
answer in this work.

Our model consisted of a particle interacting with an external
potential for some finite period of time in the past of our
measurement. The position shift was defined as the change in
position between a hypothetic control particle which does not
undergo radiation reaction, and a test particle which does include
this effect. We refer the reader to the appropriate definitions of
the models in the main chapters for the full description. The aim
here is to recall these descriptions to mind. In classical
electrodynamics, we treated the Lorentz-Dirac force, the classical
radiation reaction force, as a perturbation. This is in keeping with
the reduction in order interpretations of the theory, with the
treatment of interactions in the perturbative description of quantum
field theory, and with the fact that the Lorentz-Dirac force is a
physically small effect. We demonstrated in Chapter
\ref{classicalshiftchapter} that the classical position shift can be
given by\footnote{The quantities and factors in the equations in
this chapter are those defined in the main sections of this work.
The equations numbers of the quoted results are the original
equations numbers in the work.}
\begin{equation}\tag{\ref{cshift}}
\delta x^i_C = -\int^0_{-\infty} dt {\cal F_{\rm LD}}^j \( \f{\partial
x^j}{\partial p^i} \)_t \,.
\end{equation}
This is a fairly short and simple expression and suggests a more
general rule in addition to the case of the Lorentz-Dirac force.
This is indeed the case and analysis of our working in Chapter
\ref{classicalshiftchapter} demonstrates that the above expression
can be used for the position shift of a general perturbative
force\footnote{For another force, ${\cal F_{\rm LD}}$ in
(\ref{cshift}) would of course need to be replaced by the equivalent
expression for the new force.} to a Hamiltonian system within our
model's set-up.\footnote{We again stress that our discussion here is
within the limitations of the model we defined in full in Chapter
\ref{introchapter}.}

Before tackling the quantum theories, we calculated the
semiclassical expansions for the scalar and spinor field in Chapter
\ref{Semiclassical}. This work was necessary in order to later
investigate the $\hbar\to 0$ limit of the quantum theories. The
semiclassical expansions of the mode functions are dependent on the
details of the acceleration due to the external potential and we
calculated these expansions in the cases of the time-dependent
(space-independent) potential, and only for the scalar field, in the
case of the potential dependent on one of the spatial coordinates.
These potentials were chosen due to the conditions for the validity
of the semiclassical expansions. This chapter provided the ground
work for our description of the quantum fields in the external
potential.

Our first investigation into the quantum effects of radiation
reaction used the theory of quantum scalar electrodynamics. The use
of the scalar field is a good starting point for studying the
quantum effects and a useful toy model for electrodynamics, without
the complications of spin which is also absent in the classical
model. We started our investigation with the calculation of the
position expectation value of a non-radiating scalar particle, given
by
\begin{equation}
\bra{i} x^i(0) \ket{i} = \frac{i\hbar}{2}
\int\frac{d^3{\bf p}}{(2\pi\hbar)^3} f^*({\bf p})
\stackrel{\leftrightarrow}{\partial}_{p_i} f({\bf p}) = 0 \quad \forall
i=1,2,3 \,, \tag{\ref{posexpi}}
\end{equation}
where we recall that $f$ is heuristically to be regarded as the
one-particle wave function. This then served as our control
particle. We proceeded to calculate the position expectation value
of a particle which has undergone radiation reaction during the
period of acceleration and compared these two results. To order
$e^2$, we found that there are two main processes contributing to
the position shift. These are the emission and the forward
scattering, which in turn come from the one photon and zero photon
final states respectively:
\begin{align}
\delta x^i_{\rm em} &= -
\frac{i}{2}
\int \frac{d^3{\bf k}}{2k(2\pi)^3}
{\cal A}^{\mu *}({\bf p},{\bf k})
\stackrel{\leftrightarrow}{\partial}_{p_i} {\cal A}_\mu({\bf p},{\bf
k}) \,, \tag{\ref{deltaxem}} \\
\tag{\ref{deltaxfor}}
 \delta x^i_{\rm for} &= -\hbar \partial_{p_i} \Re {\cal F}({\bf p})
\,,
\end{align}
written in terms of the emission amplitude ${\cal A}$ and the
forward scattering amplitude ${\cal F}$. The semiclassical
expansions of the mode functions, described in Chapter
\ref{Semiclassical} enabled us to calculate the amplitudes for
these processes in the `classical' $\hbar\to 0$ limit. For the
emission amplitude we performed the calculation in the case of a
time-dependent (space-independent) external potential and also in
the case of a potential dependent on only one of the spatial
coordinates. These two cases gave the same result, namely
\begin{equation}
{\cal A}^\mu({\bf p},{\bf k})  =  -e \int_{-\infty}^{+\infty} d\xi\,
\frac{dx^\mu}{d\xi}\chi(\xi)\,e^{ik\xi}\,,\tag{\ref{cut-off}}
\end{equation}
Calculation of the resulting position shift due to the emission
process produced
\begin{equation} \tag{\ref{qemshift}}
\delta x^i_{\rm em} = - \int^0_{-\infty}
dt\, {\bf {\cal F}}^{j}_{\rm LD} \pd{x^j}{p^i}{t} \,.
\end{equation}
In other words the position shift due to the emission process in the
$\hbar\to 0$ limit of quantum scalar electrodynamics is equal to the
classical position shift. Any difference between the classical and
quantum measurements would thus need to arise from the forward
scattering effects.

We thus proceeded to calculate the position shift due to forward
scattering, for the case of the time-dependent potential. The
forward scattering amplitude results from the one-loop interaction
and we calculated the divergent contribution from these effects.
However, this divergence was then subsequently found to be cancelled
by the contribution from the counter term due to the renormalisation
of the mass. These divergent expressions were both of order
$\hbar^{-2}$ and thus, from the formula shown above in
(\ref{deltaxfor}), they would contribute at order $\hbar^{-1}$ to
the position shift. All remaining contributions from the forward
scattering amplitude were shown to be imaginary at order
$\hbar^{-1}$ in ${\cal F}$ and thus at order $\hbar^0$ in the
position shift. The reader will recall that only the real part of
the forward scattering amplitude is present in the position shift
formula. As a result the position shift contribution due to forward
scattering is zero. We can consequently conclude that the quantum
position shift for this model is equal to the classical position
shift. This would at first sight appear to imply that there are no
differences in the treatment of radiation reaction between the
classical and quantum theories, at least within the confines of the
models and in the $\hbar\to 0$ limit. However, the theoretical paths
along which we travelled for these calculations have significant
differences. In order to further expand on the similarities and
differences, we turned to an alternative, but equivalent,
description of the radiation reaction effect based on the Green's
functions of the electromagnetic field (Chapter
\ref{greenschapter}).

The key to the Green's function description of radiation reaction is
the decomposition of the particle's retarded electromagnetic field
into `regular' and `singular' components. The retarded Green's
function, $G_-$, is decomposed into the regular and singular Green's
functions, given respectively as
\begin{align}
G_R=\f{1}{2} \[ G_- - G_+ \] \,, \tag{\ref{GReg}}\\
G_S=\f{1}{2} \[ G_- + G_+ \] \,, \tag{\ref{GSing}}
\end{align}
where $G_+$ is the advanced Green's function. The singular field is
regarded as a generalisation of the Coulomb field for a static
particle and is similarly singular (hence the name) on the world
line of the particle. The regular Green's function, which solves the
homogeneous wave equation, has been shown to be entirely responsible
for the radiation reaction effect.\footnote{See, for example,
\cite{Poisson}.} In Chapter
\ref{greenschapter} we showed that the emission contribution to
the position shift can be rewritten as
\begin{align}
\delta x_{\rm em} & = \int d^4x' d^4x''\,
\partial_{p^i} j_{\rho''}(x) G_R^{\rho''\nu'}(x-x')j_{\nu'}(x')\nonumber \\
  & = \int
d^4x\, \partial_{p^i}j^\mu(x)A_{{\rm R}\mu}(x)
\,, \tag{\ref{standard0}}
\end{align}
i.e. in terms of the regular Green's function or regular field,
$A_R$. As we demonstrated that the quantum position shift was
entirely due to the emission amplitude contribution, the appearance
of $A_R$ sheds some light on the connection between the theories.
The calculation leading to (\ref{standard0}) started from the
formula for the emission contribution to the position shift
(\ref{deltaxem}) and the emission amplitude obtained using the
semiclassical expansion (\ref{cut-off}). We demonstrated in the
scalar work that this amplitude result was obtained for a potential
dependent on only one of the space-time coordinates. The position
shift result will, however, hold for any potential that can be shown
to produce the amplitude (\ref{cut-off}). The limitations of the
semiclassical expansion dictated the use of the potentials
mentioned, but given the form of the amplitude and its relation to
the amplitude for a classical field, one would expect that the
result may well be true for more general potentials, should an
appropriate semiclassical method be applied. This possibility poses
a question for future work.

Some major differences between the classical and quantum approaches
come from the analysis of the forward scattering contribution, which
however does still contain similarities. In the classical theory,
the singular field is regarded as an infinite correction to the
mass, and thus removed in a process of renormalisation. Thus in
fact, the classical theory involves a divergent self-energy `forward
scattering' effect, removed by mass renormalisation, a process not
normally associated with classical theories. Many people would think
only of quantum theories when hearing the word renormalisation, but
in both theories the mass renormalisation can be considering as
arising from an infinite self-interaction effect. So far we have
talked of the similarities. However, the quantum self-interaction as
described by the one-loop process involves effects not present in
the classical theory at all, such as contributions from the virtual
antiparticles. In the calculation of the forward scattering in
Chapter
\ref{scalarshiftchapter} we decomposed the forward scattering into
particle and antiparticle loop contributions. The particle loop was
further decomposed when we analysed the low photon energy portion of
the loop. It is this effect which is analogous to that present in
the classical theory - the high photon energy limit and the
antiparticle loop processes do not have classical counterparts. As
this description hints, we found in Chapter
\ref{greenschapter} that the low photon energy particle loop
contribution can be rewritten in terms of the singular field $A_S$
to produce
\begin{equation}
-\hbar \partial_{p^i} {\rm Re} \, {\cal F}^{<}({\bf p}) = \int
d^4x\,
 \partial_{p^i} j^\mu (x) A_{S\mu}(x) \,, \tag{\ref{singcont}}
\end{equation}
which is an analogous expression to that for the emission
contribution. We thus see that those elements of the quantum process
with classical counterparts effectively give the same results as
found in the classical theory. The total forward scattering
contribution is zero after mass renormalisation and it is clear that
the quantum mass renormalisation is not the same as the classical
case, but naturally renormalises all the quantum contributions
including the antiparticle loop.

The results in Chapter \ref{greenschapter} do not change those of
the previous chapter. They are simply a rewriting of some parts of
the calculation in terms of different quantities. The results do
however give a clearer picture of the similarities and differences
between the classical and quantum treatment of radiation reaction.

Having succeeded in comparing the classical position shift with the
scalar quantum position shift, we turned our attention to the more
accurate quantum model of the spinor field. It is this field which
is used in the standard theory of quantum electrodynamics and we
thus repeated our investigation for the spinor QED. We investigated
the case of the spinor wave packet having travelled through a
time-dependent potential. The position expectation value of the
control particle was found, to order $\hbar^0$, to give the same
expression as previously found in the case of the scalar field:
\begin{align} \tag{\ref{nonradpev2}}
 \langle x^i \rangle |_{t=0} = \frac{i\hbar}{2}
\int\frac{d^3{\bf p}}{(2\pi\hbar)^3} f^*({\bf p})
\stackrel{\leftrightarrow}{\partial}_{p_i} f({\bf p}) \,.
\end{align}
However, as an aside, we did note that spin effects, related to
spin-orbit coupling, can be observed to have an effect at order
$\hbar$. Such spin effects would still not however alter our
measured position shift due to their presence in both the control
and test particle calculations. The addition of the phenomenon of
spin to the model is one major difference between the classical
electrodynamics theory and spinor QED, although it should be noted
that the spin is present in the Dirac equation for the spinor field
and is not technically a result of the quantisation of that field.
Proceeding, we found that as per the scalar field, the contributions
to the position shift can be split into emission and forward
scattering contributions. In fact, the formula for the position
shift, written in terms of the amplitudes for these processes was
calculated to give the same answer as the scalar field:
\begin{align}\tag{\ref{emshift}}
 \delta x^i_{\rm em} =& -\f{i}{2} \int \f{d^3 {\bf k}}{(2\pi)^3 2k_0}
{\cal A}_\mu^*({\bf p},{\bf k})\stackrel{\leftrightarrow}{\partial}_{p_i}{\cal A}^\mu({\bf p},{\bf k}) \\
\delta x^i_{\rm for} =& -\hbar\partial_{p_i}\Re{\cal F}({\bf p})
\,. \tag{\ref{forshift}}
\end{align}
Again, the effects of the spin, including any spin transport effects
producing a difference between the initial and final spin states,
did not come into play at lowest order.

Despite the more complicated nature of the Dirac spinor field in
comparison with the scalar field, the interaction Hamiltonian for
the spinor field
\begin{align}\tag{\ref{spinorHamInt}}
{\cal H}_I = e : \bar{\psi}\slashed{A}\psi : \,,
\end{align}
is simpler than the scalar case. We used the spinor interactions and
semiclassical spinor solutions to proceed to calculate the emission
and forward scattering amplitudes. Many of the features of these
calculations were analogous to those found in the previous scalar
work. In fact, the result of the emission amplitude calculation in
the $\hbar\to 0$ limit gave the same result as the emission
amplitude for the scalar field, viz
\begin{align}
{\cal A}^\mu (p,k) =& -e \int d\xi \f{d x^\mu}{d\xi}
\chi(\xi) e^{ik\xi} \,.
\end{align}
Using either the direct calculation in Chapter
\ref{scalarshiftchapter} or the Green's function decomposition
method in Chapter \ref{greenschapter}, we arrive at the result that
the emission contribution to the quantum position shift for the
spinor field is equal to the classical position
shift.\footnote{Throughout this summary, we imply the $\hbar\to 0$
limit when talking about the quantum position shift results.} So far
the result of changing field has merely been to change the
intermediate calculations, rather than the final result. The pattern
of the calculation for the forward scattering amplitude for the
spinor field was initially similar to that for the scalar case. For
example, the particle and antiparticle loop contributions were
calculated and for the case of the particle loop it was again
necessary to check the $\hbar$ order of the infrared divergences and
analyse the low-energy contribution. The renormalisation of the mass
via the counterterm again removed the order $\hbar^{-1}$
contributions to the position shift. However, unlike the scalar
case, we had to additionally consider order $\hbar$ terms in the
semiclassical expansion. As the forward scattering amplitude is at
order $\hbar^{-2}$ and thus its contribution to the position shift
at order $\hbar^{-1}$, any order $\hbar$ effects in the expansion
potentially contribute at order $\hbar^0$ to the position shift and
consequently remain when the classical limit is taken. After
renormalisation, some of these terms in the amplitude were indeed
still present and after some integration and simplification we were
able to show that the `correction' to the forward scattering
amplitude is given by
\begin{align}
{\cal F}^R({\bf p}) &= -\f{e^2}{16\pi^2p_0m\hbar} \int dt
   |\phi_p(t)|^2 \, \itbd{\xi}\cdot{\bf p}\times\dot{\bf
   p} \,. \tag{\ref{FRcorrection}}
\end{align}
This result then gives an additional contribution to the position
shift when compared with either the quantum scalar case or the
classical result. The position shift for the spinor quantum position
shift was given at the end of Chapter
\ref{spinorshiftchapter} by
\begin{align}
\delta x = -\int_{-\infty}^0 dt\,{\cal F}^j_{\rm LD}\,\left(
\frac{\partial x^j}{\partial p^i}\right)_t
+\frac{e^2}{16\pi^2mp_0}\partial_{p^i} \int dt\,|\phi_p(t)|^2\,
\itbd{\xi}\cdot({\bf p}\times \dot{\bf p}) \,. \tag{\ref{spinorshift}}
\end{align}
The interpretation of this extra term was analysed in the last
subsection of that chapter. We found that the correction was
entirely the result of the renormalised vertex correction that is
produced by the one-loop process. In other words it is the result of
the correction to the coupling to the external field produced at the
one-loop level. As we described in our discussion at the end of
Chapter \ref{spinorshiftchapter}, the vertex correction is
responsible for the well known anomalous magnetic moment. This
correction is not simply a result of the addition of spin. The spin
is present in the Dirac equation prior to quantisation and produces
the prediction that the g-factor of the magnetic moment\footnote{The
magnetic moment due to the intrinsic angluar momentum from the spin
${\bf s}$ is given by $\mu =-g \,e\, {\bf s}/(2m)$.} is equal to
$2$. The anomalous magnetic moment is the correction to the g-factor
due to quantisation, starting at the one-loop level. The effect to
the position shift noted above is of similar origin.

Given the correction produced by quantisation the natural question
to ask is whether or not this effect can be measured. This is of
course an interesting question, and indeed any measurements
improving the accuracy of tests of radiation reaction would be
beneficial to our understanding of the phenomenon. The smallness of
the radiation reaction force was stated as one of the reasons that
it was frequently ignored and it is also one which hampers accurate
testing of the theories. It is also worth noting however, that the
purpose of the work presented here was to study the classical limit
of QED. This naturally leads us to consider further work and the
possibility of analysing the effect at higher orders in $\hbar$.
Indeed Higuchi and Walker are currently investigating the $\hbar$
correction to the Larmor formula
\cite{HiguchiWalker}, which would have some influence on such an extension to
this work. In addition, spin effects can be shown to be orders of
magnitude larger than the self-force at low-energies (see for
example
\cite{Hammond}) and consequently they should be considered when
predictions for possible experiments are made. Such investigations
thus present a natural extension for future investigation and would
aid the understanding of the current results by adding additional
context. They would also require further investigation into the
semiclassical expansion at higher orders, or an alternative method
for such expansions to include other more general external
potentials.

Additional directions in which this work can be extended include the
possibility of investigating quantum radiation reaction in a curved
space setting. In the introduction in Chapter
\ref{introchapter}, we presented a brief summary of the theory of
radiation reaction in curved space and noted that this is an area of
great current interest. Much of the interest is focused on the
effects of gravitational radiation reaction. It would be of great
interest to extend the current work to consider quantum
electromagnetic radiation reaction in curved space. In curved space
this could be linked to investigations of the radiation produced by
the expansion of space-time (see for example \cite{Yamamoto}).
Further work could then attempt to grapple with a quantum treatment
of gravitational radiation and gravitational radiation reaction. Due
to the fact that the self-force is fundamental to our full
understanding of dynamics and even on a classical level involves
many of the concepts which usually define the complications of
quantum theories, such as self-interaction and renormalisation, it
may provide a useful avenue in which to obtain further knowledge of
quantum fields in curved space and ultimately, signals towards the
ever elusive theory of quantum gravity. For the purposes of working
in curved space, the Green's function decomposition approach may
well be more suited to adaptation for curved space given the methods
used in both classical radiation reaction in curved
space\footnote{See the earlier introduction and the much more
detailed review by Poisson in
\cite{Poisson2}.} and also in the treatment of quantum fields in
curved space.\footnote{See, for example, \cite{Wald}.}

The main focus of future work is therefore to build on the work
presented here, using it as a base upon which to generalise the
results presented. The generalisations mentioned above and in the
main text include extensions to higher orders in $\hbar$, extensions
to more general external potentials and extensions to curved space
and radiation reaction in other fields. The work presented here has
provided a solid base for future investigation and has given us new
insight into the similarities and differences between the classical
and quantum treatments of radiation reaction. The author hopes that
the reader has found this report to be interesting, to answer the
some of questions posed about radiation reaction, and perhaps to
advance further questions in the reader's mind to be answered in
future.

\appendix

\chapter{Semiclassical Spinor Identities} \label{spinoridappendix}

In this appendix we derive a set of identities for combinations of
the time-dependent semiclassical spinors, expanded up to ${\cal
O}(\hbar)$. These identities can then be used, for example, in the
evaluation of the spinor combinations in the forward scattering
loops.

\section{Summary of semiclassical expansions}
Firstly we quote the semiclassical spinors derived in the
semiclassical chapter. For the particle spinors $u$, all
energy-momenta are $\tilde{p}=(E_p,\tilde{\bf p})$ where
\begin{align}
\tilde{\bf p} &= {\bf p}-{\bf V}(t) & E_p&=\sqrt{\tilde{\bf
p}^2+m^2} \,,
\end{align}
and for the antiparticle spinors $v$, we have
$\tilde{p}_+=(E_{p_+},\tilde{\bf p}_+)$ where
\begin{align}
\tilde{\bf p}_+ &= {\bf p}+{\bf V}(t) & E_p&=\sqrt{\tilde{\bf
p}_+^2+m^2} \,.
\end{align}
With this in mind, when there is no ambiguity, we drop the momenta
subscripts. We recall
\begin{align}
u_\alpha(p,t) &= \sqrt{\f{E+m}{2m}} \[ \(1+i\hbar g\) \begin{pmatrix} Us_\alpha \\
\Sigma U s_\alpha
\end{pmatrix}  -i\hbar \f{E+m}{\(2E_p\)^2} \begin{pmatrix} -\Sigma \dot{\Sigma} U s_\alpha \\
\dot{\Sigma} U s_\alpha \end{pmatrix} \] \,,
\end{align}
\begin{multline}
\bar{u}_\alpha(p,t) = \sqrt{\f{E+m}{2m}} \[ \(1-i\hbar
g\)\begin{pmatrix} s^\dagger_\alpha U^\dagger & -s^\dagger_\alpha
U^\dagger \Sigma
\end{pmatrix}  \right. \\ \left. +i\hbar \f{E+m}{\(2E_p\)^2} \begin{pmatrix}
-s^\dagger_\alpha U^\dagger \dot{\Sigma} \Sigma & -s^\dagger_\alpha
U^\dagger \dot{\Sigma} \end{pmatrix} \] \,,
\end{multline}
\begin{align}
v_\alpha(p,t) &= \sqrt{\f{E+m}{2m}} \[ \(1-i\hbar g\) \begin{pmatrix} \Sigma U s_\alpha \\
Us_\alpha
\end{pmatrix}  +i\hbar \f{E+m}{\(2E_p\)^2} \begin{pmatrix} \dot{\Sigma} U s_\alpha \\
 -\Sigma \dot{\Sigma} U s_\alpha\end{pmatrix} \] \,,
 \end{align}
\begin{multline}
\bar{v}_\alpha(p,t) = \sqrt{\f{E+m}{2m}} \[ \(1+i\hbar g\)\begin{pmatrix}
s^\dagger_\alpha U^\dagger \Sigma & -s^\dagger_\alpha U^\dagger
\end{pmatrix} \right. \\ \left. -i\hbar \f{E+m}{\(2E_p\)^2} \begin{pmatrix}
s^\dagger_\alpha U^\dagger \dot{\Sigma} & s^\dagger_\alpha U^\dagger
\dot{\Sigma} \Sigma \end{pmatrix} \] \,,
\end{multline}
with the following (full) notation:
\begin{align}
\Lambda_p(t) &= \f{\itbd{\sigma}\cdot\tilde{\bf p}(t)\times\dot{\tilde{{\bf p}}}}{E_p+m} \,, \\
U_p(t) &= T \( \exp \[ -i\int^t_0 d\tau
\f{\Lambda_p(\tau)}{2E_p(\tau)} \] \) \,, \\
\Sigma_p(t) &= \f{\itbd{\sigma}\cdot\tilde{\bf p}}{E_p+m} \,, \\
\dot{\Sigma}_p(t) &= \f{\itbd{\sigma}}{E_p+m}\cdot \( \dot{\tilde{{\bf p}}}(t) - \f{\dot{E}_p) \tilde{\bf p}}{E_p+m}\) \,,
\\
g_p(t) &= \int^t_{0} d\tau
\f{\dot{\tilde{p}}^2(\tau)}{8E_p^3(\tau)} \,,\\
\end{align}
and similarly for $\tilde{p}_+$. We note that $\Lambda$, $\Sigma$
and $\dot{\Sigma}$ are Hermitian, whilst $U$ is unitary.

\section{Summary of useful identities}\label{Sigmaids} The following are identities
involving some of the terms above which are useful for the
calculation of the spinor identities. For the purpose of the summary
we use the time-dependent energy momentum $(E,{\bf p})$.
\begin{align}
\Sigma^2 &= \f{E-m}{E+m} \,, \\
\Sigma\dot{\Sigma} &=
\f{m\dot{E}}{(E+m)^2}+\f{i\itbd{\sigma}\cdot{\bf p}\times\dot{\bf
p}}{(E+m)^2} \,, \\
\Sigma\dot{\Sigma}\Sigma &= \f{\dot{E}\itbd{\sigma}\cdot{\bf
p}}{(E+m)^2} - \f{\(E-m\)\itbd{\sigma}\cdot\dot{\bf p}}{(E+m)^2} \,, \\
\dot{\Sigma}+ \Sigma\dot{\Sigma}\Sigma &=
\f{2m\itbd{\sigma}\cdot\dot{\bf p}}{(E+m)^2} \,,
\end{align}
\begin{align}
\Sigma \itbd{\sigma} &= \f{{\bf p}}{E+m}
+\f{i\itbd{\sigma}\times{\bf p}}{E+m} \,, \\
\itbd{\sigma} \Sigma &= \f{{\bf p}}{E+m}
-\f{i\itbd{\sigma}\times{\bf p}}{E+m} \,, \\
\Sigma \itbd{\sigma} \Sigma &= \f{2{\bf p} \itbd{\sigma}\cdot{\bf
p}}{(E+m)^2} - \f{{\bf p}^2 \itbd{\sigma}}{(E+m)^2} \,, \\
\dot{\Sigma} \Sigma \itbd{\sigma} \Sigma &=
\f{{\bf p}\cdot\dot{\bf p} {\bf p} - {\bf p}^2 \dot{\bf p}}{(E+m)^3}
-\f{2i \itbd{\sigma}\cdot{\bf p}\times\dot{\bf p} {\bf p}}{(E+m)^3}
- \f{i{\bf p}^2}{(E+m)^3} \itbd{\sigma}\times \( \dot{\bf
p}-\f{\dot{E} {\bf p}}{E+m} \) \,.
\end{align}

\section{Zeroth order spinor identities}
Below are the standard zeroth order spinor identities showing the
normalisation we have used for the spinors:
\begin{align}
\bar{u}_\alpha^{(0)} (p) u_\alpha^{(0)} (p) &= 1 \,, \\
\bar{u}_\alpha^{(0)} (p) \gamma^0 u_\alpha^{(0)} (p) &= \f{E}{m} \,, \\
\bar{u}_\alpha^{(0)} (p) \itbd{\gamma} u_\alpha^{(0)} (p) &=
\f{{\bf p}}{m} \,, \\
u_\alpha^{(0)} (p) \bar{u}^{\alpha (0)} (p) &= \f{\gamma\cdot p
+m}{2m} \,.
\end{align}

\begin{align}
\bar{v}_\alpha^{(0)} (p) v_\alpha^{(0)} (p) &=
 -1 \,, \\
\bar{v}_\alpha^{(0)} (p) \gamma^0 v_\alpha^{(0)} (p) &= \f{E}{m} \,, \\
\bar{v}_\alpha^{(0)} (p) \itbd{\gamma} v_\alpha^{(0)} (p) &=
\f{{\bf p}}{m} \,, \\
v_\alpha^{(0)} (p) \bar{v}^{\alpha (0)} (p) &= \f{\gamma\cdot p
-m}{2m} \,.
\end{align}

\section{Equal time spinor identities}
Here we present the identities to ${\cal O}(\hbar)$ for both the
particle (`positive energy') and antiparticle (`negative energy')
semiclassical spinors in turn, evaluated with the same
(time-dependent) momenta at equal time. The identities will be
useful in the calculation of the outer spinors in the combinations
found in the forward scattering contributions. All momenta are given
by $p$ at, say, time $\bar{t}$. Consequently for simplicity of
notation we shall drop the explicit $p$ subscripts and time
arguments. Similarly we can treat $s_\alpha(\bar{t}) = s_\alpha
U_p(\bar{t})$ as the two-spinor for the direction of the spin at
that time and, as all times are equal, drop the argument notation.

\subsection{Particle equal time
spinors}\label{partequaltimesection} The first set of identities are
for the particle spinors.
\subsubsection{}
\begin{align}
 & \quad \bar{u}_\alpha (p) u_\alpha (p) \nol &= \f{E+m}{2m}  \[ \(1-i\hbar g\) \begin{pmatrix}
s^\dagger_\alpha & -s^\dagger_\alpha \Sigma \end{pmatrix} \right.
\nol & \left. \qquad \qquad \qquad \times
\(
\( 1+i\hbar g\) \begin{pmatrix} s_\alpha \\ \Sigma s_\alpha
\end{pmatrix} -i\hbar \f{E+m}{(2E)^2} \begin{pmatrix} -\Sigma
\dot{\Sigma} s_\alpha \\ \dot{\Sigma} s_\alpha \end{pmatrix} \) \right. \nol
& \qquad \qquad \left. +i\hbar \f{E+m}{(2E)^2} \begin{pmatrix}
-s^\dagger_\alpha
\dot{\Sigma} \Sigma & -s^\dagger_\alpha
\dot{\Sigma} \end{pmatrix}  \begin{pmatrix} s_\alpha \\ \Sigma s_\alpha
\end{pmatrix}
\] \nol
&= \f{E+m}{2m} \[ s^\dagger_\alpha\(1-\Sigma^2\)s_\alpha -i\hbar
\f{E+m}{(2E)^2} s^\dagger_\alpha\( -\Sigma\dot{\Sigma} -
\Sigma\dot{\Sigma} + \dot{\Sigma}\Sigma + \dot{\Sigma}\Sigma
\)s_\alpha \] \nol
&= \f{E+m}{2m} \[ \f{2m}{E+m} -i\hbar \f{E+m}{2E^2}s^\dagger_\alpha
\(-\f{2i\Lambda}{E+m}\)s_\alpha \] \,, \nonumber
\end{align}
which leads to
\begin{align}
 \bar{u}_\alpha (p) u_\alpha (p) &= 1-\f{\hbar}{2m}\f{\itbd{\xi}\cdot{\bf
p}\times\dot{\bf p}}{E^2} \,.
\end{align}

\subsubsection{}\label{ug0uidsection}
\begin{align}
& \quad \bar{u}_\alpha (p)\gamma^0 u_\alpha (p)\nol  &= \f{E+m}{2m}
\[
\(1-i\hbar g \) \begin{pmatrix} s^\dagger_\alpha & s^\dagger_\alpha
\Sigma \end{pmatrix} \right. \nol & \qquad \qquad \qquad \times \left. \(
\( 1+i\hbar g\) \begin{pmatrix} s_\alpha \\ \Sigma s_\alpha
\end{pmatrix} -i\hbar \f{E+m}{(2E)^2} \begin{pmatrix} -\Sigma
\dot{\Sigma} s_\alpha \\ \dot{\Sigma} s_\alpha \end{pmatrix} \) \right. \nol
& \qquad \qquad \left. +i\hbar \f{E+m}{(2E)^2} \begin{pmatrix}
-s^\dagger_\alpha
\dot{\Sigma} \Sigma & s^\dagger_\alpha
\dot{\Sigma} \end{pmatrix} \begin{pmatrix} s_\alpha \\ \Sigma s_\alpha
\end{pmatrix}
\] \nol
&= \f{E+m}{2m} \[ s^\dagger_\alpha\(1+\Sigma^2\)s_\alpha -i\hbar
\f{E+m}{(2E)^2} s^\dagger_\alpha \(
-\Sigma\dot{\Sigma}+\Sigma\dot{\Sigma}+\dot{\Sigma}\Sigma-\dot{\Sigma}\Sigma
\)s_\alpha \] \nol
&= \f{E+m}{2m}\f{2E}{E+m}\,,  \nonumber
\end{align}
and hence we have
\begin{align}\label{ug0uid}
 \bar{u}_\alpha (p)\gamma^0 u_\alpha (p) &= \f{E}{m} \,.
\end{align}
We note that there is no order $\hbar$ term remaining.
\subsubsection{}
\begin{align}
& \quad \bar{u}_\alpha (p)\itbd{\gamma} u_\alpha (p)\nol  &=
\f{E+m}{2m}
\[
\(1-i\hbar g \) \begin{pmatrix} s^\dagger_\alpha\Sigma\itbd{\sigma}
& s^\dagger_\alpha \itbd{\sigma}
\end{pmatrix} \right. \nol & \qquad \qquad \qquad \times \left.
\(
\( 1+i\hbar g\) \begin{pmatrix} s_\alpha \\ \Sigma s_\alpha
\end{pmatrix} -i\hbar \f{E+m}{(2E)^2} \begin{pmatrix} -\Sigma
\dot{\Sigma} s_\alpha \\ \dot{\Sigma} s_\alpha \end{pmatrix} \) \right. \nol
& \qquad \qquad  \left. +i\hbar \f{E+m}{(2E)^2}
\begin{pmatrix} s^\dagger_\alpha \dot{\Sigma}\itbd{\sigma} &
-s^\dagger_\alpha
\dot{\Sigma}\Sigma\itbd{\sigma} \end{pmatrix}  \begin{pmatrix} s_\alpha \\ \Sigma s_\alpha
\end{pmatrix}
\] \,. \nonumber
\end{align}
Multiplying the matrices, we have
\begin{multline}
 \bar{u}_\alpha (p)\itbd{\gamma} u_\alpha (p) = \f{E+m}{2m} \[ s^\dagger_\alpha
\(
\Sigma\itbd{\sigma} +
\itbd{\sigma}\Sigma \) s_\alpha \right. \\ \left. -i\hbar \f{E+m}{(2E)^2}
s^\dagger_\alpha \( \dot{\Sigma}\Sigma\itbd{\sigma}\Sigma -
\Sigma\itbd{\sigma}\Sigma\dot{\Sigma} +
\itbd{\sigma}\dot{\Sigma}-\dot{\Sigma}\itbd{\sigma} \)s_\alpha \]
\,, \nonumber
\end{multline}
which using the identities in section \ref{Sigmaids} becomes
\begin{multline}
= \f{E+m}{2m} \[ \f{2{\bf p}}{E+m}  -i\hbar \f{E+m}{(2E)^2}
s^\dagger_\alpha \( -\f{4i \itbd{\sigma}\cdot{\bf p}\times\dot{\bf
p} {\bf p}}{(E+m)^3} \right. \right. \\ \left.
\left. -
\(\f{2i{\bf p}^2}{(E+m)^3}  + \f{2i}{E+m}\)
\itbd{\sigma}\times
\( \dot{\bf p}-\f{\dot{E} {\bf p}}{E+m}\) \)s_\alpha \] \,.
\end{multline}
Thus
\begin{align}
\bar{u}_\alpha (p)\itbd{\gamma} u_\alpha (p) &= \f{{\bf p}}{m} - \f{\hbar}{2mE} \( \f{\itbd{\xi}\cdot{\bf
p}\times\dot{\bf p}\,{\bf p}}{E(E+m)} +\itbd{\xi}\times \(\dot{\bf
p}-\f{\dot{E}{\bf p}}{E+m} \) \) \,.
\end{align}

\subsubsection{Inner spinors}
The last identity for the `particle' spinors we present here is used
for the inner spinors at equal time.
\begin{align}
& \quad  u_\alpha (p) \bar{u}^\alpha (p)\nol &= \sum_\alpha
\f{E+m}{2m}
\[ \(1+i\hbar g\) \begin{pmatrix} s_\alpha \\ \Sigma s_\alpha
\end{pmatrix} \right. \nol & \qquad \qquad \qquad \times \left.
\[ \(1-i\hbar g \) \begin{pmatrix} s^\dagger_\alpha &
-s^\dagger_\alpha \Sigma \end{pmatrix} + i\hbar \f{E+m}{(2E)^2}
\begin{pmatrix} -s^\dagger_\alpha \dot{\Sigma} \Sigma &
-s^\dagger_\alpha \dot{\Sigma} \end{pmatrix} \] \right. \nol &
\qquad \qquad \left. -i\hbar \f{E+m}{(2E)^2} \begin{pmatrix}
-\Sigma\dot{\Sigma}s_\alpha \\ \dot{\Sigma} s_\alpha \end{pmatrix}
\begin{pmatrix} s^\dagger_\alpha & -s^\dagger_\alpha \Sigma \end{pmatrix} \] \nol
&= \sum_\alpha \f{E+m}{2m}
\[ \begin{pmatrix} s_\alpha s^\dagger_\alpha & - s_\alpha
s^\dagger_\alpha \Sigma \\ \Sigma s_\alpha s^\dagger_\alpha &
-\Sigma s_\alpha s^\dagger_\alpha \Sigma \end{pmatrix} +i\hbar
\f{E+m}{(2E)^2}
\begin{pmatrix} -s_\alpha s^\dagger_\alpha \dot{\Sigma}\Sigma &
-s_\alpha s^\dagger_\alpha \dot{\Sigma} \\
-\Sigma s_\alpha s^\dagger_\alpha \dot{\Sigma}\Sigma & - \Sigma
s_\alpha s^\dagger_\alpha \dot{\Sigma} \end{pmatrix} \right.
\nol & \left. -i\hbar \f{E+m}{(2E)^2} \begin{pmatrix}
-\Sigma\dot{\Sigma}s_\alpha s^\dagger_\alpha & \Sigma\dot{\Sigma}
s_\alpha s^\dagger_\alpha \Sigma \\ \dot{\Sigma} s_\alpha
s^\dagger_\alpha & -\dot{\Sigma} s_\alpha s^\dagger_\alpha \Sigma
\end{pmatrix} \] \nol
&= \f{E+m}{2m} \[ \begin{pmatrix} I & -\Sigma \\ \Sigma & -\Sigma^2
\end{pmatrix} -i\hbar \f{E+m}{(2E)^2}
\begin{pmatrix} \dot{\Sigma}\Sigma-\Sigma\dot{\Sigma} & \dot{\Sigma}
+ \Sigma\dot{\Sigma}\Sigma \\ \dot{\Sigma} +
\Sigma\dot{\Sigma}\Sigma & \Sigma\dot{\Sigma}-\dot{\Sigma}\Sigma
\end{pmatrix} \] \nol
&= \f{1}{2m} \[ \begin{pmatrix} E+m & -\itbd{\sigma}\cdot{\bf p} \\
\itbd{\sigma}\cdot{\bf p} & -E+m \end{pmatrix} -i\hbar
\f{1}{(2E)^2} \begin{pmatrix} -2i\itbd{\sigma}\cdot{\bf
p}\times\dot{\bf p} & 2m\itbd{\sigma}\cdot\dot{\bf p} \\
2m\itbd{\sigma}\cdot\dot{\bf p} & 2i\itbd{\sigma}\cdot{\bf
p}\times\dot{\bf p} \end{pmatrix} \] \,. \nonumber
\end{align}
Using the gamma matrices, we can rewrite this equation as
\begin{align}
u_\alpha (p) \bar{u}^\alpha (p) &=
\f{\gamma\cdot p +m}{2m} +
\f{\hbar}{m(2E)^2} \gamma^5
\itbd{\gamma}\cdot{\bf p}\times\dot{\bf p} -\f{i\hbar}{(2E)^2}
\gamma^0 \itbd{\gamma}\cdot\dot{\bf p} \,.
\end{align}

\subsection{Antiparticle equal time spinors}
We now repeat these identities for the antiparticle spinors.
\subsubsection{}
\begin{align}
 & \quad \bar{v}_\alpha (p) v_\alpha (p)\nol &= \f{E+m}{2m}  \[ \(1+i\hbar g\) \begin{pmatrix}
s^\dagger_\alpha \Sigma & -s^\dagger_\alpha \end{pmatrix} \right.
\nol & \quad \quad \quad \times \left. \(
\( 1-i\hbar g\) \begin{pmatrix} \Sigma s_\alpha \\ s_\alpha
\end{pmatrix} +i\hbar \f{E+m}{(2E)^2} \begin{pmatrix}\dot{\Sigma} s_\alpha  \\ -\Sigma
\dot{\Sigma} s_\alpha \end{pmatrix} \) \right. \nol
& \quad \quad \left. -i\hbar \f{E+m}{(2E)^2} \begin{pmatrix}
s^\dagger_\alpha
\dot{\Sigma} & s^\dagger_\alpha
\dot{\Sigma} \Sigma \end{pmatrix}  \begin{pmatrix} \Sigma s_\alpha \\ s_\alpha
\end{pmatrix}
\] \nol
&= \f{E+m}{2m} \[ s^\dagger_\alpha\(\Sigma^2-1\)s_\alpha +i\hbar
\f{E+m}{2E^2} s^\dagger_\alpha\( \Sigma\dot{\Sigma} -
 \dot{\Sigma}\Sigma
\)s_\alpha \] \nol
&= \f{E+m}{2m} \[ -\f{2m}{E+m} +i\hbar \f{E+m}{2E^2}s^\dagger_\alpha
\f{2i\Lambda}{E+m}s_\alpha \] \,. \nonumber
\end{align}
Thus we obtain
\begin{align}
\bar{v}_\alpha (p) v_\alpha (p) &= -1-\f{\hbar}{2m}\f{\itbd{\xi}\cdot{\bf
p}\times\dot{\bf p}}{E^2} \,.
\end{align}

\subsubsection{}
\begin{align}
& \quad \bar{v}_\alpha (p)\gamma^0 v_\alpha (p) \nol &= \f{E+m}{2m}
\[
\(1+i\hbar g \) \begin{pmatrix} s^\dagger_\alpha \Sigma &
s^\dagger_\alpha \end{pmatrix} \right. \nol & \qquad \qquad \qquad
\left. \times
\(
\( 1-i\hbar g\) \begin{pmatrix}\Sigma s_\alpha \\ s_\alpha
\end{pmatrix} +i\hbar \f{E+m}{(2E)^2} \begin{pmatrix} \dot{\Sigma} s_\alpha \\ -\Sigma
\dot{\Sigma} s_\alpha \end{pmatrix} \) \right. \nol
& \qquad \qquad \left. -i\hbar \f{E+m}{(2E)^2} \begin{pmatrix}
s^\dagger_\alpha
\dot{\Sigma} & -s^\dagger_\alpha
\dot{\Sigma} \Sigma \end{pmatrix} \begin{pmatrix} \Sigma s_\alpha \\ s_\alpha
\end{pmatrix}
\] \nol
&= \f{E+m}{2m} \[ s^\dagger_\alpha\(\Sigma^2+1\)s_\alpha +i\hbar
\f{E+m}{(2E)^2} s^\dagger_\alpha \(
\Sigma\dot{\Sigma}-\Sigma\dot{\Sigma}-\dot{\Sigma}\Sigma+\dot{\Sigma}\Sigma
\)s_\alpha \] \nol
&= \f{E+m}{2m}\f{2E}{E+m} \,. \nonumber
\end{align}
We therefore once again have
\begin{align}
 \bar{v}_\alpha (p)\gamma^0 v_\alpha (p) &= \f{E}{m} \,,
\end{align}
with no order $\hbar$ term.

\subsubsection{}
\begin{align}
&\quad \bar{v}_\alpha (p)\itbd{\gamma} v_\alpha (p)\nol &=
\f{E+m}{2m}
\[
\(1+i\hbar g \) \begin{pmatrix} s^\dagger_\alpha \itbd{\sigma} &
s^\dagger_\alpha\Sigma\itbd{\sigma}
\end{pmatrix} \right. \nol & \qquad \qquad \qquad \left. \times
\(
\( 1-i\hbar g\) \begin{pmatrix} \Sigma s_\alpha \\ s_\alpha
\end{pmatrix} +i\hbar \f{E+m}{(2E)^2} \begin{pmatrix} \dot{\Sigma} s_\alpha \\ -\Sigma
\dot{\Sigma} s_\alpha \end{pmatrix} \) \right. \nol
& \qquad  \qquad \left. -i\hbar \f{E+m}{(2E)^2}
\begin{pmatrix} -s^\dagger_\alpha
\dot{\Sigma}\Sigma\itbd{\sigma} &
s^\dagger_\alpha \dot{\Sigma}\itbd{\sigma} \end{pmatrix}
\begin{pmatrix} \Sigma s_\alpha \\ s_\alpha
\end{pmatrix}
\] \,, \nonumber
\end{align}
which gives
\begin{multline}
 \bar{v}_\alpha (p)\itbd{\gamma} v_\alpha (p) = \f{E+m}{2m} \[ s^\dagger_\alpha \(
\itbd{\sigma}\Sigma +
\Sigma\itbd{\sigma} \) s_\alpha \right. \\ \left. +i\hbar \f{E+m}{(2E)^2}
s^\dagger_\alpha \(
\itbd{\sigma}\dot{\Sigma}-\dot{\Sigma}\itbd{\sigma} + \dot{\Sigma}\Sigma\itbd{\sigma}\Sigma
- \Sigma\itbd{\sigma}\Sigma\dot{\Sigma}
 \)s_\alpha \]
\,. \nonumber
\end{multline}
Using the identities presented in section \ref{Sigmaids}, we find
\begin{align}
 \bar{v}_\alpha (p)\itbd{\gamma} v_\alpha (p) &= \f{E+m}{2m}
\[
\f{2{\bf p}}{E+m} +i\hbar
\f{E+m}{(2E)^2} s^\dagger_\alpha \[ -\f{4i \itbd{\sigma}\cdot{\bf
p}\times\dot{\bf p} {\bf p}}{(E+m)^3} \right. \right. \nol & \left.
\left.
\quad \quad -
\(\f{2i{\bf p}^2}{(E+m)^3} + \f{2i}{E+m}\)
\itbd{\sigma}\times
\( \dot{\bf p}-\f{\dot{E} {\bf p}}{E+m}\) \]s_\alpha \] \,, \nonumber
\end{align}
and so
\begin{align}
\bar{v}_\alpha (p)\itbd{\gamma} v_\alpha (p) &= \f{{\bf p}}{m} + \f{\hbar}{2mE} \( \f{\itbd{\xi}\cdot{\bf
p}\times\dot{\bf p}\,{\bf p}}{E(E+m)} +\itbd{\xi}\times \(\dot{\bf
p}-\f{\dot{E}{\bf p}}{E+m} \) \) \,.
\end{align}

\subsubsection{Inner spinors}
\begin{align}
& \quad v_\alpha (p) \bar{v}^\alpha (p) \nol &= \sum_\alpha
\f{E+m}{2m}
\[ \(1-i\hbar g\) \begin{pmatrix} \Sigma s_\alpha \\ s_\alpha
\end{pmatrix} \right. \nol & \quad \quad \quad \times \left.
\[ \(1+i\hbar g \) \begin{pmatrix} s^\dagger_\alpha \Sigma &
-s^\dagger_\alpha \end{pmatrix} - i\hbar
\f{E+m}{(2E)^2}
\begin{pmatrix} s^\dagger_\alpha \dot{\Sigma} &
s^\dagger_\alpha \dot{\Sigma} \Sigma
\end{pmatrix} \] \right. \nol
& \quad \quad \left. +i\hbar \f{E+m}{(2E)^2} \begin{pmatrix}
\dot{\Sigma} s_\alpha \\ -\Sigma\dot{\Sigma}s_\alpha \end{pmatrix}
\begin{pmatrix} s^\dagger_\alpha \Sigma & -s^\dagger_\alpha \end{pmatrix} \] \nol
&= \sum_\alpha \f{E+m}{2m}
\[ \begin{pmatrix} \Sigma s_\alpha s^\dagger_\alpha \Sigma & - \Sigma s_\alpha s^\dagger_\alpha \\ s_\alpha
s^\dagger_\alpha \Sigma & -s_\alpha s^\dagger_\alpha
\end{pmatrix} -i\hbar
\f{E+m}{(2E)^2}
\begin{pmatrix} \Sigma s_\alpha s^\dagger_\alpha \dot{\Sigma} &
\Sigma s_\alpha s^\dagger_\alpha \dot{\Sigma}\Sigma \\
 s_\alpha s^\dagger_\alpha \dot{\Sigma} &  s_\alpha s^\dagger_\alpha \dot{\Sigma}\Sigma \end{pmatrix} \right.
\nol & \quad  \left. +i\hbar \f{E+m}{(2E)^2} \begin{pmatrix}
 \dot{\Sigma} s_\alpha s^\dagger_\alpha \Sigma & -\dot{\Sigma} s_\alpha
s^\dagger_\alpha
\\ -\Sigma\dot{\Sigma} s_\alpha s^\dagger_\alpha \Sigma & \Sigma\dot{\Sigma}s_\alpha s^\dagger_\alpha
\end{pmatrix} \] \,. \nonumber
\end{align}
Hence,
\begin{align}
&\quad \quad v_\alpha (p) \bar{v}^\alpha (p) \nol
 &= \f{E+m}{2m} \[ \begin{pmatrix} \Sigma^2 & -\Sigma \\
\Sigma & -I
\end{pmatrix} -i\hbar \f{E+m}{(2E)^2}
\begin{pmatrix} \Sigma\dot{\Sigma}-\dot{\Sigma}\Sigma & \dot{\Sigma}
+ \Sigma\dot{\Sigma}\Sigma \\ \dot{\Sigma} +
\Sigma\dot{\Sigma}\Sigma & \dot{\Sigma}\Sigma-\Sigma\dot{\Sigma}
\end{pmatrix} \] \nol
&= \f{1}{2m} \[ \begin{pmatrix} E-m & -\itbd{\sigma}\cdot{\bf p} \\
\itbd{\sigma}\cdot{\bf p} & -E-m \end{pmatrix} -i\hbar
\f{1}{(2E)^2} \begin{pmatrix} 2i\itbd{\sigma}\cdot{\bf
p}\times\dot{\bf p} & 2m\itbd{\sigma}\cdot\dot{\bf p} \\
2m\itbd{\sigma}\cdot\dot{\bf p} & -2i\itbd{\sigma}\cdot{\bf
p}\times\dot{\bf p} \end{pmatrix} \] \,, \nonumber
\end{align}
leading to
\begin{align}
 v_\alpha (p) \bar{v}^\alpha (p) &= \f{\gamma\cdot p -m}{2m} - \f{\hbar}{m(2E)^2}
\gamma^5
\itbd{\gamma}\cdot{\bf p}\times\dot{\bf p} -\f{i\hbar}{(2E)^2}
\gamma^0 \itbd{\gamma}\cdot\dot{\bf p} \,.
\end{align}

\section{Split time spinor identities}
The following identities are for when the time-dependent momenta are
evaluated at different times. The identities are the expansion of
the zeroth order spinors under the transformation
$(t,t')\to(\bar{t},\eta)$ to order $\hbar$, where
\begin{align}
t &= \bar{t} - \f{\hbar}{2}\eta \,, \\
t' &= \bar{t} + \f{\hbar}{2} \eta \,.
\end{align}
All momenta are $p$ and thus we drop the subscripts. Un-primed terms
are evaluated at $t$, primed terms evaluated at $t'$ and barred
terms evaluated at $\bar{t}$. Using this transformation, we have the
following:
\begin{align}
U^\dagger(t)U(t') & \to I - \f{i\hbar\eta}{2\bar{E}}
\bar{U}^\dagger(\bar{t})
\Lambda (\bar{t}) U(\bar{t}) + {\cal O}(\hbar^2) \\
\Sigma(t)\Sigma(t') &\to \f{\bar{\bf p}^2}{\bar{E}+m}
+i\hbar\eta\bar{\Lambda} \,,
\end{align}
where the energy-momenta on the right hand side are evaluated at
$\bar{t}$.

\subsection{Particle split time spinors}

\subsubsection{}
\begin{align}
\bar{u}^{(0)}_\alpha (p,t) u^{(0)}_\alpha (p,t')
&= \f{\sqrt{(E+m)(E'+m)}}{2m} \begin{pmatrix} s^\dagger_\alpha
U^\dagger & -s^\dagger_\alpha U^\dagger \Sigma \end{pmatrix}
\begin{pmatrix} U' s_\alpha \\ \Sigma' U' s_\alpha \end{pmatrix}
\nol
&= \f{\bar{E}+m}{2m} s^\dagger_\alpha U^\dagger \( 1 - \Sigma
\Sigma'  \) U' s_\alpha \nol
&= \f{\bar{E}+m}{2m} s^\dagger_\alpha U^\dagger \( \f{2m}{\bar{E}+m}
- \f{i\hbar\eta \bar{\Lambda}}{\bar{E}+m} \) U' s_\alpha \nol &=
s^\dagger_\alpha U^\dagger U' s_\alpha - \f{i\hbar\eta}{2m}
s^\dagger_\alpha \bar{U}^\dagger \bar{\Lambda} \bar{U} s_\alpha \nol
&= s^\dagger_\alpha s_\alpha -\f{i\hbar\eta}{2\bar{E}}
s^\dagger_\alpha \bar{U}^\dagger \bar{\Lambda} \bar{U} s_\alpha
-\f{i\hbar\eta}{2m} s^\dagger_\alpha \bar{U}^\dagger
\bar{\Lambda} \bar{U} \nol
&= 1 - \f{i\hbar\eta}{2m} \f{\bar{\itbd{\xi}}\cdot\bar{\bf
p}\times\dot{\bar{\bf p}}}{\bar{E}} \,.
\end{align}

\subsubsection{}
\begin{align}
\bar{u}^{(0)}_\alpha (p,t) \gamma^0 u^{(0)}_\alpha (p,t')
&=  \f{\sqrt{(E+m)(E'+m)}}{2m} \begin{pmatrix} s^\dagger_\alpha
U^\dagger & s^\dagger_\alpha U^\dagger \Sigma \end{pmatrix}
\begin{pmatrix} U' s_\alpha \\ \Sigma' U' s_\alpha \end{pmatrix}
\nol
&= \f{\bar{E}+m}{2m} s^\dagger_\alpha U^\dagger \( 1 + \Sigma
\Sigma'  \) U' s_\alpha \nol
&= \f{\bar{E}+m}{2m} s^\dagger_\alpha U^\dagger \(
\f{2\bar{E}}{\bar{E}+m} + \f{i\hbar\eta \bar{\Lambda}}{\bar{E}+m} \)
U' s_\alpha \nol &= \f{\bar{E}}{m}\(1 -\f{i\hbar\eta}{2\bar{E}}
s^\dagger_\alpha \bar{U}^\dagger \bar{\Lambda} \bar{U} s_\alpha\)
+\f{i\hbar\eta}{2m} s^\dagger_\alpha \bar{U}^\dagger
\bar{\Lambda} \bar{U} \nol
&=\f{\bar{E}}{m} \,.
\end{align}

\subsubsection{}
\begin{align}
& \quad \bar{u}^{(0)}_\alpha (p,t) \itbd{\gamma} u^{(0)}_\alpha
(p,t')\nol &=
\f{\sqrt{(E+m)(E'+m)}}{2m} \begin{pmatrix} s^\dagger_\alpha
U^\dagger\Sigma\itbd{\sigma} & s^\dagger_\alpha U^\dagger
\itbd{\sigma} \end{pmatrix}
\begin{pmatrix} U' s_\alpha \\ \Sigma' U' s_\alpha \end{pmatrix}
\nol
&= \f{\bar{E}+m}{2m} s^\dagger_\alpha U^\dagger \( \Sigma
\itbd{\sigma} + \itbd{\sigma}\Sigma' \) U' s_\alpha \nol &=
\f{\bar{E}+m}{2m} s^\dagger_\alpha U^\dagger \(
\(\bar{\Sigma}-\f{\hbar\eta}{2}\dot{\bar{\Sigma}}\)
 \itbd{\sigma} + \itbd{\sigma}\(\bar{\Sigma}+\f{\hbar\eta}{2}\dot{\bar{\Sigma}}\)\) U' s_\alpha \nol
&= \f{\bar{E}+m}{2m} s^\dagger_\alpha U^\dagger \( \bar{\Sigma}
\itbd{\sigma} + \itbd{\sigma}\bar{\Sigma} -\f{\hbar\eta}{2}
\(\dot{\bar{\Sigma}} \itbd{\sigma} -
\itbd{\sigma}\dot{\bar{\Sigma}} \)  \) U' s_\alpha \nol
&= \f{\bar{E}+m}{2m} s^\dagger_\alpha U^\dagger \f{2\bar{\bf
p}}{\bar{E}+m} U' s_\alpha - \f{\bar{E}+m}{2m}
\f{\hbar\eta}{2} s^\dagger_\alpha
\bar{U}^\dagger \f{2i}{\bar{E}+m}\itbd{\sigma} \times \(
\dot{\bar{{\bf p}}} - \f{\dot{\bar{E}} \bar{\bf p}}{\bar{E}+m} \)
\bar{U} s_\alpha \nol
&= \f{\bar{\bf p}}{m} \( 1- \f{i\hbar\eta}{2\bar{E}} s^\dagger
\bar{U}^\dagger \bar{\Lambda} \bar{U} s_\alpha \) -
\f{i\hbar\eta}{2m} s^\dagger \bar{U}^\dagger \itbd{\sigma} \times \(
\dot{\bar{{\bf p}}} - \f{\dot{\bar{E}} \bar{\bf p}}{\bar{E}+m} \)
\bar{U} s_\alpha \nol
&= \f{\bar{\bf p}}{m} -\f{i\hbar\eta}{2m} \(
\f{\bar{\itbd{\xi}}\cdot\bar{\bf p}\times \dot{\bar{\bf p}} \bar{\bf
p}}{\bar{E}\(\bar{E}+m\)}
 + \bar{\itbd{\xi}} \times \(
\dot{\bar{{\bf p}}} - \f{\dot{\bar{E}} \bar{\bf p}}{\bar{E}+m} \)
\) \,.
\end{align}

\subsubsection{Inner spinors}
\begin{align}
& \quad u^{(0)}_\alpha (p,t) \bar{u}^{(0)\alpha} (p,t') \nol &=
\sum_\alpha
\f{\sqrt{(E+m)(E'+m)}}{2m} \[
\begin{pmatrix} U s_\alpha \\ \Sigma U s_\alpha \end{pmatrix}
\begin{pmatrix} s^\dagger_\alpha {U^\dagger}' & -s^\dagger_\alpha {U^\dagger}' \Sigma' \end{pmatrix}
\] \nol
&= \sum_\alpha \f{\bar{E}+m}{2m} \begin{pmatrix} U s_\alpha
s^\dagger_\alpha {U^\dagger}'  &
  -U s_\alpha s^\dagger_\alpha {U^\dagger}'\Sigma'   \\
\Sigma U s_\alpha s^\dagger_\alpha {U^\dagger}' & -\Sigma U
s_\alpha s^\dagger_\alpha {U^\dagger}' \Sigma'
\end{pmatrix} \nol
&= \f{\bar{E}+m}{2m} \begin{pmatrix} U  {U^\dagger}'  &
  -U  {U^\dagger}'\Sigma'   \\
\Sigma U {U^\dagger}' & -\Sigma U {U^\dagger}' \Sigma'
\end{pmatrix} \,. \nonumber
\end{align}
Expanding the elements in terms of $\hbar$, starting with the
unitary operators we have
\begin{align}
u^{(0)}_\alpha (p,t) \bar{u}^{(0)\alpha} (p,t')
 &=
\f{\bar{E}+m}{2m}
\[
\begin{pmatrix} I & -\Sigma' \\ \Sigma & -\Sigma\Sigma'
\end{pmatrix}
+\f{i\hbar\eta}{2\bar{E}} \begin{pmatrix}
\bar{\Lambda} & -\bar{\Lambda}\bar{\Sigma} \\ \bar{\Sigma}\bar{\Lambda} & -\bar{\Sigma}\bar{\Lambda}\bar{\Sigma}
\end{pmatrix} \] \nol
&= \f{\bar{E}+m}{2m} \[ \begin{pmatrix} I & -\bar{\Sigma} \\
\bar{\Sigma} & -\bar{\Sigma}^2 \end{pmatrix} -\f{\hbar\eta}{2}
\begin{pmatrix} 0 &
\dot{\bar{\Sigma}} \\ \dot{\bar{\Sigma}} &
\bar{\Sigma}\dot{\bar{\Sigma}}-\dot{\bar{\Sigma}}\bar{\Sigma}
\end{pmatrix} \right. \nol & \left. +\f{i\hbar\eta}{2\bar{E}} \begin{pmatrix}
\bar{\Lambda} & 0 \\ 0 & \f{\bar{\bf p}^2
\bar{\Lambda}}{\(\bar{E}+m\)^2} \end{pmatrix} +
\f{i\hbar\eta}{2\bar{E}}\f{i \bar{\bf p}\times
\( \bar{\bf p}\times\dot{\bar{{\bf p}}} \) }{(\bar{E}+m)^2} \cdot \begin{pmatrix} 0 &
\itbd{\sigma} \\ \itbd{\sigma} & 0 \end{pmatrix} \] \nol
&= \f{\itbd{\gamma}\cdot\bar{\bf p} +m }{2m} -\f{\hbar\eta}{2m.2}
\gamma^0 \itbd{\gamma} \cdot \( \dot{\bar{{\bf
p}}}-\f{\dot{\bar{E}}\bar{\bf p}}{\bar{E}+m} + \f{\bar{\bf
p}\bar{\bf p}\cdot\dot{\bar{{\bf p}}} - \dot{\bar{{\bf p}}} \bar{\bf
p}^2}{\bar{E}\(\bar{E}+m\)} \) \nol &
+\f{i\hbar\eta}{2m}\(\bar{E}+m\)
\begin{pmatrix} \f{\bar{\Lambda}}{2\bar{E}} & 0 \\ 0 & \bar{\Lambda}
\( \f{\bar{\bf p}^2}{2\bar{E}(\bar{E}+m)^2} - \f{1}{\bar{E}+m} \)
\end{pmatrix} \,, \nonumber
\end{align}
which finally leads to
\begin{align}
u^{(0)}_\alpha (p,t) \bar{u}^{(0)\alpha} (p,t') &=
\f{\itbd{\gamma}\cdot\bar{\bf p} +m }{2m}
-\f{i\hbar\eta}{2\bar{E}2m}\gamma^5 \itbd{\gamma}\cdot \bar{{\bf
p}}\times\dot{\bar{{\bf p}}} - \f{\hbar\eta}{2\bar{E}.2} \gamma^0
\itbd{\gamma}\cdot \dot{\bar{{\bf p}}} \,.
\end{align}

\subsection{Antiparticle Split time spinors}

\subsubsection{}
\begin{align}
\bar{v}^{(0)}_\alpha (p,t) v^{(0)}_\alpha (p,t')
&= \f{\sqrt{(E+m)(E'+m)}}{2m} \begin{pmatrix} s^\dagger_\alpha
U^\dagger \Sigma & -s^\dagger_\alpha U^\dagger \end{pmatrix}
\begin{pmatrix} \Sigma' U' s_\alpha \\ U' s_\alpha \end{pmatrix}
\nol
&= \f{\bar{E}+m}{2m} s^\dagger_\alpha U^\dagger \( \Sigma
\Sigma' -1 \) U' s_\alpha \nol
&= \f{\bar{E}+m}{2m} s^\dagger_\alpha U^\dagger \(
-\f{2m}{\bar{E}+m} + \f{i\hbar\eta \bar{\Lambda}}{\bar{E}+m} \) U'
s_\alpha \nol &= -s^\dagger_\alpha U^\dagger U' s_\alpha +
\f{i\hbar\eta}{2m} s^\dagger_\alpha \bar{U}^\dagger \bar{\Lambda}
\bar{U} s_\alpha \nol &= -s^\dagger_\alpha s_\alpha
+\f{i\hbar\eta}{2\bar{E}} s^\dagger_\alpha \bar{U}^\dagger
\bar{\Lambda} \bar{U} s_\alpha +\f{i\hbar\eta}{2m} s^\dagger_\alpha
\bar{U}^\dagger
\bar{\Lambda} \bar{U} \nol
&= -1 + \f{i\hbar\eta}{2m} \f{\bar{\itbd{\xi}}\cdot\bar{\bf
p}\times\dot{\bar{\bf p}}}{\bar{E}} \,.
\end{align}

\subsubsection{}
\begin{align}
\bar{v}^{(0)}_\alpha (p,t) \gamma^0 v^{(0)}_\alpha (p,t')
&=  \f{\sqrt{(E+m)(E'+m)}}{2m} \begin{pmatrix} s^\dagger_\alpha
U^\dagger \Sigma & s^\dagger_\alpha U^\dagger \end{pmatrix}
\begin{pmatrix} \Sigma' U' s_\alpha \\ U' s_\alpha \end{pmatrix}
\nol
&= \f{\bar{E}+m}{2m} s^\dagger_\alpha U^\dagger \(  \Sigma
\Sigma' + 1  \) U' s_\alpha \nol
&= \f{\bar{E}+m}{2m} s^\dagger_\alpha U^\dagger \(
\f{2\bar{E}}{\bar{E}+m} + \f{i\hbar\eta \bar{\Lambda}}{\bar{E}+m} \)
U' s_\alpha \nol &= \f{\bar{E}}{m}\(1 -\f{i\hbar\eta}{2\bar{E}}
s^\dagger_\alpha \bar{U}^\dagger \bar{\Lambda} \bar{U} s_\alpha\)
+\f{i\hbar\eta}{2m} s^\dagger_\alpha \bar{U}^\dagger
\bar{\Lambda} \bar{U} \nol
&=\f{\bar{E}}{m} \,.
\end{align}

\subsubsection{}
\begin{align}
& \quad \bar{v}^{(0)}_\alpha (p,t) \itbd{\gamma} v^{(0)}_\alpha
(p,t') \nol &=
\f{\sqrt{(E+m)(E'+m)}}{2m} \begin{pmatrix} s^\dagger_\alpha
U^\dagger & s^\dagger_\alpha U^\dagger\Sigma\itbd{\sigma}
\itbd{\sigma} \end{pmatrix}
\begin{pmatrix} \Sigma' U' s_\alpha \\ U' s_\alpha \end{pmatrix}
\nol
&= \f{\bar{E}+m}{2m} s^\dagger_\alpha U^\dagger \(
\itbd{\sigma}\Sigma' + \Sigma \itbd{\sigma} \) U' s_\alpha \nol &=
\f{\bar{E}+m}{2m} s^\dagger_\alpha U^\dagger \(
\itbd{\sigma}\(\bar{\Sigma}+\f{\hbar\eta}{2}\dot{\bar{\Sigma}}\) +
\(\bar{\Sigma}-\f{\hbar\eta}{2}\dot{\bar{\Sigma}}\)
 \itbd{\sigma}\) U' s_\alpha \nol
&= \f{\bar{E}+m}{2m} s^\dagger_\alpha U^\dagger \(
\itbd{\sigma}\bar{\Sigma} + \bar{\Sigma} \itbd{\sigma}
+\f{\hbar\eta}{2} \(
\itbd{\sigma}\dot{\bar{\Sigma}}+\dot{\bar{\Sigma}} \itbd{\sigma}
 \)  \) U' s_\alpha \nol
&= \f{\bar{E}+m}{2m} s^\dagger_\alpha U^\dagger \f{2\bar{\bf
p}}{\bar{E}+m} U' s_\alpha \nol & \quad + \f{\bar{E}+m}{2m}
\f{\hbar\eta}{2} s^\dagger_\alpha
\bar{U}^\dagger \(\f{-2i}{\bar{E}+m}\)\itbd{\sigma} \times \(
\dot{\bar{{\bf p}}} - \f{\dot{\bar{E}} \bar{\bf p}}{\bar{E}+m} \)
\bar{U} s_\alpha \nol
&= \f{\bar{\bf p}}{m} \( 1- \f{i\hbar\eta}{2\bar{E}} s^\dagger
\bar{U}^\dagger \bar{\Lambda} \bar{U} s_\alpha \) -
\f{i\hbar\eta}{2m} s^\dagger \bar{U}^\dagger \itbd{\sigma} \times \(
\dot{\bar{{\bf p}}} - \f{\dot{\bar{E}} \bar{\bf p}}{\bar{E}+m} \)
\bar{U} s_\alpha \,. \nonumber
\end{align}
Thus
\begin{align}
\bar{v}^{(0)}_\alpha (p,t) \itbd{\gamma} v^{(0)}_\alpha
(p,t') &= \f{\bar{\bf p}}{m} -\f{i\hbar\eta}{2m} \(
\f{\bar{\itbd{\xi}}\cdot\bar{\bf p}\times \dot{\bar{\bf p}} \bar{\bf
p}}{\bar{E}\(\bar{E}+m\)}
 + \bar{\itbd{\xi}} \times \(
\dot{\bar{{\bf p}}} - \f{\dot{\bar{E}} \bar{\bf p}}{\bar{E}+m} \)
\) \,.
\end{align}

\subsubsection{Inner spinors}
\begin{align}
& \quad v^{(0)}_\alpha (p,t) \bar{v}^{(0)\alpha} (p,t') \nol &=
\sum_\alpha
\f{\sqrt{(E+m)(E'+m)}}{2m} \[
\begin{pmatrix} \Sigma U s_\alpha \\ U s_\alpha \end{pmatrix}
\begin{pmatrix} s^\dagger_\alpha {U^\dagger}' \Sigma' & -s^\dagger_\alpha {U^\dagger}' \end{pmatrix}
\] \nol
&= \sum_\alpha \f{\bar{E}+m}{2m} \begin{pmatrix} \Sigma U s_\alpha
s^\dagger_\alpha {U^\dagger}' \Sigma'  &
  -\Sigma U s_\alpha s^\dagger_\alpha {U^\dagger}'   \\
U s_\alpha s^\dagger_\alpha {U^\dagger}'\Sigma' & -U s_\alpha
s^\dagger_\alpha {U^\dagger}'
\end{pmatrix} \nol
&= \f{\bar{E}+m}{2m} \begin{pmatrix} \Sigma U {U^\dagger}'
\Sigma'  &
  -\Sigma U {U^\dagger}'   \\
U  {U^\dagger}'\Sigma' & -U  {U^\dagger}'
\end{pmatrix} \,. \nonumber
\end{align}
Expanding the unitary matrices to order $\hbar$,
\begin{align}
& \quad v^{(0)}_\alpha (p,t) \bar{v}^{(0)\alpha} (p,t') \nol
 &=
\f{\bar{E}+m}{2m}
\[
\begin{pmatrix} \Sigma\Sigma' & -\Sigma \\ \Sigma' & -I
\end{pmatrix}
+\f{i\hbar\eta}{2\bar{E}} \begin{pmatrix}
\bar{\Sigma}\bar{\Lambda}\bar{\Sigma} & -\bar{\Sigma}\bar{\Lambda} \\ \bar{\Lambda}\bar{\Sigma} & -\bar{\Lambda}
\end{pmatrix} \] \nol
&= \f{\bar{E}+m}{2m} \[ \begin{pmatrix} \bar{\Sigma}^2 &
-\bar{\Sigma} \\ \bar{\Sigma} & -I \end{pmatrix} +\f{\hbar\eta}{2}
\begin{pmatrix}
\bar{\Sigma}\dot{\bar{\Sigma}}-\dot{\bar{\Sigma}}\bar{\Sigma} &
\dot{\bar{\Sigma}} \\ \dot{\bar{\Sigma}} & 0
\end{pmatrix} \right. \nol & \quad \left. +\f{i\hbar\eta}{2\bar{E}} \begin{pmatrix}
-\f{\bar{\bf p}^2
\bar{\Lambda}}{\(\bar{E}+m\)^2} & 0 \\ 0 & -\bar{\Lambda} \end{pmatrix}
-
\f{i\hbar\eta}{2\bar{E}}\f{i \bar{\bf p}\times
\( \bar{\bf p}\times\dot{\bar{{\bf p}}} \) }{(\bar{E}+m)^2} \cdot \begin{pmatrix} 0 &
\itbd{\sigma} \\ \itbd{\sigma} & 0 \end{pmatrix} \] \nol
&= \f{\itbd{\gamma}\cdot\bar{\bf p} -m }{2m} +\f{\hbar\eta}{2m.2}
\gamma^0 \itbd{\gamma} \cdot \( \dot{\bar{{\bf
p}}}-\f{\dot{\bar{E}}\bar{\bf p}}{\bar{E}+m} + \f{\bar{\bf
p}\bar{\bf p}\cdot\dot{\bar{{\bf p}}} - \dot{\bar{{\bf p}}} \bar{\bf
p}^2}{\bar{E}\(\bar{E}+m\)} \) \nol &
\quad +\f{i\hbar\eta}{2m}\(\bar{E}+m\)
\begin{pmatrix} \bar{\Lambda}
\( \df{1}{\bar{E}+m} - \df{\bar{\bf p}^2}{2\bar{E}(\bar{E}+m)^2} \) & 0 \\ 0 & -\df{\bar{\Lambda}}{2\bar{E}}
\end{pmatrix} \,. \nonumber
\end{align}
Therefore, we have
\begin{align}
v^{(0)}_\alpha (p,t) \bar{v}^{(0)\alpha} (p,t') &=
\f{\itbd{\gamma}\cdot\bar{\bf p} +m }{2m}
-\f{i\hbar\eta}{2\bar{E}2m}\gamma^5 \itbd{\gamma}\cdot \bar{{\bf
p}}\times\dot{\bar{{\bf p}}} + \f{\hbar\eta}{2\bar{E}.2} \gamma^0
\itbd{\gamma}\cdot \dot{\bar{{\bf p}}} \,.
\end{align}

\section{Summary of semiclassical spinor
identities}\label{spinoridsummary}
 In this section, for ease of practical use, we collect together the spinor identities derived in
the previous sections. All terms on the right hand side are
evaluated at time $\bar{t}$ with energy-momenta
 $\tilde{p} = (E_p,\tilde{\bf p})$ and
$\tilde{p}_+=(E_{p_+},\tilde{\bf p}_+)$ for the particle and
antiparticle identities respectively. These results are quoted in
the main body of this work, when needed for the evaluation of the
spinor combinations found in the forward scattering loops.

\begin{align}
\bar{u}_\alpha (p,t) u_\alpha (p,t') &=
 1-\f{\hbar}{2m}\f{\itbd{\xi}\cdot\tilde{\bf p}\times\dot{\tilde{{\bf p}}}}{E_p^2}
-\f{i\hbar\eta}{2m}\f{\itbd{\xi}\cdot\tilde{\bf p}\times\dot{\bf
p}}{E_p}
+{\cal O}(\hbar^2) \,, \\
\bar{u}_\alpha (p,t) \gamma^0 u_\alpha (p,t') &= \f{E_p}{m} +
{\cal O}(\hbar^2) \,, \\
\bar{u}_\alpha (p,t) \itbd{\gamma} u_\alpha (p,t') &=
\f{\tilde{\bf p}}{m} - \f{\hbar}{2mE_p} \( \f{\itbd{\xi}\cdot\tilde{\bf
p}\times\dot{\tilde{{\bf p}}}\,\tilde{\bf p}}{E_p(E_p+m)}
+\itbd{\xi}\times
\(\dot{\tilde{{\bf p}}}-\f{\dot{E}_p\tilde{\bf p}}{E_p+m} \) \) \nol & \quad -
\f{i\hbar\eta}{2m} \(
\f{\itbd{\xi}\cdot\tilde{\bf p}\times\dot{\tilde{{\bf p}}}\,\tilde{\bf p}}{E_p(E_p+m)}
+\itbd{\xi}\times \(\dot{\tilde{{\bf p}}}-\f{\dot{E}_p\tilde{\bf
p}}{E_p+m}
\) \) +
{\cal O}(\hbar^2) \,, \\
u_\alpha (p,t) \bar{u}^\alpha (p,t') &= \f{\gamma\cdot \tilde{p}
+m}{2m} +
\f{\hbar}{m(2E_p)^2} \gamma^5
\itbd{\gamma}\cdot\tilde{\bf p}\times\dot{\tilde{{\bf p}}} -\f{i\hbar}{(2E_p)^2}
\gamma^0 \itbd{\gamma}\cdot\dot{\tilde{{\bf p}}} \nol & \quad -
\f{i\hbar\eta}{2m2E_p} \gamma^5
\itbd{\gamma}\cdot\tilde{\bf p}\times\dot{\tilde{{\bf p}}} -\f{\hbar\eta}{2E_p.2}
\gamma^0 \itbd{\gamma}\cdot\dot{\tilde{{\bf p}}} + {\cal
O}(\hbar^2) \,.
\end{align}

\begin{align}
\bar{v}_\alpha (p,t) v_\alpha (p,t') &=
 -1-\f{\hbar}{2m}\f{\itbd{\xi}\cdot\tilde{{\bf p}}_+\times\dot{\tilde{{\bf p}}}_+}{E_{p_+}^2}
+\f{i\hbar\eta}{2m}\f{\itbd{\xi}\cdot\tilde{{\bf p}}_+\times\dot{\bf
p}}{E_{p_+}}
+{\cal O}(\hbar^2) \,, \\
\bar{v}_\alpha (p,t) \gamma^0 v_\alpha (p,t') &= \f{E_{p_+}}{m} +
{\cal O}(\hbar^2) \,, \\
\bar{v}_\alpha (p,t) \itbd{\gamma} v_\alpha (p,t') &=
\f{\tilde{{\bf p}}_+}{m} \nol & \quad + \f{\hbar}{2mE_{p_+}} \( \f{\itbd{\xi}\cdot{\bf
p}\times\dot{\tilde{{\bf p}}}_+\,\tilde{{\bf
p}}_+}{E_{p_+}(E_{p_+}+m)} +\itbd{\xi}\times
\(\dot{\tilde{{\bf p}}}_+-\f{\dot{E}_{p_+}\tilde{{\bf p}}_+}{E_{p_+}+m} \) \) \nol &  -
\f{i\hbar\eta}{2m} \(
\f{\itbd{\xi}\cdot\tilde{{\bf p}}_+\times\dot{\tilde{{\bf p}}}_+\,\tilde{{\bf p}}_+}{E_{p_+}(E_{p_+}+m)}
+\itbd{\xi}\times \(\dot{\tilde{{\bf p}}}_+-\f{\dot{E}_{p_+}{\bf
p}}{E_{p_+}+m}
\) \) +
{\cal O}(\hbar^2) \,, \\
v_\alpha (p,t) \bar{v}^\alpha (p,t') &= \f{\gamma\cdot \tilde{p}_+
-m}{2m}  -
\f{\hbar}{m(2E_{p_+})^2} \gamma^5
\itbd{\gamma}\cdot\tilde{{\bf p}}_+\times\dot{\tilde{{\bf p}}}_+ -\f{i\hbar}{(2E_{p_+})^2}
\gamma^0 \itbd{\gamma}\cdot\dot{\tilde{{\bf p}}}_+ \nol & \quad -
\f{i\hbar\eta}{2m2E_{p_+}} \gamma^5
\itbd{\gamma}\cdot\tilde{{\bf p}}_+\times\dot{\tilde{{\bf p}}}_+ +\f{\hbar\eta}{2E_{p_+}.2}
\gamma^0 \itbd{\gamma}\cdot\dot{\tilde{{\bf p}}}_+ + {\cal
O}(\hbar^2) \,.
\end{align}

The following identity is for equal momenta zeroth order spinors:
\begin{align}
\bar{u}_\alpha^{(0)} (p) \gamma^5 \itbd{\gamma} u_\alpha^{(0)} (p) &=
- \( \itbd{\xi} + \f{\itbd{\xi}\cdot\tilde{\bf p} \, \tilde{\bf
p}}{m(E+m)}
\) \,.
\end{align}

\chapter{Interaction Hamiltonian for the Scalar field}\label{IntHam}

In this appendix we calculate the interaction Hamiltonian for the
complex scalar field. The result is in contrast with those cases,
such as the spinor field, where the interaction Hamiltonian and
Lagrangian are the negative of each other. For simplicity, let us
use natural units ($\hbar=1$) here. Consider the classical
Lagrangian density for a charged scalar field interacting with
electromagnetic field and in the presence of a background potential
$V$:
\begin{equation}
{\cal L} = ({\cal D}_\mu\varphi)^\dagger {\cal D}^\mu \varphi -
m^2\varphi^\dagger\varphi \,,
\end{equation}
where ${\cal D}_\mu \varphi = (D_\mu + ieA_\mu)\varphi$ and $D_\mu
\varphi = (\partial_\mu + iV_\mu)\varphi$.
This can be written
\begin{equation}
{\cal L} = ({\cal D}_0\varphi)^\dagger {\cal D}_0\varphi - (D_i
\varphi)^\dagger D_i \varphi - m^2\varphi^\dagger\varphi +
ie(\varphi^\dagger D_i\varphi - (D_i\varphi)^\dagger\cdot \varphi)
A_i - e^2A_iA_i
\varphi^\dagger\varphi \,,
\end{equation}
where the indices $i$ are summed over $i=1,2,3$.  The canonical
conjugate momentum densities are
\begin{align}
\pi_\varphi &= \frac{\partial{\cal L}}{\partial \dot{\varphi}}
= ({\cal D}_0 \varphi)^\dagger = \dot{\varphi}^\dagger -
i(V_0+eA_0)\varphi^\dagger \nol
\pi_{\varphi^\dagger} &= \frac{\partial{\cal L}}{\partial \dot{\varphi}^\dagger}
= {\cal D}_0\varphi = \dot{\varphi} + i(V_0+eA_0)\varphi \,.
\end{align}
Hence
\begin{align}
\dot{\varphi} &= \pi_{\varphi^\dagger} - i(V_0 + eA_0)\varphi \nol
\dot{\varphi}^\dagger &= \pi_{\varphi} + i(V_0+
eA_0)\varphi^\dagger \,.
\end{align}
Thus, the Hamiltonian density is
\begin{align}
{\cal H} & =  \pi_\varphi \dot{\varphi} + \pi_{\varphi^\dagger }
\dot{\varphi} - {\cal L} \nonumber \\
& =  \pi_\varphi (\pi_{\varphi^\dagger } - i(V_0+eA_0)\varphi)
+ \pi_{\varphi^\dagger }(\pi_\varphi^\dagger  + i(V_0+eA_0)\varphi^\dagger )\nonumber \\
& \quad -\pi_\varphi\pi_{\varphi^\dagger } + (D_i\varphi)^\dagger
D_i\varphi + m^2\varphi^\dagger \varphi - ie(\varphi^\dagger
D_i\varphi - (D_i\varphi)^\dagger \cdot\varphi)A_i +
e^2A_iA_i\varphi^\dagger
\varphi
\nonumber \\
& =  \pi_\varphi\pi_{\varphi^\dagger } + (D_i\varphi)^\dagger
D_i\varphi + m^2\varphi^\dagger \varphi
\nonumber \\
& \quad + i(\varphi^\dagger \pi_{\varphi^\dagger } -
\varphi\pi_{\varphi})(V_0+eA_0) - ie(\varphi^\dagger D_i\varphi -
(D_i\varphi)^\dagger
\cdot
\varphi) + e^2 A_iA_i\varphi^\dagger \varphi\,.
\end{align}
Hence we can decompose the Hamiltonian density into free and
interacting parts ${\cal H} = {\cal H}_0 + {\cal H}_I$ with
\begin{align}
{\cal H}_0 & =   \pi_\varphi\pi_{\varphi^\dagger } +
(D_i\varphi)^\dagger D_i\varphi + m^2\varphi^\dagger
\varphi + i(\varphi^\dagger \pi_{\varphi^\dagger }-\varphi\pi_{\varphi})V_0 \\
{\cal H}_I & =   ie(\varphi^\dagger \pi_{\varphi^\dagger } -
\varphi\pi_{\varphi})A_0
- ie(\varphi^\dagger D_i\varphi - (D_i\varphi)^\dagger
\cdot
\varphi)A_i + e^2 A_iA_i\varphi^\dagger \varphi \,.
\end{align}
Hamilton's equations with $H_0 = \int d^3{\bf x}{\cal H}_0$ read
\begin{align}
\dot{\varphi} & =  \frac{\delta H_0}{\delta\pi_\varphi}
= \pi_{\varphi^\dagger } - iV_0\varphi\,,\nonumber \\
\dot{\pi}_{\varphi^\dagger } & =  -
\frac{\delta H_0}{\delta\varphi^\dagger }
= D_i D_i\varphi - m^2\varphi - iV_0\pi_{\varphi^\dagger } \,,
\nonumber
\end{align}
and their conjugates. These equations can be rewritten as
\begin{align}
& (D_\mu D^\mu + m^2)\varphi = 0  \\
& \pi_{\varphi^\dagger } = D_0\varphi \,,
\end{align}
and their conjugates, as expected.

In the interaction picture, $\varphi$ obeys the Hamilton's equations
with $H_0$, so we can let $\pi_\varphi = (D_0\varphi)^\dagger $ and
$\pi_{\varphi^\dagger } = D_0\varphi$.  Then
\begin{equation}
{\cal H}_I =  ie(\varphi^\dagger D_\mu\varphi -
(D_\mu\varphi)^\dagger \cdot\varphi)A^\mu + e^2A_iA_i
\varphi^\dagger \varphi\,.
\end{equation}
The na\"{\i}ve interaction Hamiltonian density is
\begin{equation}
{\cal H}_I^{\rm naive} = ie(\varphi^\dagger D_\mu\varphi -
(D_\mu\varphi)^\dagger \cdot \varphi)A^\mu - e^2A_\mu A^\mu
\varphi^\dagger \varphi\,.
\end{equation}
Overall, the difference is ${\cal H}_I - {\cal H}_I^{\rm naive} =
e^2A_0A_0\varphi^\dagger
\varphi$.

\chapter{Reference: Dirac representation
matrices}\label{diracrefs}

In this appendix we give the matrix representations of the Pauli,
alpha, beta and gamma matrices frequently employed in mathematical
discussions on spin and in the Dirac equation. We present the
Pauli-Dirac or Standard representations here. These are the
representations used in the calculations in this work and thus they
are repeated here as a reference for the reader.

\section{Pauli Matrices}
The three Pauli spin matrices are given by
\begin{align}
\sigma_1 &=  \begin{pmatrix} 0 & 1 \\ 1 & 0 \end{pmatrix}  \,, &
\sigma_2 &=  \begin{pmatrix} 0 & -i \\ i & 0 \end{pmatrix} \,, &
\sigma_3 &=  \begin{pmatrix} 1 & 0 \\ 0 & -1 \end{pmatrix} \,. &
\end{align}
These matrices can easily be seen to have the following eigenvalues
and eigenvectors: \newline
\begin{center}
\begin{tabular}{|c|c|c|c|c|c|c|}
\hline
\textbf{Pauli Matrix} &
\multicolumn{2}{|c|}{$\sigma_1$} &
\multicolumn{2}{|c|}{$\sigma_2$} &
\multicolumn{2}{|c|}{$\sigma_3$} \\ \hline
\textbf{Eigenvalue} & 1 & -1 & 1 & -1 & 1 & -1 \\ \hline & & & & & &  \\
\textbf{Eigenvector} &
$\begin{pmatrix} 1 \\ 0 \end{pmatrix}$ & $\begin{pmatrix} 0 \\ 1
\end{pmatrix}$ & $\begin{pmatrix} 1 \\ 1\end{pmatrix}$ &
$\begin{pmatrix} -1 \\ 1 \end{pmatrix}$ & $\begin{pmatrix} -i \\ 1
\end{pmatrix}$ &
$\begin{pmatrix} i \\ 1 \end{pmatrix}$ \\ & & & & & & \\
\hline
\end{tabular}
\end{center}

\section{Alpha, Beta, Gamma Matrices}

\subsection{Alpha, Beta Matrices}
The alpha and beta matrices, frequently used in the non-covariant
form of the Dirac equation and its standrad derivation from the
assumption that the equation of motion is first order, are given in
the standard representation below:

\begin{eqnarray}
\beta = \( \begin{array}{cc} I_2 & 0 \\ 0 & -I_2 \end{array} \)
,\,\,
\alpha_i = \( \begin{array}{cc} 0 & \sigma_i \\ \sigma_i & 0 \end{array}
\)\,.
\end{eqnarray}
\begin{eqnarray}
\beta&=& \( \begin{array}{cccc} 1 & 0 & 0 & 0 \\ 0 & 1 & 0 & 0 \\ 0 & 0 & -1 & 0 \\ 0 & 0 & 0 & -1 \end{array} \) \\
\alpha_1 &=& \( \begin{array}{cccc} 0 & 0 & 0 & 1 \\ 0 & 0 & 1 & 0 \\ 0 & 1 & 0 & 0 \\ 1 & 0 & 0 & 0 \end{array} \) \\
\alpha_2 &=& \( \begin{array}{cccc} 0 & 0 & 0 & -i \\ 0 & 0 & i & 0 \\ 0 & -i & 0 & 0 \\ i & 0 & 0 & 0 \end{array} \) \\
\alpha_3 &=& \( \begin{array}{cccc} 0 & 0 & 1 & 0 \\ 0 & 0 & 0 & -1 \\ 1 & 0 & 0 & 0 \\ 0 & -1 & 0 & 0 \end{array}
\)\,.
\end{eqnarray}

\subsection{Gamma Matrices}
The gamma matrices, from the covariant form of the Dirac equation
and the Feynman slash notation, are given in terms of the alpha and
beta matrices by
\begin{equation}
\gamma^0 = \beta ,\, \gamma^i = \beta\alpha_i \,.
\end{equation}
In the standard representation, these matrices are therefore
\begin{align}
\gamma^0 &= \begin{pmatrix} 1 & 0 & 0 & 0 \\ 0 & 1 & 0 & 0 \\ 0 & 0 & -1 & 0 \\ 0 & 0 & 0 & -1 \end{pmatrix} \\
\gamma_1 &= \begin{pmatrix} 0 & 0 & 0 & 1 \\ 0 & 0 & 1 & 0 \\ 0 & -1 & 0 & 0 \\ -1 & 0 & 0 & 0 \end{pmatrix} \\
\gamma_2 &= \begin{pmatrix} 0 & 0 & 0 & -i \\ 0 & 0 & i & 0 \\ 0 & i & 0 & 0 \\ -i & 0 & 0 & 0 \end{pmatrix} \\
\gamma_3 &= \begin{pmatrix} 0 & 0 & 1 & 0 \\ 0 & 0 & 0 & -1 \\ -1 & 0 & 0 & 0 \\ 0 & 1 & 0 & 0 \end{pmatrix} \,. \\
\end{align}

Finally, the $\gamma^5$ matrix is defined by
\begin{equation}
\gamma^5 \equiv i \gamma^0\gamma^1\gamma^2\gamma^3 \,.
\end{equation}
Thus, we have in this representation
\begin{equation}
\gamma^5 = \begin{pmatrix} 0 & 0 & 1 & 0 \\ 0 & 0 & 0 & 1 \\ 1 & 0 & 0 & 0 \\ 0 & 1 & 0 & 0 \end{pmatrix} \,. \\
\end{equation}

\backmatter


\end{document}